%% file: main.tex
\documentclass[]{scrartcl}
\usepackage[utf8]{inputenc}
\usepackage{authblk}
\usepackage[numbers,sort&compress]{natbib}
\usepackage{hyperref}
  \hypersetup{colorlinks,linkcolor=[rgb]{0.2,0.2,0.75},citecolor=[rgb]{0.3,0.5,0.5},urlcolor=[rgb]{0.4,0.4,0.6},breaklinks}
\usepackage{doi}
\usepackage{amsmath}
\usepackage{graphicx}
\usepackage{fancyhdr}
\usepackage{xcolor}
\usepackage{soul}
\usepackage{bm}
\usepackage{siunitx} 
\usepackage[margin=10pt,font=small,labelfont=bf,labelsep=endash]{caption}
\usepackage[numbib,nottoc,notlof,notlot]{tocbibind} 
\usepackage{tikz} 

\usepackage[symbol]{footmisc}

\definecolor{lime}{HTML}{A6CE39}
\DeclareRobustCommand{\orcidicon}{
	\hspace{-2.7mm}
	\begin{tikzpicture}
	\draw[lime, fill=lime] (0,0) 
	circle [radius=0.16] 
	node[white] {{\fontfamily{qag}\selectfont \tiny ID}};
	\draw[white, fill=white] (-0.0625,0.095) 
	circle [radius=0.007];
	\end{tikzpicture}
	\hspace{-2.7mm}
}
\newcommand{\orcid}[1]{\href{https://orcid.org/#1}{\orcidicon}}

\title{\Large The JOREK non-linear extended MHD code and applications to large-scale instabilities and their control in magnetically confined fusion plasmas}

\author[1]{\footnotesize M Hoelzl\orcid{0000-0001-7921-9176}} 
\author[2,3]{GTA~Huijsmans}                                
\author[4]{SJP~Pamela\orcid{0000-0001-8854-1749}}          
\author[2]{M~Bécoulet}                                     
\author[2]{E~Nardon}                                       
\author[5]{FJ~Artola\orcid{0000-0001-7962-1093}}           
\author[6]{B~Nkonga}                                       
\author[7]{CV~ Atanasiu}                                   
\author[1]{V~Bandaru\orcid{0000-0003-4096-1407}}           
\author[6]{A~Bhole}                                        
\author[8]{D~Bonfiglio\orcid{0000-0003-2638-317X}}         
\author[1,9]{A~Cathey\orcid{0000-0001-7693-5556}}          
\author[10]{O~Czarny\orcid{0000-0002-0827-7843}}           
\author[2]{A~Dvornova}                                     
\author[1]{T~Fehér\orcid{0000-0003-2095-4349}\footnote{A number preceded by ``*'' corresponds to the affiliation at the time of contributing to this work.}}            
\author[4]{A~Fil\orcid{0000-0001-5755-4440}}               
\author[11]{E~Franck}                                      
\author[12]{S~Futatani\orcid{0000-0001-5742-5454}}         
\author[13]{M~Gruca\orcid{0000-0001-7443-2250}}            
\author[14]{H~Guillard}                                    
\author[15]{JW~Haverkort\orcid{0000-0001-5028-5292}}       
\author[16,1]{I~Holod}                                     
\author[17]{D~Hu}                                          
\author[18]{SK~Kim\orcid{0000-0002-0701-8962}}             
\author[3]{SQ~Korving\orcid{0000-0002-0395-3494}}          
\author[19]{L~Kos\orcid{0000-0002-1790-7093}}              
\author[20]{I~Krebs\orcid{0000-0001-9275-8991}}            
\author[21]{L~Kripner\orcid{0000-0001-8014-7296}}          
\author[2]{G~Latu}                                         
\author[2]{F~Liu}                                          
\author[1]{P~Merkel$^*$}                                      
\author[1]{D~Meshcheriakov$^*$}                               
\author[1,9]{V~Mitterauer}                                 
\author[1]{S~Mochalskyy\orcid{0000-0002-0421-7560v}}       
\author[2]{JA~Morales\orcid{0000-0002-3140-0504}}          
\author[*1,22,23]{R~Nies\orcid{0000-0002-9508-1223}}       
\author[1,9]{N~Nikulsin\orcid{0000-0003-1861-1777}}        
\author[1]{F~Orain$^*$}                                       
\author[24]{J~Pratt\orcid{0000-0003-2707-3616}}            
\author[1,9,25]{R~Ramasamy\orcid{0000-0002-1300-2043}}     
\author[26]{P~Ramet\orcid{0000-0002-6179-9819}}            
\author[2]{C~Reux}                                         
\author[1]{K~S{\"a}rkim{\"a}ki}                            
\author[1]{N~Schwarz\orcid{0000-0002-0574-1233}}           
\author[1]{P~Singh Verma}                                  
\author[4]{SF~Smith\orcid{0000-0003-2319-0356}}            
\author[27]{C~Sommariva\orcid{0000-0001-7676-3811}}        
\author[1]{E~Strumberger}                                  
\author[3]{DC~van Vugt\orcid{0000-0002-1108-3927}$^*$}        
\author[3]{M~Verbeek$^*$}                                     
\author[20]{E~Westerhof\orcid{0000-0002-0749-9399}}        
\author[1,9]{F~Wieschollek\orcid{0000-0002-6111-543X}}     
\author[28]{J~Zielinski\orcid{0000-0002-5683-9468}}        

\affil[1]{\scriptsize Max Planck Institute for Plasma Physics, Boltzmannstr. 2, 85748 Garching b. M., Germany}
\affil[2]{CEA, IRFM, 13108 Saint-Paul-Lez-Durance, France}
\affil[3]{Eindhoven University of Technology, P.O. Box 513, 5600 MB Eindhoven, The Netherlands}
\affil[4]{CCFE, Culham Science Centre, OX14 3DB, UK}
\affil[5]{ITER Organization, 13067 St. Paul Lez Durance Cedex, France}
\affil[6]{Université Côte d'Azur \& Inria Sophia-Antipolis Méditerranée, CNRS, LJAD, Nice, France}
\affil[7]{National Institute for Laser, Plasma and Radiation Physics, Atomistilor 409, P.O. BoxMG-36, 077125 Magurele-Bucharest, Romania}
\affil[8]{Consorzio RFX-CNR, ENEA, INFN, Università di Padova, Acciaierie Venete SpA. I-35127 Padova, Italy}
\affil[9]{Department of Physics, Technical University Munich,  James-Franck-Str. 1, 85748 Garching b. M., Germany}
\affil[10]{FRAMATOME, 10 Rue Juliette Récamier, 69456 Lyon Cedex, France}
\affil[11]{Inria Nancy Grand Est and IRMA Strasbourg, France}
\affil[12]{Universitat Politècnica de Catalunya, Barcelona, Spain}
\affil[13]{Institute of Plasma Physics and Laser Microfusion, P.O. Box 49, PL-00-908 Warsaw 49, Poland}
\affil[14]{Inria Sophia-Antipolis Méditerranée \& Université Côte d’Azur, Inria, CNRS, LJAD, Nice, France}
\affil[15]{Delft University of Technology, Delft, The Netherlands}
\affil[16]{Max Planck Computing and Data Facility, Boltzmannstr. 2, 85748 Garching b. M., Germany}
\affil[17]{School of Physics, Beihang University, Beijing, 100191, China}
\affil[18]{Department of Nuclear Engineering, Seoul National University, Gwanak 08826, Seoul, Korea}
\affil[19]{LECAD laboratory, Mech.Eng., University of Ljubljana, A\v{s}ker\v{c}eva 6, SI-1000 Ljubljana,  Slovenia}
\affil[20]{DIFFER - Dutch Institute for Fundamental Energy Research, De Zaale 20, 5612 AJ Eindhoven, The Netherlands}
\affil[21]{Institute of Plasma Physics, Prague, Czech Republic}
\affil[22]{Department of Astrophysical Sciences, Princeton University, Princeton, NJ, 08543, USA}
\affil[23]{Princeton Plasma Physics Laboratory, Princeton, NJ, 08540, USA}
\affil[24]{Georgia State University, Department of Physics and Astronomy, Atlanta Georgia, 30303, USA}
\affil[25]{Max-Planck/Princeton Research Center for Plasma Physics}
\affil[26]{LaBRI, INRIA Bordeaux Sud-Ouest, Université de Bordeaux, 351, cours de la Libération, 33405 Talence Cedex, France}
\affil[27]{Ecole Polytechnique Fédérale de Lausanne (EPFL), Swiss Plasma Center (SPC), CH-1015 Lausanne, Switzerland}
\affil[28]{University of Saskatchewan, S7N 5E2 Saskatoon, Canada}

\makeatletter
\renewcommand\maketitle{
   \begin{center}
     {\Huge\sffamily\bfseries\@title\par\vspace{0.3em}}
     {\scshape\normalsize\@author}
   \end{center}
}
\makeatother

\newcommand{\pderiv}[2]{\frac{\partial #1}{\partial #2}}       

\newcommand{\showcomment}[1]{}   
\newcommand{\coloredchange}[1]{\color{black}{#1}} 

\newcommand{\revised}[1]{{\color{orange}\coloredchange{#1}}}

\def\noreprint{\showcomment{\color{teal}\scriptsize Original figure -- no re-print permission required.}}
\def\reprintaps{\showcomment{\color{teal}\scriptsize APS owned -- no re-print permission required.}}
\def\reprintiop{\showcomment{\color{teal}\scriptsize IOP owned -- no re-print permission required.}}
\def\reprintccby{\showcomment{\color{teal}\scriptsize Published under CC BY 3.0 -- no re-print permission required.}}

\def\reprintaipOK{\showcomment{\color{teal}\scriptsize AIP owned -- re-print permission was obtained.}}
\def\reprintiaea{\showcomment{\color{teal}\scriptsize IAEA owned -- re-print permission was obtained from Miriam Edvardsen.}}

\def\reprintunpub{\showcomment{\color{red}\scriptsize In the submission/publication process -- will need to ask respective publishers later on.}}

\def\Elmt{e}

\def\VecA{\mathbf{A}}
\def\VecV{\mathbf{V}}
\def\VecVm{\mathbf{v}}
\def\dens{\rho}
\def\pres{p}

\def\VecAS{\VecA^{\star}}
\def\VecVS{\VecV^{\star}}
\def\densS{\dens^{\star}}
\def\presS{\pres^{\star}}

\def\SaS{\text{A}^{\star}}
\def\SvS{\text{v}^{\star}}

\def\VecB{\mathbf{B}}
\def\VecE{\mathbf{E}}
\def\VecJ{\mathbf{J}}
\def\pvec{u}

\def\coefD{\text{D}}
\def\coefR{\eta}
\def\coefV{\nu}  
\def\coefK{\kappa}

\def\TensD{\underline{\pmb{\text{D}}}}
\def\TensV{\underline{\pmb{\tau}}}
\def\TensK{\underline{\pmb{\kappa}}}
\def\Id{\underline{\pmb{\text{I}}}}

\begin{document}

\maketitle

\input{00_abstract}

{\scriptsize
\tableofcontents}

\input{01_introduction}

\input{02_models}

\input{03_numerics}

\input{04_verification}
\input{05_elms}

\input{06_disruptions}

\input{07_other}

\input{08_summary}

\addtocontents{toc}{\protect\setcounter{tocdepth}{0}}

\input{09_acknowledgements}

{\scriptsize
\bibliography{main}}

\input{10_appendix}

\end{document}

%% file: 00_abstract.tex
\section*{Abstract}

JOREK is a massively parallel fully implicit non-linear extended MHD code for realistic tokamak X-point plasmas. It has become a widely used versatile simulation code for studying large-scale plasma instabilities and their control and is continuously developed in an international community with strong involvements in the European fusion research program and ITER organization. This article gives a comprehensive overview of the physics models implemented, numerical methods applied for solving the equations and physics studies performed with the code. A dedicated section highlights some of the verification work done for the code. A hierarchy of different physics models is available including a free boundary and resistive wall extension and hybrid kinetic-fluid models. The code allows for flux-surface aligned iso-parametric finite element grids in single and double X-point plasmas which can be extended to the true physical walls and uses a robust fully implicit time stepping. Particular focus is laid on plasma edge and scrape-off layer (SOL) physics as well as disruption related phenomena. Among the key results obtained with JOREK regarding plasma edge and SOL, are deep insights into the dynamics of edge localized modes (ELMs), ELM cycles, and ELM control by resonant magnetic perturbations, pellet injection, as well as by vertical magnetic kicks. Also ELM free regimes, detachment physics, the generation and transport of impurities during an ELM, and electrostatic turbulence in the pedestal region are investigated. Regarding disruptions, the focus is on the dynamics of the thermal quench and current quench triggered by massive gas injection (MGI) and shattered pellet injection (SPI), runaway electron (RE) dynamics as well as the RE interaction with MHD modes, and vertical displacement events (VDEs). Also the seeding and suppression of tearing modes (TMs), the dynamics of naturally occurring thermal quenches triggered by locked modes, and radiative collapses are being studied.

\scriptsize
{\color{gray}
\textbf{Keywords:} Magneto-hydrodynamics, MHD, extended MHD, reduced MHD, particle in cell, PiC, magnetic confinement fusion, equilibrium, Tokamak, ITER, plasma, plasma instabilities, edge localized modes, ELMs, ELM cycles, ELM types, disruption, vertical displacement event, VDE, tearing modes, pellets, ablation, free boundary, vertical kicks, ELM suppression, ELM mitigation, resonant magnetic perturbations, RMPs, ELM pacing, pellet ablation, disruption mitigation, massive material injection, massive gas injection, shattered pellet injection, relativistic particles, runaway electrons, tungsten, detachment, burn-through, finite elements, Bezier, weak form, Galerkin method, QH-mode, Edge Harmonic Oscillations (EHO), bootstrap current, resistive wall modes, tearing mode seeding, thermal quench, current quench, eddy currents, halo currents, implicit time stepping, sparse matrix, iterative solver, preconditioning, stellarator, ITG}

\normalsize

%% file: 01_introduction.tex
\section{Introduction}\label{:intro}

The present article provides a comprehensive overview of the non-linear extended MHD code JOREK, which is among the leading simulation codes worldwide for studying large scale plasma instabilities and their control in realistic divertor tokamaks. The article provides a detailed description of the physics models, numerical methods, and physics applications of the code.

In the existing literature, Refs.~\cite{Huysmans2007,Czarny2008} already describe some aspects of the numerical methods and physics models of the JOREK code, which has been extended significantly since these articles were published. Ref.~\cite{Huijsmans2015} contains an overview of modelling activities worldwide regarding ELMs and ELM control based on many different simulation codes, and Refs.~\cite{Pamela2018IAEA,Hoelzl2018,Hoelzl2018EPS} provide a partial overview of JOREK activities regarding plasma edge and scrape off layer. The present article, in contrast, aims to give a comprehensive description of the code and its applications, with a particular focus on recent developments.

In the present Section, we describe the motivation for the research activities (Subsection~\ref{:intro:motiv}) followed by a very brief review of (extended) magnetohydrodynamics (Subsection~\ref{:intro:mhd}) and some words on the historic development of JOREK (Subsection~\ref{:intro:history}).

The rest of the article is organized as follows: Section~\ref{:code:models} provides a detailed overview of the physics models available in JOREK and Section~\ref{:code:numerics} describes the numerical methods employed for solving the equations. Selected tests performed for code verification are shown in Section~\ref{:verification}. After this ``technical'' part, a detailed picture is drawn of the physics studies and validation activities performed in particular in the fields of plasma edge and scrape off layer physics (Section~\ref{:applic:edge}) as well as disruption physics (Section~\ref{:applic:core}). Further code applications are described in Section~\ref{:applic:other}. Each Section contains a brief outlook towards further plans and developments. Finally, a concise summary is provided in Section~\ref{:summary:conclusions}.

The support received from many entities and useful discussions with various scientists are acknowledged in Section~\ref{:ack}. Additional details on the coordinate systems, finite element basis, normalization of quantities, and time stepping scheme are provided in Appendices~\ref{:app:coord}--\ref{:app:tstep}.

\subsection{Motivation and challenges}\label{:intro:motiv}

Among the obstacles, which need to be overcome on the path towards a magnetic confinement fusion power plant, large scale plasma instabilities may well be the most critical one. A plasma configuration suitable for harvesting energy needs to have good confinement properties, however, a reliable control\footnote{The term ``control'' is in this article is meant to include both avoidance and control strategies.} of plasma instabilities is equally important. Robust predictions of the properties of such instabilities and of effective control methods are urgently needed
\begin{enumerate}
    \item to provide input to the ITER design, where it can still be influenced (e.g., the disruption mitigation system),
    \item to prepare a robust, efficient, and successful exploitation of ITER across all phases of the planned operation, and
    \item to answer critical questions regarding the design of a successful DEMO reactor.
\end{enumerate}

Revealing the underlying physics processes of plasma instabilities and developing control mechanisms constitutes a major challenge for experiments, theory, and modelling. While the suitability of control techniques for present devices may be tested in a straight forward manner experimentally, their applicability to future machines, with plasma parameters very different both quantitatively (e.g., Lundquist number) and qualitatively (e.g., large amount of fusion-born fast particles) from present machines, needs to be ensured by developing truly predictive capabilities. In such a holistic approach based on fundamental plasma theory, experimental studies across devices, and numerical simulations of the plasma dynamics, the computational models play a key role. Simulation codes can provide the capability to predict the relevant processes in future devices after being carefully validated against theory predictions and experiments first. Activities with the JOREK code ultimately aim at reaching that goal. 

Key challenges in this respect are the immense scale separations in both time and space of the involved processes, the intrinsic highly non-linear multiphysics nature of the problem, and the complicated magnetic topology of divertor plasmas. Magneto-hydrodynamic (MHD) models have become a very robust and reliable framework for describing large-scale plasma instabilities. And via numerous extensions beyond the classical MHD, more and more effects can be captured accurately in the simulations. Worldwide, a number of specialized simulation codes for calculating non-linear MHD dynamics in magnetically confined tokamak and stellarator plasmas have been developed in the past years and decades including BOUT++~\cite{Dudson2009}, JOREK (this article and Refs.~\cite{Huysmans2007,Czarny2008,Huysmans2009}), MEGA~\cite{Todo1998}, M3D~\cite{Park1999}, M3D-C$
^1$~\cite{Jardin2007,Jardin2012,Ferraro2016}, NIMROD~\cite{Glasser1999,Sovinec2004}, and XTOR~\cite{Lutjens2008,Luetjens2010} (listed alphabetically, not a complete list).

Besides the challenges imposed by the multi-scale nature already mentioned, in particular the large number of different physical effects, which need to be treated consistently and which are mutually interacting in a highly non-linear way, requires simulation codes that can capture this rich multi-physics behaviour in a reliable way. In a typical mitigated disruption scenario, for instance, the dynamics of magnetic islands, the ablation of (shattered) pellets, the reconnection of the plasma leading to a stochastic state, the fast losses of thermal energy along magnetic field lines, the radiative losses by partly ionized impurities, the generation and transport of runaway electrons (REs), the interaction of REs with the MHD modes and the electromagnetic interaction of the plasma with conducting structures in the device may all play an important role simultaneously. Developing the capability to describe the non-linear interaction of all these processes is necessary for unravelling the complete physics picture and becoming truly predictive regarding the dynamics in future machines. At the same time, simpler models are needed to allow faster access to larger parameter studies. JOREK is a advanced simulation framework for studying large-scale instabilities in magnetized plasmas. It offers such a hierarchy from simple and fast to very complex and computationally demanding models.

\subsection{Extended Magnetohydrodynamics (MHD)}\label{:intro:mhd}

This article does not give a complete overview of magnetohydrodynamics (MHD) and its computational treatment. We mention only key features in this Section, which are directly relevant as context for this article. For literature on MHD, in particular the References~\cite{GoedbloedI,GoedbloedII,SchnackBook,BiskampBook,FreidbergBook,ZohmBook,JardinBook} are recommended.

Magneto-hydrodynamics developed first by H. Alfv{\'e}n in 1942~\cite{Alfven1942} describes a magnetized plasma as an electrically conducting fluid. In the ideal MHD model, the plasma is assumed to be perfectly conducting. Ideal MHD can describe certain stability limits in tokamak plasmas well (e.g., see References~\cite{Gohil1988,HuysmansEPS1995,Connor1998,Snyder2004} for type-I ELMs). However, 3D non-linear simulations need to be based on resistive extended MHD models, which include anisotropic heat conduction, plasma resistivity, diamagnetic flows, finite Larmor radius effects, neoclassical physics, source/sink terms, two-fluid effects, neutrals, impurities, sheath boundary conditions, and many more effects depending on the addressed problem. A certain class of models includes also electron inertia effects~\cite{Abdelhamid2015}.

Tokamak plasmas are typically in approximate force balance $\nabla p \approx \mathbf{j} \times \mathbf{B}$, where $p$ denotes (the isotropic component of) the pressure, $\mathbf{j}$ the plasma current vector, and $\mathbf{B}$ the magnetic field vector. The stability of this equilibrium state determines whether the plasma will remain in this equilibrium state or is prone to instabilities. This is traditionally studied by linearizing the equations and analyzing the eigenvalue spectrum of the system along with the associated eigenvectors. However, linear growth rates may be affected dramatically by background flows and non-ideal plasma effects, which are not always accounted for in linear codes. Also, non-linear dynamics cannot be predicted from the linear stability analysis in general, and linearly stable eigenmodes might become non-linearly unstable at sufficiently large ``seed perturbation'' amplitudes (e.g., neoclassical tearing modes). As a result, predicting the full consequences of plasma instabilities is only possible by employing advanced non-linear models. Solving such models in realistic geometries typically is only possible numerically.
MHD involves very different time scales: The Alfv{\`e}n time $\tau_A=a \sqrt{\mu_0\,m_i\,n_i}/B$ is about $0.3\mathrm{\mu s}$ for ITER like parameters, where $a$ denotes the minor radius of the plasma, $\mu_0$ the vacuum permeability, $m_i$ the ion mass, and $n_i$ the ion density. On the other hand, the resistive time scale $\tau_R=\mu_0\,a^2/\eta$, where $\eta$ denotes the plasma resistivity, is $\gg 1\mathrm{s}$ for ITER like parameters. Plasma instabilities typically develop on mixed time scales of tens of $\mu\mathrm{s}$ to tens of $\mathrm{ms}$. The resistive time scale of the ITER vacuum vessel is around $\tau_W=500  \mathrm{ ms}$ slowing down some instabilities to that time scale (e.g. axisymmetric resistive wall modes). 
Consequently, the relevant time scales for large scale instabilities are two to six orders of magnitude longer than the Alfv{\'e}n time. The frequencies of fast magneto-acoustic waves propagating in the plane orthogonal to the magnetic field are typically even two to three orders of magnitude larger than the Alfv{\'e}n frequency,
thus constituting the most challenging time scale in the system. The so-called reduced MHD model, described in Section~\ref{:code:models:base:reduced}, eliminates the fast waves from the model to facilitate its numerical solution.

In spatial dimensions, a similarly challenging splitting of scales can be observed. While the size of the whole system typically is in the range of several meters (minor radius of $2\mathrm{m}$ in ITER), the resistive skin depth is given by $\sqrt{2\eta/(\mu_0\,\omega)}$ at a given frequency $\omega$, which can easily drop into the mm or even sub-$\mathrm{mm}$ range at the low resistivity of large fusion devices (which decreases strongly with temperature) -- a separation by four orders of magnitude. The strong increase of this scale separation towards larger (and at the same time hotter) fusion devices is a particular challenge for the modelling.

Anisotropic heat conduction is another particularly challenging physics aspect to be dealt with in MHD simulations. While the transport coefficients across field lines determined by neoclassical or turbulent processes typically are in the range of $1\mathrm{m^2/s}$, the heat transport along field lines by electrons can reach values of $10^{10}\mathrm{m^2/s}$ in hot plasmas~\cite{Spitzer1953,Malone1975}. Avoiding overly restrictive time scales, numerical instabilities, or a pollution of cross-field transport by errors in the parallel transport is a significant challenge for the numerical treatment.

Magnetohydrodynamics is strictly valid only when the plasma is sufficiently collisional, and many important kinetic effects are not reflected by the MHD equations. However, a large number of corrections (e.g., effective parallel heat diffusion coefficients~\cite{Malone1975}) and extra terms (e.g., two-fluid effects, or a consistent evolution of the bootstrap current~\cite{Sauter2002}) allow to apply MHD outside its original boundaries. In many cases, the full MHD equations can be further simplified to eliminate the fast magneto-sonic waves from the system, reducing the separation of time scales. A significant number of reduced MHD models with different levels of approximation exist (e.g., References~\cite{Strauss1983,Strauss1997}), which lower the number of physical variables in the system. JOREK presently has several different reduced MHD models (the one described in Section~\ref{:code:models:base} with and without parallel velocity; a reduced MHD model suitable for stellarator applications is in development) and a full MHD model (Section~\ref{:code:models:fullMHD}) implemented for tokamak
configurations along with numerous physics extensions.

\subsection{Historic development of the JOREK code}\label{:intro:history}

The development of a first version ``JOREK 1'' was started by G.T.A. Huysmans in 2002 at CEA/IRFM and is described in Ref.~\cite{Huysmans2005}. Applications of the JOREK 1 code include the current hole problem, the stability of external kink modes in X-point plasmas ~\cite{Huysmans2005}, the first nonlinear ELM simulations ~\cite{Huysmans2007} and the application of RMP fields ~\cite{Nardon2007}. The JOREK 1 code was based on so-called generalised, h-p refinable, finite elements ~\cite{Oden1998}. However, in practice, the p refinement, i.e., adapating the order of the finite elements was never used. Therefore it was decided to change the finite elements to cubic Bezier finite elements, an extension to the iso-parametric bicubic Hermite elements which are succesfully applied in the HELENA equilibrium code ~\cite{Huysmans1990}. The code ``JOREK 2'', which has been developed since 2006, is first described in the references~\citep{HuysmansASTER,HuysmansASTER2,Czarny2008} and has successively evolved into the presently existing JOREK code, which is described in the article at hand. As major changes in version 2, an iterative solver, and a $G^1$ continuous finite element formulation had been implemented. \revised{$G^1$ continuity refers to both values and real-space gradients being continuous throughout the computational domain but without continuity in the gradients in the local finite element coordinates.} The present JOREK code is being further developed continuously regarding physics models, numerical methods, and applications as shown in this article. The JOREK website~\cite{jorekeu} contains some regularly updated information. The article at hands intends to give a complete overview of the code including references to all original publications which go more into detail than possible here.

%% file: 02_models.tex
\section{Physics models}\label{:code:models}

\begin{figure}
\centering
  \includegraphics[height=0.25\textwidth]{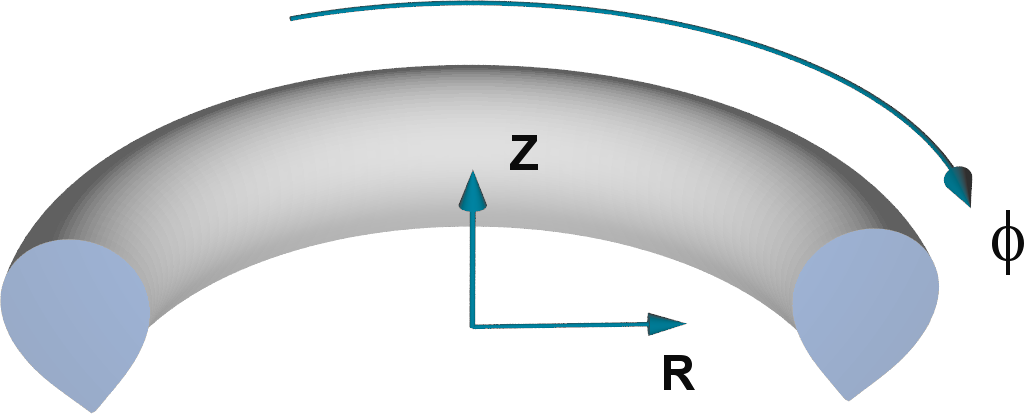}
\caption{The base cylindrical coordinate system used in JOREK. Refer to Appendix~\ref{:app:coord} for details. \noreprint}
\label{fig:cylcoord}
\end{figure}

This section describes the physics models and corresponding extensions available in JOREK. Before turning towards these models, the coordinate systems used in JOREK are introduced briefly.

\subsection{Coordinate systems}\label{:code:models:coord}

The base cylindrical coordinate system $(R,Z,\phi)$ is given by $x=R~\mathrm{cos} \phi$, $y=-R~\mathrm{sin} \phi$, $z=Z$, where $(x,y,z)$ denotes Cartesian coordinates (Figure~\ref{fig:cylcoord}). Thus, $\phi$ is oriented clockwise if viewed from the top. According to the definitions in Ref.~\cite{Sauter2013}, the JOREK conventions correspond to a COCOS number of 8. To describe the Bezier elements, the coordinates $R$ and $Z$ are expanded in the same Bezier basis functions, that are also used for the expansion of the physics variables (``isoparametric''). This introduces a local $(s,t,\phi)$ coordinate system inside each grid element. See Sections~\ref{:code:numerics:discretization} and Appendix~\ref{:app:coord} for more details on the discretization.

\subsection{Grad-Shafranov solver}\label{:code:models:equil}
JOREK has a built-in Grad-Shafranov equilibrium solver which uses the same finite element grid and representation of the variables used in the nonlinear time evolution. This guarantees that the discrete initial state used in MHD equations accurately satisfies the initial equilibrium force balance, avoiding any initial discontinuous behaviour. JOREK can solve both fixed boundary equilibria and, through the coupling to the STARWALL code, free boundary equilibria (see Section~\ref{:code:models:freebound}).

The solver requires the profiles of pressure (provided by temperature and density separately, since they are needed for the initial conditions) and $FF'$. These profiles are provided as functions of the normalized poloidal flux $\Psi_N=(\Psi-\Psi_\text{axis})/(\Psi_\text{bnd}-\Psi_\text{axis})$,
either via a simple analytical function, or via a numerical representation. Here $\Psi_\text{axis}$ and $\Psi_\text{bnd}$ denote the values of the poloidal magnetic flux at the magnetic axis and on the boundary of the plasma domain, respectively. In addition, the poloidal flux $\Psi$ on the boundary of the computational domain needs to be specified by by a numerical list of (R,Z,$\Psi$) points (or by coefficients for analytical moments for simpler cases). This input can be extracted, for instance, from ``geqdsk'' files or from equilibria created with the CLISTE code. Starting from an initial guess, the equilibrium is determined iteratively by Picard or Newton iterations to a specified accuracy.
After solving the GS equation on the initial finite element grid, the solution is typically used to create a new grid aligned to the equilibrium flux surfaces. The GS equation is solved a second time on this new grid, providing the accurate initial conditions for the time evolution part.  

After the equilibrium calculation, all physical variables are initialized consistently to it: the poloidal flux $\Psi$ is directly taken from the equilibrium solution; the toroidal current density is calculated directly from $\Psi$ via the current definition equation; density and temperature are initialized according to the specified profiles. All velocity related quantities (velocity stream function, vorticity, and parallel velocity) are initialized to be zero unless background rotation profiles are prescribed. In case of sheath boundary conditions, the parallel velocity is initialized to the ion sound speed at divertor targets\footnote{Simulations with sheath boundary conditions typically need to be run axi-symmetrically for a short while, such that the parallel flows in the scrape-off layer (SOL) can establish a steady state. Non-axisymmetric Fourier modes are added then, once SOL flows have equilibrated.}.

\subsection{Base MHD model}\label{:code:models:base}
The MHD model is formulated as a set of normalized equations for the
evolution of the magnetic potential ($\VecA$), mean velocity
($\VecV$), total density ($\rho$) and total pressure ($p$). The
equations are normalized with respect to the central mass density
$\rho_0$ and the vacuum permeability $\mu_0$ such that $\mu_0$ does
not appear explicitly. Length scales are not normalized, while the
time is normalized by a factor\footnote{In case of ITER with minor
  radius $a\approx2\,\mathrm{m}$ and magnetic field amplitude on axis
  $B_0\approx5\,\mathrm{T}$,
  $\tau_\text{norm}\approx2.5\,\tau_{A}$. Note that $\sqrt{\mu_0\,\rho_0}$ does not have the dimension of a time and it is therefore more exact to say that the numerical value of $\tau_\text{norm}$ is $\sqrt{\mu_0\,\rho_0}$.} $\tau_\text{norm}=\sqrt{\mu_0\,\rho_0}$
which is typically close to the Alfv{\'e}n time
$\tau_A=a\sqrt{\mu_0\,\rho_0}/B_0$. The total pressure and
mass density are normalized by $\mu_0$ and $\rho_0$
respectively. Details on the normalization of further quantities are
given in Appendix~\ref{:app:norm}. The normalized equations are
written as

\begin{align}
\pderiv{\VecA}{t} &= -\VecE - \nabla \Phi , \label{eq:mhd:A}\\
\rho\pderiv{\VecV}{t} &= -\rho\VecV\cdot\nabla\VecV - \nabla p +
\VecJ\times\VecB + \nabla\cdot\TensV 
+\mathbf{S}_{\VecV},
\label{eq:mhd:v} \\
\pderiv{\rho}{t} &=
-\nabla\cdot(\rho\,\VecV)
+\nabla\cdot(\TensD\nabla\rho)
+S_\rho,\label{eq:mhd:rho}
\\
\pderiv{p}{t} &= -\VecV\cdot\nabla p
- \gamma p\nabla\cdot\VecV
+ \nabla\cdot(\TensK\nabla T) + (\gamma -1)\TensV:\nabla\VecV + S_{\pres} 
\label{eq:mhd:p}\end{align}
The total pressure is defined by the ideal gas law $p= p_i + p_e = \rho\,T$ ($\mu_0$
disappears due to normalization).  This pressure is the sum of
electron ($p_e$) and ion ($p_i$) pressures. Electron and ion pressures are then 
assumed in some of the models to be half of the total pressure\revised{, e.g., where the diamagnetic terms are calculated. The model extension described in Section~\ref{:code:models:twotemp} allows to separately evolve variables for the electron and ion temperatures instead.}
The magnetic field vector\revised{, $\VecB$,} and the current vector\revised{, $\VecJ$,} are defined as:
\begin{equation}\label{eq:EBJ}
 \VecB = F\nabla\phi + \nabla\times\VecA
 \quad \text{ and } \quad \VecJ = \nabla\times\VecB, 
\end{equation}
Here, the toroidal flux function $F\left(\psi\right) \equiv R B_\phi$ is not essential for the model but is added for numerical reasons. $F\left(\psi\right)$ is constant in time and typically taken from the initial Grad-Shafranov equilibrium such that the initial vector potential in the poloidal plane is zero. This does not constrain $\mathbf{B}$ in any way since the magnetic vector potential $\VecA$ takes into account the (arbitrarily large) time evolution and perturbations from the equilibrium. 

\revised{Ohm’s law includes resistivity and drift-ordered (diamagnetic) terms \cite{Hazeltine_and_Meiss}}:
 \begin{equation}\label{eq:Eprim}
    \VecE = -  \VecV\times \VecB + \coefR( \VecJ -\VecJ_{\star})  +
    F_0\frac{\delta^*}{\rho}(\nabla_\perp p_i - \nabla_\parallel p_e)
   \end{equation} 
The resistivity $\coefR$ \revised{with} Spitzer temperature dependence and the diamagnetic coefficient $\delta^*$ are given by 
 \begin{align}
\coefR&=\coefR_0\cdot(T/T_0)^{-3/2} \\
\delta^*&=m_\text{ion}/(e F_0 \sqrt{\mu_0\rho_0})
\end{align}
with the constant parameter $\coefR_0$ and initial plasma core temperature ($T_0$).  $m_\text{ion}$ is the ion mass and $e$ the elementary charge. The constant $F_0 = R_0 B_{\phi0}$ is defined as the major radius at the geometric centre times the vacuum toroidal field. This constant appears due to the definition of the $\delta^*$, for consistency with the reduced MHD model described below. A more accurate modeling of the diamagnetic effect is possible with the two pressures extension described in Section~\ref{:code:models:twotemp}.
The term $\VecJ_{\star} = j_S\,\mathbf{e}_\phi$
denotes a toroidal current source term. It can be used to preserve the
original current profile $j_0$ approximately throughout the
simulation, if one chooses $j_S=j_0$\footnote{For a more consistent
treatment, a loop voltage can also be applied at the computational
boundary.}.  The current source
term is also used to model a consistently evolving bootstrap
current~\cite{Pamela2017}. In that case, the initial current profile
needs to include the initial bootstrap current correctly and the
current source term takes the form: $j_S=j_0+j_B-j_{B,0}$, where
$j_{B,0}$ denotes the initial bootstrap current and $j_{B}$ is the
bootstrap current corresponding to the self-consistent profiles during
time evolution. For the calculation of the bootstrap current, the expressions of
Refs.~\cite{Sauter1999,Sauter2002} are used.

In the MHD model shown in Equations~(\ref{eq:mhd:A}--\ref{eq:mhd:p}), the gauge still needs to be defined. In the JOREK full MHD model (Section~\ref{:code:models:fullMHD}), the Weyl Gauge, $\Phi = 0$, is used. That implies that the toroidal component of the magnetic vector potential changes with time even in steady state.

The tokamak plasma evolves in a low collisionality
 regime and the associated viscous stress tensor ($\TensV$) is decomposed into
 three main parts 
 \begin{equation} 
 \TensV \simeq  \TensV_{f} + \TensV_{neo} +  \TensV_{gv} 
 \end{equation} 
 These components model the Newtonian-fluid type, neoclassical and
 gyro-viscous effects respectively. 
 \revised{The separation in the different physical effects contained in the stress tensor does not imply a possible double counting but rather a representation on different time scales and directions. Following \cite{Callen_2009}, the parallel viscous stress is included as a Braginskii viscous stress acting on a fast timescale and a residual neoclassical stress on a slower collisional time scale.}
 The Newtonian stress tensor
 ($\TensV_{f}$) is decomposed into the parallel and the perpendicular
 directions to the magnetic field and the associated coefficients of
 viscosity are $\coefV_\parallel$ and $\coefV_{\perp}$.
According to the Chew-Goldberger-Low formulation \cite{Chew1956}, the parallel stress tensor $\TensV_{f\|}$ for arbitrary collisionality in a magnetized plasma is written as \cite{SchnackBook}:
   \[
   \TensV_{f\|} =
   3\coefV_{\parallel}\left ( \mathbf{b}\otimes\mathbf{b} - \frac{1}{3}\Id
   \right )\otimes\left ( \mathbf{b}\otimes\mathbf{b} - \frac{1}{3}\Id
   \right ) \nabla\VecV
\]   
The coefficient $\coefV_{\parallel}$ is modelled as a spatial constant but such a  dependency can be changed easily. The explicit formulation of the perpendicular tensor  $\TensV_{f\perp}$ can be found in Ref.~\cite{SchnackBook}.
The associated coefficient $\coefV_{\perp}$ is typically chosen to have the same temperature dependence as $\eta$ in order to keep the magnetic Prandtl number spatially nearly constant (except for the weaker density dependency).
\begin{equation}
  \coefV_{\perp} =\coefV_{\perp 0}\cdot(T/T_0)^{-3/2}
\end{equation} 
where $\coefV_{\perp 0}$ is a constant parameter and $T_0$ is the initial  plasma core temperature.
The neoclassical viscous tensor ($\TensV_{neo}$) is determined by a  heuristic
formulation~\cite{Gianakon2002}  
\begin{equation}
    \nabla\cdot{\TensV_{neo} } =\dens \coefV_{neo}
    \frac{\|\VecB\|^2}{\|\VecB_\theta\|^2} \left (\mathbf{b}_{\theta}\otimes\mathbf{b}_{\theta}\right )
    \left ( \VecV - \VecV_{neo}  \right )
    \quad \text{ where } \quad
 \mathbf{b}_{\theta} = \frac{\VecB_\theta}{\|\VecB_\theta\|}   
\end{equation}
with $\VecB_\theta = \VecB - (\VecB\cdot\mathbf{e}_\phi) \mathbf{e}_\phi$ the poloidal magnetic field. The
neoclassical coefficient $\coefV_{neo}$ and  velocity $\VecV_{neo}$ are given functions of the
temperature and the magnetic field (see Refs.~\cite{Orain2013,OrainPhD}). 
In magnetized plasmas it is usual to assume gyro-viscous cancellation~\cite{Hazeltine1985,SchnackBook} caused by the finite Larmor-radius effect. Therefore, to enforce gyro-viscous cancellation, the gyro-viscous stress tensor $\TensV_{gv}$ is modeled as
 \begin{equation}\label{eq:gyroV:T}
   \nabla\cdot{\TensV_{gv} } =  \dens\left (
 \pderiv{\mathbf{v}^*_i}{t} + \VecV\cdot\nabla\mathbf{v}^*_i +
 \mathbf{v}^*_i\cdot\nabla{\mathbf{v}_{||}}
 \right )
 \end{equation}
where $\mathbf{v}_{dia,i}$ is the ion diamagnetic drift velocity defined in equation~\eqref{eq:Vreduced} and $\mathbf{v}_{||}$ the parallel velocity.
\revised{The terms involving the ion-diamagnetic heat flux (and the associated cancellation with the density convection due to the ion diamagnetic drift \cite{Schnack2013}) have not been implemented to avoid the possible destabilisation of ITG modes. These terms are essential to study the interaction of MHD instabilities with underlying ITG turbulence but this is left for future applications.}

The heat diffusion tensor $\TensK$ is decomposed parallel and perpendicular to the magnetic field
 \begin{equation} 
 \TensK = \coefK_{\parallel} \mathbf{b}\otimes \mathbf{b}
        + \coefK_\bot\left ( \Id - \mathbf{b}\otimes \mathbf{b}
        \right )
        \label{eq:heat_diff_tens}
        \end{equation} 
Note that the factor $(\gamma-1)$ in the heat diffusion terms is
absorbed in the coefficients $\coefK_{\parallel} $ and $\coefK_\bot$. Here, $\gamma$ is the ratio of specific heats (usually $\gamma=5/3$).
The vector $\mathbf{b}=\mathbf{B}/B$ denotes the unit vector in the direction of the magnetic field. Radial profiles of $\coefK_\bot$ are usually specified in an ad-hoc manner to mimic the background transport that cannot be captured with the present model. For instance, low values are set in the pedestal region to model the transport barrier. The parallel heat diffusion coefficient $\coefK_{\parallel}$ is implemented with  Spitzer-H{\"a}rm~\cite{Spitzer1953} temperature dependency according to
\begin{equation}
\coefK_{\parallel}=\coefK_{||,0}\cdot(T/T_0)^{5/2},
\end{equation}
where the central value $\coefK_{||,0}$ is calculated according to the Spitzer-H{\"a}rm
formula. An optional parameter
$\coefK_{||,\text{max}}$ can be specified to account for the 
heat flux limit~\cite{Malone1975} in a simplified way by ensuring that the parallel heat conductivity cannot exceed this maximum
value. Realistic anisotropies even beyond
$\coefK_{||}/\coefK_\bot\approx10^{10}$ can be handled without producing large spurious perpendicular transport provided a grid is used that is aligned to the equilibrium flux surfaces (see Section~\ref{:verification:anisotropy}). The particle diffusion tensor $\TensD$ has an analogous form to expression \eqref{eq:heat_diff_tens} although the parallel component ($D_\parallel$) is usually not used as the parallel particle transport is dominated by convection. The profile of $D_\perp$ is also specified by ad-hoc profiles reflecting underlying small-scale turbulence that is not included in the MHD model.

The source term in the momentum equation contains the contribution of the diffusion and the source of density, as well as specific source $\mathbf{S}_{\mathbf{m}} $  of momentum 
\begin{equation}
 \mathbf{S}_{\VecV} =  \mathbf{S}_{\mathbf{m}} -   \left ( \nabla\cdot(\TensD\nabla\rho) + S_\rho \right )\VecV, 
\end{equation}
The source terms in the pressure equation contains the Ohmic
heating term, thermal energy source and particles source effects. 
\begin{equation} 
 S_{\pres}  = (\gamma - 1)\left ( \left( \VecE + \VecV\times\VecB \right )\cdot\VecJ  +
S_E  - \VecV \cdot\mathbf{S}_{\mathbf{m}} + \frac{\VecV\cdot\VecV}{2}(S_\rho + \nabla\cdot(\TensD  \nabla\rho)) \right ),
\end{equation} 
where $S_E$ is the thermal energy source and $S_\rho$ is the particle source. Sources are
typically specified as radial profiles.\\
Given equations~(\ref{eq:mhd:A}--\ref{eq:mhd:p}), a proper mathematical treatment of this system should specify the functional spaces where the solutions are sought for. Moreover, since JOREK uses a finite element method, a weak form of the equation is preferred which implies to define basis and test functions. In the following, for brevity, the ``$dV$'' in all volume integrals is omitted.
Thus let $\cal{V}_\VecA,\cal{V}_\VecV,\cal{V}_\dens$, ${\cal{V}}_p$ and 
$\cal{V}^*_\VecA,\cal{V}^*_\VecV,\cal{V}^*_\dens$, ${\cal{V}}^*_p$ be the chosen function spaces for the basis and test functions respectively, a weak form of the MHD problem will be reformulated as: Find $( \VecA, \VecV, \dens, T)$ in 
${\cal{V}}_\VecA \times {\cal{V}}_\VecV \times {\cal{V}}_\dens \times {\cal{V}}_p$ such that, for any
test functions $( \VecAS, \VecVS, \densS, \presS)$ in 
 ${\cal{V}}^*_\VecA \times {\cal{V}}^*_\VecV \times {\cal{V}}^*_\dens \times {\cal{V}}^*_p$, we have:
\begin{align}
  \int \pderiv{\VecA}{t}\cdot\VecAS &=
  -\int \rule{0mm}{4mm}
  \VecE   \cdot\VecAS, \label{eq:mhdW:A}\\
  \int \dens\pderiv{\VecV}{t} \cdot\VecVS &=
  -\int \left ( \rule{0mm}{4mm} \dens\VecV\cdot\nabla\VecV + \nabla \pres -
\VecJ\times\VecB - \nabla\cdot\TensV - \mathbf{S}_{\VecV} \right ) \cdot\VecVS, \label{eq:mhdW:v} \\
\int \pderiv{\dens}{t}\densS &=
-\int \left ( \rule{0mm}{4mm} \nabla\cdot(\dens\,\VecV)  - \nabla\cdot(\TensD  \nabla\rho)- S_\dens\right ) \densS,\label{eq:mhdW:rho}
\\
\int \pderiv{\pres}{t}\presS &=
-\int \left ( \rule{0mm}{4mm}  \VecV\cdot\nabla \pres
+ \gamma \pres\nabla\cdot\VecV
- \nabla\cdot(\TensK\nabla T) - (\gamma -1)\TensV:\nabla\VecV- S_{\pres} \right )\presS\label{eq:mhdW:p}
\end{align}
where $\pres=T\,\dens$. 
The identity $\VecJ\times\VecB = -
\nabla\left (\frac{\VecB\cdot\VecB}{2} \right ) + \nabla\cdot\left (
\VecB\otimes\VecB\right ) $ and integration by parts is used to avoid computation of second order derivatives.
Equations (\ref{eq:mhdW:A}) and (\ref{eq:mhdW:v}) are vector equations. For numerical purpose, each of them must be transformed into three scalar equations by projecting the vectors onto some basis. 

Following the representation of $\VecA$ and $\VecV$ in the basis
$ \left(\mathbf{e}_\text{R}, \mathbf{e}_\text{Z}, \mathbf{e}_\varphi \right)$,
and
$\left( \mathbf{e}_\text{R},\mathbf{e}_\text{Z}, \VecB \right)$ respectively, the basis for the vector-test function $\VecVS$ and $\VecAS$ in the  weak formulation (\ref{eq:mhdW:A})-(\ref{eq:mhdW:p}) is chosen as:
\begin{align}
\VecA &=  A_R \mathbf{e}_\text{R} + A_Z \mathbf{e}_\text{Z} + \frac{1}{R} A_3 \mathbf{e}_\varphi \\
\VecV &=  V_R \mathbf{e}_\text{R} + V_Z \mathbf{e}_\text{Z} + V_\parallel \VecB \\
\VecAS &= a^* \left( \mathbf{e}_\text{R},  \mathbf{e}_\text{Z}, \mathbf{e}_\varphi \right) \\
\VecVS &= v^* \left(\mathbf{e}_\text{R},  \mathbf{e}_\text{Z}, \VecB \right)
\end{align}

where $a^*$ and $v^*$ represent the scalar test functions as defined in Section~\ref{:code:numerics:discretization:Bezier}.
This choice of projection in the parallel direction ensures on the discrete level that the Lorentz force $\VecV^*\cdot(\VecJ\times\VecB)$ is
exactly vanishing.

\subsubsection{Reduced MHD}\label{:code:models:base:reduced}

In order to reduce computational requirements, one often employs reduced MHD models, which eliminate fast magnetosonic waves while retaining the relevant physics \cite{guillard:hal-01145009,Strauss1983,Strauss1997,Jardin2012}. The removal of fast magnetosonic waves, the fastest waves in the system, allows one to use larger time steps due to \revised{relaxing} the CFL condition. Even when implicit time integration methods are used, and the CFL condition is no longer a hard limit, using time steps that are large compared to the shortest time scale can lead to poor accuracy \cite{Jardin2012,Kruger1998}. In addition, reduced MHD has less unknowns compared to full MHD, which decreases the computational costs and memory requirements for simulations.

Reduced MHD, as first introduced by Greene and Johnson \cite{Greene1961}, and later developed by Kadomtsev, Pogutse and Strauss \cite{Kadomtsev1974,Strauss1976}, relied on ordering in a small parameter, often taken to be the inverse aspect ratio. The ordering itself is a system of several approximations and assumptions involving the ordering parameter that allows one to determine the relative order (in terms of the ordering parameter) of any quantity with respect to any other quantity of the same dimension. In this context, terms corresponding to fast magnetosonic waves have a higher order than the terms that one wants to keep, allowing the fast wave terms to be dropped. Naturally, there are many choices one can make in the ordering assumptions, depending on which physical effects one wants to keep, all of which result in different reduced equations \cite{Strauss1997,Strauss1976,Strauss1977,Strauss1980,Kruger1998}. The ideas of reduced MHD have also found use in astrophysics, where toroidal geometry cannot be assumed, and thus the inverse aspect ratio cannot be used as an ordering parameter \cite{Oughton2017}.

Starting in the 1980s, a new ansatz-based approach was introduced by Park \emph{et al} \cite{Park1980}, where an ansatz form that eliminates fast magnetosonic waves is used for the velocity and terms of all orders are kept \revised{(eliminating partially the fluid compression)}. Their reduced model corresponds to ideal MHD in the incompressible limit and was used to resolve internal kink modes in a cylindrical geometry, something that ordering-based reduced MHD could not do. Izzo \emph{et al} used a similar ansatz in their study \cite{Izzo1985}. Later papers also adopt an ansatz for the magnetic field that eliminates field compression \cite{Breslau2009,Franck2015}. The ansatz approach allows one to make less assumptions and keep more physical effects, while generally resulting in more complicated equations than the ordering approach. Thus, while keeping more physics, the various terms in the equations of ansatz-based reduced MHD are harder to interpret due to their complexity. In addition, without an ordering parameter, error estimation becomes much more difficult. \revised{The ansatz method allows to conserve energy exactly on the equation level. Some of the ordering-based models also conserve energy, however it is then often necessary to keep selected higher order terms to ensure that. The ansatz method thus is an alternative way of eliminating fast magnetosonic waves, which makes energy conservation easier. At the same time, in comparison to models which use $\epsilon=a/R$ for the ordering, the ansatz based reduced MHD model is still fairly accurate in the spherical tokamak limit as shown in Ref.~\cite{Pamela2020} by comparison to full MHD.}

The 
reduced MHD model used in JOREK is derived following the ansatz-based approach. In this approach, instead of the whole  functional spaces used in full MHD, the variables are constrained to 
lie in a subset of these spaces and the equations are established by a Galerkin 
truncation. Another way to present this procedure is to say that an ansatz is postulated for some variables. The ansatz considered here
assumes that the time dependent part of the magnetic potential is dominated by the toroidal component. The ansatz for the magnetic field is deduced by approximating $B_\phi$ as the vacuum $F_0/R$ toroidal field.
\begin{equation}\label{eq:ABreduced}
  \mathbf{B}=\frac{F_0}{R}\mathbf{e}_\phi + \frac{1}{R}\nabla\psi\times\mathbf{e}_\phi
\quad \Longrightarrow \quad\VecA = \psi\nabla\phi \end{equation}
where $F_0$ is constant in space as well as time and $\mathbf{e}_\phi$ is
the normalized toroidal basis vector. In the weak formulation (\ref{eq:mhdW:A}), this corresponds to defining 
${\cal{V}}_\VecA =\{\VecA : \exists \psi \in H^2 s.t. ~\VecA=\psi \nabla \phi \}$. The ansatz (\ref{eq:ABreduced}) implies that the velocity cannot be arbitrary. Indeed, taking
the cross product of equation
(\ref{eq:mhd:A}) with $\nabla\phi$, after substituting equations \eqref{eq:Eprim} and \eqref{eq:ABreduced} for $\VecE$ and $\VecA$ and neglecting resistivity and the poloidal component of $\VecB$, we obtain:
\begin{equation}\label{eq:Vreduced}
\mathbf{v}_P=\underbrace{-R\nabla  u\times\mathbf{e}_\phi}_{\equiv\mathbf{v}_E}
-\underbrace{(\delta^*\,R/\rho)\,\nabla \pres_i\times\mathbf{e}_\phi}_{\equiv\mathbf{v}_\text{dia,i}}.
\end{equation}
where $u$ is defined as $\Phi/F_0$ and $\mathbf{v}_P=(\mathbf{e}_\phi\times \mathbf{v}) \times \mathbf{e}_\phi$ denotes the poloidal component of the velocity. In this expression, $\mathbf{E}\times\mathbf{B}$ effects are captured by the first term,
and the ion diamagnetic drift
velocity by the second one. Given this expression, we can define the  
approximation space for the $\mathbf{v}_E$ poloidal component of the velocity variable as 
$ {\cal{V}}_{\mathbf{v}_E}=\{ \mathbf{v} : \exists u \in H^2 s.t. ~\mathbf{v}=-R \nabla u \times\mathbf{e}_\phi\}$
and according to Ritz-Galerkin method, it is natural to choose the velocity test functions in the poloidal  direction in the same space:  
\[
\VecVS_{\text{E}} = -R \nabla \SvS\times\mathbf{e}_\phi
\]
A first version of the reduced MHD model can be obtained using the definition of these spaces. 
However, for many problems, flows are not purely poloidal and one must take into account flows 
along the magnetic field lines. Therefore an improved version of the reduced MHD model used in JOREK defines the velocity approximation space by: 
\begin{equation}\label{eq:Vreduced2}
\VecV = \VecVm_E + \mathbf{v}_\text{dia,i}
+\underbrace{ \revised{v_{||}}\mathbf{B}}_{\equiv\mathbf{v}_{||}} 
\end{equation}

This reduced MHD model is thus characterized by a magnetic 
potential defined by a single scalar function ($\psi$) and a velocity field defined by the two scalars functions ($\pvec$ and \revised{$v_{||}$}). As  
done for $\VecVm_{\text{E}}$, it will be natural to define the parallel test functions 
using Ritz-Galerkin recipe as $ \SvS\VecB$ for some scalar $\SvS$. 

It is important to point out here that, even if a constant $F_0$ is used, instead of the function $F$ used in the computation of the Grad-Shafranov equilibrium (i.e. assuming that $RB_\phi = \mathrm{const}$), the reduced model preserves the equilibrium. Indeed, when focusing on the momentum equation, we can prove that all the terms associated with the function $F$ (from the initial Grad-Shafranov equilibrium) are in the kernel of the momentum projectors: $
  \SvS\VecB$ and $-R \nabla \SvS\times\mathbf{e}_\phi$.  This is a
  direct consequence of the fact that \\ $R \nabla
  \SvS\times\mathbf{e}_\phi = R^2\nabla\times(\SvS\nabla\phi) $,
  $\quad\VecB\cdot(\VecJ\times\VecB)=0\;$ and $\;\nabla\times (\nabla F^2/2) =
  0$. \revised{Note that, although the ansatz \eqref{eq:ABreduced} neglects toroidal field compression by using a constant $F_0$ instead of the flux function $F(\psi)$ for $RB_\phi$, the Shafranov shift is retained due to the use of the solution of the Grad-Shafranov equation with nonzero $FF'$ as the initial condition for $\psi$.}\\
  Summarizing, the reduced MHD is defined by the magnetic and velocity
 ansatz given by equations (\ref{eq:ABreduced}--\ref{eq:Vreduced}) respectively. The weak form for the reduced MHD equations after integration by parts can be directly derived from the general expressions (\ref{eq:mhdW:A}--\ref{eq:mhdW:p}) taking into account 
 the present definition of the functional spaces. We detail 
 in the sequel the expression of the magnetic potential and momentum 
 equations. For the magnetic potential, it is convenient to use 
 $\SaS\,\VecB$ as test function and we obtain 
 \begin{displaymath}
\int \pderiv{\VecA}{t}\cdot\VecB \; \SaS =
  -\int \left ( \rule{0mm}{4mm}
  \VecE  - F_{0}\nabla \pvec \right ) \cdot\VecB \;  \SaS
\end{displaymath}
that gives the problem: Find $\psi \in H^2$ such that for any $\SaS \in H^2$ we have: 
\begin{align}\label{eq:rmhdW:A}
\int \frac{1}{R^2}\pderiv{\psi}{t} \SaS 
 & = - \int \left (\frac{1}{R} [u,\psi]
  +\frac{\coefR}{R^2}(j - j_{\star})  
  - \frac{F_0}{R^2}\pderiv{u}{\phi} 
 + \frac{\delta_*}{2\rho}\nabla p\cdot \VecB 
 \right )\SaS
\end{align}
where we have introduced the Poisson bracket 
$[f,g]=\nabla f\times \nabla g \cdot \mathbf{e}_\phi$. Note, that the poloidal current component has been neglected in the resistive term. To establish the momentum equation, we first use the expression of the velocity (\ref{eq:Vreduced2}) together with our definition of the gyro-viscous tensor (\ref{eq:gyroV:T}) to obtain the equation: 
\begin{equation}
 \rho\pderiv{\VecVm}{t} 
 = -\rho\VecVm\cdot\nabla\VecVm
 -\rho\VecVm_i^*\cdot\nabla\VecVm_{\text{E}}
 - \nabla p +
\VecJ\times\VecB + \nabla\cdot(\TensV_f +\TensV_{neo}) 
+ \mathbf{S}_{\VecV},
\label{eq:mhdR:v}   
\end{equation}
where $\VecVm= \mathbf{v}_{||}+\mathbf{v}_{\text{E}}$. Now, using successively $\SvS\,\VecB$ and $\VecV_{\text{E}}^*$ as test functions allows to obtain two scalar equations for the parallel and poloidal components of the velocity: 
\begin{align}
  \int \dens\pderiv{\VecVm}{t} \cdot\VecB \; \SvS &=
  \int \left (
 \frac{\VecVm\cdot\VecVm}{2}\nabla\dens + \dens \VecVm \times\mathbf{w} 
 - \nabla p + \nabla\cdot(\TensV_f +\TensV_{neo}) 
+ \mathbf{S}_{\VecV} \right )\cdot\VecB \SvS
 \nonumber  \\
  & \hspace*{3cm}\rule{0mm}{4mm}
  - \int \rho(\mathbf{v}_i^*\cdot\nabla\VecVm_{\text{E}}) \cdot\VecB \SvS 
   + \text{BT}_{v_\parallel}, \label{eq:rmhdW:vpar} \\
  \int \dens\pderiv{\VecVm}{t} \cdot \VecV_{\text{E}}^* &=
  \int \left ( \rule{0mm}{4mm} \frac{\VecVm\cdot\VecVm}{2R^2}\nabla(R^2\dens) + \dens \VecVm \times\mathbf{w}
 - \nabla \pres  + \nabla\cdot(\TensV_f +\TensV_{neo}) + \mathbf{S}_\VecV\right )  \cdot\VecV_{\text{E}}^* \nonumber \\
& \hspace*{3cm} 
  - \int \rho(\mathbf{v}_i^*\cdot\nabla\VecVm_{\text{E}}) \cdot \VecV_{\text{E}}^*
+ \int v^* \VecB\cdot\nabla j + \text{BT}_{v_{\text{E}}},
  \label{eq:rmhdW:v}
\end{align}
where we have used the relation $\VecVm\cdot\nabla\VecVm= \nabla\frac{\VecVm\cdot\VecVm}{2} - \VecVm\times\mathbf{w}$  
introducing the 
vorticity $\mathbf{w}\equiv\nabla\times\VecVm$ and $j\equiv -R\,J_\phi$.
The equations for density and temperature are similar to the full MHD context, but with the prescribed velocity and magnetic field expressions
\begin{align}  
\int \pderiv{\dens}{t}\densS &=
-\int \left ( \rule{0mm}{4mm} \nabla\cdot(\dens\,\VecVm) + \mathbf{v}_i^*\cdot\nabla\dens
- S_\dens\right ) \densS -\int \TensD\nabla\dens\cdot\nabla\densS + \text{BT}_{\dens}, \label{eq:rmhdW:rho}
\\
\int \pderiv{\pres}{t}\presS &=
-\int \left( \rule{0mm}{4mm}  \VecVm\cdot\nabla \pres
+ \gamma \pres\nabla\cdot\VecVm -(\gamma -1)(\TensV_f +\TensV_{neo}):\nabla\VecVm - S_{\pres} \right)\presS  \nonumber  \\
  & \hspace*{4cm}
  \left . \rule{0mm}{4mm}-\int \TensK \nabla T \cdot\nabla\presS + \text{BT}_{\pres}  \right . \label{eq:rmhdW:p}
\end{align}
To derive the final formulation of the pressure equation, gyro-viscous cancellation assumptions have been used. The boundary integrals appearing after the integration by parts are indicated by the symbol BT and defined as
\begin{align}
\text{BT}_{v_\parallel} &=  -\oint v^* \frac{\dens \VecVm\cdot\VecVm}{2}\VecB \cdot  d\mathbf{S} \\
\text{BT}_{v_E} &= -\oint v^*\left( \nabla\left(\frac{R^2\dens\VecVm\cdot\VecVm}{2}\right) -\VecJ\times\VecB \right)\times\nabla\phi \cdot d\mathbf{S} \\
\text{BT}_{\dens} &=  \oint \densS\TensD \nabla\dens \cdot  d\mathbf{S} \label{eq:BT:rho} 
\\
\text{BT}_{\pres} &=  \oint \presS\TensK \nabla T \cdot  d\mathbf{S} \label{eq:BT:pres}
\end{align}
Note that in this derivation, once the ansatz and projection functions are defined, there are no approximations on geometry. I.e., the reduced MHD derived here is not an aspect ratio expansion of the full MHD model.

These equations involve some high order derivatives whose computations can be alleviated by the introduction of intermediate
variables: the toroidal current density ($j$) and 
toroidal vorticity ($\omega$) satisfying the following partial
differential equations 
\begin{align}
j &= \Delta^*\Psi \equiv R^2 \nabla\cdot\left(\frac{1}{R^2}\nabla_\text{pol} \Psi\right),\label{eq:red:currdef} \\
\omega &= \Delta_\text{pol} u \equiv \nabla\cdot\nabla_\text{pol} u.\label{eq:red:vortdef}
\end{align}
Here, $\nabla_\text{pol} u$ denotes the gradient in the R-Z plane. The reduced MHD base model consists of seven scalar physical quantities as variables, see Table~\ref{tab:variables}. Five variables are evolved in time (``five field model''),
while $\omega$ and $j$ are
coupled to $u$ and $\Psi$ by definition equations\footnote{Via the definition equations, $\omega$ and $j$ are projected to the $G^1$ continuous Bezier basis, while expressing them directly in terms of $u$ and $\Psi$ would correspond to a discontinuous representation.}. The evolution and definition equations
are solved simultaneously at every time step in a fully implicit numerical scheme (see
Section~\ref{:code:numerics:tstep}).

\begin{table}
\centering
\begin{tabular}{ r | l | l }
Symbol & Description \\
\hline
$\Psi$ & Poloidal magnetic flux & $=R\;\mathbf{A}\cdot\mathbf{e}_\phi$ with $\mathbf{A}$ the vector potential \\
$u$ & Velocity stream function & $=\Phi/F_0$ with $\Phi$ the electric potential \\
$j$ & Toroidal current density & $=-R\,\mathbf{j}\cdot\mathbf{e}_\phi=\Delta^*\Psi$ \\
$\omega$ & Toroidal vorticity & $=\Delta_\text{pol}u$ \\
$\rho$ & Mass density & $=n_e m_\text{ion}$ for singly charged ions \\
$T$ & Temperature & $\equiv T_e+T_i$ in the single temperature model \\
$v_{||}$ & Parallel velocity & $=\mathbf{v}_{||}\cdot\mathbf{B}/B^2$
\end{tabular}
 \caption{Scalar physics variables of the base reduced MHD model. $\omega$ and $j$ are derived variables, connected to $u$ and $\Psi$ by definition equations.}
 \label{tab:variables}
\end{table}

\subsubsection{Boundary conditions}

Boundary conditions can be set in a flexible way. 
By default, all variables are kept constant in time on the computational domain boundary, wherever the latter is aligned to a flux surface (Dirichlet). Where flux surfaces are intersecting the boundary (e.g., in the divertor region or for grids extended to the true physical wall) sheath boundary conditions are applied as commonly done in divertor physics codes~\cite{StangebyBook}. The poloidal flux, current density, electric potential, and vorticity are kept fixed at the boundary, while the parallel velocity is forced to be equal to the ion sound speed. For the density no Dirichlet condition is forced and the boundary term \eqref{eq:BT:rho} is not included. Not including the latter boundary term in the finite element method naturally implies that $\TensD \nabla\dens \cdot \mathbf{n}=0$ and therefore the perpendicular ion flux to the boundary is purely convective $\mathbf{\Gamma}\cdot\mathbf{n}=\dens \VecV\cdot\mathbf{n}$. The evolution of the boundary temperatures are constrained by the following B.C.s for the normal ion and electron heat fluxes to the boundary
\begin{align}
\mathbf{q}_i\cdot\mathbf{n}\equiv & \left(\frac{\dens}{2}\,\VecV\cdot\VecV+\frac{\gamma}{\gamma-1} \, \rho \, T_i\right)\VecV\cdot\mathbf{n} -\frac{\TensK_i}{\gamma-1} \,\nabla T_i  \cdot\mathbf{n}=\gamma_\textrm{sh,i}\,\rho\, T_i\, \VecV\cdot\mathbf{n},\\
\mathbf{q}_e\cdot\mathbf{n}\equiv & \frac{\gamma}{\gamma-1}  \,\rho \,T_e\,\VecV\cdot\mathbf{n} -\frac{\TensK_e}{\gamma-1} \nabla T_e  \cdot\mathbf{n}=\gamma_\textrm{sh,e}\,\rho\, T_e\, \VecV\cdot\mathbf{n},
\end{align}
where $\gamma_\textrm{sh,i}\sim $ 2-3 and $\gamma_\textrm{sh,e}\sim$ 5-6 are the ion and electron sheath transmission factors ~\cite{StangebyBook}. For the single temperature model ($T_e=T_i$), the two latter expressions can be added to find the total heat flux equation
\begin{equation}
\mathbf{q}\cdot\mathbf{n}\equiv \left(\frac{\dens}{2}\,\VecV\cdot\VecV+\frac{\gamma}{\gamma-1}  \rho T\right)\VecV\cdot\mathbf{n} -\frac{\TensK}{\gamma-1}\nabla T  \cdot\mathbf{n}=\gamma_\textrm{sh}\,\rho\, T_e\, \VecV\cdot\mathbf{n},
\end{equation}
 
where $\gamma_\textrm{sh}$ is the total sheath transmission factor that has typical values of 7-8. The latter B.C.s are expressed in the form $\TensK \nabla T  \cdot\mathbf{n}=-(c_b-1)\,\rho\, T\, \VecV\cdot\mathbf{n}$ and replaced in the boundary term \eqref{eq:BT:pres} in order to implement them as natural B.C.s. For the electrons $c_{b,e} = (\gamma-1)(\gamma_\textrm{sh,e}-1)$, for the ions $c_{b,i} = (\gamma-1)(\gamma_\textrm{sh,i}-\gamma-1)$ and the total heat flux  $c_b = (\gamma-1)(\gamma_\textrm{sh}/2-\gamma/2-1)$. Note that sheath boundary conditions are applied on the whole boundary of the computational domain if grids extended to the physical first wall are used (see Section~\ref{:code:numerics:discretization:grids}). In case of free boundary simulations, the Dirichlet condition on the plasma current density and poloidal flux is removed, and a natural boundary condition is implemented instead, like described in Section~\ref{:code:models:freebound}. Further extensions of the boundary conditions have been developed for particular applications, e.g., a limitation of the current density to the ion saturation current ~\cite{artola2020COMPASS}.

\subsubsection{Properties of the reduced MHD model}

Since a significant number of reduced MHD models with very different properties have been proposed in literature, some confusion exists regarding their capabilities. We explain a few key features of our reduced MHD model in the following. A recent discussion of reduced and full MHD models is also provided by Ref.~\cite{Graves2019}, and Section~\ref{:code:models:outlook} shows reduced MHD models for stellarator configurations yet to be implemented, including a detailed discussion of the conservation properties. 

In Ref.~\cite{Franck2015}, it is shown that the model implemented in JOREK is energy conserving as consequence of the full MHD being energy conserving and the ansatz based approach being used. This is strictly valid only for the single-fluid model, where diamagnetic drift effects are excluded, since the gyro-viscous cancellation is not exactly energy conserving~\cite{Zeiler1988,Pamela2020}. \revised{The non-conservation introduced by gyro-viscous cancellation is of the order of $\delta^*p_m/(\rho_0\,v_A)$, where $p_m$ and $v_A$ denote the magnetic pressure and Alfv\'en velocity, respectively. As long as $\delta^*$ is small, the error should be acceptable. The energy conservation test shown in the right panel of Figure~\ref{fig:energy_conservation} supports a good energy conservation also when diamagnetic drift effects are included.
Note that, even without diamagnetic drift, energy conservation is only exact in the limit of continuous time, and introducing a finite time step also introduces a small error in the energy conservation. Another source of error is the truncation of the toroidal Fourier series (see Section \ref{:code:numerics:discretization:four}).} More formally, for a simplified version of the reduced MHD model, it has been shown in Ref.~\cite{guillard:hal-01145009} that reduced MHD models are a valid approximation of the full MHD model, i.e., the solutions of the full MHD system converge to the solutions of an appropriate reduced model.

\revised{Linear momentum is not exactly conserved locally by the reduced MHD models even in the continuous limit. This becomes obvious when one acknowledges that each of the three Cartesian components of the momentum equation govern momentum conservation in the respective Cartesian direction, and all three must be satisfied individually in order for linear momentum to be conserved. However, there are only two scalar velocity variables, $u$ and $v_{||}$, in the reduced MHD model, and only two scalar equations governing these variables, namely \eqref{eq:rmhdW:vpar} and \eqref{eq:rmhdW:v}, remain in the reduced model. Thus, it is impossible to locally conserve all three components of linear momentum in reduced MHD, except for some special cases. Globally, the $z$-component of momentum is always conserved, as can be seen by letting $\SvS = \ln R$ (which gives $\VecVS_{\text{E}} = \nabla z$) in equation \eqref{eq:rmhdW:v}. One can also ensure global conservation of the $x$ and $y$ components of linear momentum by excluding the $n=1$ term from the Fourier series, which forces the $x$ and $y$ components of the integrated total momentum to zero due to symmetry. This topic is considered in more detail in Ref.~\cite{Nikulsin2020}. The full MHD model is conserving momentum exactly on the equation level, comparisons presented in Section~\ref{:verification} between both models provide confidence to some degree that the introduced error does not affect linear and non-linear dynamics significantly.}
  


The presented reduced MHD model satisfies $\nabla\cdot\mathbf{J}=0$. In fact it can be shown that equation~\eqref{eq:rmhdW:v} is identical to the weak form of $\nabla\cdot\mathbf{J}=0$. This is demonstrated by applying the cross product $\times\nabla\phi$ to equation \eqref{eq:mhdR:v} in order to obtain the poloidal current density   
\begin{align}
    F_0 \mathbf{J}_\text{pol} = &-j \mathbf{B}_\text{pol} - R^2\nabla p\times\nabla\phi  \nonumber\\ 
    &-R^2\left(  \rho\pderiv{\VecVm}{t} 
  +\rho\VecVm\cdot\nabla\VecVm
  +\rho\VecVm_\text{dia,i}\cdot\nabla\VecVm_{\text{E}}
 - \nabla\cdot(\TensV_f +\TensV_{neo}) 
- \mathbf{S}_{\VecV}   \right)\times\nabla\phi 
\label{eq:RMHD_poloidal_curr}
\end{align}
Then using the latter expression in $\int v^*\nabla\cdot\left(\mathbf{J}_\phi + \mathbf{J}_\text{pol}\right) = 0$ and applying integration by parts, equation~\eqref{eq:rmhdW:v} is recovered. As it can be inferred from equation~\eqref{eq:RMHD_poloidal_curr}, even if the toroidal field is fixed in time, the poloidal currents exist in this model and evolve according to the momentum equation and conservation of current.
The ansatz~\eqref{eq:ABreduced} together with the projection operator $e_\phi \cdot \nabla \times$ projects out, i.e. removes, the poloidal currents from the system of equations. This does not imply that the poloidal currents are neglected in the model, but rather their contribution to the toroidal field is dropped. The poloidal currents can be calculated, a posteriori, from \eqref{eq:RMHD_poloidal_curr}.
As mentioned above, the reduced ideal MHD momentum equation is consistent with the Grad-Shafranov equation, even in the absence of poloidal currents. The force balance $\VecJ\times\VecB = \nabla p$ appears as $\left[\psi,j \right] - \left[ R^2,p \right]$ in the momentum equation. Substituting $j = FF'\left(\psi \right) + R^2 p' \left( \psi \right)$ shows that the two terms in the pressure balance cancel.
The very successful benchmarks of VDE simulations between the JOREK reduced MHD model and the M3D-C$^1$ and NIMROD full MHD models shows the validity of this approach: the agreement regarding plasma dynamics and halo currents is excellent (see Section~\ref{:applic:core:vdes}).

As mentioned above, reduced MHD models aim to eliminate fast magnetosonic waves, the fast propagation of which can impose restrictive CFL limits for explicit methods and significantly increase the stiffness of the problem for implicit methods. In the JOREK reduced MHD model presented here, the fast magnetosonic waves are eliminated by the velocity ansatz \eqref{eq:Vreduced2}. In Ref. \cite{Nikulsin2019}, it is shown that any velocity field can be decomposed into an $\VecE\times\VecB$ term, a field-aligned flow term and a perpendicular fluid compression term\footnote{The perpendicular fluid compression term is mostly responsible for plasma compressional motion orthogonal to the background vacuum field, but some such compression is allowed already by the $\VecE\times\VecB$ term. See Ref. \cite{Nikulsin2019} and the references therein for more detailed discussion.}, which are responsible for Alfv\'en waves, slow magnetosonic and fast magnetosonic waves, respectively. The $\VecE\times\VecB$ and field-aligned flow terms are the first and third terms, respectively, in the velocity ansatz \eqref{eq:Vreduced2}, whereas the \revised{perpendicular} fluid compression term is not present in the ansatz.

Finally, it is important to note that the reduced MHD model presented here cannot correctly reproduce pressure-driven modes under all circumstances. In particular, the $1/1$ internal kink mode at nonzero $\beta$ is affected. As shown in Ref.~\cite{Graves2019}, the term associated with fast magnetosonic waves in the energy functional can be written as
\begin{equation*}
    \frac{1}{2}\int_V \left|\delta B_\parallel + \delta p/B_0\right|^2 dV,
\end{equation*}
where $\VecB = \VecB_0 + \delta\VecB$ and $p = p_0 + \delta p$; the quantities with a '0' subscript correspond to the equilibrium, and quantities with a '$\delta$' prefix are perturbations. Since in our reduced MHD model, we have set $\mathbf{e}_\phi\cdot\delta\VecB = 0$, we have $\delta B_\parallel = \VecB_0\cdot\delta\VecB/B_0 \approx 0$, whereas in full MHD simulations, one often finds that $B_0\,\delta B_\parallel \approx -\delta p$. Thus, in the case of pressure-driven instabilities, the term above contributes a stabilizing effect due to the $\delta p$ contribution not being cancelled by $B_0\,\delta B_\parallel$ \cite{Graves2019}. Traditional ballooning modes in the plasma edge are an exception to this rule. As shown in Ref \cite{Graves2019}, the Mercier criterion is modified as $4\epsilon\alpha(1-q^2) > s^2 q^2 + \alpha^2$, where $\alpha = -2q^2R_0p'/B_0^2$ is the ballooning parameter and $s = rq'/q$ is the shear. Since $q \gg 1$ and $\alpha\sim 1$ in the plasma edge, ballooning modes are largely unaffected, which can be seen in various benchmarks (Section~\ref{:verification:edge}). In Figure~\ref{fig:FMHD_KINK_BETA}, the linear growth rates for the internal kink mode are shown. For $\beta$ near zero, the mode is mostly current driven, and the stabilizing effect discussed above is negligible. However, as $\beta$ increases, the accuracy of reduced MHD quickly deteriorates. Alternate reduced MHD models, such as that by Kruger \emph{et al} \cite{Kruger1998}, can better capture most pressure-driven modes due to the incorporation of the constraint $B_0\delta B_\parallel = -\delta p$ into their model.

\subsubsection{Related models and extensions}

The various extensions available for the reduced MHD model are described in the following, Sections~\ref{:code:models:rmps}--\ref{:code:models:REfluid}. Note, that also a simplified version of the reduced MHD model is available, where the parallel momentum equation and the variable $v_{||}$ have been eliminated, while the rest of the description shown above remains unchanged. This corresponds to a drop of slow magneto-sonic waves. The full MHD model of JOREK is described briefly in Section~\ref{:code:models:fullMHD}. Formulations of reduced as well as full MHD appropriate for stellarators yet to be implemented are shown in Section~\ref{:code:models:outlook}.

\subsection{External magnetic perturbations}\label{:code:models:rmps}

For simulations of resonant magnetic perturbations (RMPs) used typically for ELM mitigation or suppression, a 3D poloidal flux perturbation at the boundary of the computational domain can be ramped up during the simulation~\cite{OrainPhD}. The perturbation needs to be pre-calculated by an external code (vacuum assumption for the boundary condition). This approach has widely been used for previous studies of RMPs (see Section~\ref{:applic:edge:rmp}) with the drawback that the magnetic field perturbation at the computational boundary cannot evolve consistently. Using the free boundary extension (see Section~\ref{:code:models:freebound}), RMPs can alternatively be described fully consistently from 3D coils~\cite{MitterauerSchwarzArtolaHoelzlCoils}.

\subsection{Separate electron and ion temperatures}\label{:code:models:twotemp}

An extension for treating electron and ion temperatures separately~\cite{PamelaPhD} introduces one additional variable to evolve the both temperatures independently in time. In particular, different parallel heat diffusion coefficients can be used for the species, allowing to capture the temperature evolution across an ELM cycle more accurately. The parallel heat conductivity does not only influence the non-linear evolution of the plasma considerably, but also affects linear stability properties (neglected in most stability codes).

\subsection{Neutrals}\label{:code:models:neutrals}

A model extension is available to include a neutral particle fluid in the simulations, a development originally started in Ref.~\cite{Reux38thEPS}. The present version was derived and implemented in Ref.~\cite{FilPhD}.
One additional physics variable was introduced to describe the distribution of neutrals across the computational domain. In this model, the neutral transport is purely diffusive. Ionization and recombination terms, as well as radiative loss terms are implemented. Recycling boundary conditions at the divertor targets have recently been implemented~\cite{SmithEPS2019,HuijsmansEPS2019}. The model is used for deuterium massive gas injection or shattered pellet injection simulations (see Sections~\ref{:applic:core:disr}) as well as detachment studies (see Section~\ref{:applic:edge:detachment}). A kinetic treatment of neutrals is also possible using the framework described in Section~\ref{:code:models:particles} with first applications on the way.

\subsection{Impurities}\label{:code:models:impurities}

For the modelling of impurities, several options exist. As a particularly simple model to incorporate some impurity effects, a radiative loss term can be switched on in the reduced MHD model with neutrals (Section~\ref{:code:models:neutrals}). Losses are then calculated under the assumption of a spatially and temporally constant background impurity distribution with a prescribed radiative cooling rate.

For a more realistic description, a model exists where impurities are treated as an additional fluid species~\cite{Nardon2017,HuEPS2018,Hu2020}. This model is applied to massive gas or shattered pellet injection simulations (see Section~\ref{:applic:core:disr}), but also to radiative collapse simulations (see Section~\ref{:applic:core:impurities}). One additional variable is introduced to describe the impurity density distribution. All impurities are assumed to be convected together with the main plasma independently of the charge state, and the impurities are assumed to be in coronal equilibrium. The latter assumption may lead, at least in certain cases, to an underestimation of energy dissipation by impurities. For example, for an axisymmetric benchmark case on impurity dynamics~\cite{Lyons_2019}, JOREK (with its coronal equilibrium model) predicts a roughly two times slower thermal collapse than M3D-C$^1$ and NIMROD which have a more advanced model tracking the density of each impurity charge state. The coronal equilibrium assumption also results in an instantaneous change in the ionization state according to the electron temperature, resulting in difficulties in treating the ionization energy and the corresponding recombination radiation which would not be present in a self-consistently evolving non-equilibrium model. To avoid such artificial recombination radiation, we currently treat the ionization energy as a potential energy. A more advanced model, going beyond the coronal equilibrium assumption, is presently in preparation (see Section~\ref{:code:models:outlook}).

Also kinetic particles (Section~\ref{:code:models:particles}) can be used to describe impurities. This has already been applied to study the transport of tungsten during ELM crashes (see Section~\ref{:applic:edge:elms}) as well as the sputtering and SOL transport of tungsten (see Section~\ref{:applic:core:impurities}).

\subsection{Pellets}\label{:code:models:pellet}

Several pellet ablation models are available in JOREK both for pellets consisting of the same material as the main plasma (``Deuterium pellets'') and for pellets consisting of a different material (``impurity pellets'', e.g., Argon or Neon). These ablation models are combined with the neutrals model (Section~\ref{:code:models:neutrals}) or the impurity model (Section~\ref{:code:models:impurities}), respectively.

For the particle source corresponding to pellet ablation, scaling laws from various Neutral Gas Shielding type of models in a Maxwellian plasma are implemented~\cite{Parks1978,Sergeev2006,Bosviel2020EPS,Parks2020}. The main idea behind these models is that the ablation rate naturally adapts such that the incoming heat flux from the ambient plasma is almost fully absorbed by the ablation cloud surrounding the pellet. Literature provides scaling laws for ablation rates for various pellet materials (including mixtures) which have been obtained by fitting numerical results of gas dynamics simulations~\cite{Sergeev2006,Bosviel2020EPS,Parks2020}. 

Ablated atoms are deposited \textit{via} a volumetric source term of the form:
\begin{equation}
S_n \propto \left[ 0.5 - 0.5 \tanh \left( \frac{(R - R_p)^2 + (Z - Z_p)^2}{\delta r_c} \right) \right] \left[ 0.5 - 0.5 \tanh \left( \frac{\phi - \phi_p}{\delta \phi_c} \right) \right],
\end{equation}
where $R_p$ and $Z_p$ are the pellet location and $\delta r_c$ and $\phi_c$ characterize the poloidal and toroidal extension of the ablation cloud. The parameters $\delta r_c$ and $\delta \phi_c$ determine the width of smoothing of the source profile in poloidal and toroidal direction. The pellet is presently assumed to move along a straight line with constant velocity, and its particle content (and physical size) is evolved according to the ablation.
The toroidal extension $\delta\phi_c$ of the ablation cloud in simulations is typically far larger than in reality due to limited toroidal resolution, but tests shown in Ref.~\cite{Futatani2014} for a Deuterium pellet found that for a sufficiently small $\delta\phi_c$, JOREK results converge, i.e. MHD dynamics becomes independent of $\delta\phi_c$. For impurity pellets, the same may however not be true because of the radiative loss term, which scales like $n_{imp}n_{e}$, i.e. like $n_{imp}^2$ in regions where the impurity density $n_{imp}$ is large, such that the total power radiated in the ablation cloud scales like $1/\delta\phi_c$. For shattered pellet injection simulations~\cite{Hu2018}, the model described above is applied for each individual shard. Input parameters allow specifying the shard size distribution, the averaged velocity and velocity spread of the shards.

\subsection{Free boundary and resistive walls}\label{:code:models:freebound}

Via a coupling~\cite{Hoelzl2012B} to the STARWALL code~\cite{Merkel2015}, JOREK is capable of free boundary simulations. In the Greens functions approach applied here, STARWALL discretizes the conducting structures by triangles (thin-wall approximation), while the vacuum region surrounding the plasma and conducting structures is not discretized. The JOREK-STARWALL coupling is then performed via a natural boundary condition at the edge of the JOREK computational domain \revised{that replaces the Dirichlet boundary conditions for the poloidal flux and current density}. In a boundary integral, that arises from partial integration (see Section~\ref{:code:numerics:discretization:weakform}) of the current definition equation, and that vanishes for fixed boundary simulations, the tangential magnetic field is expressed in terms of the poloidal flux values and the currents in the conducting structures as shown in detail in Ref.~\cite{Hoelzl2012B}. The B.C.\ is a Neumann type condition for the magnetic vector potential, which results from the analytical solution of the vacuum field given by the Green's functions. In terms of the response matrices, the magnetic field has the form
\begin{align*}
    \textbf{B}\times\textbf{n} = \underline{\textbf{M}_{vac}}\,\mathbf{B}\cdot\mathbf{n} + \underline{\textbf{M}_{I}} \text{I},
\end{align*}
where $\mathbf{n}$ is the normal vector to the boundary, $\underline{\textbf{M}_{vac}}$ is the vacuum response matrix and $\underline{\textbf{M}_{I}}$ is a matrix that calculates the contribution of the wall and coil currents ($\text{I}$). The evolution of the wall currents is calculated with resistor-inductor circuit equations that arise for each of the discretized wall elements. The ``response matrices'', which allow to calculate the evolution of the wall currents and the tangential magnetic field at the JOREK boundary, are calculated by STARWALL only once in the beginning of a simulation. Since they are only dependent on the JOREK grid and wall geometry, response matrices may even be re-used for further simulations if the geometry remains the same. The response matrices are written out by STARWALL into a file and read by JOREK using MPI I/O in both codes. In JOREK, the matrices are used to evolve wall currents in time and to implement the natural boundary condition. The coupling between plasma and wall currents is implemented in a fully implicit way that is entirely consistent with the time evolution of the intrinsic JOREK equations. The dimensionality of the sparse matrix system is not increased in spite of this fully implicit approach compared to fixed boundary simulations since the implicit values of the wall currents are analytically eliminated from the system~\footnote{Note, that the natural boundary condition, however, leads to a less sparse matrix structure for boundary degrees of freedom. Local interactions between neighboring grid nodes are replaced by global interactions on the boundary. To ensure efficient load balancing also for such simulations, the domain decomposition is slightly adapted compared to fixed boundary simulations.}.

JOREK-STARWALL allows to choose a fixed or free boundary mode independently for each toroidal harmonic. This is sometimes used to keep fixed boundary conditions for the axisymmetric $n=0$ component, while a free boundary treatment is applied to non-axisymmetric $n\ne0$ components. For simulations, where also the $n=0$ component is treated to be free, the equilibrium solver has been extended to free boundary cases~\cite{Hoelzl2012B} and has recently been updated for Newton iterations to enhance convergence (using the methods described in Ref.~\cite{heumann2017finite}).
Magnetic field coils have been implemented self-consistently in STARWALL, including time varying coil currents and their interaction with conducting \revised{structures~\cite{HoelzlSchwarzArtolaCoils} and recently also arbitrary 3D coils have been implemented in a self-consistent way.}~\cite{MitterauerSchwarzArtolaHoelzlCoils} (see also Section~\ref{:verification:freeb}). This allows to include active coils (poloidal field coils, RMP coils, etc.) and passive coils (Mirnov coils, saddle coils, etc.) consistently in JOREK-STARWALL simulations. A functionality is available also, which allows to create a free boundary equilibrium for a given fixed boundary case, by automatically determining appropriate coil currents~\cite{ArtolaPhD}.

\begin{figure}
\centering
  \includegraphics[width=0.7\textwidth]{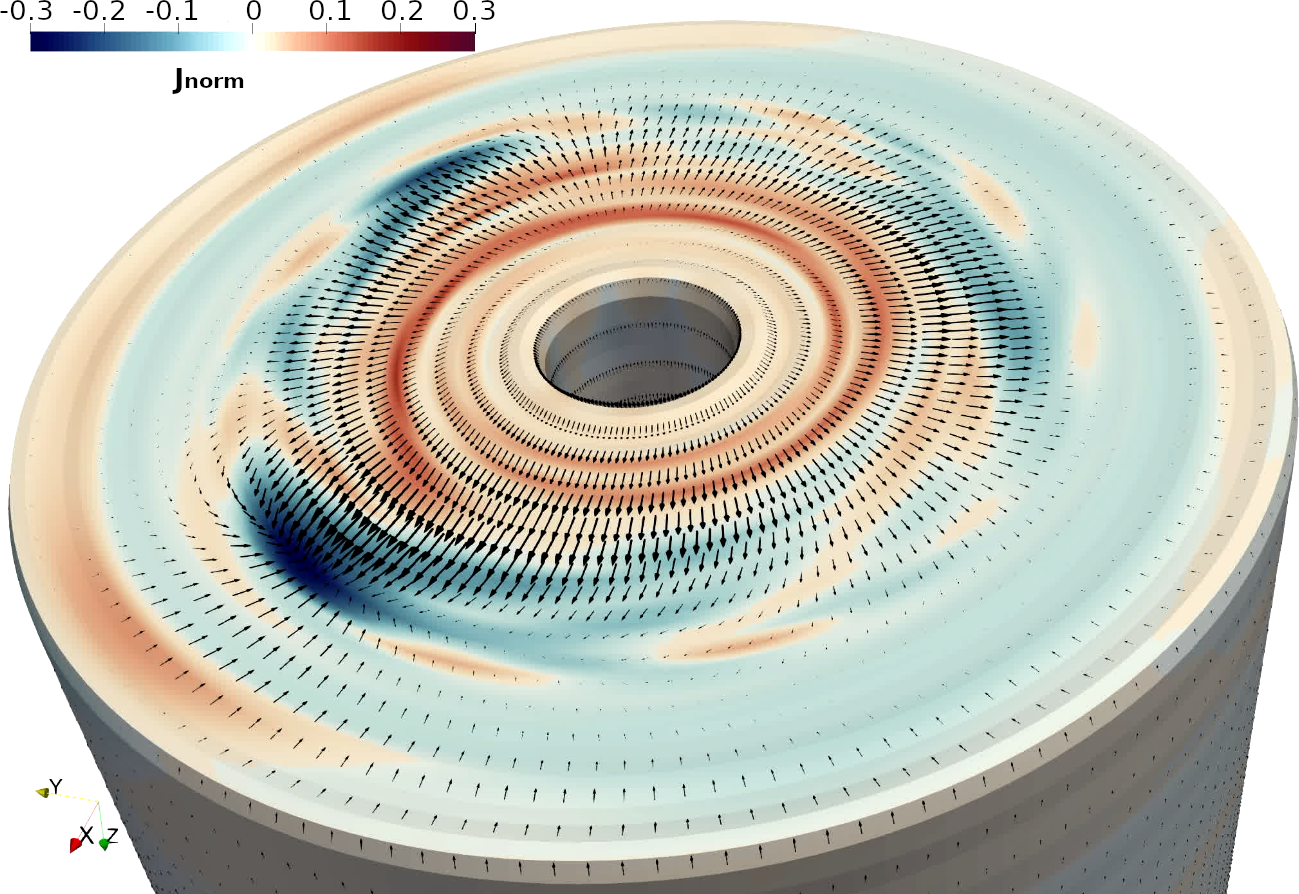}
\caption{Bottom view of the thin wall used for the 3D VDE benchmark case published in Ref.~\cite{Artola2020B} showing the distribution of source/sink (halo) currents (black arrows). The colors indicate the perpendicular current density into the wall respectively out of it (in a.u.). The B.C. for the electric potential was $\Phi=0$ here such that the wall acts ideally conducting for the halo currents. \noreprint}
\label{fig:source_sink}
\end{figure}

Via the derivation shown in Refs.~\cite{Zakharov2015,Atanasiu2016,Atanasiu2018}, plasma currents flowing directly into conducting structures or out of them (current sharing between plasma and wall), can be treated consistently with the STARWALL formalism. The respective derivation for JOREK-STARWALL including the interaction of eddy and halo currents are shown in Ref.~\cite{ArtolaPhD}. However, the implicit coupling of wall source/sink (halo currents) with the plasma electric potential has not been implemented yet. For walls with a low poloidal path resistance, the usual JOREK B.C. for the electric potential ($\Phi=0$) gives the correct distribution of halo currents as demonstrated in \cite{Krebs2020}. The formalism derived in ~\cite{Zakharov2015,Atanasiu2016,Atanasiu2018} has been implemented as a post-processing tool to calculate wall forces and to visualize the source/sink currents (see Figure~\ref{fig:source_sink}). This tool has been validated as well for 3D walls with holes. A formalism for treating ferromagnetic components in a thin wall model is shown in Ref.~\cite{Atanasiu2021} but integration in JOREK-STARWALL hasn't been approached so far.

Some numerical limitations had originally restricted JOREK-STARWALL to moderate toroidal and poloidal resolutions. In particular, STARWALL was originally purely OpenMP parallelized, the coupling terms inside JOREK were treated OpenMP parallel only, and the fairly large and sparse ``response matrices'' were duplicated across all MPI ranks in JOREK. Within the project described in Ref.~\cite{Mochalskyy2017}, an MPI parallelization of STARWALL, a hybrid MPI+OpenMP parallelization of the coupling terms in JOREK, parallel input/output for the response matrices in both codes as well as a distributed storage of the response matrices across the MPI ranks were implemented including the distributed matrix-matrix operations. With these developments, high resolution cases are possible now with an excellent performance. For a verification of free boundary simulations, see Section~\ref{:verification:freeb}.

\subsection{Kinetic particles}\label{:code:models:particles}

For a number of applications, such as the transport and interaction of fast particles, impurities and neutrals with the MHD fluid, the main fluid model(s) in JOREK have been extended with a kinetic particle module.  Particles are followed in the time dependent 3D magnetic and electric fields given on the cubic finite element grid. For the fast ions, impurities and neutrals, the well-known Boris scheme is used. The full orbit of the particles is followed both in real $(R,Z,\phi)$ space and in the local coordinates $(s,t,\phi)$ of each element. At every particle step, the Boris scheme is applied in $(R,Z, \phi)$ coordinates including a correction for the toroidal geometry~\cite{Delzanno2013}. The updated local element coordinates are found by Newton iterations. Following the particles in real space has the advantage that the crossing of a finite element boundary does not influence the Boris scheme. The change of element of a particle position is handled in the update of the local element coordinates. For the rare case a particle crosses many element boundaries, an efficient RTree algorithm is used to find the particles element. The Boris scheme combined with a higher order interpolation of the magnetic and electric fields in time assures accurate particle trajectories with good energy conservation (see Section~\ref{:verification:particles}). Multiple particle species can be traced within one simulation allowing for example simulations combining neutral deuterium, heavy impurities and fast ions.
For the simulation of relativistic runaway electrons, a relativistic particle tracer (both full orbit or guiding center) was implemented~\cite{Sommariva2017,SommarivaPhD} with excellent energy and momentum conservation properties (see Section~\ref{:verification:particles}). This model was applied to study various aspects of runaway electron dynamics with a test particle approach (Section~\ref{:applic:core:res}).
In addition to the particle following, the charge state of the particles and the ionisation, recombination and radiation rated are consistently evolved in time according to the background fluid properties, using the OPEN-ADAS rate coefficients including all charge states. A model for particle collisions with the background fluid or with projected moments of the kinetic particles, following Refs.~\cite{Takizuka1977,Homma2011,Homma2013} has been implemented and successfully bench-marked with the cases in Refs.~\cite{Homma2011,Homma2013}. The collision model includes the thermal force, relevant for the movement of impurities upwards relative to the temperature gradient in on open field lines. To model the main source of impurities, the sputtering of divertor/first wall material by incoming particles and MHD fluid has been implemented using the sputtering yields from Refs.~\cite{Eckstein2002,Eckstein2007,Eckstein2011} and bench-marked with the results from Ref.~\cite{Kallenbach2005}.
The first application, with test particles, studied the transport of tungsten impurities during (simulations of) ELM crashes (see Section~\ref{:applic:edge:elms}), tungsten sputtering during ELMs and scrape-off layer transport of tungsten (see Section~\ref{:applic:core:impurities}).
For the modelling of the interaction of the MHD fluid and particles, coupling terms have been implemented. These terms appear mostly as additional explicit source terms on the right hand side of the fluid equations. To describe the excitation of MHD instabilities driven by fast particles, both the coupling schemes based on the pressure and the current have been implemented. To include the neutral and impurity physics, density, momentum and energy sources resulting from ionisation, recombination and radiation have been added. 
From the particle distribution, the source terms are calculated by projecting the moments of the particle distribution onto the finite element representation. For the mass density source $S_\rho$:
\begin{align}
S_\rho\left( \mathbf{x} \right) &= \int{m_i f \left(\mathbf{x},\mathbf{v}\right) d\mathbf{v}} \\
\tilde{S_\rho} \left( \mathbf{x} \right) &= \sum_{ijk} p_{\rho,ijk} H_{ij}\left(s,t \right) H_{\phi,k}\left(\phi \right)
\label{eq:projection}
\end{align}
Where $p_{\rho,ijk}$ are the finite element expansion coefficients to be determined by the projection.
The projection uses the weak form with the same testfunction as the finite element basis functions $v^* = H_{ij} H_{\phi,k}$:
\begin{equation}
\int{v^* \tilde{S_\rho} \left(\mathbf{x}\right) d\mathbf{x} } = \int{v^* S_\rho \left( \mathbf{x} \right) d\mathbf{x}} =\sum_i v^*\left(\mathbf{x}\right) m  w_i \delta \left(\mathbf{x} - \mathbf{x_i} \right)
\label{eq:projection_weak}
\end{equation}
Here, the integral over the particles becomes a sum over all particles weighted by the finite element basis functions. The system of equations \eqref{eq:projection_weak} is factorised only once and solved at particle projection. The projection results can be smoothed by solving instead:
\begin{equation}
    \int{v^* \left(1 + \alpha \nabla^4 \right) \tilde{S_\rho} \left(\mathbf{x}\right)  d\mathbf{x} } = \int{v^* S_\rho \left( \mathbf{x} \right) d\mathbf{x} } 
    \label{eq:projection_smoothing}
\end{equation}
Typical values for $\alpha$ are, depending on the application, of the order of $10^{-10}$ to $10^{-12}$. The projection is required at every time step of the main fluid part. The number of particle steps for each fluid step varies from 100-1000 for fast particles to order 1-10 for slow heavy impurities. The sources are either a time integrated source (typically required for good conservation properties for particle/energy sources), or at one given time (for fast particle pressures). 

\subsection{Relativistic electron fluid}\label{:code:models:REfluid}

A relativistic electron fluid model is available~\cite{Bandaru2019A}, which allows to simulate generation, transport, and losses of runaway electrons (REs). One additional variable is introduced to describe the spatial distribution of the RE density. Most of the relevant primary and secondary generation mechanisms have been implemented, and successfully been benchmarked against lower-dimensional codes. This fluid approach does not capture some kinetic aspects (i.e., accurate treatment of the energy spectrum), but allows to study the mutual non-linear interaction between REs and MHD. In that sense, the approach is complementary to the RE test particle model described in the previous Section, which captures kinetic effects but does not account yet for the back-reaction of the REs to the plasma. The model has so far been applied to the interaction of REs with internal kinks, vertical displacement events, and tearing modes as well as RE beam termination studies (Section~\ref{:applic:core:res}). An extension has very recently been developed, which takes into account the effect of REs onto the radial force balance of the plasma~\cite{Bandaru2020B}. The implementation of the interaction terms with the impurity fluid is under development.

\subsection{Full MHD model}\label{:code:models:fullMHD}

A full MHD model suitable for production simulations is available in JOREK~\cite{HaverkortPhD,Haverkort2016,Pamela2020} and can be used for many applications. The model was, nevertheless, only used for few physics applications so far, since its final robust implementation has only been completed recently. Several important extensions available for the reduced MHD model have now been implemented in the full MHD model already.

As discussed in Section~\ref{:verification}, benchmarks of the reduced and full MHD models of JOREK show that the reduced MHD model is capturing the key physics very well under many conditions while reducing computational costs. However, it is also shown that for certain types of instabilities such as the internal kink it is necessary to use the full MHD model as also has been found in analytical calculations~\cite{Graves2019}. Furthermore, VDE benchmarks revealed an overall excellent agreement between the reduced MHD JOREK model and full MHD codes (Section~\ref{:verification:freeb}). However, the toroidal variation in the plasma current is not reflected.

The full MHD model has recently been extended for sheath boundary conditions and numerical stabilization terms were implemented~\cite{HuijsmansNkongaFullMHD}. The full-MHD physics model implemented in Ref.~\cite{Pamela2020} includes plasma flows like the reduced-MHD model, with diamagnetic terms, neoclassical poloidal friction, and toroidal rotation. The bootstrap current source has also been included. A neutrals fluid model like used for MGI and SPI disruption simulations, has not been implemented yet and will be addressed in future studies.

\subsection{Electrostatic fluid turbulence model}\label{:code:models:itg}
JOREK code is very flexible for the implementation of many other physical models not only related to MHD. The realistic geometry, global equilibrium obtained from the Grad-Shafranov solver, flux-aligned grid, numerical methods, sparse matrix solver are common for many applications of the JOREK code. An electrostatic turbulence  model has been implemented into JOREK which allows to study ion temperature gradient (ITG) driven turbulence in realistic tokamak geometry including SOL and X-point. Both fluid  ~\cite{Garbet2001,Zielinski44thEPS,Zielinski2020} and full kinetic orbits approaches are under development. 
Benchmarks with standard CYCLONE case ~\cite{Dimits2000,Merlo2018} for fluid and kinetic approaches have already confirmed that the implementation is capturing the growth rates of the instabilities in simplified configurations accurately~\cite{Zielinski44thEPS,Zielinski2020}. The model has recently been extended to model ITGs in X-point geometry with SOL for realistic JET and COMPASS discharges (see Section~\ref{:applic:other:itgturbulence}).
 
\subsection{Fully kinetic electrostatic model}\label{:code:models:kin}
 The particle framework described in Section~\ref{:code:models:particles} has been used to implement a fully kinetic (electrostatic) model (i.e., no fluid part) of the plasma with full orbit ions and adiabatic electrons. In this case, there is only one equation to solve for the electric potential $\Phi$ in terms of ion density $n_i$:
 \begin{align}
     N > 0 \; &: \; \: \Phi_N\left(\mathbf{x}\right) = n_{i,N}\left(\mathbf{x}\right) \\
     N = 0 \; &: \; \; \nabla \cdot \left( \frac{m_e T_e}{e^2 B^2} \nabla_\perp \Phi \right)
     + \left(\Phi - \left< \Phi \right> \right> = \frac{n_i - n_{i0}}{n_{e0}}
     \label{eq:model:kinetic}
 \end{align}
 where N is the toroidal harmonic, $n_{i0}$ and $n_{e0}$ the initial ion and electron density. Due to the full orbit ions, there is no ion polarisation density in Eq.~\eqref{eq:model:kinetic}. This equation is solved as a slightly modified form of the projection operator from Eq.~\eqref{eq:projection_weak}. The full orbit, full-f model has been successfully benchmarked against the linear ITG growth rates and frequencies from \cite{Merlo2018} and against the zonal flow frequencies and damping from \cite{Biancalani2017}, see Section~\ref{:applic:other:itgturbulence}. The model can be used with any of the JOREK finite element grids, allowing ITG simulations in arbitrary X-point geometry, including the open field line region (see Section~\ref{:applic:other:itgturbulence}).
 
\subsection{Outlook}\label{:code:models:outlook}

Various extensions to the models are in preparation, and only some of them can be mentioned here. Present plans involve a more modular structure of the physics models to simplify combining arbitrary extensions when needed. Sheath boundary conditions will be further refined, in particular allowing for outflows with Mach numbers larger than one. A fully consistent neoclassical model for the plasma resistivity is being implemented.

Two types of impurity models going beyond the coronal equilibrium assumption are under development: a fluid model involving one additional continuity equation per impurity charge state (or bundles of charge states), and a model in which impurities are treated as a set of particles,  the charge of each particle being evolved individually.

Regarding the RE fluid model, it is planned to evolve in time an average parallel momentum of the particles, allowing to capture the effect of the REs onto the radial force balance, which leads to a major-radial shift between the flux surfaces and the RE drift-orbit surfaces in the absence of thermal pressure.

An effort is presently taken to implement an energy conserving current-coupling scheme for the kinetic particles and the MHD fluid allowing to study the energy exchange of MHD modes with super-sonic particles. Related to that, kinetic MHD or simplified gyrokinetic models like described in Refs.~\cite{Lanthaler2019,Lu2019A} are presently evaluated and might be implemented in JOREK.

In preparation of a 3D extension aiming to simulate stellarator plasmas, a hierarchy of reduced and full MHD models with good energy and momentum conservation properties has been derived in a form suitable for stellarators~\cite{Nikulsin2019,Nikulsin2020}. The reduced model with some further improvements is presently being implemented into JOREK. Further extensions of the numerical methods, in particular regarding spatial discretization and solver, are necessary to achieve well resolved stellarator simulations and are evolving in parallel (see Section~\ref{:code:numerics:outlook}).

%% file: 03_numerics.tex
\section{Numerical methods}\label{:code:numerics}

The JOREK code is largely written in Fortran with a few core routines in C and C++. It has only few library dependencies allowing to port the code easily to new machines. Some aspects of the code development are described in Section~\ref{:code:numerics:management}. The numerical methods used in the code are shown in the following, in particular the spatial and temporal discretization, numerical stabilization schemes, the construction of the sparse matrix system, the iterative solver with its preconditioner, and the hybrid parallelization of the code are described.

\subsection{Spatial discretization}\label{:code:numerics:discretization}

\begin{figure}
\centering
  \includegraphics[height=0.25\textwidth]{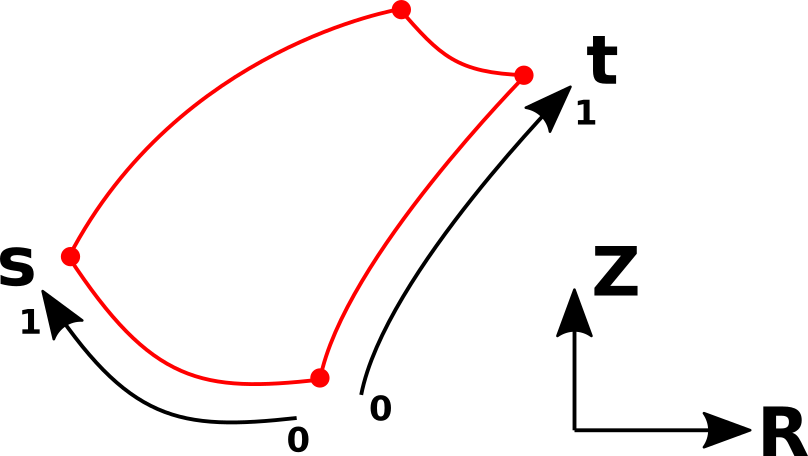}
\caption{Single 2D Bezier element in R-Z space. The element-local s-t coordinate system is also shown. The mapping from s-t into the R-Z space is expressed in Bernstein polynomials. The same basis functions are also used for physical variables like the temperature (iso-parameteric). \noreprint}
\label{fig:bezier-st}
\end{figure}

JOREK solves the equations of the respective physics models in weak form on a $G^1$ continuous 2D isoparametric Bezier finite element grid combined with a toroidal Fourier expansion~\cite{HuysmansASTER,Czarny2008}. The Bezier basis and the finite element grids constructed from them are shown in Subsections~\ref{:code:numerics:discretization:Bezier} and~\ref{:code:numerics:discretization:grids}, respectively. The toroidal Fourier expansion is described in Subsection~\ref{:code:numerics:discretization:four}.

\subsubsection{Bezier basis}\label{:code:numerics:discretization:Bezier}

\begin{figure}
\centering
  \includegraphics[width=0.35\textwidth]{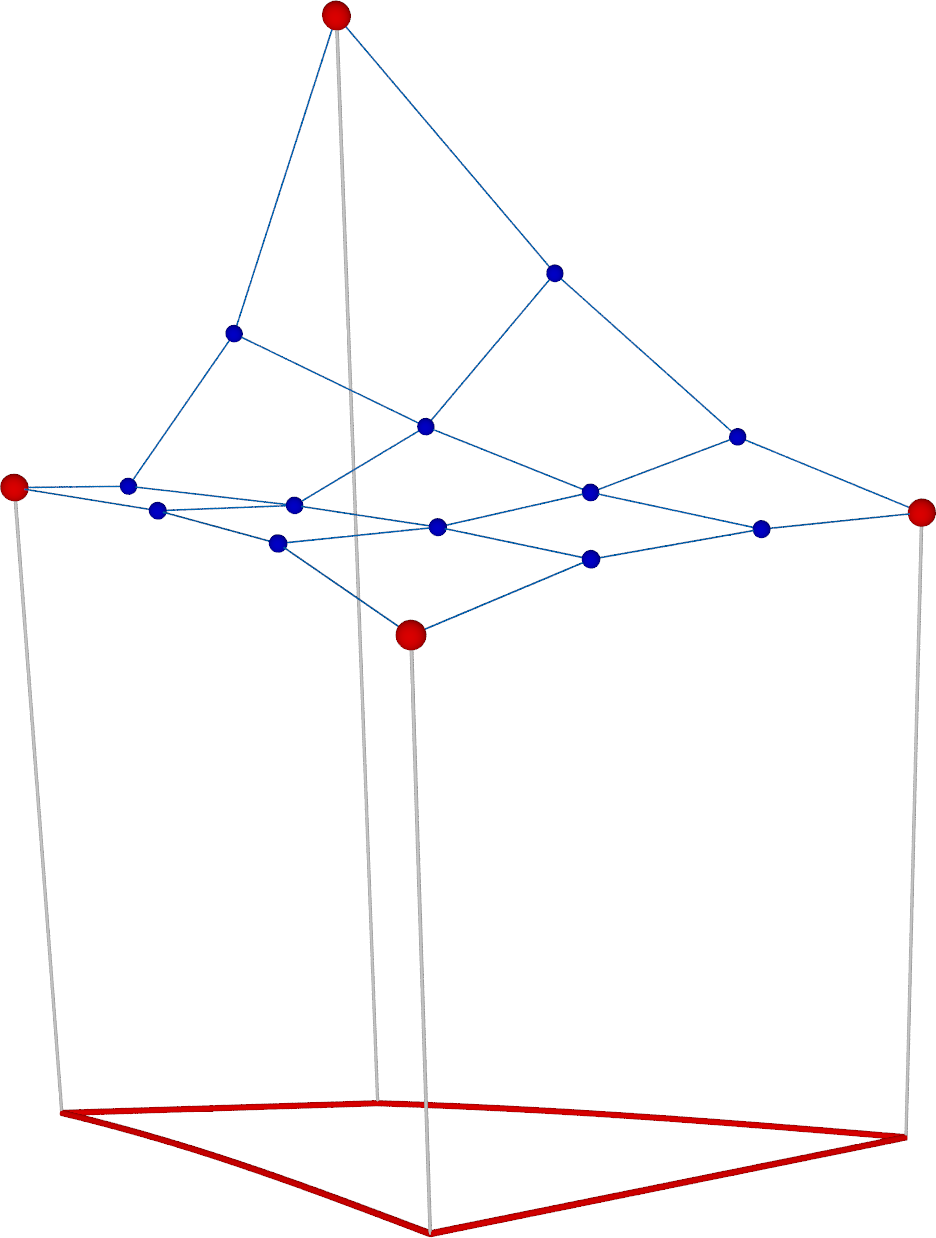}
  \includegraphics[width=0.35\textwidth]{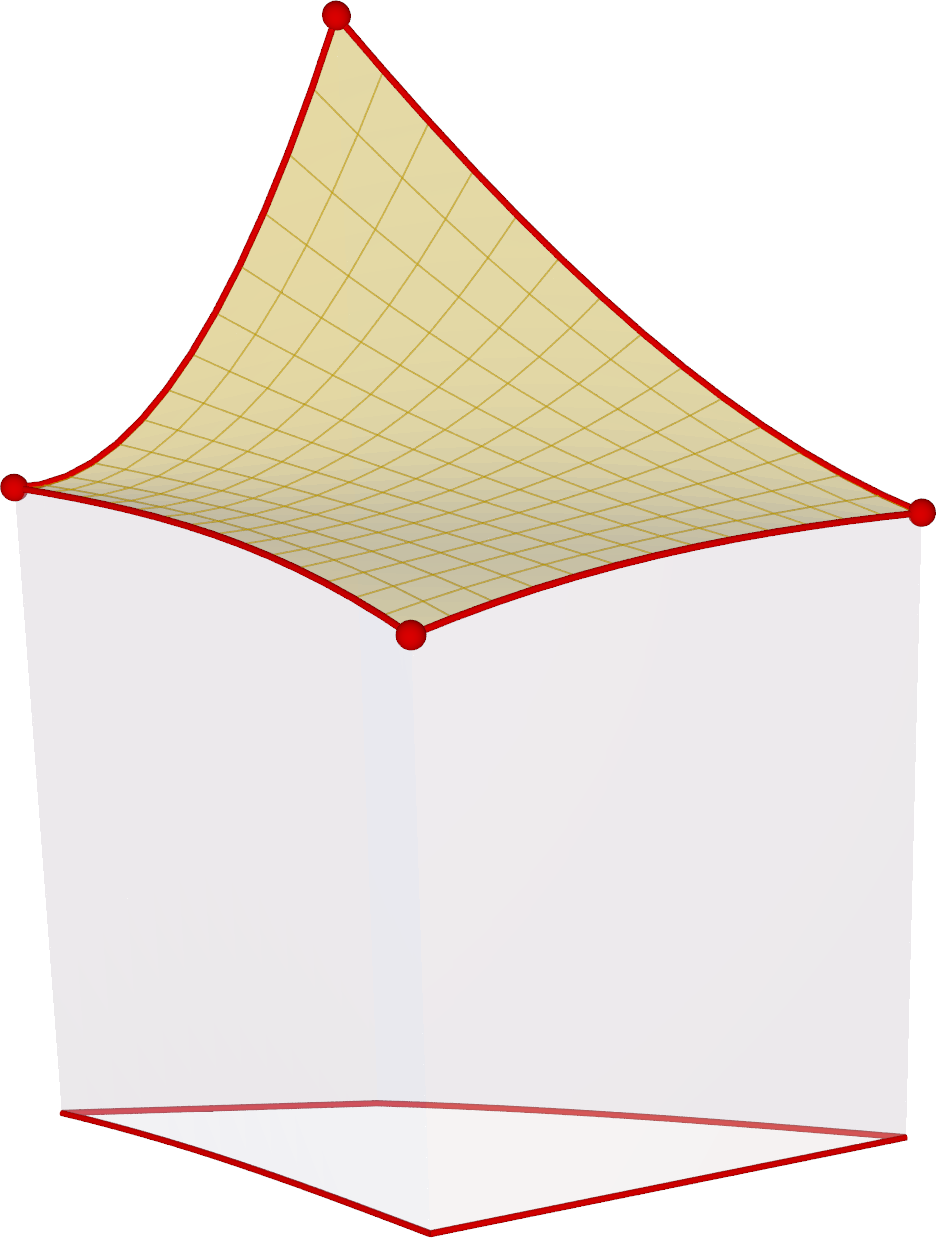}
\caption{A Bezier element is shown in 3D space. The left plot represent the control points of the elements (blue). The red points are the grid nodes. The right plot depicts the resulting surface. The red structure at the bottom of the plots represents the element in (R,Z) space. Within each element, a local (s,t) coordinate system is defined (Figure~\ref{fig:bezier-st}). The elevation represents the spatial distribution of a physical variable, e.g., the temperature. \noreprint}
\label{fig:beziersingle}
\end{figure}

Two-dimensional third order Bernstein polynomials are defined by
\begin{equation}\label{eq:BezierA}
  B_{i,j}^3(s,t)=B_i^3(s)\;B_j^3(t)\hspace{20mm}i,\;j\;=\;0\;\dots\;3,
\end{equation}
with
\begin{equation}\label{eq:BezierB}
  B_i^3(s)=\frac{3!}{i!(3-i)!}\;s^i (1-s)^{3-i}
\end{equation}
and where $0\leq s\leq 1$ and $0\leq t \leq 1$ denote the element-local coordinates, which take values of $0$ respectively $1$ at the four element vertices (see Figure~\ref{fig:bezier-st} and Appendix~\ref{:app:coord}). All physical quantities, and also the R and Z positions of the elements themselves (i.e., the mapping from element local to global coordinates) are expressed in this basis:
\begin{equation}
\mathbf{X}(s,t) = \sum_{i=0}^3\sum_{j=0}^3 \mathbf{P}_{i,j}\;B_{i,j}^3(s,t)
\end{equation}
with $\mathbf{X}$ being an N-dimensional vector containing R, Z, and all interpolated  variables of the model (e.g., R, Z, $\Psi$, T, $\rho$, etc). Here, $\mathbf{P}_{i,j}$ denotes the Bezier control points in the N-dimensional space (see Figure~\ref{fig:beziersingle}). For all physical variables, this 2D expression is multiplied with the toroidal Fourier basis (next Section), while for R and Z, axisymmetry is assumed. The coordinates $s$ and $t$ are orthogonal to the toroidal coordinate $\phi$\footnote{An ongoing 3D extension of the code will generalize the mapping to be non-axisymmetric, i.e., R and Z will also be expanded in the Fourier basis. In addition, the orthogonality of $s$ and $t$ with respect to $\phi$ will be given up depending on the choice of coordinates. See Sections~\ref{:code:models:outlook} and~\ref{:code:numerics:outlook}.}.

\begin{figure}
\centering
  \includegraphics[width=0.35\textwidth]{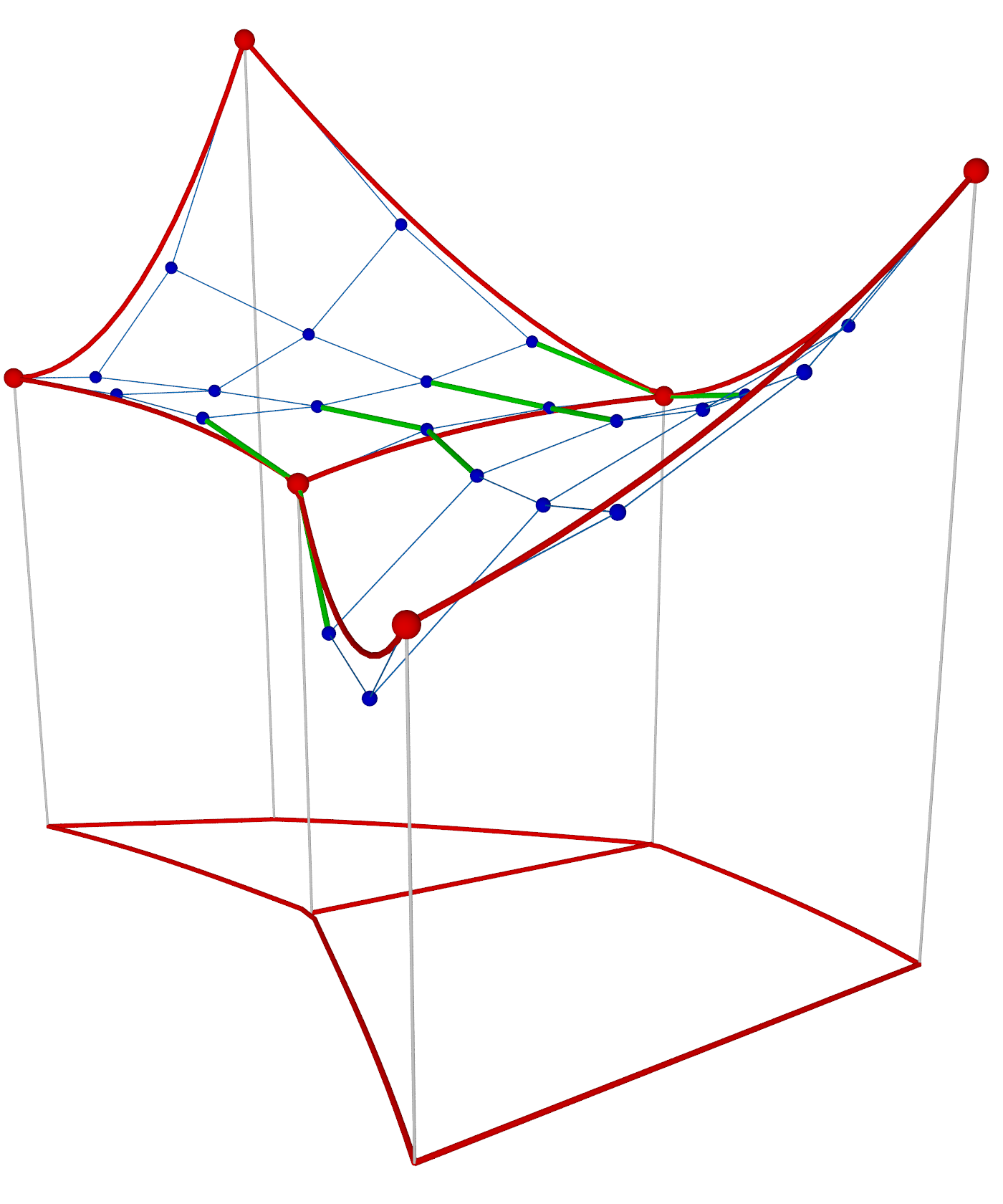}
  \includegraphics[width=0.35\textwidth]{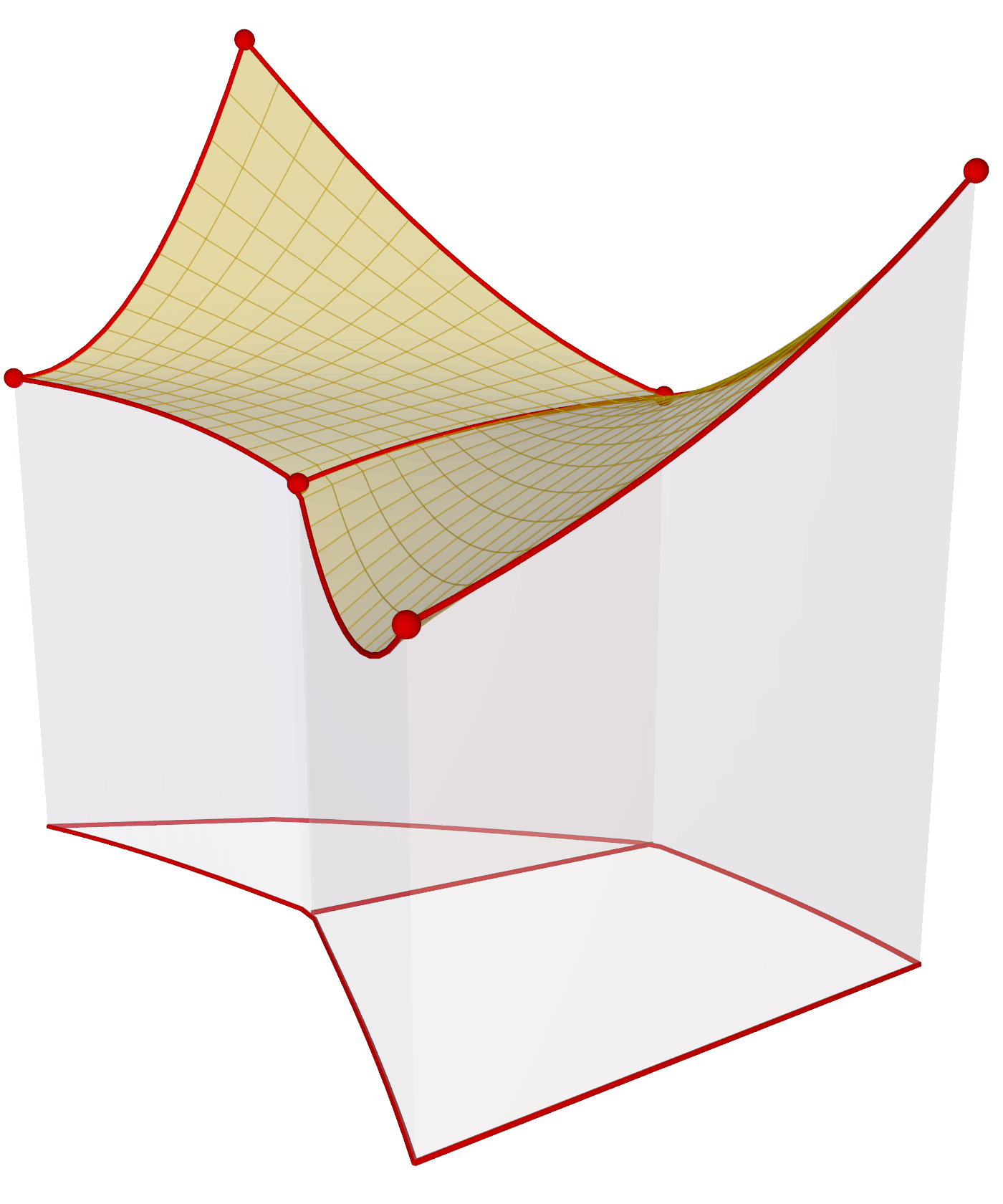}
  \includegraphics[width=0.35\textwidth]{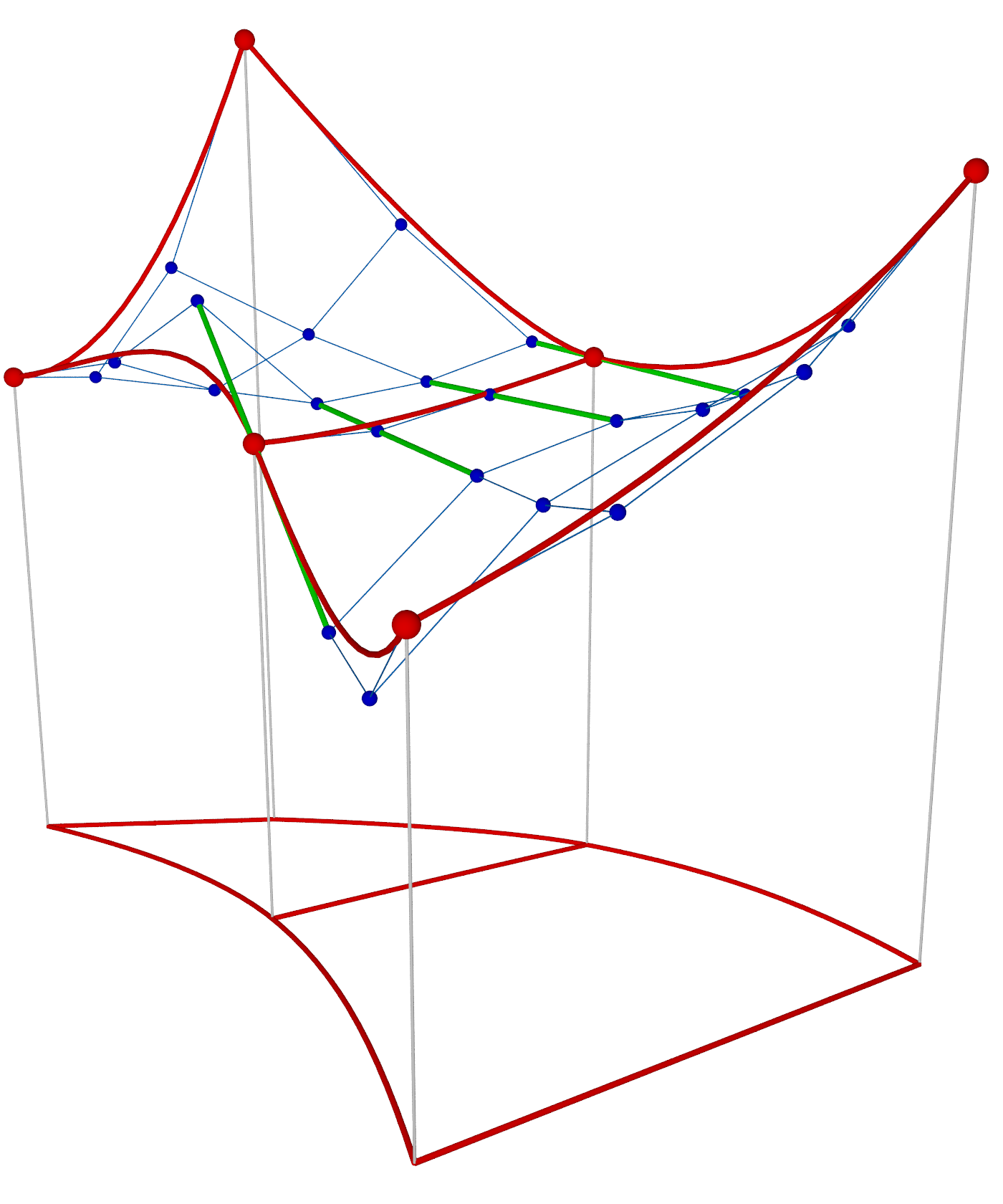}
  \includegraphics[width=0.35\textwidth]{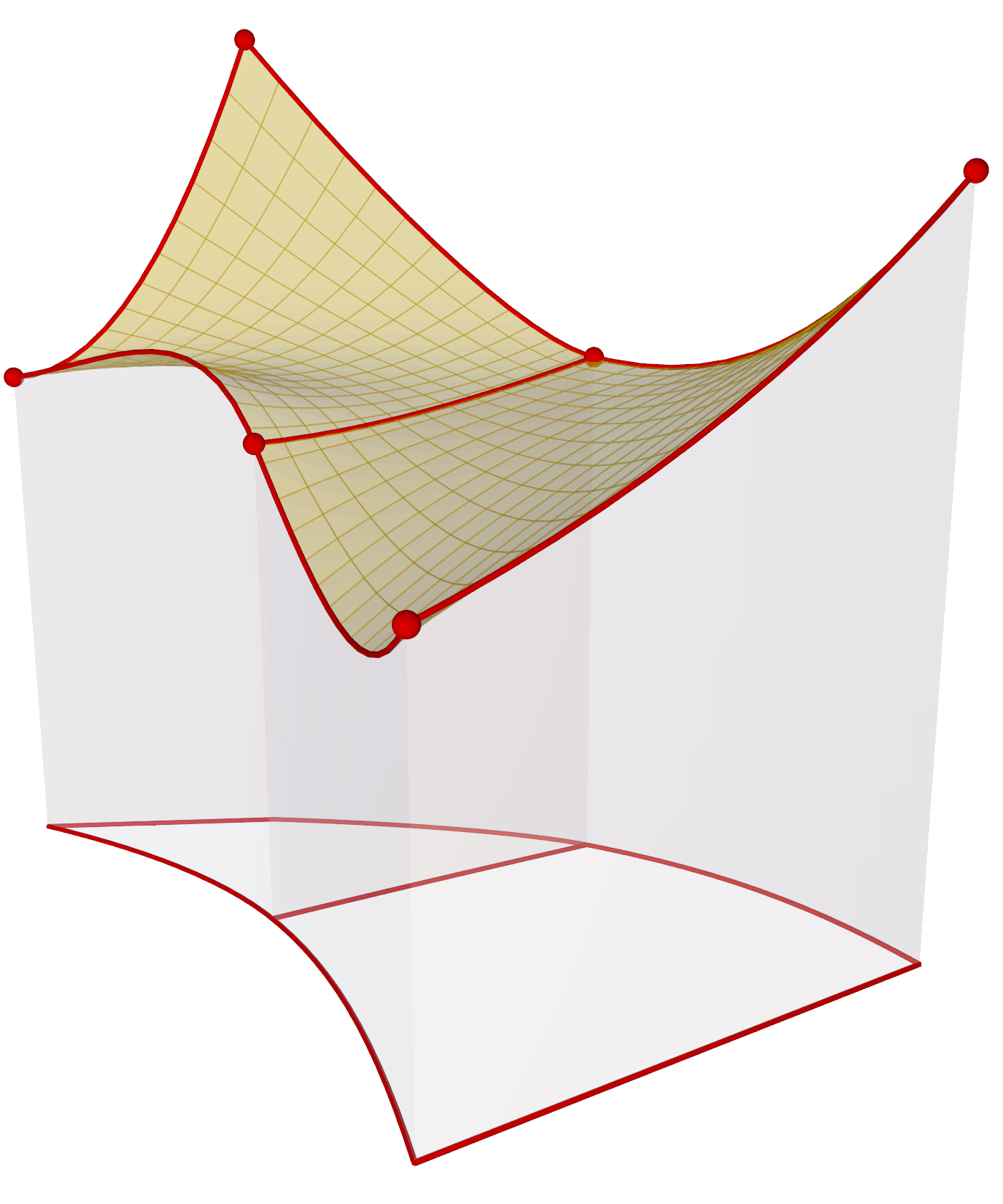}
\caption{Two neighbouring Bezier elements are shown. The left plots represent the control points of the elements and the right plots depict the resulting surface. In the \textbf{plots at the top}, the neighboring Bezier elements share the control points at the common boundary such that values are continuous across the boundary between the elements ($G^0$ continuity). However, since the control points around the boundary are not aligned to each other, i.e., green lines exhibit bends, derivatives are not continuous across the element boundary. The \textbf{plots at the bottom} correspond to the discretization used in JOREK, where both values and derivatives are continuous across the element boundaries ($G^1$). Here, the control points at the element boundaries are aligned reducing the degrees of freedom, i.e., the green lines do not exhibit bends. \noreprint}
\label{fig:bezier}
\end{figure}

Since first order continuity in real space ($G^1$), i.e. continuity of the values and the gradients, is applied as constraint between neighboring elements, corresponding control points of adjacent elements need to lie on a common line in the N-dimensional space through the vertex location, which reduces the number of degrees of freedom (see Figure~\ref{fig:bezier}). Effectively, only four degrees of freedom $p_k$, $u_k$, $v_k$, and $w_k$
remain per node $k$ and for each component of the vector $X$, which are shared by all elements connected to the respective node. These degrees of freedom are linked to the value, s-derivative, t-derivative, and s-t cross-derivative at the location of the grid node. The relation of the control points with the coefficients $p_k$, $u_k$, $v_k$, and $w_k$ is explicitly shown in Ref.~\cite{Czarny2008}. Since metric tensor and Jacobian differ between elements, each element $e$ has specific scale factors $d_{k,l}^{e}$ for each degree of freedom $l$ at each node $k$ to guarantee $G^1$ continuity. These scale factors are a geometric grid property and therefore time independent and identical for each physical quantity.
For any given quadrangular element $\Elmt$ of the mesh, a  physical variable, e.g., temperature $T$, is expanded in the $G^1$ continuous basis in the following way (the Fourier expansion in toroidal direction is omitted here for clarity):
\begin{equation}
\left . T(s,t) \rule{0mm}{6mm}\right |_{e} = \sum_{k=1}^{n_\text{vert}}\sum_{l=1}^{n_\text{dofs}}T_{k,l}H_{k,l}(s,t)\,d_{k,l}^{e}
\end{equation}
where $n_\text{vert}=4$ is the number of vertices per element,
$n_\text{dofs}=4$ is the number of degrees of freedom per vertex. $H_{k,l}$ are the sixteen basis functions written as a product of 1D basis functions in s and t: $H_{k,l}(s,t)=H_k(s)\,H_l(t)$. The four 1D basis functions $H_k(s)$ are constructed as linear combinations of Bernstein polynomials to satisfy:
\begin{itemize}
\item $H_1(s=0)=1$, $H'_1(s=0)=0$, $H_1(s=1)=0$, $H'_1(s=1)=0$
\item $H_2(s=0)=0$, $H'_2(s=0)=1$, $H_2(s=1)=0$, $H'_2(s=1)=0$
\item $H_3(s=0)=0$, $H'_3(s=0)=0$, $H_3(s=1)=1$, $H'_3(s=1)=0$
\item $H_4(s=0)=0$, $H'_4(s=0)=0$, $H_4(s=1)=0$, $H'_4(s=1)=1$
\end{itemize}
i.e., the well-known cubic Hermite finite elements. This way, the coefficients $T_{k,l}$ become node properties (nodal formulation) which all elements containing the respective node are sharing, while the coefficients $d_{k,l}^{e}$ corresponding to the same node are different for each element to guarantee the $G^1$ continuity.

$G_1$ continuity is not strictly enforced at the grid axis and in the direct vicinity of the X-point(s) in a flux surface aligned grid due to the specific topology at these points (more than four elements share a common node). Although this is typically not an issue in the simulations, an implementation of strict $G^1$ continuity also at these special points is on its way by locally combining basis functions in an appropriate way (see Section~\ref{:code:numerics:outlook}). In a few cases, the higher order FE formulation can produce ``overshoots'' leading to zero or negative values of density and temperature, in particular in the presence of strong convection. For that reason, numerical stabilization has been implemented as shown in Section~\ref{:code:numerics:stab}. The cubic Bezier element formulation is a generalisation of the isoparametric cubic Hermite elements \cite{Huysmans1990}, allowing a local refinement of each element in 2 or 4 sub elements \cite{Czarny2008}.

\subsubsection{Finite element grids}\label{:code:numerics:discretization:grids}

\begin{figure}
\centering
  \includegraphics[width=0.405\textwidth]{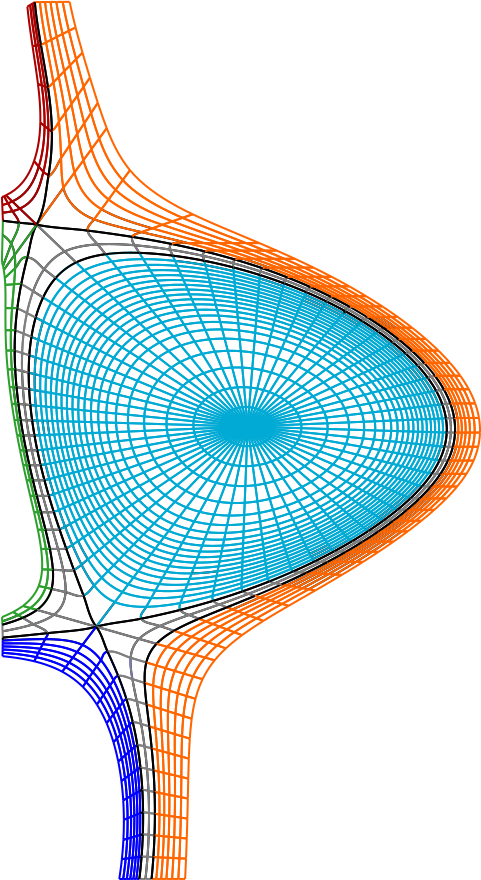}
  \includegraphics[width=0.27\textwidth]{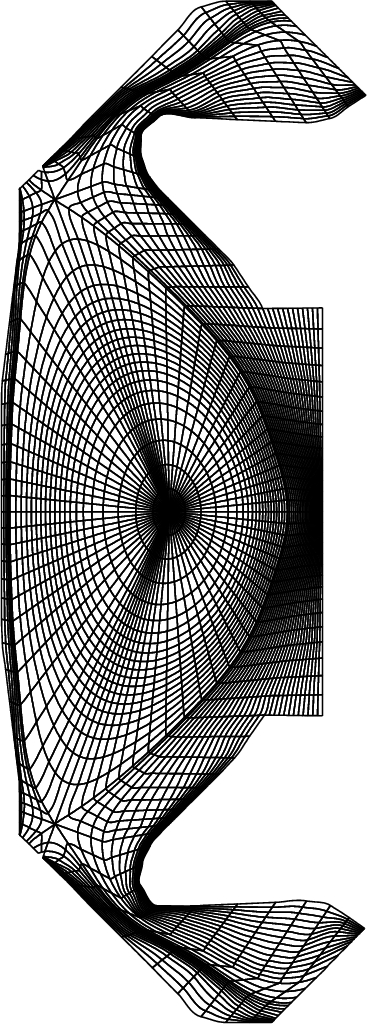}
\caption{JOREK example grids shown at reduced resolution for clarity. \textbf{(Left)} Flux-surface aligned grid~\cite{Gruca2020} for a TCV  equilibrium with two X-points (the lower X-point is active). The part in the closed flux region is shown in light blue, the two separatrices in black, the grid between the separatrices in grey, the lower private flux region in dark blue, the upper private flux region in red, the outer open flux region in orange, and the inner open flux region in green. \textbf{(Right)} Flux surface aligned X-point grid extended to the true first wall (using the methods described in Ref.~\cite{Pamela2019}) for a MAST Upgrade case with super-X divertor. \noreprint}
\label{fig:xpointgrid}
\end{figure}

All data structures and routines in JOREK are implemented in a general way suitable for unstructured grids. Nevertheless, most grids in use are effectively flux surface aligned grids which are structured within the respective topological domains. Usually, a JOREK simulation is started by calculating the equilibrium on an initial grid, which is not flux surface aligned. Based on this equilibrium, a flux surface aligned grid is then constructed. For numerical accuracy, using such a flux surface aligned grid is in general beneficial (see Section~\ref{:verification:anisotropy}), while the code can be used also with non-aligned grids where appropriate. The grid types available include simple rectangular grids, simple polar grids, polar grids with a rectangular central region. These grids are mostly used as initial grid and only occasionally for the time evolution, e.g., in case of VDEs, where the grid would not remain aligned during the simulation anyways.

Flux surface aligned grid generators are available for plasmas without X-point, with an upper X-point, a lower X-point, or two X-points. Additionally, an extension of grids to the true first walls is possible as described in~\cite{Pamela2019}. Grid patches (which are not flux-surface aligned any more) are added in the far scrape-off layer to cover the space between a flux-surface aligned X-point grid and plasma facing components. Figure~\ref{fig:xpointgrid} contains two examples for a double X-point grid and a double X-point grid extended to the true physical wall (both are shown at reduced resolutions for clarity).

\subsubsection{Toroidal Fourier series}\label{:code:numerics:discretization:four}

As mentioned before, the poloidal discretization in Bezier finite elements is combined with a toroidal Fourier expansion for the physical variables. The toroidal real Fourier harmonics included in a JOREK simulation are selected by the number of harmonics $n_\text{tor}$, where cosine and sine components are counted separately -- since $n=0$ does not have a sine component, $n_\text{tor}$ must be an uneven number. In addition, the periodicity $n_\text{period}$ can be specified. The latter parameter allows to enforce periodicity in the torus after a fraction $1/n_\text{period}$ of a full toroidal turn. This effectively means, that instead of $n=0,1,2,\dots$, the harmonics $n=0,1\cdot n_\text{period},2\cdot \,n_\text{period},\dots$ are included in a simulation. The number of toroidal planes $n_\text{plane}$ for the integration along $\phi$ direction needs to be chosen large enough to avoid aliasing. Typically, $n_\text{plane}\ge 2(n_{tor}-1)$ is sufficient, i.e., 4 planes per period. While fast Fourier transforms performed during the matrix construction using the FFTW library (see Section~\ref{:code:numerics:matrix}) can handle arbitrary integer numbers, best performance is usually achieved, when $n_\text{plane}$ is taken to be a power of two.

\subsubsection{Weak form}\label{:code:numerics:discretization:weakform}

As a consequence of using the finite element approach, the model equations are solved in weak form such that each equation is multiplied by a test function from the same space as the basis functions, and then integrated over the volume $V$ of the computational domain (Galerkin method). Since the resulting equation needs to be fulfilled for each test function, many linear relations are obtained, which form the rows of the matrix system~(see Section~\ref{:code:numerics:matrix} for more details on the matrix construction). In case of operators involving higher order derivatives, partial integration is performed e.g.,
\begin{equation}
\int_V dV\,a\,\nabla\cdot\mathbf{b} = - \int_V dV\,\nabla a \cdot \mathbf{b} - \oint_A dA\,a\,\mathbf{b}\cdot\mathbf{n}    
\end{equation}
Here, $a$ and $\mathbf{b}$ are arbitrary scalar respectively vector expressions, $A$ denotes the boundary of the computational domain and $\mathbf{n}$ the normal vector to that boundary. Depending on boundary conditions, some of the boundary integrals vanish, otherwise they are implemented. In particular cases, relations known from physical properties of the system may be plugged into the boundary integrals. In such a case, the boundary integral is expressed in terms of other variables and natural boundary conditions are obtained\footnote{Since natural boundary conditions are not enforced, it needs to be checked whether a simulation actually fulfills the original relation. If they are not satisfied well enough, smaller time steps are typically required.}. Such a natural boundary condition is used, for instance, for the JOREK-STARWALL coupling where the STARWALL expression for the magnetic field tangential to the boundary is plugged into the boundary integral of the current definition equation (see Section~\ref{:code:models:freebound}).

\subsection{Temporal discretization}\label{:code:numerics:tstep}

The physical equations expressed in the form
$\partial\mathbf{A}(\mathbf{u})/\partial t=\mathbf{B}(\mathbf{u},t)$,
where $\mathbf{u}$ denotes the vector of physical variables, are discretized in time as
\begin{equation}\label{eq:timestepB}
    \left[(1+\xi)\left(\pderiv{\mathbf{A}}{\mathbf{u}}\right)^n-\Delta t\theta\left(\pderiv{\mathbf{B}}{\mathbf{u}}\right)^n\right]\delta\mathbf{u}^n
      =\Delta t\mathbf{B}^n+\xi\left(\pderiv{\mathbf{A}}{\mathbf{u}}\right)^n\delta\mathbf{u}^{n-1},
\end{equation}
where the second order linearised Crank-Nicolson scheme is selected by $(\theta,\,\xi)=(1/2,\,0)$ and the second order BDF2 Gears scheme is selected by $(\theta,\,\xi)=(1,\,1/2)$. The first order implicit Euler method corresponds to $(\theta,\,\xi)=(1,\,0)$ but is normally not used in practice. Here, $\delta\mathbf{u}^n=\mathbf{u}^{n+1}-\mathbf{u}^n$ denotes the change of the variables from time step $n$ to $n+1$. The definition equations for current and vorticity are directly implemented to be satisfied at time point $n+1$, the above scheme does not need to be applied.

Details, and the time stepping scheme suitable for a more general form of equations is shown in Appendix~\ref{:app:tstep}. The linearization involved in the above time stepping can be replaced by Newton iterations with a beneficial effect onto non-linear stability in some cases, as demonstrated in Ref.~\cite{Franck2015}, however this is not implemented in the present code version \revised{as these earlier tests showed that larger time steps are possible, however at an increased computational cost and memory consumption such that the overall benefits were limited.}

\subsection{Numerical stabilization}\label{:code:numerics:stab}

\begin{figure}
\centering
  \includegraphics[width=0.7\textwidth]{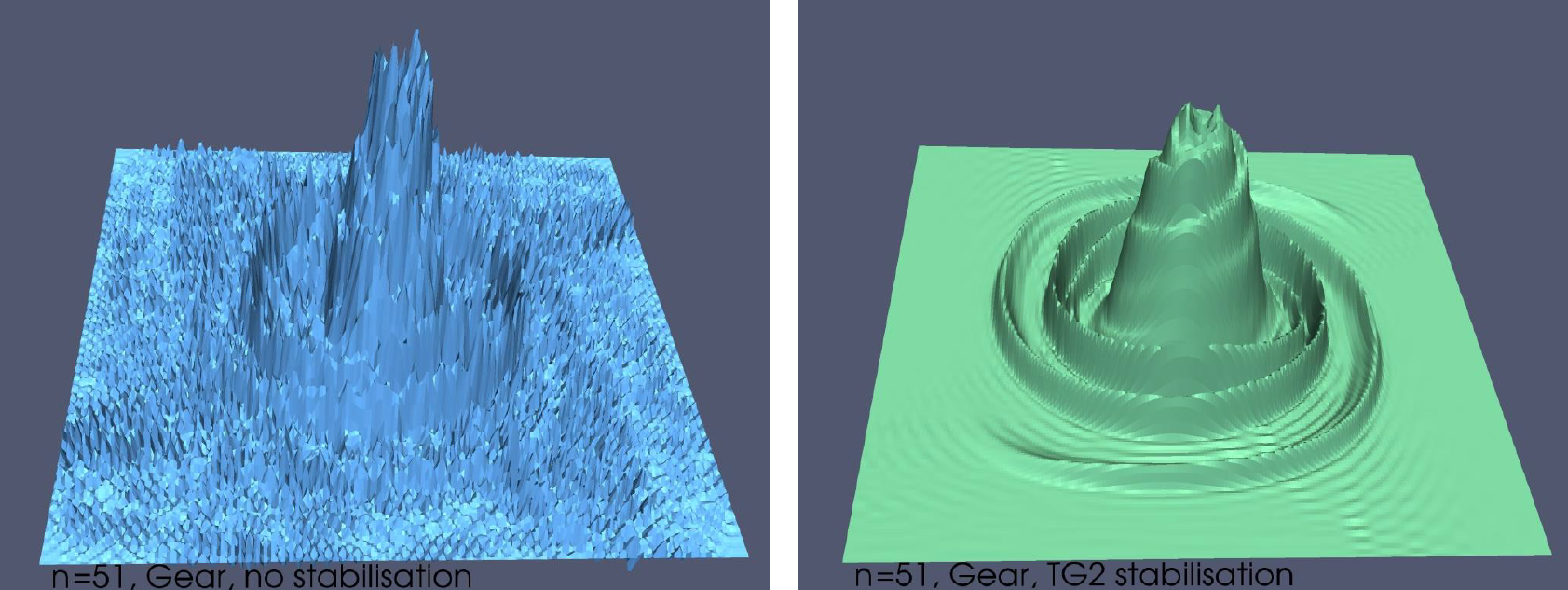}
\caption{Example of the effect of the Taylor-Galerkin stabilisation (TG2) for a simplified model for the vorticity evolution (Ref.~\cite{Hammett2008}) in 2D on a grid of 50 by 50 cubic finite elements. \noreprint}
\label{fig:tg_example}
\end{figure}

When the system considered is dominated by convection, the finite
element method may suffer from numerical instabilities as illustrated on the left hand side of Figure \ref{fig:tg_example}. The variational
multi-scale (VMS) is a general framework to avoid these numerical
artifacts~\cite{hughes2004,Billaud2011}. 
The main purpose of the VMS is to take into account the
effect of the unresolved scales onto the numerically resolved scales. The
consequence is to complement the physical model by a proper
numerical dissipative term that scales with the mesh size. The SUPG and the
\revised{Taylor}-Galerkin~\cite{Donea1984} methods are subsets of the VMS. Let us
explain with the mass conservation equation how a simplified VMS
contribution is added to the weak form. Assume the initial equation is
\[
\pderiv{\rho}{t} + \VecV\cdot\nabla\rho = -\rho\nabla\cdot\VecV
+S_\rho
\]
In order to stabilize the discretization of the convection term
$\VecV\cdot\nabla\rho$ a dissipation term is added to the weak
form to remove numerical artifacts without altering the physical \revised{dynamics} (Figure~\ref{fig:tg_example}). With this numerical dissipation the equation that is 
numerically solved now turns into the following form
\[
\pderiv{\rho}{t} + \VecV\cdot\nabla\rho = -\rho\nabla\cdot\VecV
+\nabla\cdot(\TensD_{vms} \nabla\rho)
+S_\rho.
\]
Here, the  stabilization tensor $\TensD_{vms}$, in the context of magnetized
plasma, is designed such as to take into account the strong anisotropy
of the flow
\[
\TensD_{vms} =
\coefD_{\|,vms}(\mathbf{b}\otimes\mathbf{b})
+ \coefD_{\perp,vms}
\left ( \Id - \mathbf{b}\otimes\mathbf{b}
\right )
\]
where  
\[
\coefD_{\|,vms} = \tau_{\|}^*(\VecV\cdot\mathbf{b})^2
\quad \text{ and } \quad
\coefD_{\perp,vms} = \tau_{\perp}^* (\VecV\cdot\VecV - (\VecV\cdot\mathbf{b})^2)
\]
\revised{ The scaling factors $\tau_{\|}^*$ and $\tau_{\perp}^*$ are positive
and of the order of the mesh size. More precisely, scaling factors correspond to the ratio of the mesh size over the maximum speed of the acoustic waves and thus have units of time.
This scaling ensures that, under 
mesh refinement (acoustic speeds do not change as the same physics is considered), the VMS numerical dissipation asymptotically
vanishes.}
Therefore, the modified equation numerically solved is
consistent with the initial physical model. For the Taylor-Galerkin
approach~\cite{Donea1984,Roig2007}  $\tau_{\|}^* = \tau_{\perp}^* = \delta t/2$. 
Sometimes, we directly fix $\coefD_{\|}$ and $\coefD_{\perp}$ to values that
\revised{scale} with the mesh size and the estimated background turbulence.
Similar stabilization is applied to the other equations. For the
full MHD model (Section~\ref{:code:models:fullMHD}), a more
general stabilization has been also developed~\cite{Nkonga2016}, where the hyperbolic
part (acoustic and material waves) of the system is stabilized globally,
to take into account the coupling between variables. 
In the context of the reduced MHD without diamagnetic \revised{effects}, the velocity profile is defined by
two scalar variables: the parallel velocity (\revised{$v_{||}$}) and the velocity
stream function ($u$)
\[
\VecV = -R^2\nabla u\times \nabla\phi+ \revised{v_{||}}\mathbf{B}
\]
It is then \revised{assumed} that the  parallel velocity is $\revised{v_{||}}\mathbf{B}$ and the perpendicular velocity is  $-R^2\nabla u\times \nabla\phi$. In this context, we use the approximations 
$\coefD_{\|,vms}(\mathbf{b}\otimes\mathbf{b})
\simeq \frac{\delta t}{2}\revised{v_{||}}^2\mathbf{B}\otimes\mathbf{B}$  and 
$ \coefD_{\perp,vms} \left(\Id - \mathbf{b}\otimes\mathbf{b} \right) \simeq \frac{\delta t}{2}R^4
(\nabla u\times \nabla\phi)\otimes(\nabla u\times \nabla\phi)$. The Taylor-Galerkin  stabilisation associated to the density equation can finally be written as:
\[
\pderiv{\rho}{t} + \VecV\cdot\nabla\rho = -\rho\nabla\cdot\VecV
+\frac{\delta t}{2}\nabla\cdot \left ( \revised{v_{||}}^2\mathbf{B}\mathbf{B}\cdot\nabla\dens \right )
+\frac{\delta t}{2}\nabla\cdot \left ( R^4
\nabla u\times \nabla\phi(\nabla u\times \nabla\phi)\cdot\nabla\dens
\right )
+S_\rho.
\]
The associated weak form now becomes: 
\[
\begin{array}{rcl}\displaystyle
\int \pderiv{\dens}{t}\densS &=& \displaystyle
-\int \left ( \rule{0mm}{4mm} \nabla\cdot(\dens\,\VecV)  -
\nabla\cdot(\TensD  \nabla\rho)- S_\dens\right )
\densS\\[3mm] & & \displaystyle
- \frac{\delta t}{2}\int
\revised{v_{||}}^2\mathbf{B}\cdot\nabla\dens\mathbf{B}\cdot\nabla\densS
- \frac{\delta t}{2}\int
R^4\left( (\nabla u \times\nabla\phi)\cdot\nabla\dens\right )
\left( (\nabla u\times \nabla\phi)\cdot\nabla\densS\right )
  \end{array}
  \]
Similar stabilization is applied to the other equations of the reduced MHD model. \revised{Since users perform both spatial and temporal convergence tests for production simulations, and since VMS and TG stabilization both consistently vanish in the limit of high spatial and temporal resolutions, a significant influence of the stabilization onto physics results is excluded, in particular energy and momentum conservation is not affected. Furthermore, the energy conservation diagnostics automatically evaluated during each simulation allow to check this easily. For instance in a violent massive gas injection (MGI) case, for which energy conservation is discussed in the verification section briefly, the small non-conservation observed is not associated to stabilization as separate checks confirmed. Details about numerical stabilization are omitted when discussing simulations in the results sections for that reason.}

\subsection{Matrix construction}\label{:code:numerics:matrix}

The linear system of equations to be solved in each time step is described by a sparse matrix system. Due to the locality of the Bezier basis functions, usually only one out of several thousand entries in the matrix is non-zero. The matrix is constructed in a distributed way by a domain decomposition, where each MPI task is responsible for creating the matrix entries corresponding that correspond to its respective part of the finite element grid. \revised{Note that the full matrix construction could be avoided by a matrix free evaluation of matrix vector products, however earlier tests showed a higher computational cost while the memory consumption dominated by the preconditioner isn't reduced too much.} Each MPI tasks uses OpenMP to parallelize further. The matrix contributions are calculated separately for each grid element (``element matrices''). These element matrices are calculated by carrying out the integrals in the weak form of the equations over the element-local coordinates $s$ and $t$ via Gauss quadrature. The integration in toroidal direction is either carried out by a direct summation over equidistant toroidal planes (for very small problems), or by fast Fourier transform (FFT). Vectorization and parallel scalability of the matrix construction has recently been improved~\cite{Feher2018}. Details on the matrix construction and properties are shown in Ref.~\cite{Holod2020}.

\subsection{Solver}\label{:code:numerics:solver}

Several variants are implemented in JOREK for solving the linear sparse matrix system after it has been assembled like described in the previous Section. In the simplest approach, which is usually only used for axisymmetric $n=0$ simulations, the whole matrix system is treated by a direct solver. Interfaces to the PaStiX~\cite{Henon2002}, MUMPS~\cite{MUMPS:1,MUMPS:2}, and WSMP~\cite{Gupta2002} direct sparse matrix solvers are implemented. PaStiX is mostly used in practice. PT-Scotch~\cite{Chevalier2008} or ParMETIS~\cite{Karypis2003} are used internally by the solvers, to minimize fill-in during the matrix factorization and to achieve good performance. An interface to the STRUMPACK library \cite{Ghysels2017} has been added very recently~\cite{Holod2020}. Several of the solvers offer the possibility to compress the matrix system via block low rank (BLR) compression or via hierarchically semiseparable (HSS) matrices, for which only first tests have been done with JOREK so far~\cite{NiesSolver}.

For non-axisymmetric simulations, a restarted GMRES variant~\cite{Fraysse1997} is usually applied~\cite{HuysmansASTER2}. The necessary linear operations on the sparse matrix system are implemented directly in JOREK on the distributed representation of the system of equations. Since the system is very stiff, good preconditioning is mandatory. For this purpose, a physics based approach is used. Since each linear eigenfunction in a tokamak is associated to a single toroidal mode number, the preconditioner assumes approximate decoupling of the toroidal harmonics. Consequently, if the matrix is written into blocks corresponding to the toroidal harmonics, the diagonal blocks describing the interaction of each harmonic with itself are kept while all off-diagonal blocks are dropped in the preconditioning. This block-diagonal structure of the preconditioning matrix allows to treat each block independently of the others. For that purpose, parts of the global matrix, which has been assembled in a distributed way by domain decomposition (see Section~\ref{:code:numerics:matrix}) needs to be re-distributed by an all to all communication. The complete distributed matrix remains in memory as it is needed for GMRES operations. Since this re-distribution can become inefficient for large problem sizes, an option for re-calculation of the preconditioning matrices has recently been implemented using the same routines applied for the global matrix construction~\cite{Holod2020}.

Each block matrix is solved by one or several MPI tasks on one or several compute nodes, with the number of cores inside a compute node being exploited via the built-in thread support of the sparse matrix libraries (pthreads in case of PaStiX). Note that the analysis step of the solvers is performed only once in the beginning of a simulation (or upon a restart). The LU factorization of the block matrices is only performed in the first time step of a simulation, and again, when the GMRES convergence has deteriorated too much, i.e., the number of GMRES iterations in a time step exceeds a user defined threshold. Consequently, the factorized preconditioning matrices may be re-used in several or even many consecutive time steps before an update is needed.

The solve step, on the other hand has to be carried out in each GMRES iteration of every time step. In the linear phase of a simulation, the preconditioning matrix is an extremely good approximation of the complete system such that an update is not needed often and the factorization has almost no impact onto the overall performance. However, in a highly non-linear state, the factorization may need to be updated almost every time step and dominates the overall computational costs of the solver in that case. The preconditioner in such a highly non-linear state is not approximating the complete system well any more, since the energy exchange between the toroidal harmonics becomes strong. For such situations, a recently implemented generalization of the preconditioner~\cite{Holod2020} can be of advantage, where each block matrix in the preconditioner can contain several toroidal harmonics. While this improves the approximation of the complete system by the preconditioner, it also increases the computational costs for inverting the block matrices such that the best configuration for the preconditioner is case specific. In case these ``mode groups'' are not mutually exclusive, several options exist for obtaining the combined result vector. Another recently implemented extension~\cite{Holod2020} allows to use complex matrix solvers for the preconditioner, reducing memory consumption and computational costs considerably.

When PaStiX is used, JOREK exploits a specific feature of the solver, where the connectivity graph is not considered separately for each degree of freedom (dof). Instead, square blocks of dimension $n_\text{var}$ are treated as dense blocks, where the number of variables typically is around $n_\text{var}=8$. Exploiting this structure, greatly enhances the performance of the analysis and factorization steps and can even improve the performance of the solve step considerably. Since the result vector of the matrix system contains the \textit{changes} of all dofs for all physical variables, the actual values of the variables need to be updated after the solve step. However, also the changes of the variables in the previous time step need to be stored, since those are required for some of the time stepping schemes (e.g., BDF2 Gears, see Section~\ref{:code:numerics:tstep} for more information on the time stepping).

\revised{Main limitations of the solver strategy are the high memory consumption associated to the factorized preconditioning block matrices (partly mitigated by the recently added complex matrix treatment in the preconditioner), the limited parallel scalability of the direct solver used in the preconditioning (addressed by adapting the code for better optimized or further improved solver libraries), and the deteriorating efficiency of the preconditioning in case of strong toroidal mode coupling (mitigated by the recently implemented generalization of the preconditioner to integrate the most important non-linear interactions in the approximation).}

\subsection{Parallelization}\label{:code:numerics:parallel}

\begin{figure}
\centering
  \includegraphics[width=0.5\textwidth]{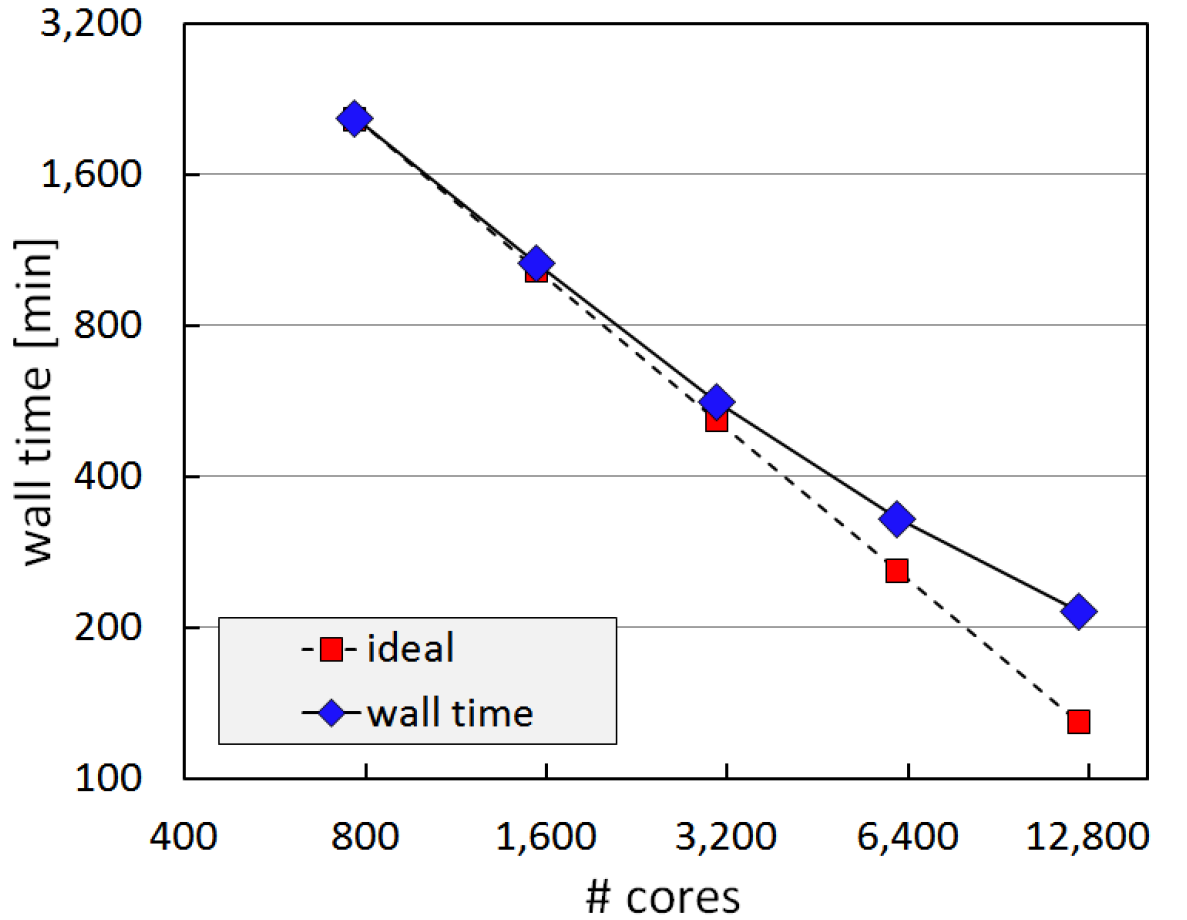}
\caption{Strong scaling of a JOREK particle simulation (with $10^9$ ions) from an entirely kinetic electro-static ITG turbulence simulation with JOREK on the Marconi-Fusion supercomputer (see Sections~\ref{:code:models:kin} and~\ref{:applic:other:itgturbulence}). \noreprint}
\label{fig:particlescaling}
\end{figure}

JOREK uses a hybrid MPI plus OpenMP parallelization. The number of MPI tasks must be a multiple of number of toroidal harmonics (cosine and sine components are not counted separately). Typically, a few MPI tasks are started per compute node and the number of OpenMP threads is adapted such that all CPU cores are exploited. The number of compute nodes used is primarily determined by the memory consumption of the simulation. As shown in Ref.~\cite{Holod2020}, already the matrix system itself requires significant storage and the LU factorization used in the preconditioner (see Section~\ref{:code:numerics:solver}) increases this memory consumption further. JOREK is mostly used on conventional architectures with Intel or AMD CPUs including Intel Xeon Phi. Exploiting GPUs is considered for the future, in particular for the kinetic particles (see Section~\ref{:code:models:particles}). Base information about the simulation like the input parameters, but also the plasma state, i.e., all degrees of freedom of all physical variables on the whole grid, is duplicated across all MPI tasks. This duplication is affordable in terms of memory consumption, and allows an efficient load balancing for instance for the kinetic particles. The communication overhead typically associated with PiC methods using a domain decomposition is avoided since particles do not need to be transferred from one MPI task to another. An example for the scaling of the kinetic particle model is shown in Figure~\ref{fig:particlescaling}.

MPI rank 0 is responsible for reading the namelist input file, broadcasting this input to the other MPI ranks, reading and writing restart files\footnote{For reading/writing large files in the free boundary extension (Section~\ref{:code:models:freebound}) and the kinetic particles framework (Section~\ref{:code:models:particles}), parallel I/O (MPI I/O and parallel HDF5, respectively) is used to achieve good performance. In case of the free boundary extension, this concerns reading the vacuum response matrices~\cite{Mochalskyy2017}. In case of the particles framework, it allows to avoid unnecessary communication when reading/writing large amounts of particle data.}, printing information to the log file, etc. The grid construction and equilibrium calculation~\ref{:code:models:equil} is only carried out on one MPI rank typically as this is not an expensive part of the simulation. \revised{This is executed} in a separate ``job'' from the time evolution such that no CPUs are idling. For (large) free boundary equilibrium calculations, several MPI ranks may be necessary due to memory requirements.

The parallelization of the matrix construction during the time stepping is done by a domain decomposition such that the system of equations is assembled in a distributed way (see Section~\ref{:code:numerics:matrix}). For the iterative solver (see Section~\ref{:code:numerics:solver}), an all to all MPI communication is carried out to extract the preconditioning matrices from the complete system (see Section~\ref{:code:numerics:solver} for details on the solver and preconditioner). Optionally the harmonic matrices can be recomputed bringing performance gains in specific limits~\cite{Holod2020}. Handling the preconditioning matrices may be slightly unbalanced since the $n=0$ block is smaller than the others (see Section~\ref{:code:numerics:matrix}). However, MPI rank 0, which is responsible for the $n=0$ block, either alone or together with a few other MPI tasks, usually anyway has other responsibilities, which can overlap with the solve on the other MPI ranks, such that this is not affecting performance in a significant way. The ``mode groups'' described in the previous section can lead to a larger imbalance, which can be compensated by adapting the number of MPI tasks assigned to each preconditioner block matrix, which is possible in a flexible way~\cite{Holod2020}. The sparse matrix libraries applied for solving the preconditioning matrices employ a hybrid parallelization internally as well. In case of PaStiX, this is done by an MPI plus pthread parallelization, for instance. The linear operations inside GMRES, e.g., sparse matrix vector products, are well balanced, since they are based on the evenly distributed global matrix.

\begin{figure}
\centering
  \includegraphics[width=0.5\textwidth]{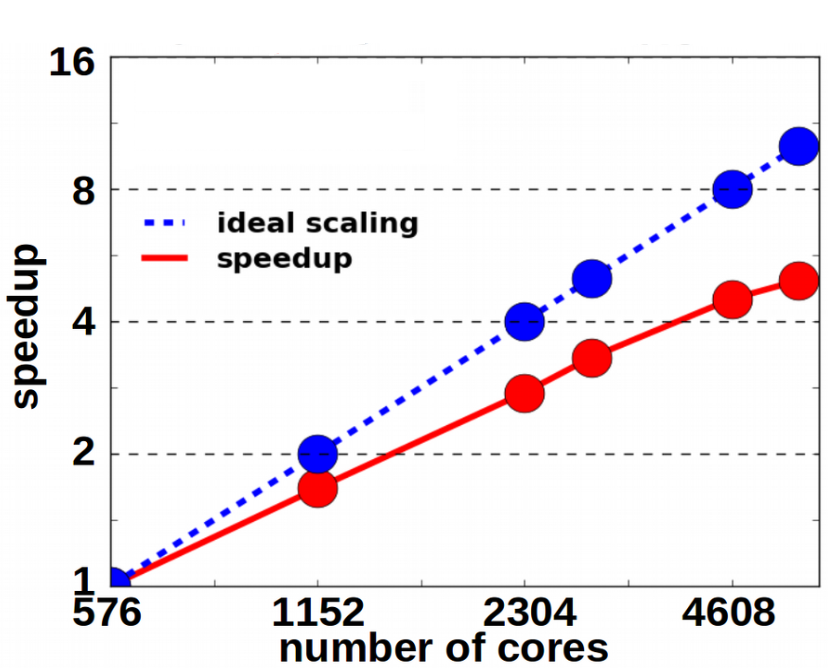}
\caption{Strong scaling test of the JOREK parallel efficiency for a moderate problem size on the Marconi-Fusion supercomputer using Intel Skylake CPUs. The setup is a JT-60SA ELM simulation (Ref.~\cite{Pamela2018IAEA}) \revised{with a moderate 2D grid resolution of about 5k grid nodes and a relatively high toroidal resolution including} $n=1\dots15$ included. For a fixed problem size, the number of compute nodes is increased from 16 (576 cores) to 160 (5760 cores). Note, that only 36 out of the 48 cores per node were used in this test. The relative efficiency is approximately 50\% when comparing the performance between 16 and 160 compute nodes, which constitutes an excellent value for a fully implicit code. In practice, the problem size considered here would not be addressed using more than 32 compute nodes, for which the relative efficiency is around 85\%. \noreprint}
\label{fig:jorekscaling}
\end{figure}

\begin{figure}
\centering
  \includegraphics[width=0.6\textwidth]{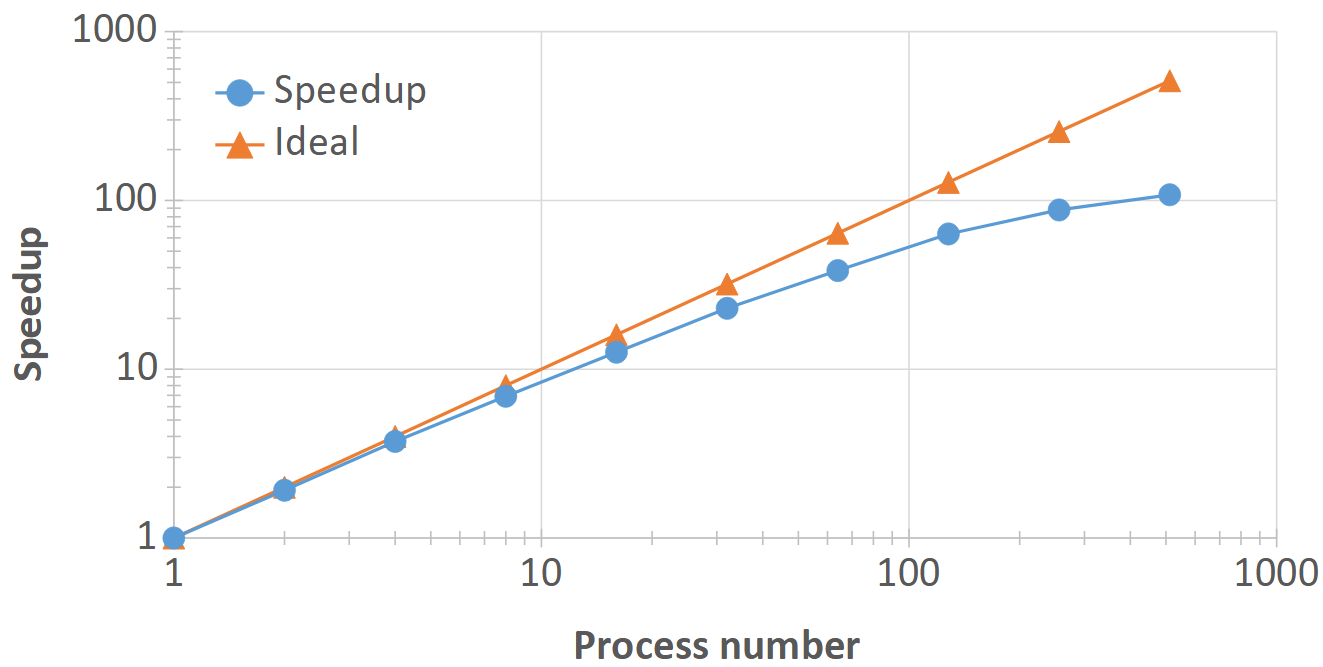}
\caption{Strong scaling test of the STARWALL parallel efficiency for a small problem size on the Marconi-Fusion supercomputer~\cite{Mochalskyy2017} \revised{with 35200 wall triangles equivalent to 17600 degrees of freedom (size chosen to fit into the memory of a single compute node). The execution time of the whole STARWALL run is given. When the number of MPI tasks is increased from 1 to 128, a speed-up of 63.3 can be obtained (parallel efficiency of 50\%). At 64 MPI tasks which would be used in practice, the parallel efficiency is around 60\%}. Larger production cases scale to a higher number of cores with a better efficiency. \revised{Note that this figure shows the scaling for response matrix calculation in STARWALL, not for the time stepping of the JOREK simulation. Calculating the STARWALL contribution to the boundary conditions in the JOREK time stepping is usually not altering the JOREK scaling significantly as this constitutes a sub-dominant and well parallelized~\cite{Mochalskyy2017} part of the computational costs except for exotic setups which combine a low plasma resolution with a very high wall resolution.} \noreprint}
\label{fig:starwallscaling}
\end{figure}

An example of the scalability of JOREK is shown in Figure~\ref{fig:jorekscaling} for a moderate problem size. Further work to improve the performance and parallel scalability of JOREK is ongoing, as shown in Section~\ref{:code:numerics:outlook}. The scalability of STARWALL is shown in Figure~\ref{fig:starwallscaling}, refer to Ref.~\cite{Mochalskyy2017} for more information about the parallelization of STARWALL and the JOREK-STARWALL coupling terms.

\subsection{Code management}\label{:code:numerics:management}

JOREK is written in Fortran 90/95 using Fortran 2003 and 2008 extensions in particular in the kinetic particles part. A few smaller core routines and interfaces are written in C and C++. JOREK presently consists of more than 250 thousand lines of source code, to which dozens of developers have contributed over the years. More than 100 pull requests are merged per year at the moment with a strongly increasing trend. More than 40 scientists presently use the code world-wide for very different purposes and a significant fraction of those is actively contributing to the development. Via modern code development techniques and a common shared code repository, a unique stable code basis is maintained shared by all users and for all code applications (The code is hosted on the Atlassian Bitbucket system operated by the ITER Organization). A dedicated Wiki contains the collaboratively written code documentation.

\revised{A particular challenge is posed by the contradictory goals of code stability and agile development. With the strongly increasing number of users and developers over the last years, it became critical to ensure that developments for specific purposes cannot break entirely different code applications.} For guaranteeing this, limited manpower was available such that a high level of efficiency was required. To achieve these goals to the best possible extent, the modular structure of the code was continuously improved and the development work flow was switched from Subversion to a Git based repository more than five years ago. The developments work flow starts by describing issues or planned new features in the Jira issue tracker, discussing about the planned implementation, creating a branch for the respective development based on the latest code version from the main development branch, implementing the changes into that branch, raising a pull request with several code-reviewers, refining the solution iteratively, and merging the modifications into the main development branch. To ensure coordination and inform users and developers, all changes to the code are discussed and communicated in regular development meetings. At each meeting, a release version of the code is created. Smaller developments are usually merged quickly, large developments are sometimes carried out in a branch over months or even years. Measures are taken to reduce the number of long-lived branches and the remaining ones are actively kept up to date with the main development branch by regularly merging in all recent changes.

To avoid introducing problems by the numerous smaller and larger developments which take place in parallel with the large number of different physics models and code applications, we introduced a framework for carrying out automatic test cases in addition to the reviewing of pull requests. Our approach to these tests is described in the following (Ref.~\cite{Latu2012arXiv} explains an older approach, which has been replaced several years ago). About 50 automatic regression tests are in place at present, which are executed automatically whenever new modifications are pushed to any branch on the central repository. These test cases can also easily be executed locally by each user. The framework is flexible enough to allow for different ``versions'' of the same regression test in different branches, a feature needed in case a bug is detected and resolved, or a modification is implemented that intentionally changes the code behaviour (e.g., improved grid construction). The data required for carrying out a regression test and checking the result is stored on a separate server and the correct version of the test is identified via hash keys. All regression tests are carried out with a large set of compiler/linker debugging options and some regression tests are using a second compiler (gFortran instead of Intel Fortran) to catch as many problems as possible.

The majority of the test cases are based on low-resolution non-linear simulations, which are restarted from an HDF5 restart file and continued for a single time step only. The result is then compared to a reference result with a specific \revised{absolute tolerance}. This has been tested extensively to capture problems reliably and the execution in the non-linear phase ensures that all physics terms influence the solution. Among the test cases, also a few specific ones exist that work differently. The simplest ones only compile the code and all diagnostic binaries for the various physics models, other tests cover the grid construction and equilibrium solver. Specifically for the kinetic particles module, also unit tests are in place.

In spite of the considerable number of test cases, not all applications of the code can be covered, however new regression tests are continuously added and are introduced for new code features already before these are merged into the main development branch. Furthermore, a generalization of the framework is foreseen to simplify testing all combinations of different use cases and model extensions.

\subsection{Outlook}\label{:code:numerics:outlook}

This section gives a brief outlook onto developments regarding the numerical methods of the JOREK code.

\begin{figure}
\centering
  \includegraphics[width=0.9\textwidth]{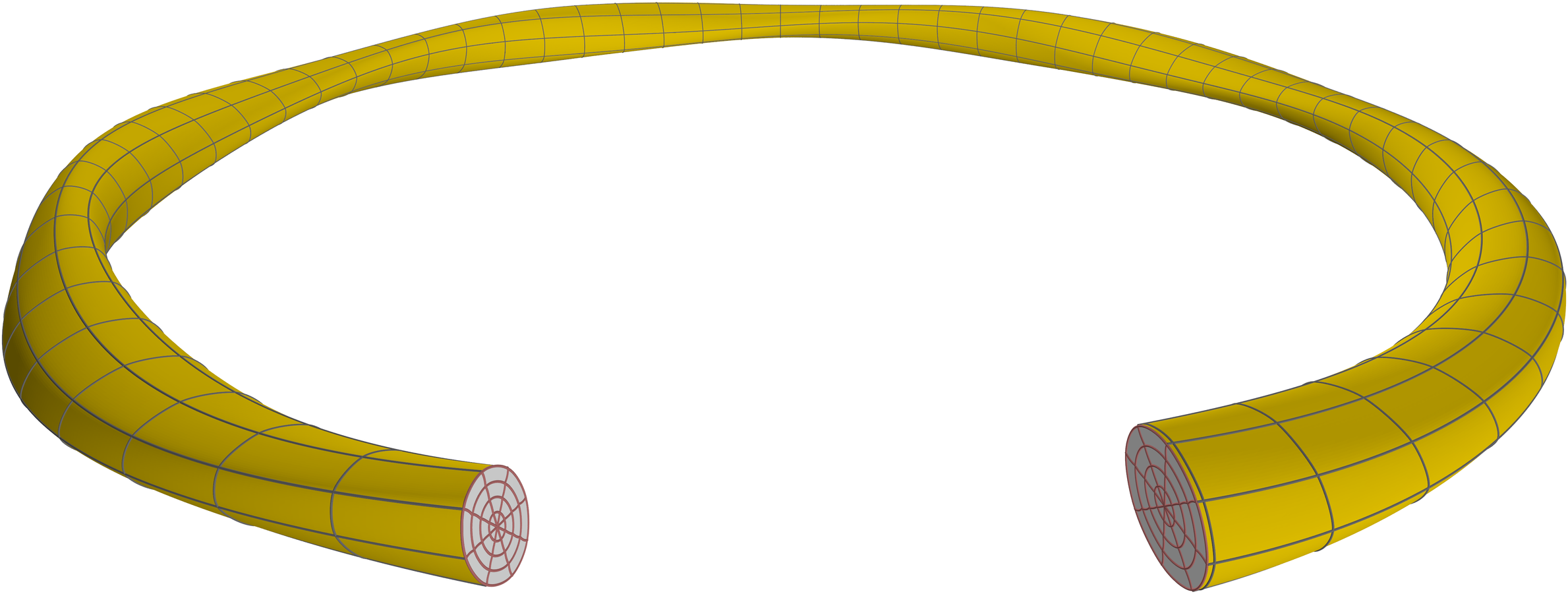}
\caption{Example of a 3D grid (shown with very low resolution for clarity), imported from the GVEC code for a 5 field period, elliptical equilibrium, similar to the W7-A stellarator. \noreprint}
\label{fig:w7a_gvec_grid}
\end{figure}

For future simulations of non-axisymmetric configurations (stellarators), in addition to the derivation of a hierarchy of MHD models (Section~\ref{:code:models:outlook}), the numerical methods also need to be adapted. In particular, the following assumptions need to be relaxed:
\begin{itemize}
\item The simulation grid is not axisymmetric any more, i.e.~R and Z are expressed in $s$ and $t$ only, but also expanded in a toroidal Fourier series. An import of such a 3D configuration from GVEC \cite{Hindenlang2019} has already been implemented, as shown in Figure~\ref{fig:w7a_gvec_grid}. 
\item The assumption that $s$ and $t$ are orthogonal to $\phi$ will not be fulfilled any more in general (for instance, in case of PEST coordinates). In addition, a $\phi=const$ surface will in general not be planar any more (for instance, in case of Boozer coordinates). This is taken into account in the ongoing implementation of the new stellarator capable models.
\item In a stellarator, the linear eigenfunctions of plasma instabilities cover a toroidal spectrum instead of being restricted to a single toroidal mode number. Consequently, the presently used preconditioner, which relies on the linear decoupling of toroidal harmonics, cannot be applied any more and solver developments are important. Work in this direction is moving forward, in particular see the ``mode groups'' in the preconditioner described in Section~\ref{:code:numerics:solver} which constitutes a first important step towards efficient stellarator simulations.
\end{itemize}

Regarding the solver, several developments are on their way building up onto previous work, this includes for instance an assessment for the use of GPUs, enhanced preconditioning via an integrated iterative refinement, the use of reduced order for the preconditioner, etc. These efforts are at early stages and are thus not described here in detail.

For the anisotropic heat transport (see also Section~\ref{:verification}), reduced order basis functions for selected quantities are being studied to further improve the numerical accuracy in particular when finite element grids are used that are not aligned to the flux surfaces, e.g., in case of VDEs.

Aiming at simplified data exchange, integrated modelling, and standardized analysis tools, the adaptation of JOREK to IMAS is on the way. The ITER Integrated Modelling \& Analysis Suite (IMAS)~\cite{imbeaux15:_desig_iter} is a scientific software framework infrastructure that orchestrates execution of integrated plasma codes. The \emph{Physics Data Model (PDM)} is a basis for coupling the plasma codes and experimental data at different space and time scales through its 60+ \emph{Interface Data Structures (IDS)}.
A widely used IDS substructure for description of computational domain is the \emph{General Grid Description (GGD)}, with the purpose to describe any $N$-dimensional numerical grid geometry and associated plasma state quantities with time slices sampled during the simulation. Generally, JOREK couples to the Equilibrium and MHD IDS for input conditions and nonlinear plasma state output that can be directly compared with other simulators that use the GGD mapping (see e.g. Ref.~\cite{penkod_JOFE_2020}). 
Integration of JOREK in IMAS is currently underway, with a few aspects and methods already available to allow incorporation of JOREK relevant data in the MHD IDS employing the GGD representation~\cite{JOREK_NENE2019}. Spatial discretisation under GGD in 3D requires B\'{e}zier finite elements for poloidal mesh discretisation (nodes and corresponding values with derivatives) combined with real Fourier series for toroidal harmonics. Temporal slices in MHD IDS are saved as flux aligned GGD and corresponding values on the grid. Time slices for GGD and values on GGD can be different to respect database size growth per simulation. Finally, stored simulation results enable analysis with unified tools and runs under integrated modelling workflows.

In the present grid structures, $G^1$ continuity is satisfied across the whole domain except for the magnetic axis and the direct vicinity of the X-point. The $G^1$ continuity at these special points will be restored by a special treatment with locally adapted basis functions.
\revised{Furthermore, a generalization of the Bezier FE formulation to higher order basis functions and higher order continuities across element boundaries is ongoing~\cite{Pamela2021}.}
The implementation of an option for triangular FEs in JOREK is being considered, operator splitting or relaxation techniques are being discussed, and further refined numerical stabilization methods such as VMS and higher order Taylor Galerkin methods are being investigated.

%% file: 04_verification.tex
\section{Verification}\label{:verification}


\begin{figure}
\centering
  \includegraphics[width=0.6\textwidth]{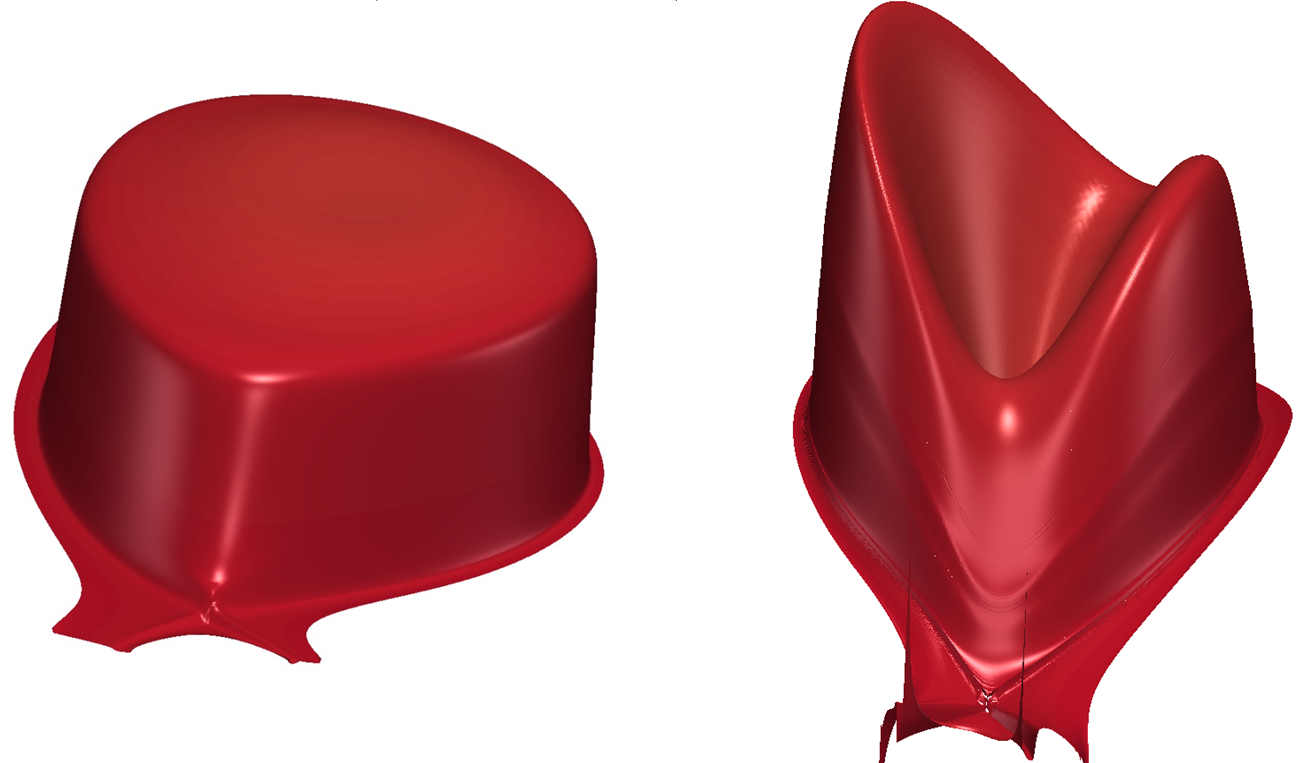}
\caption{Example of the $G^1$ continuous, i.e.,  continuous values and first derivatives in real space, profiles in the poloidal plane for \textbf{(left)} a density distribution and \textbf{(right)} the perpendicular conductive heat flux. \noreprint}
\label{fig:smooth_profiles}
\end{figure}

In this Section, selected \textit{verification} work done for the JOREK code is highlighted. This includes code-code benchmarks, comparisons to analytical solutions, comparisons between reduced and full MHD, and an assessment of energy conservation in a violent non-linear example. The emphasize here is on more recent tests, of which many have not been shown in publications and conference contributions yet. A lot more verification work can be found in the literature. E.g., the JOREK pellet model has been successfully benchmarked, for a Deuterium pellet, with a dedicated code by B. P\'egouri\'e, as can be seen in Fig.~2 of Ref.~\cite{Futatani2014}. Note that all the \textit{validation} against experiments is included along with the physics studies in Sections~\ref{:applic:edge}--\ref{:applic:other}. Figure~\ref{fig:smooth_profiles} shows an example to demonstrate that the finite element solutions are $G^1$ continuous and (if resolved well) smooth.

\subsection{Convergence properties}\label{:verification:convergence}

The convergence of the exponential growth rates with the spatial grid resolution is shown in the following. Two cases are considered in circular plasmas for simplicity: A tearing mode test-case using the full-MHD model and a ballooning mode test case with the reduced-MHD model (the so-called CBM18 equilibrium for mode number $n=20$). The grid resolution is scanned homogeneously in the radial and poloidal directions, from $(n_\text{flux},n_\text{tht})$=$(27,180)$ to $(90,600)$, where $n_\text{flux}$ and $n_\text{tht}$ are the equidistant number of radial and poloidal elements. The relative error of the growth rates scales inversely with the 5th power of the spatial resolution like $\propto(\sqrt{n_\text{flux}n_\text{tht}})^{-5}$.

\begin{figure}
\centering
  \includegraphics[width=0.6\textwidth]{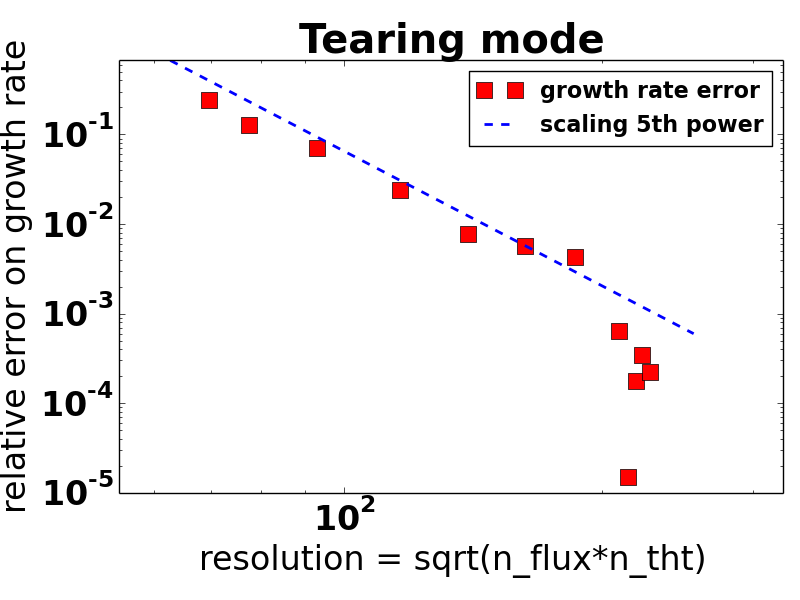}
\caption{The relative error of the growth rate of a tearing mode is plotted as a function of the spatial resolution $\sqrt{n_\text{flux}n_\text{tht}}$, for the full-MHD model. The error converges as the 5th power of the spatial resolution, as expected. Re-print from Ref.~\cite{Pamela2020}. \reprintaipOK}
\label{fig:FMHD_CONVERGENCE}
\end{figure}

\begin{figure}
\centering
  \includegraphics[width=0.6\textwidth]{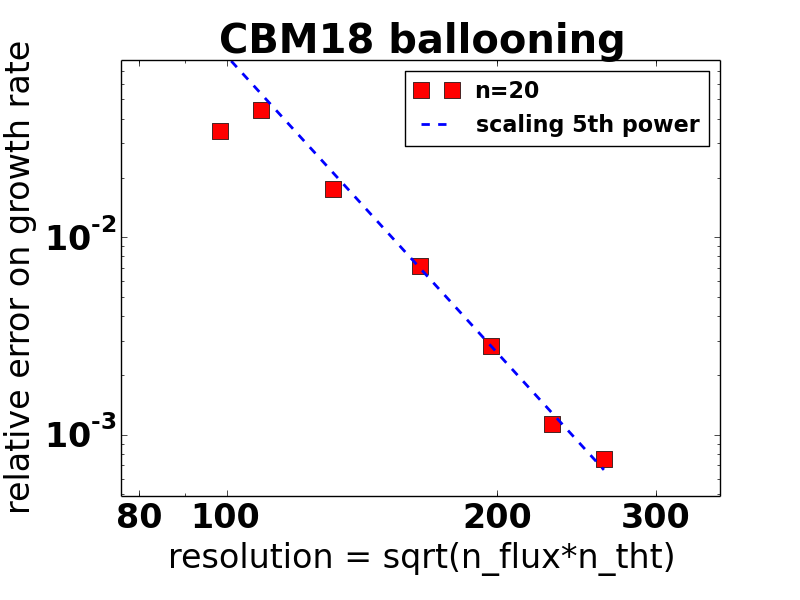}
\caption{The relative error of the growth rate of a ballooning mode n=20 is plotted as a function of the spatial resolution $\sqrt{n_\text{flux}n_\text{tht}}$, for reduced MHD. The error converges as the 5th power of the spatial resolution, as expected. \noreprint}
\label{fig:RMHD_CONVERGENCE}
\end{figure}

With finite elements, the local error is estimated as $E\propto h^p$ according to Ref.~\cite{StrangFix}, where $h$ is the element size, and $p$ is the polynomial order of the finite elements. Here, $(\sqrt{n_\text{flux}n_\text{tht}})^{-1}$ is used as an approximation of the element size $h$, and $p$=$4$ because the Bezier elements are bi-cubic. Since the value of interest in these tests are the growth rates of the toroidal modes, which are obtained by integrating the mode energies over all the elements of the simulation domain, this adds another factor $(\sqrt{n_\text{flux}n_\text{tht}})^{-1}$ to the error estimate. Hence, the error of the growth rates is expected to scale with the 5th power of the spatial resolution, $(\sqrt{n_\text{flux}n_\text{tht}})^{-5}$. 

Figure~\ref{fig:FMHD_CONVERGENCE} and Figure~\ref{fig:RMHD_CONVERGENCE} show the convergence of the growth rate error, as a function of spatial resolution. For the tearing mode, the 5th order scaling is found as expected, and beyond a high enough resolution, the error diminishes dramatically, suggesting the growth rate is already fully converged. \revised{For this case, since the error diminishes faster than the expected scaling at high resolution, the intermediate case of $\sqrt{n_\text{flux}n_\text{tht}} = 2\times 10^2$ is assumed to be the converged growth-rate.} For ballooning modes, using the CBM18 case, the growth rate convergence also follows a scaling of the expected 5th order with the reduced-MHD model. \revised{For this case, since the error still diminishes at the highest resolution, the growth-rates are fitted with the expected scaling, such that the converged growth-rate is obtained by extrapolating to infinite resolution.} With the full MHD model, only 3rd order scaling was found for the ballooning mode test case (see Ref.~\cite{Pamela2010}). The reason for this different behaviour is still being investigated.

\subsection{Anisotropic heat transport} \label{:verification:anisotropy}

\begin{figure}
\centering
  \includegraphics[width=0.53\textwidth]{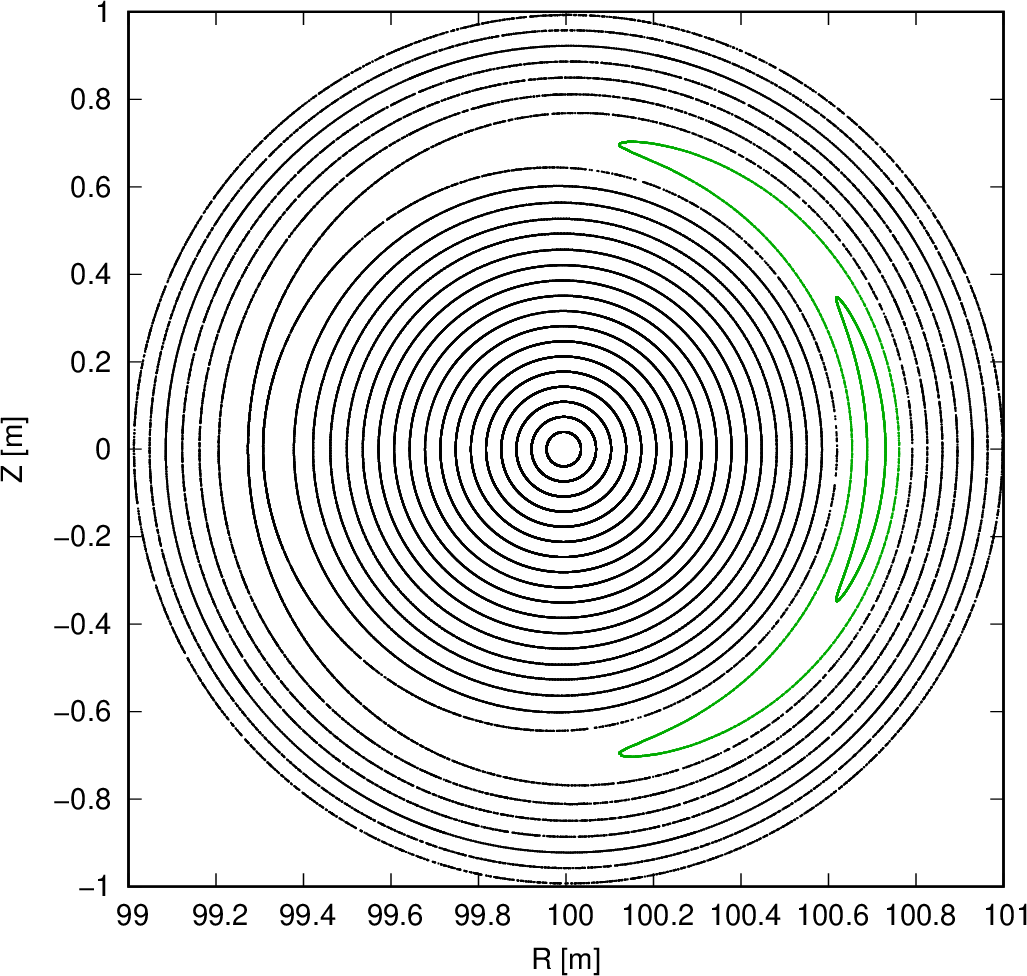}
  \includegraphics[width=0.42\textwidth]{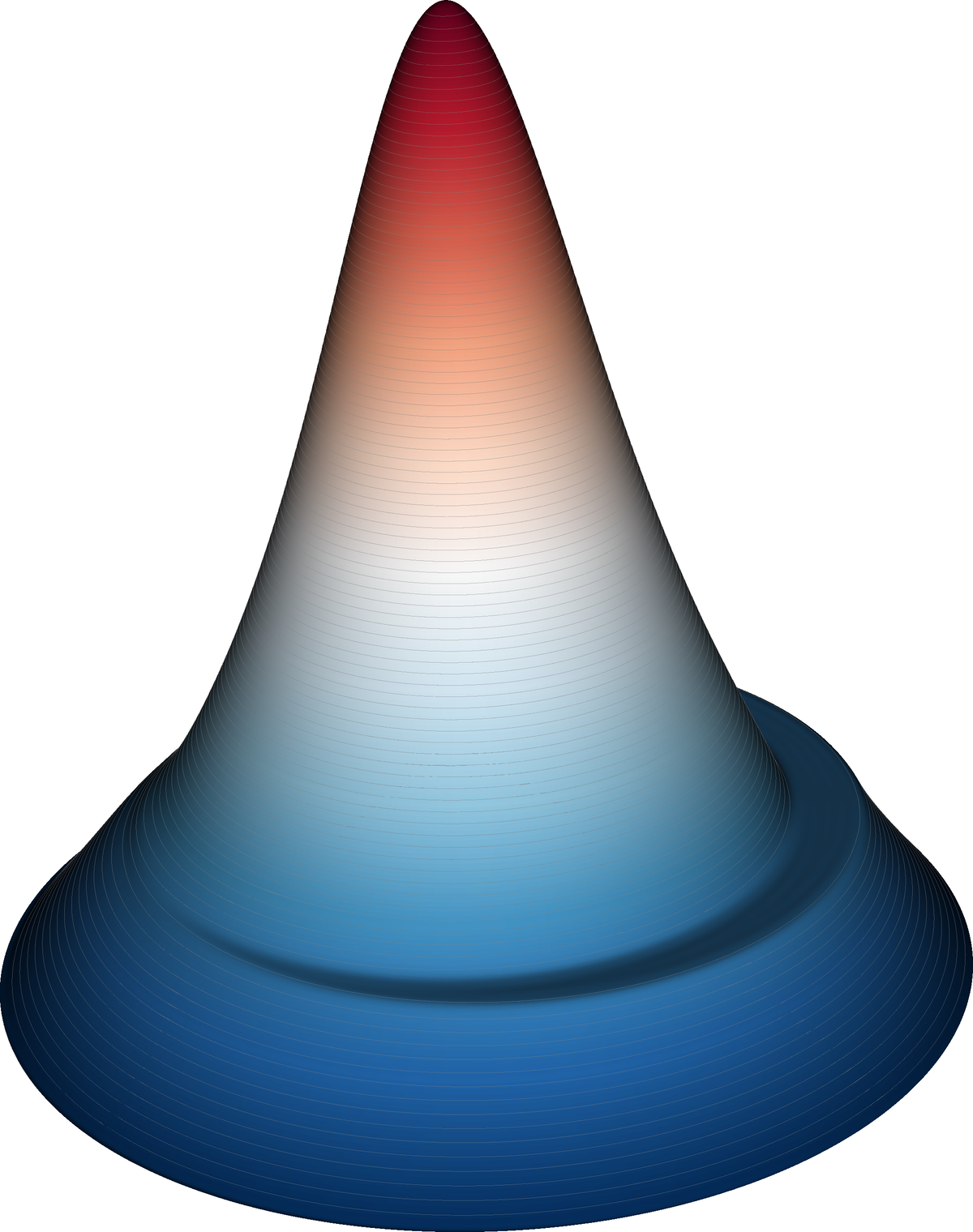}
\caption{\textbf{(left)} Poincar{\'e} plot of the magnetic field configuration used to demonstrate anisotropic heat transport in the JOREK reduced MHD model. \textbf{(right)} Temperature distribution in a fully converged simulation with a heat diffusion anisotropy of $10^{10}$. The temperature flattening inside the island is clearly visible. \noreprint}
\label{fig:poinc-anisotropy-test}
\end{figure}

The numerical treatment of anisotropic heat transport in JOREK allows to handle experimentally relevant parameters, where the diffusion coefficients along magnetic field lines and perpendicular to them differ typically by a factor $10^{8}$ to $10^{10}$. We show only a small demonstration of the properties here, based on an example with large aspect ratio circular configuration where $R=100$ and $a=1$. This simplified configuration is used to ease interpretation since it eliminates mode coupling and allows to initialize a single magnetic island without secondary islands or stochastization. The anisotropic heat diffusion equation is solved both using a polar grid (approximately aligned to the flux surfaces) and a rectangular grid (as example for a grid that is not flux surface aligned at all). A more detailed analysis is planned for the future in the context of implementing an advanced scheme following Ref.~\cite{Guenter2007} for particularly demanding cases like hot VDEs with a grid not aligned at all to the magnetic field and RE fluid simulations where parallel diffusion is sometimes used as a computationally less demanding proxy for the fast parallel advection.

Figure~\ref{fig:poinc-anisotropy-test} shows the magnetic configuration via a Poincar{\'e} plot as well as the temperature distribution obtained at high anisotropy. A pure $1/1$ magnetic perturbation is applied to create a large magnetic island. While keeping the magnetic field fixed in time and assuming a spatially uniform density, the temperature evolution equation is solved for single large time step with the implicit Euler method to determine the steady state temperature distribution. A localized Gaussian heat source is applied in the center of the plasma (zero inside the island surfaces). \revised{Details of the test case are described in Appendix~\ref{:app:anisotropy-test}.}

\begin{figure}
\centering
  \includegraphics[width=0.48\textwidth]{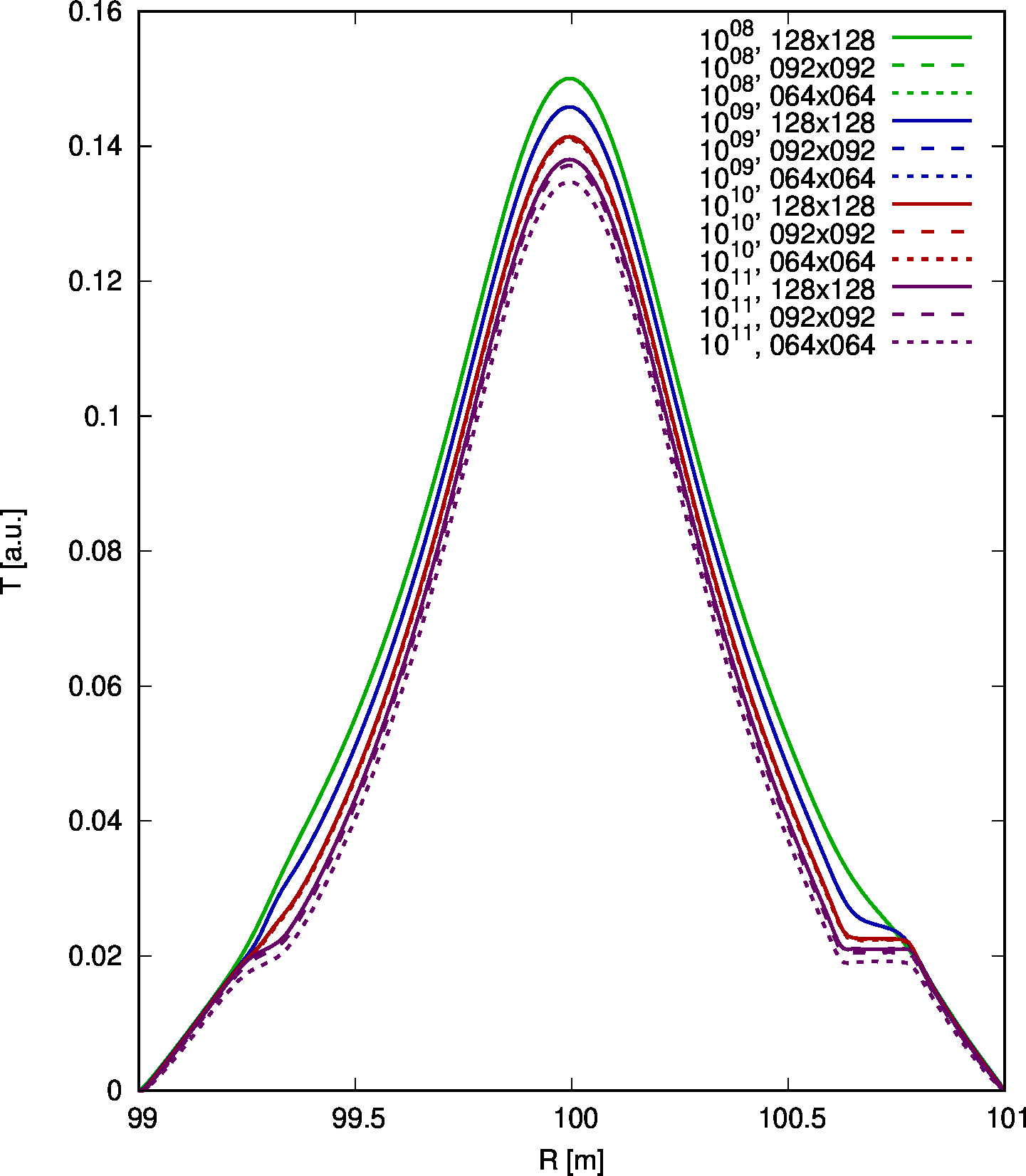}
  \includegraphics[width=0.46\textwidth]{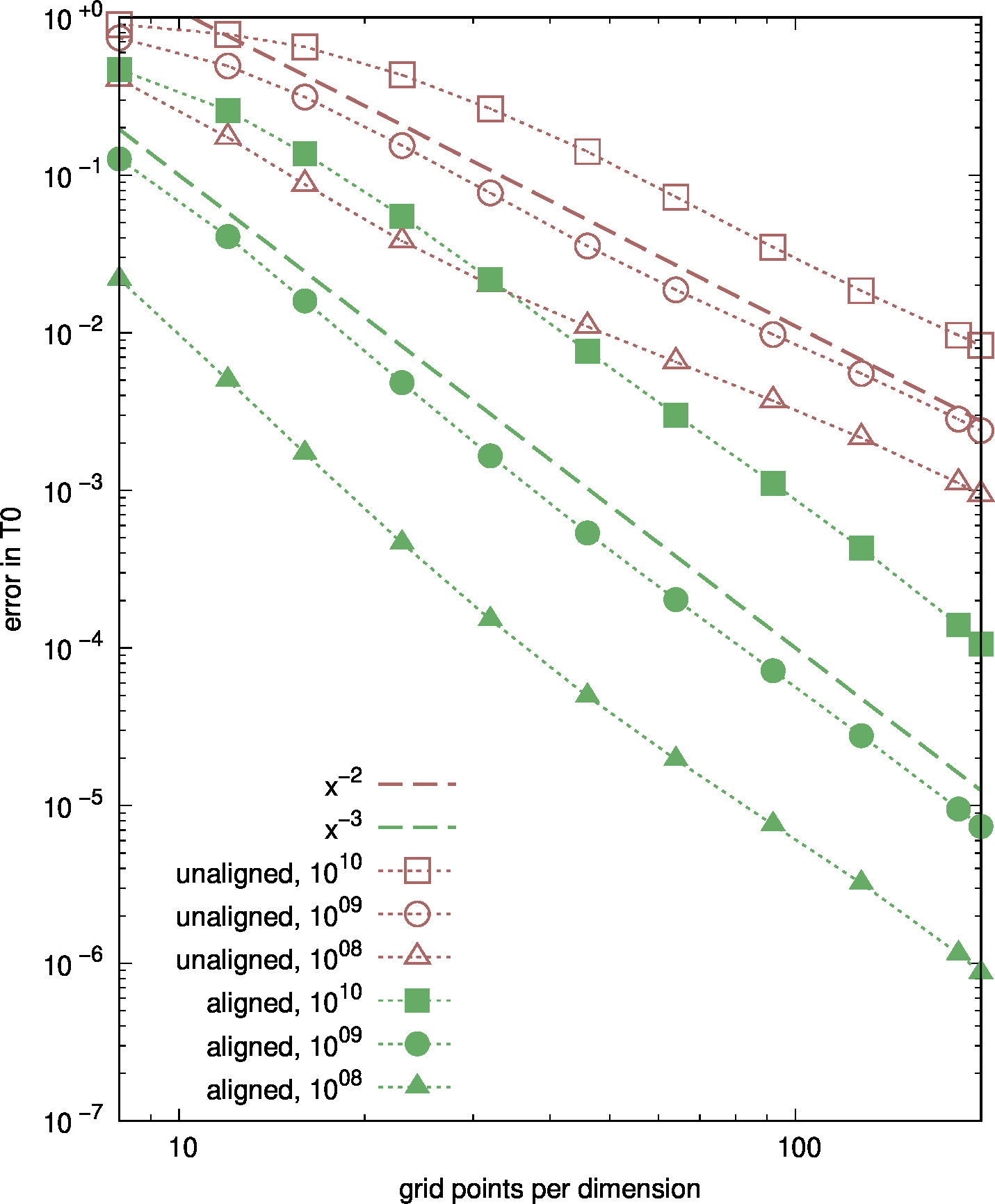}
\caption{\textbf{(left)} Temperature distribution on the midplane for the polar grid aligned to the unperturbed flux surfaces at different heat diffusion anisotropies and resolutions. The simulations shown here are done with the toroidal mode numbers $n=0\dots4$. \textbf{(right)} Convergence of the relative error in the core temperature versus the number of grid points per dimension. \noreprint}
\label{fig:anisotropy-cut-convergence}
\end{figure}

Figure~\ref{fig:anisotropy-cut-convergence} shows cuts of the temperature distribution across the mid plane at different anisotropies and different resolutions for the polar grid (left panel). It also contains a convergence study (right panel) of the central temperature value with the grid resolution for both, aligned and unaligned grids. Note that the error scales with the number of grid points to the third power in case of an approximately aligned grid and only with the second order in case of a non-aligned grid. When tolerating a relative error of 0.3\% in the core temperature, a grid resolution of 200x200 grid points allows to easily resolve anisotropy values beyond $10^{10}$ in the flux surface aligned case. With the unaligned grid, the limit is around $10^9$ for this resolution and error. The results would naturally change a bit with different mode numbers. However, the simulation grids tested here do not make use of any localized refinement, which is routinely applied in production simulations to concentrate the grid elements in the MHD active regions and resolve them better than the remainder of the plasma. Note that in case of a hot VDE, which corresponds to a highly anisotropic case that usually needs to be simulated in an unaligned grid, the ``spurious perpendicular heat transport'' becomes negligible as soon as strong 3D MHD activity is triggered and the physical transport along stochastic field lines dominates over numerical errors.

\subsection{Energy conservation}\label{:verification:energyconservation}

The set of equations \eqref{eq:mhd:A}-\eqref{eq:mhd:p} can be written in a conservative form and summed up to find the evolution of the total energy density

\begin{figure}
\centering
\begin{tabular}{cc}
 \includegraphics[height=0.4\textwidth]{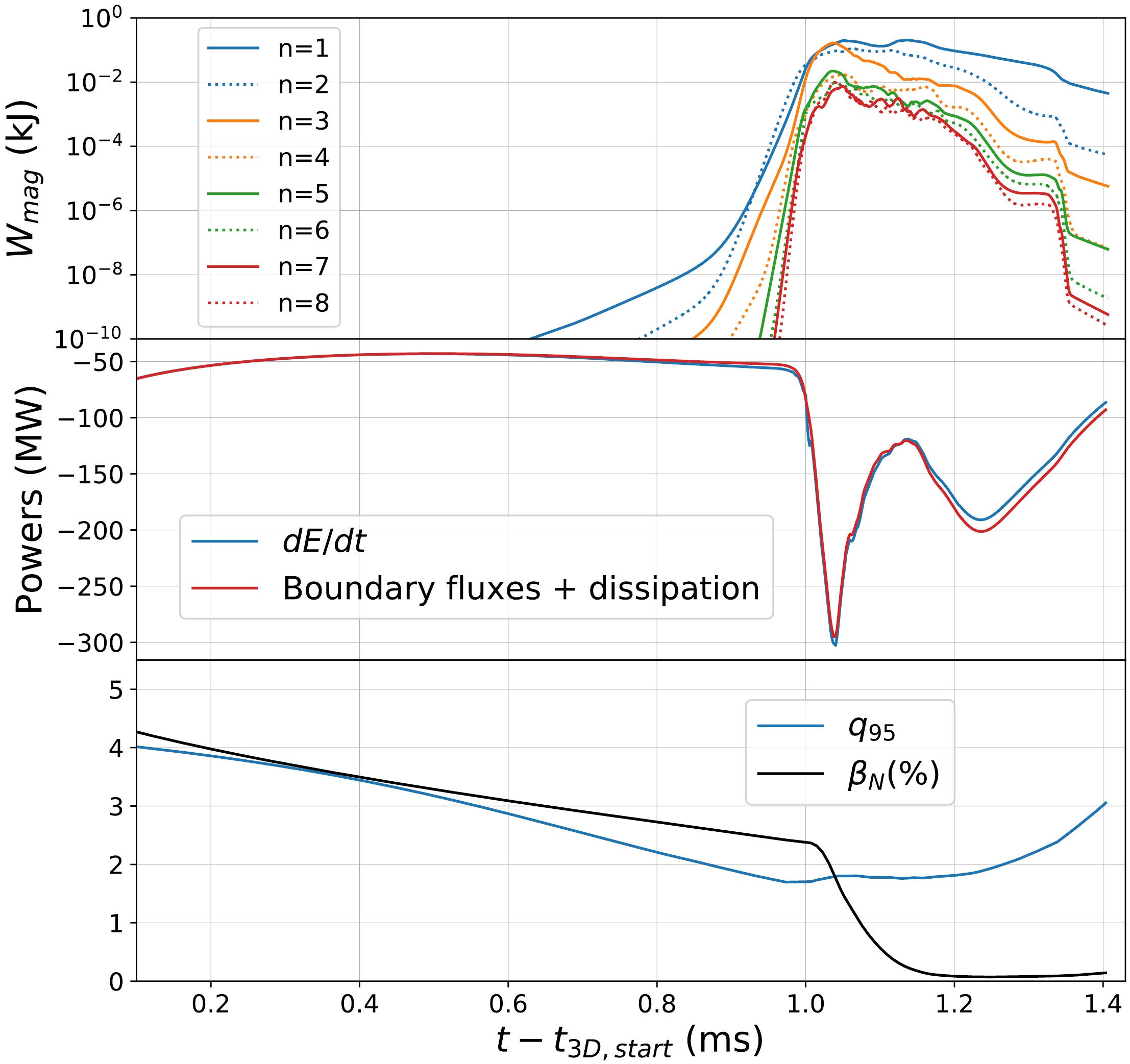}
 &
 \includegraphics[height=0.4\textwidth]{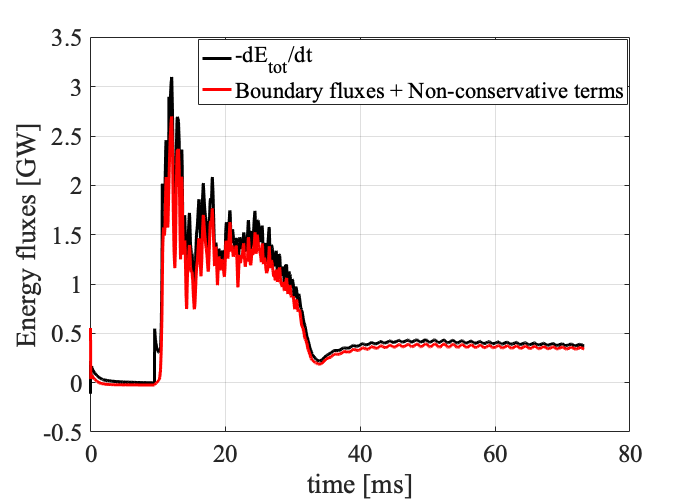}
 \end{tabular}
\caption{\revised{\textbf{(Left)}} 3D Vertical Displacement Event simulated with JOREK. (Top) Poloidal magnetic energy of the different toroidal harmonics. (Middle) Time derivative of the total plasma energy ($E=W_{mag} + W_{th}+W_{kin}$) and sum of dissipation powers and boundary fluxes (powers representing all energy losses). (Bottom) Evolution of the edge safety factor ($q_{95}$) and of $\beta_N$ during the VDE. The case is explained in detail in \cite{Artola2020B}. \revised{\textbf{(Right)} Time derivative of the total plasma energy and the sum of non-conservative terms and boundary fluxes for a 3D RMP simulation in an ITER 15~MA plasma simulated with JOREK. The simulations includes realistic ExB and diamagnetic flows and undergoes strong MHD activity between 10 and 35 ms which affects the edge confinement. The case confirms that the error in the energy conservation introduced by gyro-viscous cancellation is small as expected from analytical estimates.}
\noreprint}
\label{fig:energy_conservation}
\end{figure}

\begin{equation}
    \pderiv{e_\text{tot}}{t} + \nabla\cdot \mathbf{\Gamma}_\text{tot}=S_E
    \label{energy_conservation:total_evolution}
\end{equation}
where the total energy density is composed by the kinetic, the thermal and the magnetic energy densities
\begin{equation}
    e_\text{tot}=\frac{\rho}{2}\VecV\cdot\VecV + \frac{p}{\gamma-1} + \frac{\VecB\cdot\VecB}{2}
\end{equation}
and the total energy flux is
\begin{equation}
    \mathbf{\Gamma}_\text{tot}= \left[ \frac{\rho}{2}\VecV\cdot\VecV + \frac{p}{\gamma-1}  \right]\VecV - \frac{ \TensK\nabla T}{\gamma-1} + \VecE\times\VecB 
\end{equation}
Equation \eqref{energy_conservation:total_evolution} can be integrated over the plasma volume to obtain 
\begin{equation}
    \pderiv{E}{t} = \int S_E dV -\oint \mathbf{\Gamma}_{tot}\cdot d\mathbf{S} 
\end{equation}
where $E=\int e_{tot} dV$ is the total plasma energy, $\int S_E dV$ is the total injected power into the plasma and $\oint \mathbf{\Gamma}_{tot}\cdot d\mathbf{S}$ represents the total energy flowing through the boundary (boundary fluxes). An important test for MHD codes is to check that the previous equation is satisfied also in practice, where small errors from the numerical discretization of space and time introduce errors. This can be assessed by calculating the LHS and the RHS independently. Such a test has been done for a variety of different cases. We present this comparison here for the 3D VDE case that was used for the 3 code benchmark presented in \cite{Artola2020B}. \revised{Two tests for energy conservation are} shown in Figure~\ref{fig:energy_conservation} (powers), where the time derivative of the total energy is compared to the sum of boundary fluxes for a 3D VDE simulation \revised{and for an RMP simulation}. As the viscous and Ohmic heating terms were switched off for these particular cases, they have been calculated via postprocessing and included \revised{in $dE/dt$}. The results indicate that energy is conserved well \revised{including the RMP case with diamagnetic drift that relies on the model with gyro-viscous cancellation}. Note, that the plasma in the VDE case undergoes a very non-linear phase with stochastization of the entire domain and that all channels for energy transport contribute to the losses (anisotropic heat conduction, Poynting flux, convection, etc.). The small discrepancies at the end of the thermal quench ($t\sim 1.2$ ms) originate from the calculation of the Ohmic heating term and are under investigation. \revised{Energy is also rather well conserved, with about 90\% precision, in simulations of disruptions triggered by massive material injection (described in Section \ref{:applic:core:disr}), which include Taylor-Galerkin stabilization. The slight imbalance will be investigated. The energy balance diagnostics has been implemented recently such that the various terms involved in the kinetic, thermal and magnetic energy balances are calculated automatically for any simulation, allowing an easy and systematic check of energy conservation properties.}

\subsection{Core instabilities}\label{:verification:core}

%

\begin{figure}
\centering
 \includegraphics[height=0.3\textwidth]{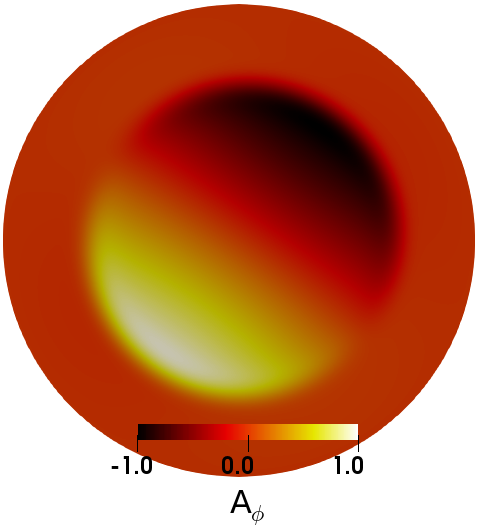}
\includegraphics[height=0.3\textwidth]{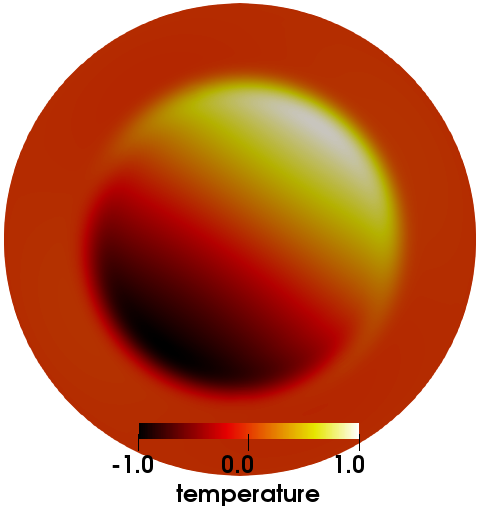}
\includegraphics[height=0.3\textwidth]{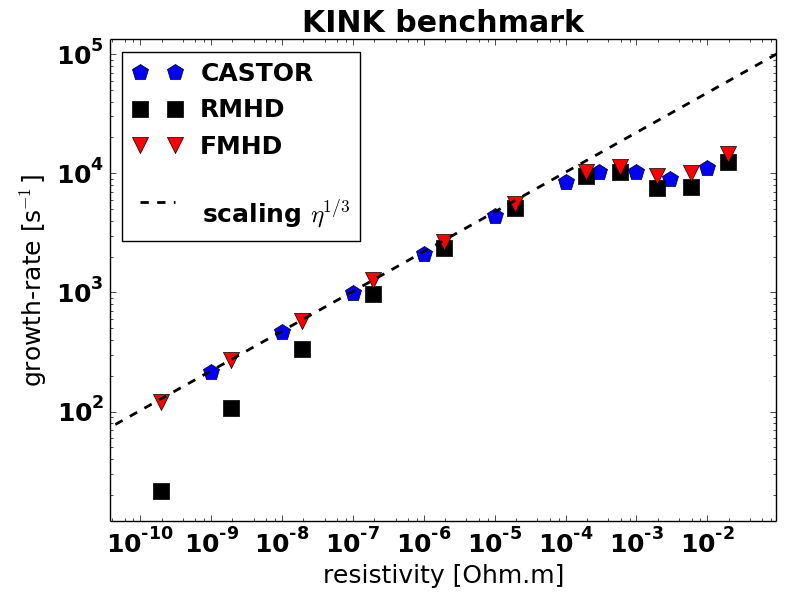}
\includegraphics[height=0.3\textwidth]{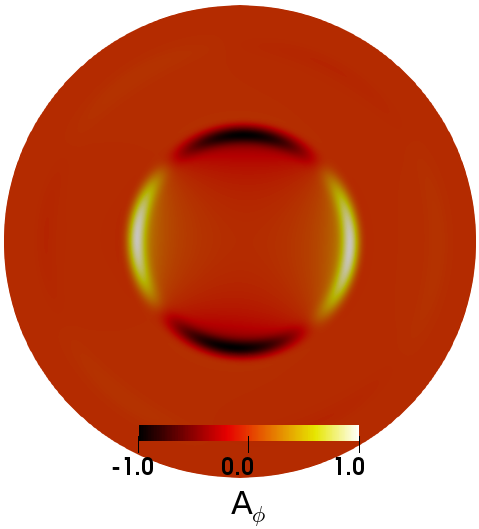}
\includegraphics[height=0.3\textwidth]{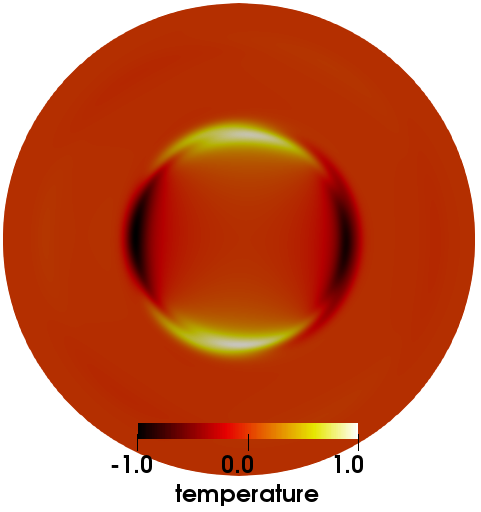}
\includegraphics[height=0.3\textwidth]{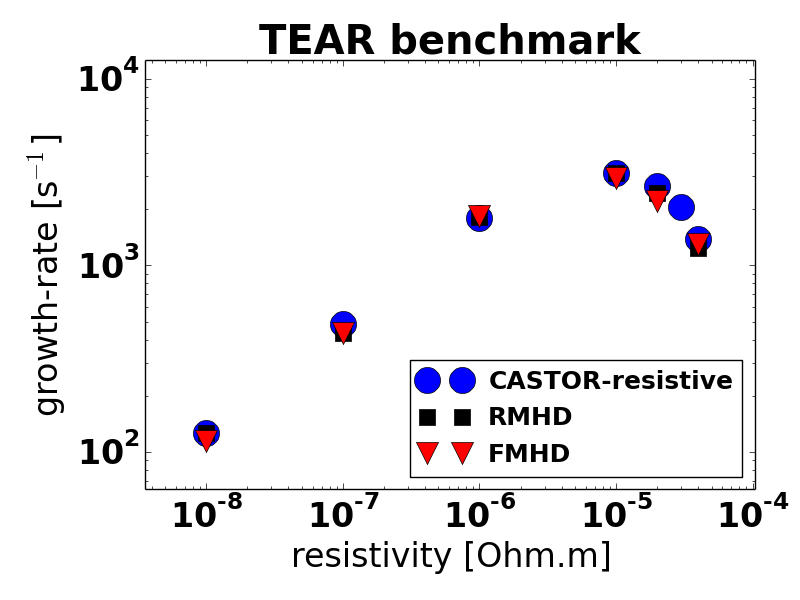}
\caption{$1/1$ kink (top) and $2/1$ tearing mode (bottom) test cases. Normalized perturbations are shown of the toroidal component of the magnetic vector potential (left) and the temperature (middle). Linear growth rates from the JOREK reduced and full MHD models are compared to results from CASTOR3D. Re-print from Ref.~\cite{Pamela2020}. \reprintaipOK}
\label{fig:FMHD_KINKTEAR}
\end{figure}

\begin{figure}
\centering
\includegraphics[height=0.3\textwidth]{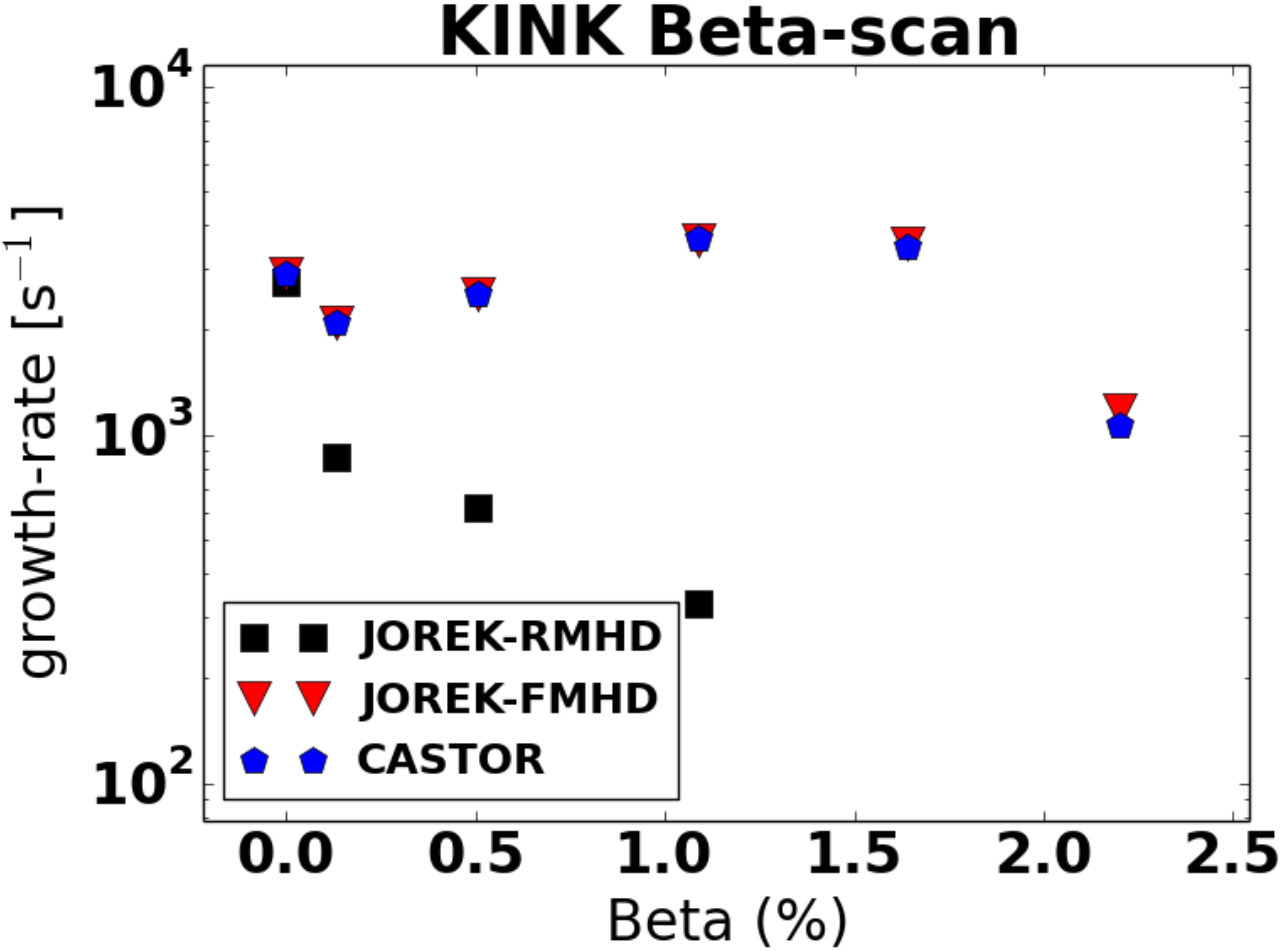}
\caption{Linear growth rates of an internal kink instability in a plasma with circular cross section as a function of normalized beta, obtained by the reduced- and full-MHD models of JOREK and the linear full-MHD code CASTOR3D.  Re-print from Ref.~\cite{Pamela2020}. \reprintaipOK}
\label{fig:FMHD_KINK_BETA}
\end{figure}

This section presents several benchmarks for core instabilities, to compare the full-MHD model against the reduced-MHD, as well as established linear MHD codes like MISHKA and CASTOR \cite{Mikhailovskii1997,Huysmans2001,Huysmans1993,Strumberger2017}. The first two linear benchmarks are a low-$\beta$ $m$=$n$=$1$ internal kink mode, and a low-$\beta$ $m$=$n$=$1$ tearing-mode. Both cases are run for a scan in resistivity. The kink mode is run with resistivity alone (without viscosity, and without particle or thermal diffusion), while the tearing mode is run including all diffusive terms, with viscosity $\nu_{_0}$=$10^{-8}\,\textnormal{kg\,m}^{-1}\,\textnormal{s}^{-1}$, particle diffusion $D_{\bot}$=$0.7\,\textnormal{m}^2\,\textnormal{s}^{-1}$, and heat conduction $\kappa_{\bot}$=$1.7$$\,\times$$10^{-8}\textnormal{kg\,m}^{-1}\,\textnormal{s}^{-1}$.

Figure~\ref{fig:FMHD_KINKTEAR} shows benchmarks for an internal kink mode and a tearing mode case respectively. Poloidal cross-sections of $n=1$ perturbed quantities are shown for the toroidal magnetic potential $A_{\phi}$ and the temperature (for the full-MHD model), and the growth rates of the modes are plotted as a function of resistivity, compared to the reduced-MHD model, and to calculations from CASTOR3D, which is also a full-MHD code \cite{Huysmans1993,Strumberger2017}.

Although the agreement between reduced-MHD and full-MHD is reasonable for both cases, the reduced-MHD model starts to deviate from the full-MHD solution at low resistivity for the internal kink mode. This is an example where reduced-MHD becomes insufficient: internal kink modes at finite-$\beta$ are not well represented like explained in Ref.~\cite{Graves2019}. Although this is a low $\beta_{_N}$=$0.4\%$ case, reduced-MHD already seems to be affected. At higher-$\beta$, the deviation becomes more pronounced as can be seen in the benchmark between the reduced- and full-MHD model of JOREK and the CASTOR3D code shown in Figure~\ref{fig:FMHD_KINK_BETA}. In this case, the linear growth rate of the internal kink instability is compared as a function of normalized beta. At finite beta, the reduced-MHD model clearly fails to reproduce the results of the full-MHD calculations.

\begin{figure}
\centering
 \includegraphics[width=0.85\textwidth]{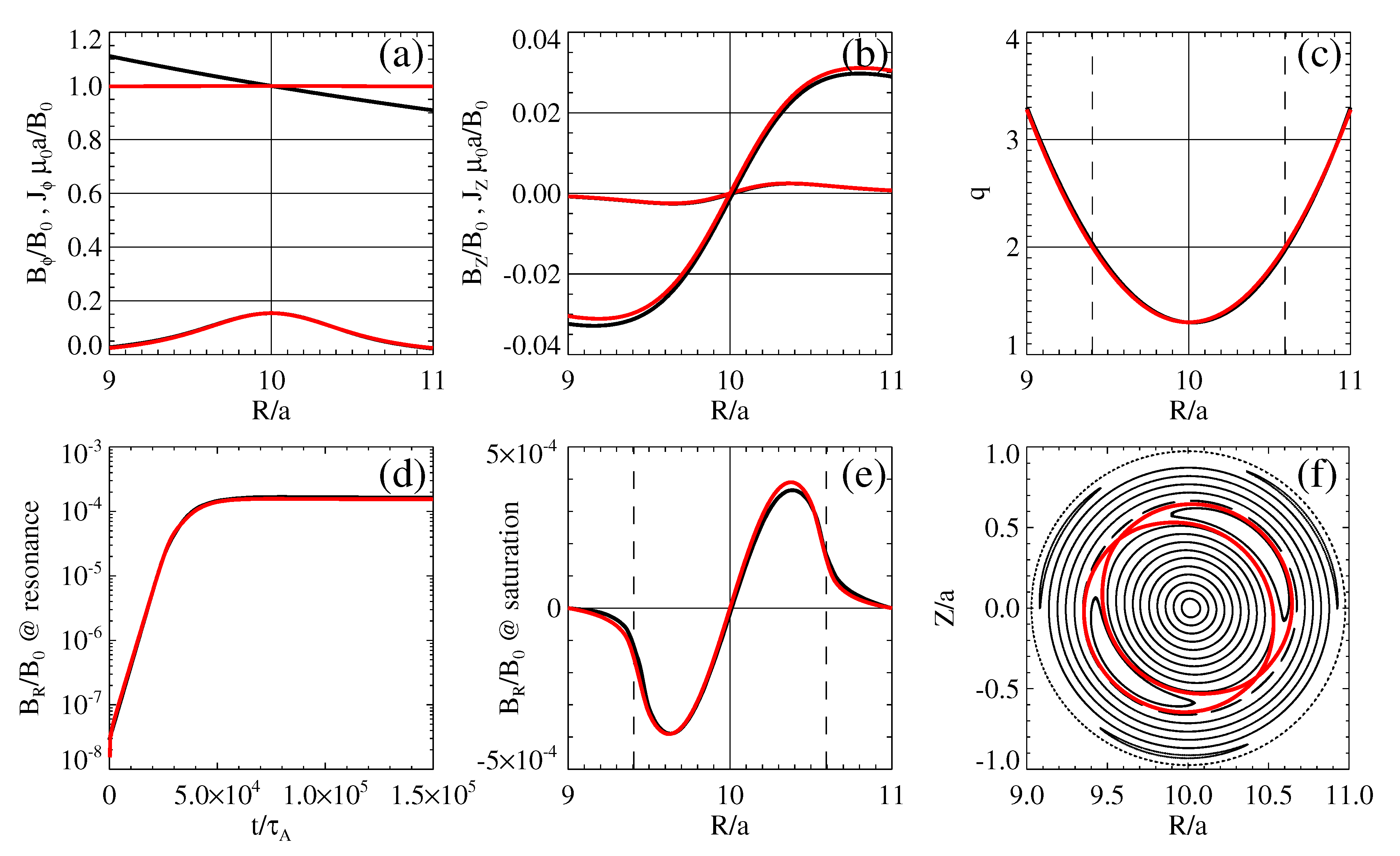}
\caption{Nonlinear verification benchmark between JOREK and the cylindrical full-MHD code Specyl. A \revised{$2/1$} tearing mode is considered in a circular tokamak with large aspect ratio $R/a=10$ in the zero-$\beta$ limit. Black and red curves are for JOREK and SpeCyl calculations, respectively. Initial equilibrium profiles along the horizontal diameter are shown in the first row: \textnormal{(a)} magnetic field and current components in the toroidal direction, \textnormal{(b)} magnetic field and current components in the $Z$ direction, and \textnormal{(c)} safety factor. The tearing mode nonlinear evolution is shown in the second row: \textnormal{(d)} temporal evolution of $B_R$ at the $q=2$ rational surface, \textnormal{(e)} $B_R$ profiles along the mid plane at in the non-linearly saturated state, and \textnormal{(f)} Poincar{\'e} plot at nonlinear saturation from JOREK, with the separatrix of the $2/1$ saturated island from SpeCyl overplotted in red. \noreprint}
\label{fig:nonlinear_benchmark_JOREK-SpeCyl}
\end{figure}

The resistive layer width for a tearing mode was simulated with JOREK visco-resistively in Ref.~\cite{Pratt2016} and compared to analytical scaling laws as well as the Phoenix code showing excellent agreement.
Furthermore, a non-linear simple tearing mode benchmark is shown in the following. JOREK is compared here to the cylindrical full-MHD code SpeCyl~\cite{Cappello1996} in simple geometry. The nonlinear verification benchmark of SpeCyl with another MHD code, PIXIE3D, is reported in Ref.~\cite{Bonfiglio2010}. Here, we consider the nonlinear saturation of a $m=2$, $n=1$ tearing mode in a circular tokamak with large aspect ratio in the zero-$\beta$ limit. The simulations are performed at Lundquist number $S=10^6$ in the limit of negligible viscosity. The results from the two codes are compared in Figure~\ref{fig:nonlinear_benchmark_JOREK-SpeCyl}. It is observed that, despite the different geometry (toroidal vs. cylindrical) and models (reduced vs. full-MHD), the two codes agree very well for this strong guide-field, large aspect-ratio problem. In particular, both the linear growth and the nonlinear saturation of the tearing mode resulting from the two codes turn out to be in \revised{very good quantitative agreement (the maximum deviation of about 5\% can be attributed to the difference between cylindrical and large-aspect ratio geometries)}.

\subsection{Edge instabilities}\label{:verification:edge}


\begin{figure}
\centering
  \includegraphics[width=0.6\textwidth]{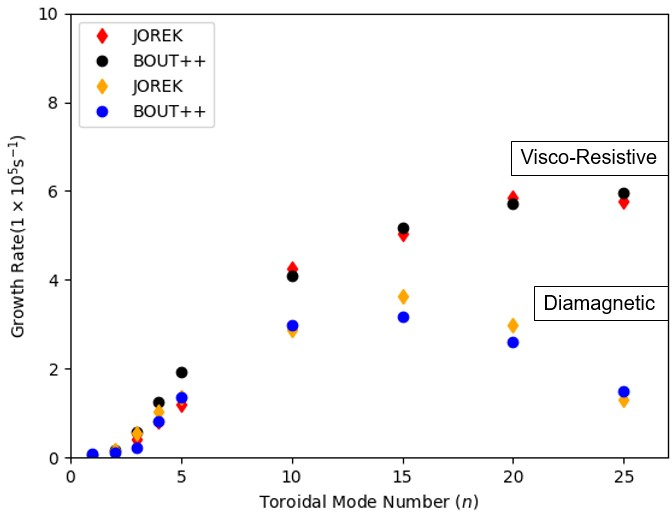}
\caption{Linear growth rates for the different toroidal mode numbers are compared between BOUT++ and JOREK for visco-resistive simulations, and for visco-resistive simulations including diamagnetic flows. \noreprint}
\label{fig:boutbench}
\end{figure}

Comparisons have been made between JOREK and BOUT++ with the aim to validate the nonlinear MHD codes. BOUT++~\cite{Dudson2009} is a framework for plasma fluid simulations, the model used for the comparison is given in Ref.~\cite{Dudson2011} and for JOREK in Section~\ref{:code:models:base}. The comparison was performed with a largely simplified geometry, a circular plasma with an aspect ratio of 3.3, a $q_{95} = 3.0$, a uniform density ($1.0\times10^{19}$ m$^{-3}$), and a hyperbolic tangent fit is used for the temperature where $T_e+T_i = 19$ keV in the core decreasing to $94$ eV at the edge of the plasma. The simulations performed have a resistivity of $6.1\times10^{-6} \Omega$m and the kinematic viscosity is 117 m$^{2}$s$^{-1}$. The differences between the two codes are detailed in~\cite{SmithPhD} and include some deviations between the actual models (equations), evolution of equilibrium quantities, numerical methods, grids and boundary conditions.
Despite some differences, the linear benchmark showed \revised{agreement within 4\%} between BOUT++ and JOREK with and without diamagnetic effects as seen from Figure~\ref{fig:boutbench}. Setting up a non-linear comparison based on this case would require further efforts in resolving some of the differences between the codes, e.g., regarding the boundary conditions and hasn't been attempted as of now.


\begin{figure}
\centering
  \includegraphics[width=0.6\textwidth]{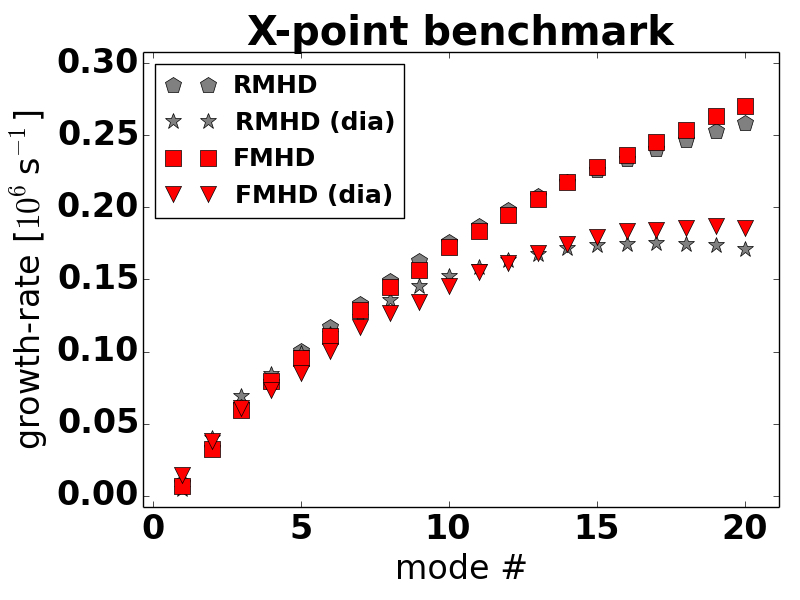}
\caption{The growth rate of the peeling-ballooning modes as a function of toroidal mode number $n$, with and without the diamagnetic effects, for both the reduced- and full-MHD models. Re-print from Ref.~\cite{Pamela2020}. \reprintaipOK}
\label{fig:FMHD_DIA}
\end{figure}

A benchmark of peeling-ballooning modes is done using an X-point JET-like plasma. Refer to Ref.~\cite{Pamela2020} for details regarding the benchmark configuration. Figure~\ref{fig:FMHD_DIA} shows the growth rate with and without diamagnetic effects, as a function of toroidal mode number, for both reduced- and full-MHD. \revised{The agreement between the two models is reasonable with deviations below 5\% (highest deviations for the largest mode numbers that are most critical in terms of resolution), particularly considering that this includes $E\times B$, parallel and diamagnetic flows in the pedestal and SOL.}

\revised{The ballooning mode rotation obtained in JOREK simulations has been successfully validated against analytical linear computations in Ref.~\cite{Morales2016}. In the laboratory frame, the $\mathbf{E}\times\mathbf{B}$ velocity should be added to the mode velocity when comparing with experimental measurements ~\cite{Becoulet2017}.}

\subsection{Scrape-off layer (SOL)}\label{:verification:sol}

The SOL modelling in JOREK at this point is still fairly simplified, while several attempts are presently on their way to improve -- via a better representation of low temperature physics in the fluid picture as well as via a kinetic treatment of neutral particles. In spite of the incomplete picture presently available, comparisons have been made between JOREK and SOLPS~\cite{XBONNIN2016} to get a first assessment of the status of the the JOREK SOL model. The comparison demonstrates the capabilities of the diffusive neutrals model~\cite{Smith2020}, which was used to perform the simulations. Full details and results are given in~\cite{SmithPhD}.

\begin{figure}
\centering
  \includegraphics[width=0.6\textwidth]{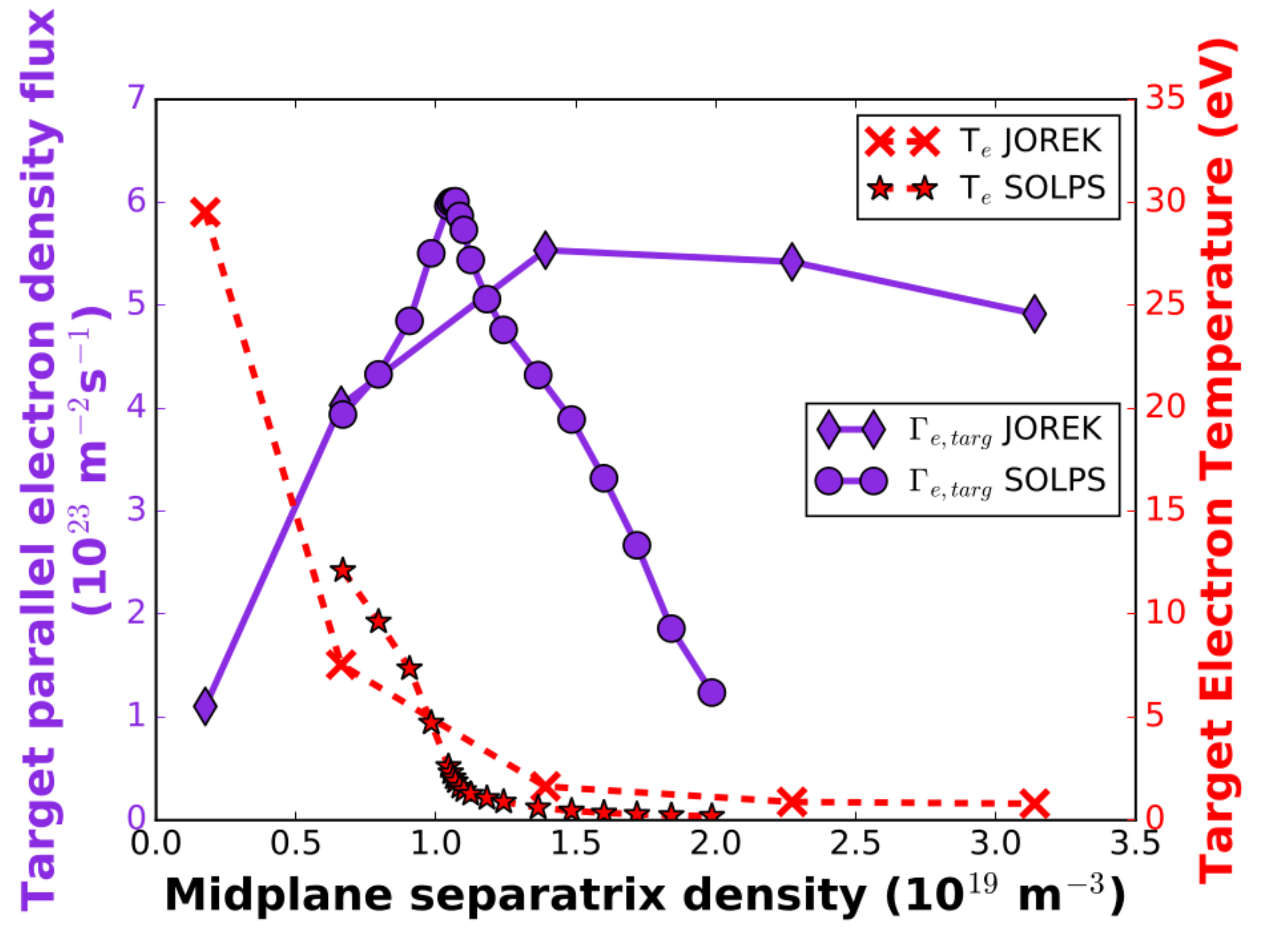}
\caption{The target parallel electron density flux and target electron temperature as a function of upstream density, comparing JOREK with SOLPS results from \cite{Moulton2017EPS} for a MAST-U Super-X H-mode case. Re-print from Ref.~\cite{Smith2020}. \reprintiaea}
\label{fig:SOLPScomparison}
\end{figure}

A double-null H-mode MAST-U Super-X case was used in an attempt to obtain a detached divertor for ELM burn-through studies. This case had a toroidal field of 0.64 T, plasma current 1 MA, $q_{95} = 7.9$, a central density and temperature ($T_{e}$ + $T_{i}$ ) of $5.2\times10^{19}$ m$^{-3}$ and 1.8 keV respectively. Parameter scans were performed and compared to the SOLPS results in \cite{Moulton2017EPS} to test the JOREK diffusive neutrals model. Figure~\ref{fig:SOLPScomparison} shows the upstream density scan. Due to the missing physics, in particularly charge exchange, it was not possible to obtain a roll-over as steep (deeply detached plasma) as the SOLPS simulations. Nevertheless, a roll-over in the target parallel electron density flux was obtained as the upstream density increased, the target electron temperature decreased to a few electron volts and the ionisation front moved from the target - indicating a detached regime was obtained.

\subsection{Runaway electron fluid}\label{:verification:refluid}

Verification of the  RE fluid model in JOREK was performed with respect to a) the conversion of thermal current into a RE beam in a simple axisymmetric plasma via comparison to a lower dimensional code, and b) regarding the linear growth of resistive kink modes. Further efforts are on their way.

\begin{figure}
\centering
\includegraphics[width=0.49\textwidth]{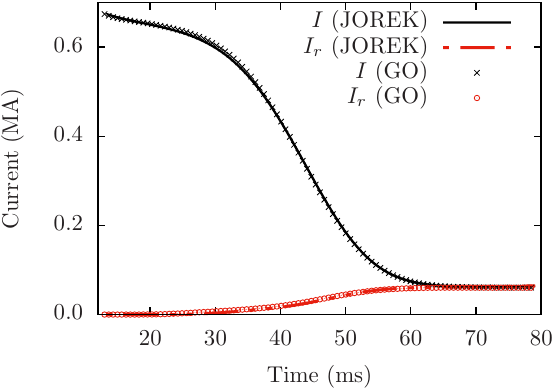}
\includegraphics[width=0.49\textwidth]{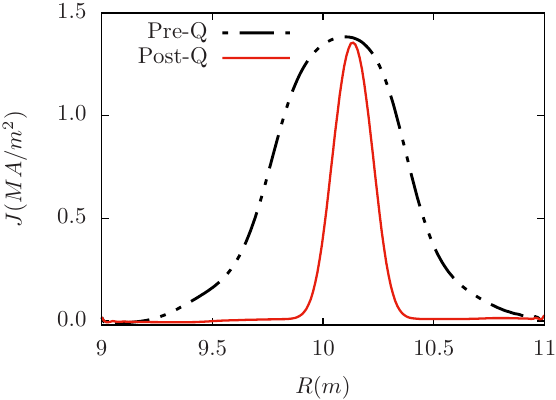}
\caption{\textbf{(Left)} Time evolution of the total plasma current $I$ and the RE current $I_{r}$ during the current quench phase. \textbf{(Right)} Midplane current density profiles before and after the current quench obtained from JOREK, showing a relatively peaked RE current profile. Re-print from Ref.~\cite{Bandaru2019A}. \reprintaps} 
\label{fig:jorek_vs_GO}
\end{figure}  

Benchmarking for the current conversion was done with the one-dimensional runaway electron code GO~\cite{Papp:2013}, by triggering an artificial thermal quench  in a large aspect ratio circular plasma ($R=10$m and $a=1$m) by imposing a large perpendicular thermal diffusivity. The parameters at the initial equilibrium state of the plasma were $I_p=\SI{0.67}{\mega\ampere}$, on-axis toroidal magnetic field $B_{\phi,0}=\SI{1}{\tesla}$, central temperature $T_0=\SI{1.7}{\kilo \electronvolt}$, central density  $n_0=\SI{1e20}{\meter^{-3}}$ and central resistivity $\eta_0=\SI{1.1e-7}{\ohm\meter}$.
The large perpendicular diffusion leads to a drop in the core temperature of the plasma to about $\SI{25}{\electronvolt}$ in a time of about $\SI{60}{\milli\second}$. Thresholds were set to initiate the Drecier generation when $E_\parallel/E_D \geq 0.01$ and the avalanching when $E_\parallel/E_c \geq 1.7$. The temperature profile evolution in GO is taken as an input from JOREK.  Figure~\ref{fig:jorek_vs_GO} (left panel) shows an excellent agreement between the result obtained with JOREK and GO for the evolution of the total and RE currents \revised{(deviations around 1\%)}. The simulations also show the often observed central peaking of RE current profile, as can be seen in the pre-quench and post-quench current density profiles shown in Figure~\ref{fig:jorek_vs_GO} (right panel). \revised{Advanced source terms that incorporate the influence of partially ionized impurities onto the RE generation~\cite{Hesslow2019} are presently being implemented and validated.}

\begin{figure}
\centering
\includegraphics[width=0.65\textwidth]{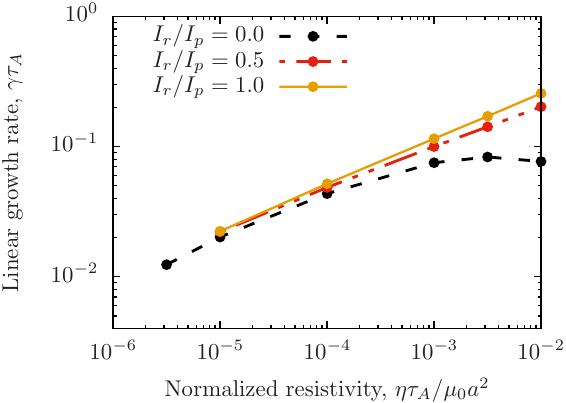}
\caption{Linear growth rate of the resistive internal kink $\left(1,1\right)$ mode as a function of the normalized resistivity at various initialized fractions of runaway current. Here, Alfv{\'e}n time $\tau_A=a\sqrt{\mu_0 \rho_0}/B$ and $\gamma$ is the growth rate in SI units. Re-print from Ref.~\cite{Bandaru2019A}. \reprintaps}
\label{fig:LGR_vs_eta_kink}
\end{figure} 

Verification with respect to linear growth of the resistive internal kink mode is done by considering that a certain fraction of the equilibrium plasma current is assumed to be carried by REs, wherein the RE current density has qualitatively the same profile as the total current. Both, thermal and RE background current densities are kept fixed in time. A large aspect ratio circular plasma ($R=\SI{10}{\meter}$ and $a= \SI{1}{\meter}$) in a fixed-boundary static equilibrium ($\bm{v}=0$) is chosen with parameters $B_{\phi,0}=\SI{1}{\tesla}$, $I_p=\SI{0.31}{\mega\ampere}$ and on-axis temperature $T_0=\SI{48}{\electronvolt}$. The equilibrium is $\left(m=1,n=1\right)$ kink unstable with the $q=1$ surface within the plasma. 
Thermal and mass diffusivities, all the sources (including RE generation) and RE advection were set to zero, while the resistivity was assumed to be temperature independent and spatially constant. Figure~\ref{fig:LGR_vs_eta_kink} shows the linear growth rate of the internal kink mode as a function of normalized resistivity (inverse Lundquist number $S^{-1}$) for the various fractions of RE current considered. It can be observed that an increase in the RE current fraction leads to a gradual recovery of the low resistivity analytical scaling $S^{-1/3}$ even at large values of the normalized resistivities. This is primarily due to the reduced effective resistivity in the presence of REs. That is, when the RE current fraction is increased, the region outside the resistive layer tends towards the ideal MHD limit in which the low-resistivity analytical scaling is valid. Such a behaviour has also been observed by Matsuyama et al~\cite{Matsuyama:2017} in a similar (but not identical) case.

\subsection{Free boundary simulations}\label{:verification:freeb}

Regarding free boundary simulations, a lot of verification work has been performed. This includes the benchmarks shown in Ref.~\cite{Hoelzl2012B} for a free boundary equilibrium and the linear growth rates of tearing modes in the presence of a conducting wall. In Ref.~\cite{Hoelzl2014}, the growth rate of resistive wall modes was compared to analytical values and a comparison of VDEs in a simplified ITER-like geometry was compared to the CEDRES++ code including realistic wall resistances. \revised{Basic verification tests (not shown here) have been performed including convergence checks, the induction of currents by the plasma in a conducting wall or passive coil, the mutual interaction between conductive structures and a comparison of an axisymmetric and non-axiymmetric coil field on the JOREK boundary with Biot-Savart~\cite{ArtolaPhD,SchwarzMaster,MitterauerSchwarzArtolaHoelzlCoils}.}
Further validation work is included in Ref.~\cite{ArtolaPhD} and some more recent and more advanced tests are shown in the following.

\begin{figure}
\centering
  \includegraphics[width=0.8\textwidth]{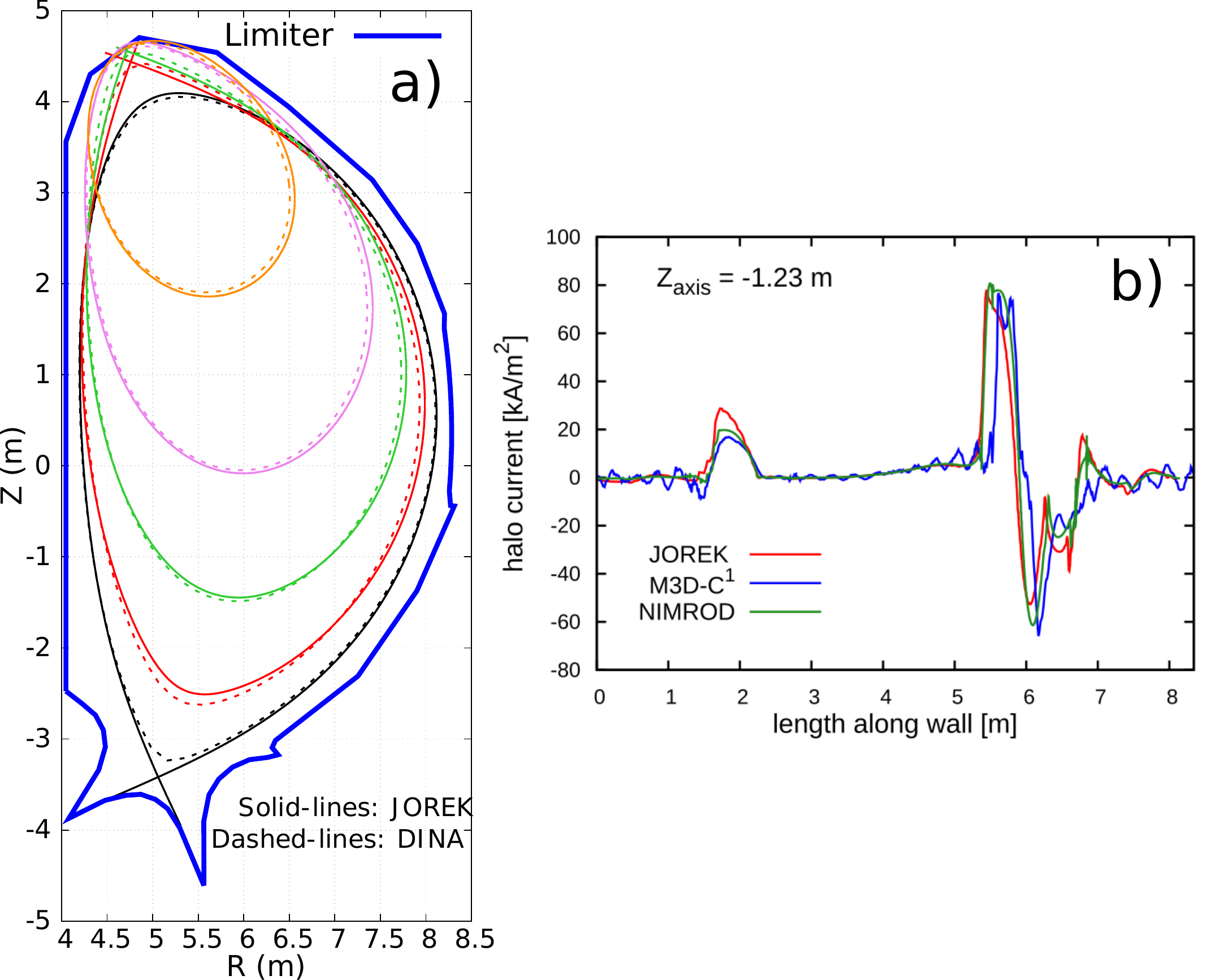}
\caption{Non-linear 2D axisymmetric benchmarks for VDEs. (a) JOREK/DINA benchmark where the evolution of the LCFS is compared \cite{ArtolaPhD}. (b) JOREK/NIMROD/M3D-C$^1$ benchmark where the normal component of the current density is compared as a function of the distance along the wall during the late stage of a VDE~\cite{Krebs2020}. Panel (b) is a re-print from Ref.~\cite{Krebs2020}. \reprintaipOK}
\label{fig:VDE_2D_nonlinear_bench}
\end{figure}
A non-linear benchmark with the DINA code \cite{ArtolaPhD} showed good agreement for the evolution of the LCFS during an ITER 15 MA axisymmetric VDE as shown in Figure~\ref{fig:VDE_2D_nonlinear_bench}a \revised{with differences between the shapes of the last closed flux surfaces of only few centimeters}.

An international benchmark~\cite{Krebs2018IAEA,Krebs2020} between the JOREK, M3D-C$^1$ and NIMROD codes for an axisymmetric VDE in an NSTX-like plasma has shown excellent agreement not only on the growth rate of the instability, the evolution of the radial and vertical position of the magnetic axis as well as of the toroidal plasma and wall currents, but also on the spatial structure of the halo currents flowing into the wall (see Figure~\ref{fig:VDE_2D_nonlinear_bench}b). The agreement is remarkable in particular, since the models implemented in the codes differ significantly (e.g., full MHD versus reduced MHD; Greens functions approach versus a direct discretization of the vacuum region with conductors in the computational domain). 
  
\begin{figure}
\centering
  \includegraphics[width=0.75\textwidth]{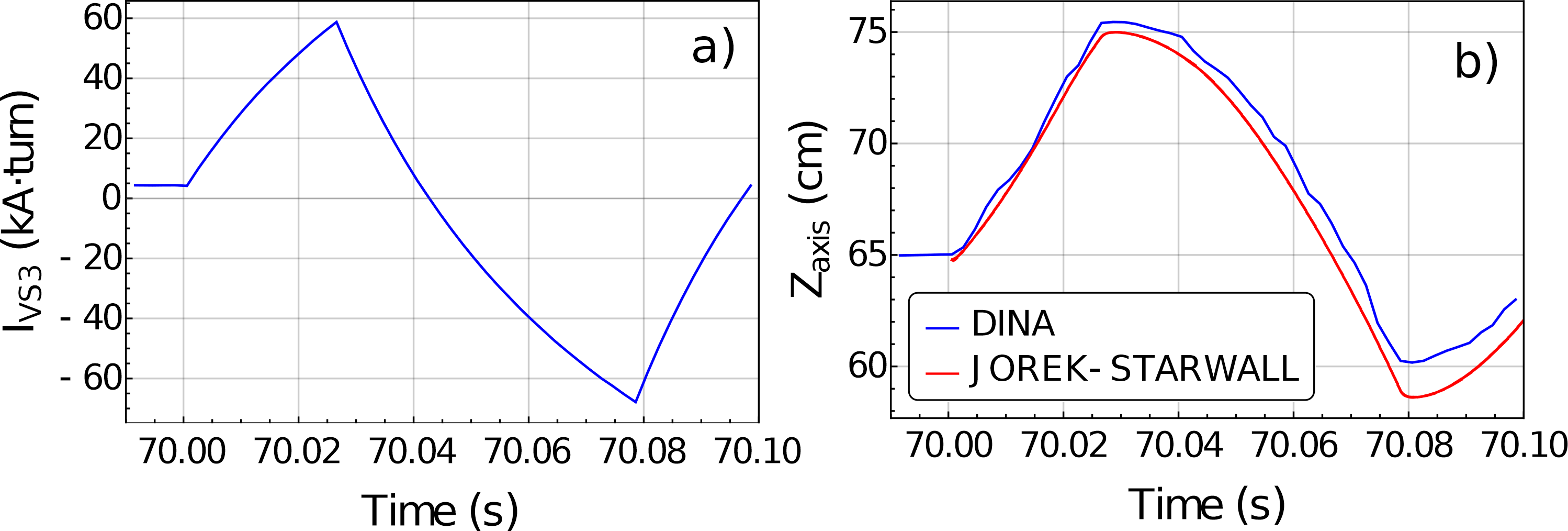}
\caption{Non-linear 2D axisymmetric benchmark for a vertical position oscillations of an ITER 7.5MA/2.65T plasma. (a) Current waveform for the ITER vertical position control coil \textit{VS3}. (b) Vertical position of the magnetic axis in DINA and in JOREK. Panel (b) is a re-print from Ref.~\cite{Artola2018}. \reprintiaea}
\label{fig:DINA_kick_bench}
\end{figure}

\revised{To test the mutual interaction between plasma, coils, and passive conductors, a benchmark was performed with the DINA code. The plasma motion resulting from a prescribed time evolution of currents flowing in coils was simulated} to verify that the plasma dynamics are captured correctly in the presence of time varying coil currents~\cite{Artola2018}. For that purpose, an ITER 7.5MA/2.65T case was chosen where \revised{the plasma position was modified in time by the prescribed coil currents} for ELM control studies. The oscillation in the vertical position is produced by applying the current waveform shown in Figure~\ref{fig:DINA_kick_bench}a in the in-vessel vertical stability coils (VS coils). \revised{To get good agreement for the plasma motion (Figure~\ref{fig:DINA_kick_bench}b), active and passive conductive structures had to be modelled accurately and the} interaction between coils, passive structures and plasma \revised{needed to be captured correctly} (see Figure~\ref{fig:kick} for the setup of the conducting structures) \revised{as the time evolving coil currents induce wall currents which further modify the overall plasma motion.} For this simulation, the ITER vacuum vessel was modelled as two thin stainless steel toroidal shells with a width of 6 cm each, the conducting OTS (outer triangular support) and DIR (divertor inboard rail) are also included.

\begin{figure}
\centering
 \includegraphics[height=0.7\textwidth]{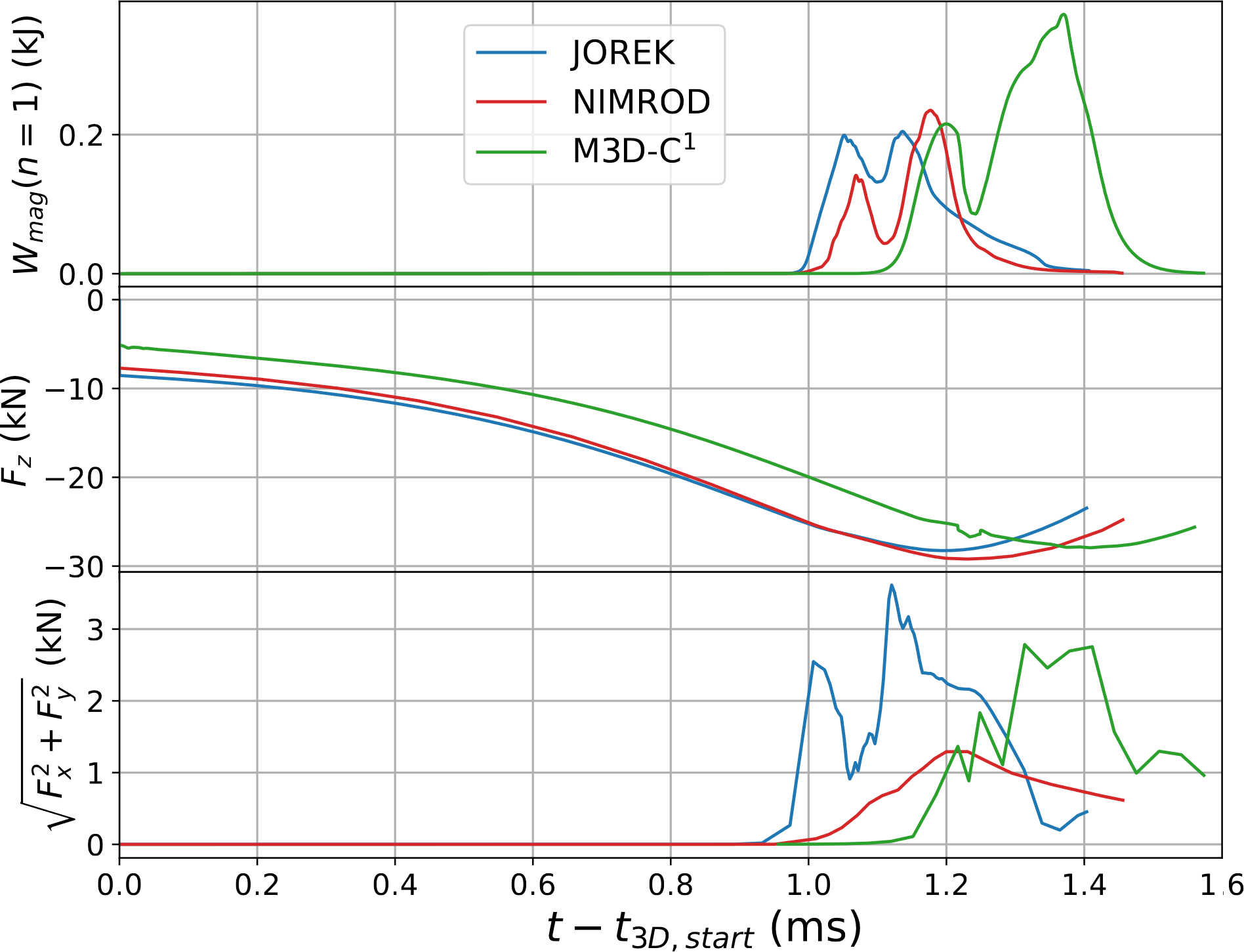}
\caption{3D Vertical Displacement Event benchmark between JOREK, NIMROD and M3D-C$^1$. The case and comparison is explained in detail in Ref.~\cite{Artola2020B}. \textbf{(Top)} Magnetic energies of the dominant modes ($n=1$ and $n=2$). \textbf{(Middle)} Thermal energy. \textbf{(Bottom)} Total horizontal wall force in JOREK, NIMROD and M3D-C$^1$. Re-print from Ref.~\cite{Artola2020B}. \reprintunpub}
\label{fig:3DVDE_bench}
\end{figure}

\begin{figure}
\centering
 \includegraphics[height=1\textwidth]{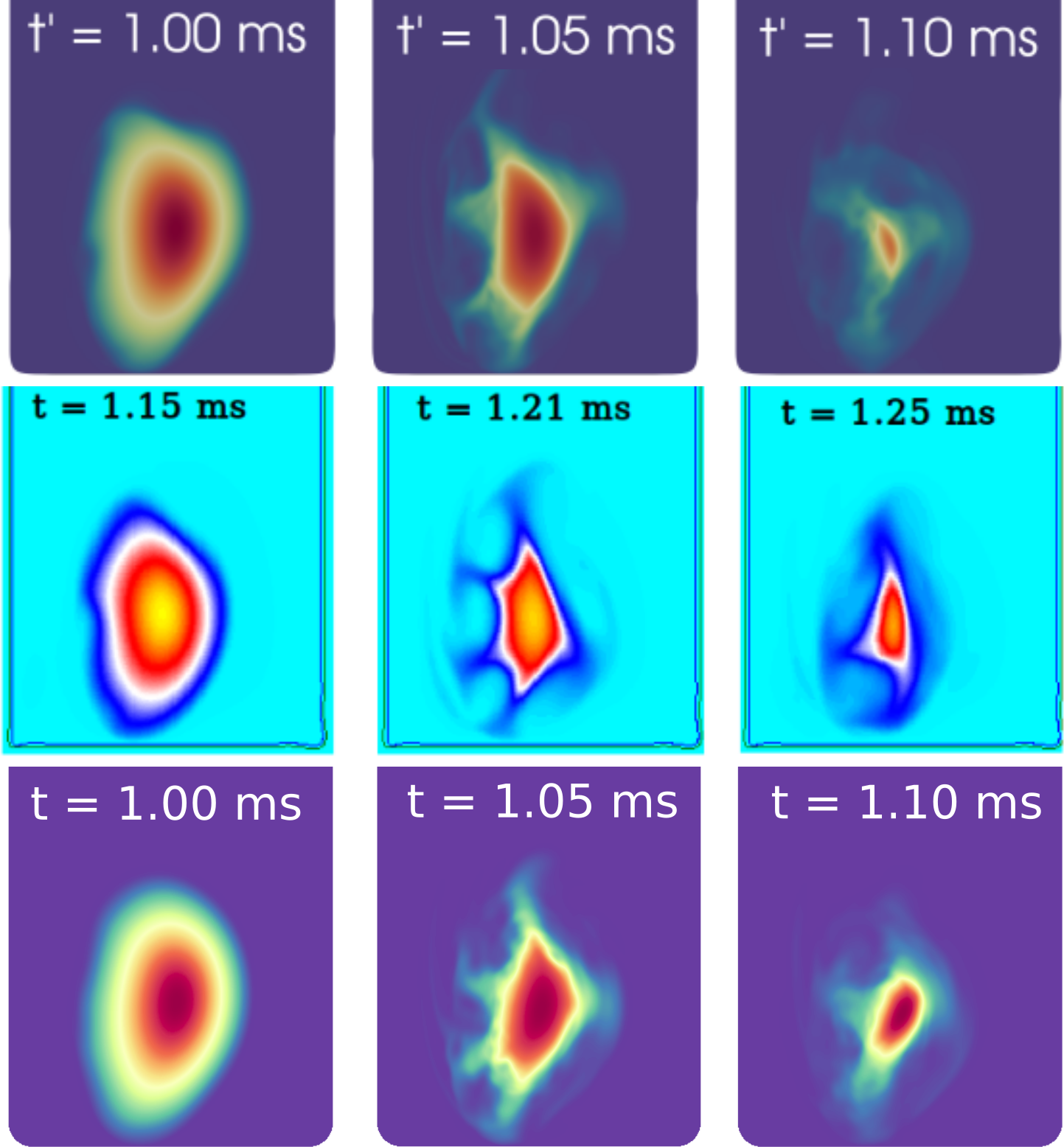}
\caption{Evolution of the pressure in the $\phi=0$ plane over time as computed by JOREK (top), M3D-C$^1$ (middle) and NIMROD (bottom) in arbitrary units. Re-print from Ref.~\cite{Artola2020B}. \reprintunpub}
\label{fig:3DVDE_bench_pressure}
\end{figure}

The international 3-code axisymmetric benchmark between the JOREK, M3D-C$^1$ and NIMROD codes shown above has been recently extended to 3D VDEs~\cite{Artola2020B}. The run was divided into two phases: an axisymmetric part for the early evolution of the VDE and a 3D simulation for the MHD active phase. The 3D run was started when the plasma became limited by the wall instead of the lower X-point. The toroidal harmonics that have been included in JOREK and NIMROD for the 3D phase are $n\in [0,10]$ and M3D-C$^1$ discretized the toroidal direction with 16 Hermite cubic elements.
The JOREK simulation was run with the reduced MHD model and the other codes employed full MHD models. In spite of pronounced differences on \revised{models for plasma and wall as well as} numerical methods, the 3-dimensional features are in very good \revised{qualitative} agreement. The three codes show that the plasma is unstable to low-n external kink modes 0.85 - 1.1 ms after the plasma becomes limited by the wall with a dominant n=1 component of similar amplitude in all codes. This  happens  when  the $q=2$ rational surface moves into the open field-line region (see Figure~\ref{fig:energy_conservation}). The development of the kink modes causes a stochastization of the magnetic field lines triggering a thermal quench with a duration of 0.14 ms in JOREK, 0.24 ms in M3D-C$^1$ and 0.20 ms in NIMROD. The evolution of the thermal pressure during the thermal quench is shown in Figure~\ref{fig:3DVDE_bench_pressure}, revealing similar filamentary structures for the three codes. Figure~\ref{fig:3DVDE_bench} shows the evolution of the magnetic energy of the \revised{$n=1$ harmonic and the vertical and horizontal wall forces in the three codes. The vertical force is in good agreement between the codes throughout the simulations except for a small shift in time} and the more sensitive horizontal force is in \revised{reasonable quantitative agreement with NIMROD showing a factor two smaller force than the other two codes. In spite of some differences, the benchmark demonstrates that the very different numerical descriptions of the resistive wall structures used in the codes lead to comparable results for such a violent 3D VDE case and that the ansatz based reduced MHD model used here for the JOREK simulation} is capturing the 3D dynamics of the wall currents, even for the large $\beta$ spherical plasma considered here. The full MHD models of M3D-C$
^1$ and NIMROD found maximum toroidal asymmetries in the plasma current ($I_p$) of 1-4\%. However such asymmetries were not found in the JOREK simulations due to the employed B.C. for the electric potential in the reduced MHD model ($\Phi=0$) .

\begin{figure}
    \centering
    \includegraphics[width=0.35\textwidth]{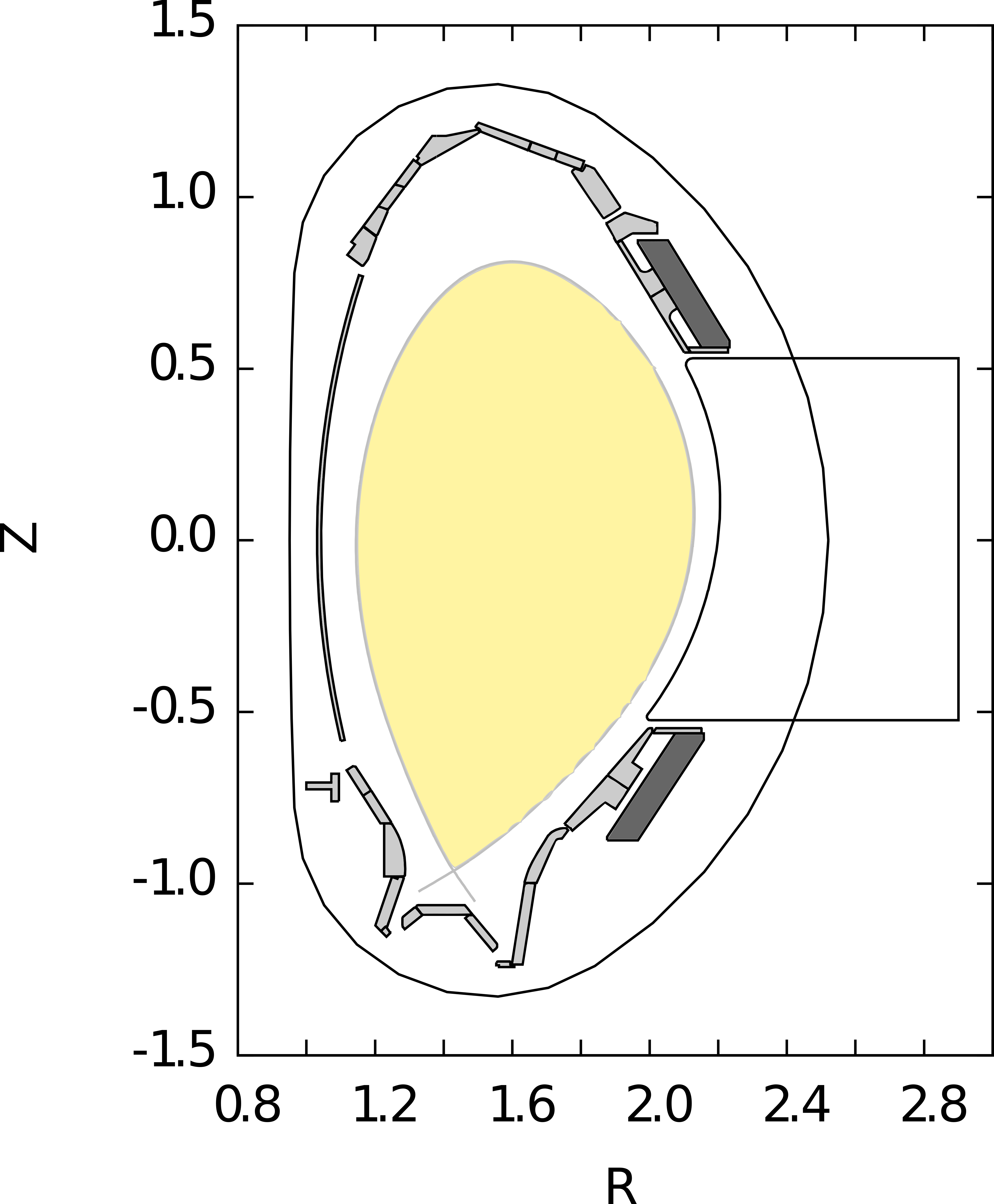}    
    \includegraphics[width=0.55\textwidth]{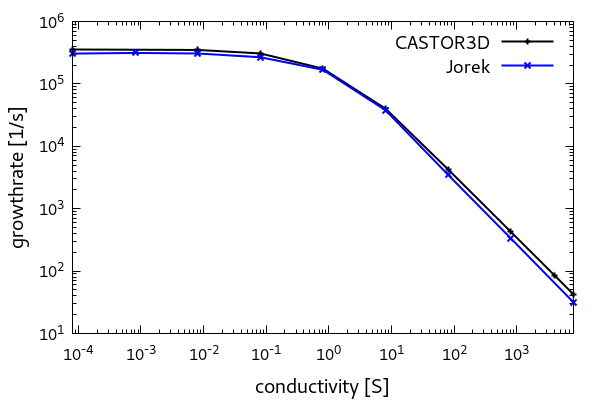}
    \caption{The figure on the left shows the conductive structures in ASDEX Upgrade including the PSL (massive structures above and below the sketched port marked in dark grey). As the conducting vacuum vessel is located at a large distance from the plasma, the PSL is needed to reduce the growth rate of the vertical motion. On the right, the growth rate for different PSL conductivities is shown. For a low conductivity, the growth rate is limited by the inertia of the plasma as the coil resistance is too high to allow for significant currents. For large conductivities, the PSL resistance determines the growth rate. \noreprint}
    \label{fig:bench_fb_castor}
\end{figure}

A benchmark with CASTOR3D was carried out to test the effect of passive coils on the plasma in an ASDEX Upgrade case where the main passive stabilization against vertical displacement is due to the Passive Stabilisation Loop (PSL). The conductivity was varied to compare the dependence of the VDE growth rate on this conductivity.
In the setup, no conducting wall was included for simplicity to only assess the coil effects. A high plasma resistivity in the SOL region was required to get agreement with CASTOR3D, which treats the SOL as part of the vacuum.
The results in Figure~\ref{fig:bench_fb_castor} show good agreement in the linear growth rates over many orders of magnitude in the PSL conductivity \revised{with largest deviations around 20\% in the limit of high PSL conductivity where small differences in numerical treatment of the PSL geometry are likely dominating the errors}. This includes the no wall limit, where the growth rate is not depending on the PSL conductivity any more, but on the plasma inertia. 

\revised{

}

\subsection{Kinetic particles}\label{:verification:particles}

\revised{
For non-relativistic particles, solvers are available for both full orbit and guiding respectively gyro centers. The well-known Boris method is used for the full orbit following. The implementation of the Boris method has been verified in detail in Ref.~\cite{vanVugtPhD}. For guiding centers, both the standard Runge-Kutta (RK4) and the variational method by Qin~\cite{Qin2009} in the modified form~\cite{KrausPhD} have been implemented and verified. For the gyro-center following, RK4 with an orbit averaged electric field is used.
The solvers have been verified following a banana orbit for 26 periods in a static tokamak equilibrium without electric field. This standard benchmark case has been described in detail in Refs.~\cite{Qin2009,KrausPhD}. The error is defined as the time averaged difference with the initial values. The scalings follow the expected order as a function of the time step. The Boris method conserves energy to a level close to machine precision. The error in the toroidal momentum scales quadratically with the time step. The RK4 method, being a fourth order method, should yield an error scaling with the 5th power of the time step, but integrated over a given time interval the error is expected to scale as the 4th power. Figure~\ref{fig:error_equilibrium_fields} shows a scaling between 4th and 5th order for the error in both the toroidal momentum as well as the energy. The variational method is expected to conserve the toroidal momentum up to machine precision. The observed error actually decreases with increasing time step. The error in the energy scales quadratically up to a time step  of 1000 gyro-periods. For larger time steps the modified Qin method is unstable (in agreement with Ref.~\cite{KrausPhD}).
As the amount of work (i.e.\ computing time) is about equal for the RK4 and Qin methods, it appears that RK4 is significantly better for all values of the time step. However the variational method has the advantage that the error is bound in time, i.e.\ the error will not increase when the integration time is increased. This is in contrast with the error in the RK4 method which tends to increase with time. In this particular case, the error of the variational integrator becomes smaller than the RK4 error for an integration time of the order of 0.1 s.   
\begin{figure}
    \centering
    \includegraphics[width=0.33\textwidth]{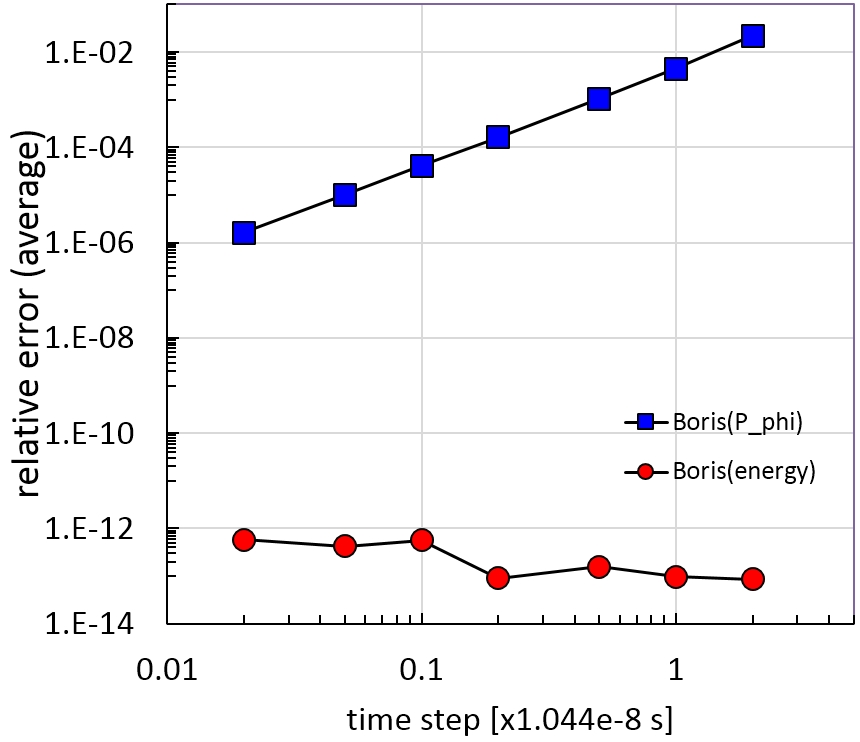} 
    \includegraphics[width=0.35\textwidth]{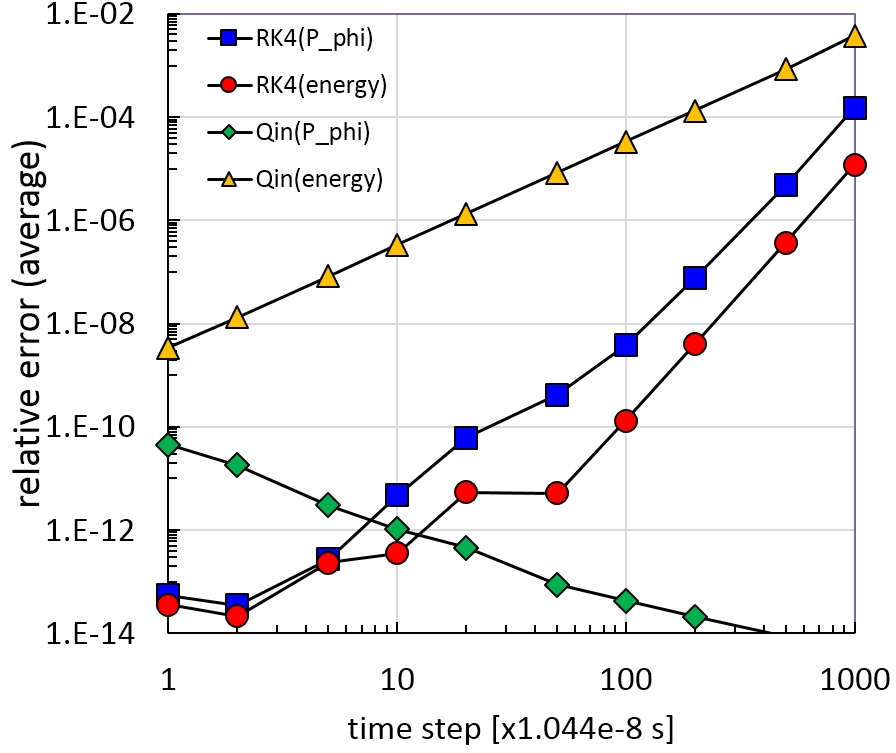}
\caption{Scaling of the relative error on energy and toroidal momentum for the Boris, RK4 and Qin solvers for single banana particle orbit (as defined in Refs.~\cite{Qin2009,KrausPhD}. The time step is normalised to the ion cyclotron frequency. The total integration time is fixed at $10^6$ (0.01s).}
\label{fig:error_equilibrium_fields}
\end{figure}
}
\subsection{Relativistic kinetic particles}\label{:verification:relativistic_particles}

\revised{
The relativistic particle tracing in JOREK has so far mainly been used to study RE confinement during plasma disruptions (Section~\ref{:applic:core:res}).
Similar studies have been done with an orbit-following code ASCOT5~\cite{Sarkimaki2020, ascot5}.
In both codes, full gyro-orbits are solved with the volume-preserving algorithm~\cite{zhang2015volume} and the (fixed time-step) guiding center orbit is solved with RK4.
The main difference between ASCOT5 and JOREK-particles is that ASCOT5 does not use finite elements for the calculation of fields; the magnetic field is interpolated in a uniform cylindrical grid with cubic splines instead.
Because both codes are being used to study transport of REs in a perturbed magnetic field, it is reasonable to verify that the codes yield equivalent results.

The thermal quench phase of a JET disruption was chosen as test case, where the stochastic field line region extended all the way from the edge to the core.
Both electric and magnetic field were included and the fields where set to be stationary in time.
JOREK postprocessing tools were used to evaluate the electromagnetic field on a cylindrical grid and the data was exported to ASCOT5.
It was verified that the grid was dense enough as not to affect the benchmark results significantly by repeating the simulations with different grid resolutions.

For the benchmark, a marker population of 5000 passing 1 MeV electrons was initialized on the outer mid plane on a fixed radial position. 
The markers were spread toroidally by sampling the initial toroidal coordinate from a uniform distribution.
The markers were traced until they exited the plasma, which was defined to occur when a marker passed a the poloidal flux value corresponding to a flux surface inside the separatrix.
The accumulation of the losses in time was used as measure for transport to compare the results.
For guiding center simulations, ASCOT5 was used to calculate the guiding center positions from the particle coordinates, and these guiding center positions were exported to JOREK to ensure that the initial transformation did not bias the results.

The results of the benchmark are shown in Fig.~\ref{fig:ascot_vs_jorek_comparison}.
The stepwise structure seen in JOREK results is due to (the simple method used here in) JOREK storing the marker status only at fixed time intervals, whereas ASCOT5 stores the exact loss-time of each marker at the end of the simulation.
In this test case, there is little difference between the full gyro-orbit and guiding center results in both codes.
Between JOREK and ASCOT5 there is a slight deviation, and it is not certain what is the cause.
One possibility is the fact that the poloidal flux in ASCOT5 is an axisymmetric quantity (the 3D components of the field are evaluated separately), and so the markers are not terminated at the same $(R,\;z)$-coordinates as can be seen in Fig.~\ref{fig:ascot_vs_jorek_comparison}~(b). Nevertheless, the results show good agreement between the different pushers in both codes.

\begin{figure}
    \centering
    \includegraphics[width=0.85\textwidth]{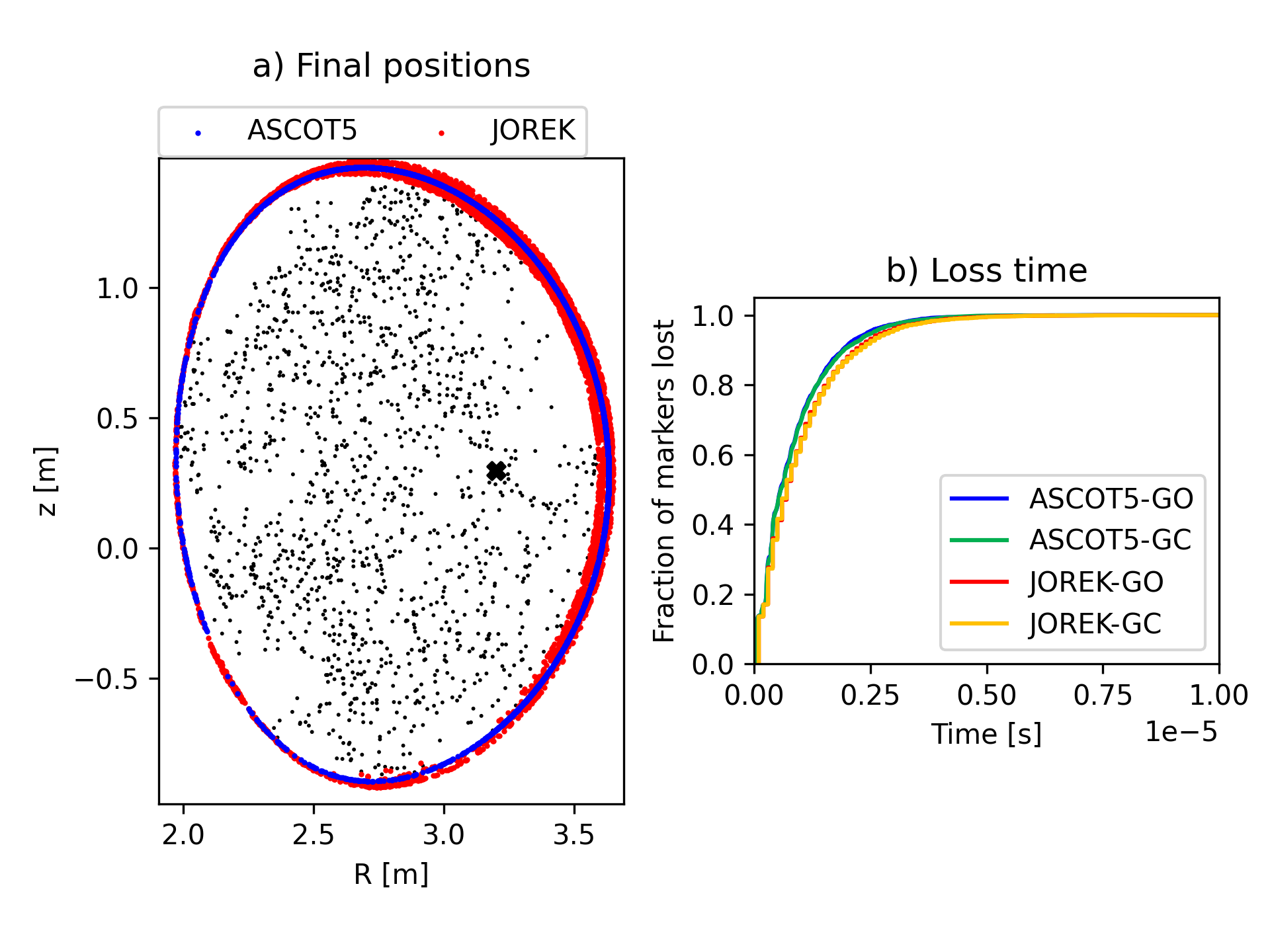}
    \caption{\revised{\textbf{(a)} The final marker position (blue and red dots) on the edge shown together with the magnetic field Poincar\'e plot (small black dots) and the position where markers where initialized (the black cross). \textbf{(b)} The cumulative losses as a function of time for each case. GO and GC refer to full gyro-orbit and guiding center simulation, respectively.} \noreprint}
    \label{fig:ascot_vs_jorek_comparison}
\end{figure}
}

%% file: 05_elms.tex
\section{Applications to ELMs and ELM control}\label{:applic:edge}

Unmitigated type-I edge localized modes (ELMs)~\cite{Zohm1996,Connor1998} cannot be tolerated in ITER at least in full current operation~\cite{Loarte2003} and even ELM types associated with reduced energy losses from the plasma and lower transient heat loads to plasma facing components, are likely not acceptable for DEMO~\cite{Wenninger2015}. Consequently, the dynamics of various ELM types, the access to ELM free regimes, and a robust understanding of ELM control methods are crucial research topics regarding the successful design and operation of future fusion devices. In this Section, JOREK simulations regarding the plasma pedestal, edge and scrape off layer are summarized focusing on recent results. Section~\ref{:applic:edge:elms} presents work on natural ELM crashes and cycles, Section~\ref{:applic:edge:pellet} contains results regarding the triggering of ELMs by pellets, Section~\ref{:applic:edge:kicks} shows the excitation of an ELM crash by a vertical magnetic kick, Section~\ref{:applic:edge:rmp} addresses the control of ELMs by RMP fields,  Section~\ref{:applic:edge:elmfree} presents work on the simulation of ELM free regimes, and Section~\ref{:applic:edge:detachment} shows recent results on an advanced modelling of SOL and divertor region as well as detachment physics. Section~\ref{:applic:edge:outlook} finally, contains a very brief outlook. \revised{For investigations of ELMs and ELM control with other non-linear MHD codes, refer to the review provided in Ref.~\cite{Huijsmans2015} and for more recent work to Refs.~\cite{Xu2016,King2017,Fil2017,Wilcox2017,Moyer2017,Canal2017,Wu2018,Ebrahimi2017,XuGS2019,HuQM2020,Li2021} and references therein.}

\subsection{Natural ELMs}\label{:applic:edge:elms}

The phenomenology of ELMs has been studied intensively since the discovery of H-modes in tokamaks~\cite{Wagner1982}. Several experimental observations like magnetic field, density and temperature perturbations localised on the Low Field Side (LFS) of the tokamak suggest that MHD instabilities are involved in ELMs phenomenology~\cite{Zohm1996,Connor1998}. In particular, linear ideal MHD theory and codes indicate that ballooning modes driven by large edge pressure gradient and kink-peeling modes driven by large edge bootstrap current, both typical for the pedestal region in H-mode scenarios, are the underlying physical instabilities of ELMs destabilisation~\cite{Zohm1996,Connor1998,Snyder2004}. However, only non-linear resistive MHD codes could explain the full non-linear dynamics of ELMs, reproducing in particular rotation of the modes, the mechanisms of density and temperature profiles relaxation, ELM cycling behaviour, heat and particle fluxes structure arriving into divertor and many other key experimental observations during ELMs. 

\begin{figure}
\centering
  \includegraphics[width=0.5\textwidth]{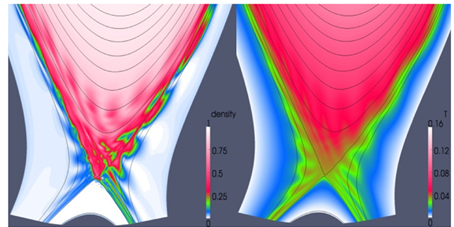}
\caption{Perturbed density and temperature in the poloidal plane at the time of the maximum of ELM magnetic perturbation for a JET-like 3MA plasma. Re-print from  Ref.~\cite{Huijsmans2013}. \reprintiaea}
\label{fig:fig_cond_convec}
\end{figure}

\begin{figure}
\centering
  \includegraphics[width=0.55\textwidth]{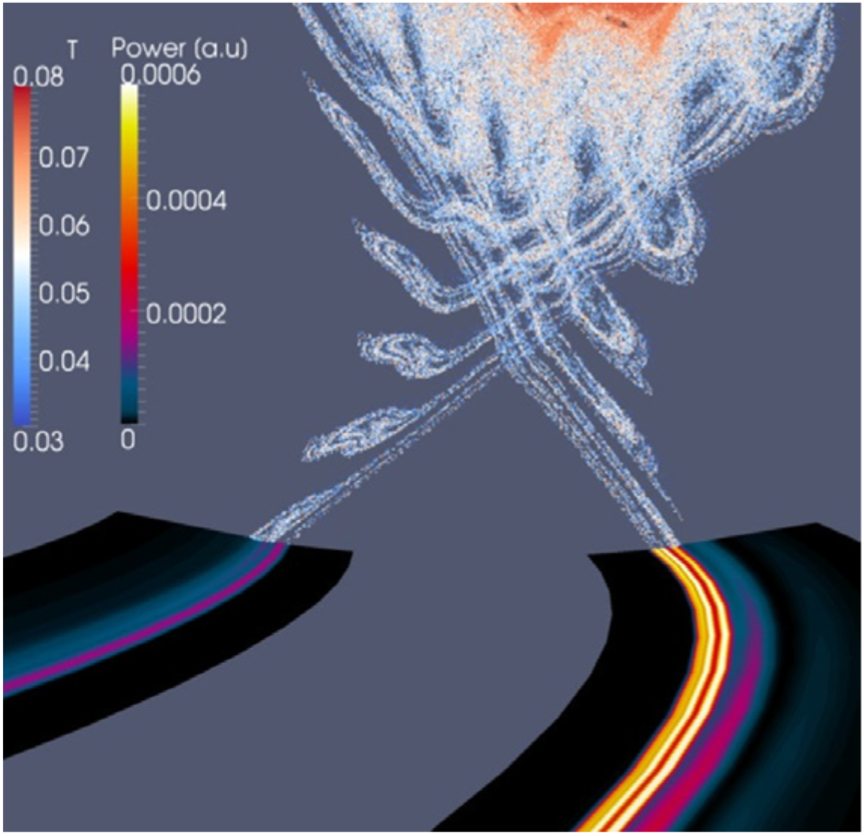}
\caption{The homoclinic tangles forming during an ELM crash in JET and the resulting strike line splitting are shown. Re-print from Ref.~\cite{Huijsmans2013}. \reprintiaea}
\label{fig:ELM-tangles}
\end{figure}

The first non-linear simulations of ELMs with JOREK were done without two fluid diamagnetic effects and are described in Refs.~\cite{Huysmans2007,Huysmans2005,Pamela2010,Huysmans2009,Huijsmans2013} for JET-scale plasmas. Indeed, it was confirmed that ELM offset is due to the destabilisation of coupled kink-peeling and medium-n ballooning modes and their linear phase is very close to the ideal MHD predictions. However, compared to the ideal linear MHD, the presence of the separatrix and plasma resistivity showed a strong stabilization of the ideal MHD external kink-peeling modes replacing them by the so-called peeling–tearing mode which is much less sensitive to edge safety factor q value compared to ideal MHD description. The different transport channels for thermal energy and particles during an ELM crash and resulting heat and particle fluxes in divertor were described in Ref.~\cite{ Huysmans2007,Huysmans2009,Huijsmans2013}. The non-linear evolution of a medium-n ballooning mode shows the formation of density filaments and, due to non-zero resistivity, causes magnetic reconnection leading to stochastic magnetic fields. Thereafter, the density evolution is mostly determined by the ExB convection cells (Figure~\ref{fig:fig_cond_convec}, left), while the temperature evolution (Figure~\ref{fig:fig_cond_convec}, right) is mainly dominated by parallel conduction in the stochastic fields formed during the ELM crash (Figure~\ref{fig:ELM-tangles}).

The density filaments are sheared off from the main plasma by a mean n=0 poloidal flow which is non-linearly induced via Maxwell stress~\cite{Huysmans2009} forming “blobs”, expelled from the main plasma towards the SOL. The amplitude of the ballooning mode is limited by this mean flow stabilising it and multiple (in time) density filaments can develop to bring the plasma below the stability boundary. The importance of poloidal flows onto ELM crashes on one hand and the generation of poloidal flows during the crash on the other hand was investigated also in Refs.~\cite{Pamela2010,PamelaPhD}. It was shown that large ELMs are mostly conductive-type, meaning relatively large losses in pedestal temperature, because the associated magnetic perturbations cause ergodisation of the edge and lead to significant conductive energy losses along perturbed field lines. The interception of the resulting homoclinic tangles with divertor plates (Figure~\ref{fig:ELM-tangles}) determine the pattern of the ELM heat flux in the divertor. Therefore explaining the splitting of the strike lines to the divertor during an ELM often observed in the experiments~\cite{Eich2017A}. The smaller ELMs are mostly convective-type with dominant density losses. These ELMs have divertor footprints determined by the radial distance travelled by plasma filaments expelled by an ELM and the loss of the plasma energy is determined by the energy stored in the expelled filaments. The simulated divertor wetted area during the ELM was shown to grow linearly with the ELM size in agreement with experimental observations. ITER simulations~\cite{Huijsmans2013} showed that for ELM losses up to 4 MJ of energy, the broadening of the wetted area is similar for conductive and convective ELMs, in spite of the different origin of the widening.

\begin{figure}
\centering
  \includegraphics[width=0.5\textwidth]{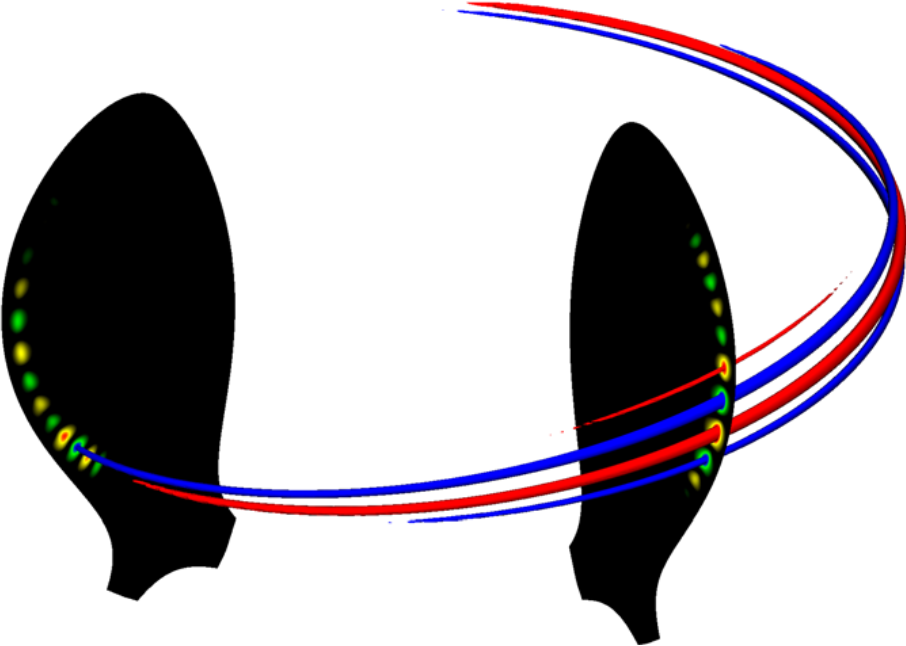}
\caption{The poloidally and toroidally localized magnetic flux perturbation in an ASDEX Upgrade simulation is shown, which has similarities to so-called solitary magnetic perturbations observed experimentally~\cite{Wenninger2012}. Re-print from Ref.~\cite{Hoelzl2012A}.  \reprintaipOK}
\label{fig:localizedELM}
\end{figure}

\begin{figure}
\centering
  \includegraphics[width=0.5\textwidth]{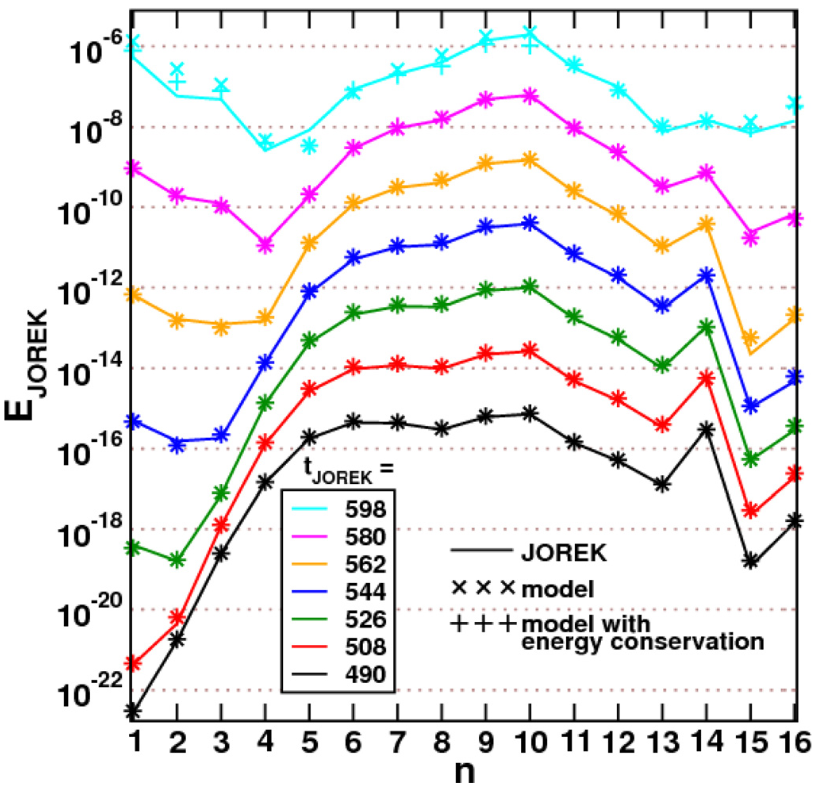}
\caption{For various time points during a JOREK simulation (times in JOREK units), the toroidal mode spectrum is shown. The linearly most unstable mode numbers around $n=10$ give rise to a broad mode spectrum in the non-linear phase at the ELM onset driven by non-linear mode coupling. In particular the linearly stable $n=1$ mode is driven to amplitudes comparable to the linearly dominant $n=10$ mode. Re-print from Ref.~\cite{Krebs2013}. \reprintaipOK}
\label{fig:ELMspectrumevolution}
\end{figure}

After the first confirmation of the generic features of full non-linear ELM dynamics using the JOREK code, extensive studies of ELMs physics in existing tokamaks (JET, ASDEX Upgrade, KSTAR, DIII-D, MAST) were performed with the aim of further validation and improvement of JOREK physical models for more confident predictive modelling for next-step machines and in particular ITER. Such extensive studies have also enabled the JOREK code to contribute to general ELM physics issues like the role of filamentary structures in ELM and ELM control \cite{Ham2020A}. In the following, we briefly describe some examples and main results. First simulations of type-I ELMs in the ASDEX Upgrade tokamak~\cite{Meyer2019} were shown in Ref.~\cite{Hoelzl2011}, where a spatial and temporal sub-structure of the ELM was observed leading to several bursts. This work pointed out the importance of including multiple toroidal harmonics in non-linear modelling to obtain a realistic dynamics of ELMs. After the early linear growth phase dominated by the most unstable medium-n ballooning mode, many other harmonics become destabilized through non-linear coupling while approaching the ELM crash. The non-linear destabilization of the low-n harmonics explained the poloidally and toroidally localized structures of the ELM crash~\cite{Hoelzl2012A} observed in experiment~\cite{Wenninger2012} as “solitary magnetic perturbations” (Figure~\ref{fig:localizedELM}).
The excitation of low-n harmonics during an ELM crash by non-linear mode coupling (Figure~\ref{fig:ELMspectrumevolution}) was studied in Refs.~\cite{Krebs2013,KrebsMaster}. \revised{Note two limitations of this study: ExB and diamagnetic background flows were not included and the toroidal resolution does not include higher harmonics of the linearly most unstable mode due to computational limitations. More recent work like~\cite{Cathey2020} has overcome both limitations.} Density filaments and their propagation into the scrape-off layer were compared qualitatively to the experiment in Ref.~\cite{Orain2016EPS}.
More recent simulations for ASDEX Upgrade were performed at realistic experimental parameters and with plasma background flows consistently included~\cite{Hoelzl2018}. Here, ELM losses, magnetic measurements~\cite{Mink2017}, and measurements for cold-front penetration~\cite{Trier2019} agree well with experiments. Based on such simulations, the role of density fluctuations onto ECE-Imaging measurements was studied~\cite{Vanovac2018A} and the excitation of parametric decay instabilities during electron cyclotron resonance heating was investigated~\cite{Hansen2020}. Further simulations confirmed the trend of decreasing dominant toroidal mode numbers with an increasing edge safety factor like experimentally observed~\cite{Mink2018}.

\begin{figure}
\centering
  \includegraphics[width=0.5\textwidth]{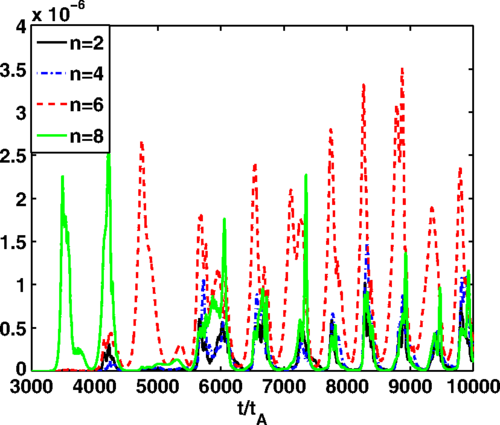}
\caption{High frequency multi-ELM cycles were \revised{obtained with JOREK} when background plasma flows were consistently included. Re-print from Ref.~\cite{Orain2015}. \reprintaps}
\label{fig:ELMcycles}
\end{figure}

The two fluid diamagnetic (for electron and ions), neoclassical and toroidal background plasma flows \revised{were implemented and applied to ELM simulations in Ref.~\cite{Orain2015} entailing an important} improvement of the physical model. A stabilizing effect of poloidal flows and in particular diamagnetic effects known from the ideal MHD model onto high-n ballooning modes was confirmed. Moreover, after the first ELM crash, poloidal flows tend to damp the otherwise continuing ballooning mode turbulence, so the pedestal profiles can be rebuilt again by heating and particle sources until they again reach the stability limit producing the following ELM crash. The numerically obtained multi-cycles ELMs exhibit similarities to high frequency ELMs in experiments (Fig.~\ref{fig:ELMcycles}). Moreover, the diamagnetic drifts were found to yield a near-symmetric ELM power deposition on the inner and outer divertor target plates, consistent with experimental measurements~\cite{Eich2005}.

\revised{The ballooning mode rotation obtained in JOREK simulations has been successfully validated against analytical linear computations in Ref.~\cite{Morales2016}. In successive studies the explanation of the rotating structures in the pedestal region before type-I ELM crashes and in the inter-ELM periods (ELM precursors) observed in the KSTAR tokamak~\cite{Park2019} was proposed~\cite{Becoulet2017,KimSK2020}. The two fluid diamagnetic effects and toroidal rotations included in the model were found to be the most important factors in explaining the experimentally observed rotating structures~\cite{Becoulet2017}.}

\begin{figure}
\centering
  \includegraphics[width=0.5\textwidth]{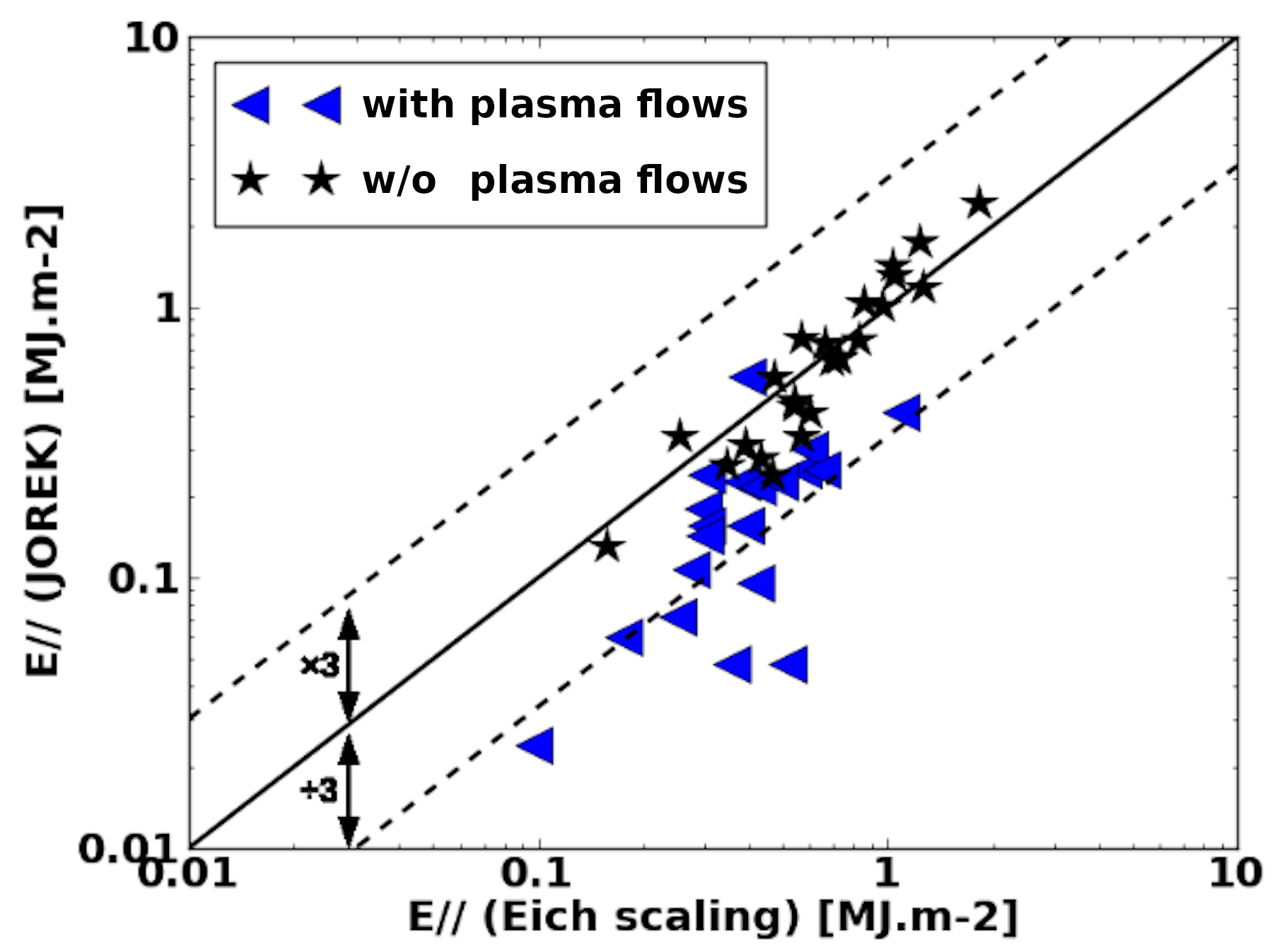}
\caption{The parallel heat fluence to the divertor targets is plotted versus the value predicted by the Eich scaling. Good agreement is obtained in general, while the heat fluence is underestimated in tendency, when plasma background flows are taken into account (blue triangles). Note a misprint in the original publication, which was corrected in the Figure shown here: the labels for the black and blue points had accidentally been swapped both in Figures 4 a and b of Ref.~\cite{Pamela2017}. Re-print with minor adaptations (key) from Ref.~\cite{Pamela2017}. \reprintiaea}
\label{fig:ELMheatfluence}
\end{figure}

Simulations aiming to capture ELMs in the JET tokamak~\cite{Litaudon2017} as realistically as possible were shown in Reference~\cite{Pamela2011}. The dynamics of filaments, divertor heat fluxes and the influence of collisionality onto the ELM size was studied and precursor modes \revised{were shown in ELM simulations}. Successively, Ref.~\cite{Pamela2015} addressed simulations for a series of JET ITER-like wall (ILW) discharges to perform quantitative comparisons with the experiments. It was shown that the accuracy of the pre-ELM equilibria used for the simulations as well as realistic parallel heat conductivity are both critical for reproducing experimentally observed ELM energy losses. Divertor peak heat fluxes were still underestimated and the ELM duration overestimated in the simulations. Ref.~\cite{Pamela2017} took more accurate pre-ELM equilibrium reconstructions as basis for the simulations and also investigated the impact of $\mathbf{E}\times\mathbf{B}$ and diamagnetic background flows onto the ELM dynamics. Very good agreement with the experimental scaling law~\cite{Eich2017A} for the energy fluence to the divertor targets was obtained in general (Figure~\ref{fig:ELMheatfluence}), although the values were slightly underestimated when $\mathbf{E}\times\mathbf{B}$ and diamagnetic background flows were taken into account, in particular for configurations corresponding to small ELM sizes in the experiment. Agreement regarding ELM energy losses and divertor peak heat fluxes improved significantly compared to Ref.~\cite{Pamela2015}, also here with some deviation when stabilizing background flows are taken into account. Resolving the remaining deviations might require simulations of complete ELM cycles. Ref.~\cite{Pamela2017} also demonstrated, that the threshold for the onset of edge instabilities in non-linear JOREK simulations is in better agreement with experimental observations than predictions from linear codes, which typically neglect many non-ideal effects.

\begin{figure}
\centering
  \includegraphics[width=0.8\textwidth]{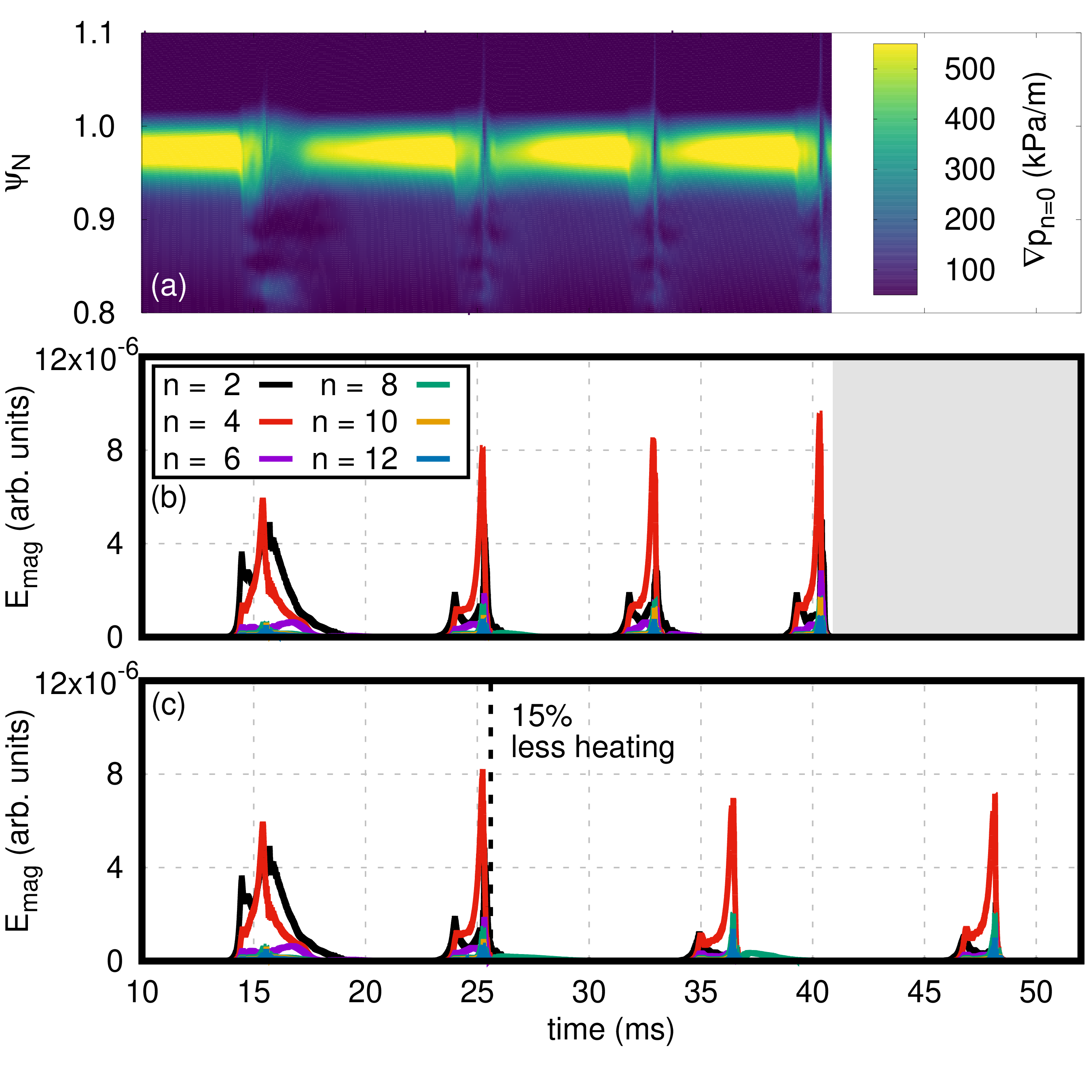}
\caption{Results from a simulation of several type-I ELM cycles in ASDEX Upgrade adapted from Ref.~\cite{Cathey2020}. (a) shows the evolution of the toroidally averaged outer midplane pressure gradient versus time and normalized poloidal flux. (b) contains the evolution of the magnetic energies of the $n=2,4,...,12$ toroidal harmonics included in the simulation. Finally, (c) shows the magnetic energies of a simulation with 15\% less input heating power starting from $25.2~\mathrm{ms}$. The ELM repetition frequency is seen to be directly proportional to the input heating power. The first ELM crash is determined by random seed perturbations and therefore shows different dynamics from the following ones, which are fully self-consistent. Strong precursor-like modes which already affect the pedestal gradient are observed prior to the very fast crashes. \noreprint}
\label{fig:typeIcycle}
\end{figure}

Using the JOREK code, realistic simulations of multiple type-I ELM cycles were recently obtained for the first time~\cite{Cathey2020} (Fig.~\ref{fig:typeIcycle}). The inclusion of pressure-gradient driven diamagnetic drifts is imperative to obtain the cyclical dynamics. The simulated ASDEX Upgrade plasma corresponds to a moderate triangularity and high pedestal density. Simulations for different plasma parameters are the subject of ongoing efforts. The simulations find that with increasing heating power, the ELM frequency rises, consistent to experimental observations of type-I ELMs~\cite{Zohm1996}. A precursor-like mode activity is observed before the violent onset of the ELM crash. It was found that the interaction between this precursor-like mode and the background plasma is responsible for the explosive ELM onset. The first ELM crash simulated was observed to be different from the subsequent ELMs because of differences in the non-axisymmetric seed perturbations that exist before each ELM crash. For the first ELM, these are simply noise-level perturbations, while for the subsequent ELMs the seed perturbations are consistent with the prior existence of an ELM. This observation stresses the importance of simulating the entire ELM cycle -- particularly for predictive simulations. These simulations were also used as basis for investigations regarding the scattering of ion cyclotron waves by filaments and ELMs~\cite{Zhang2020} and for recent studies of pellet ELM triggering (see Section~\ref{:applic:edge:pellet}).

The regime of small ELMs is another topic of current research with the JOREK code. Experiments with small ELM host a quasi-continuous power exhaust and, with appropriate plasma shaping, can avoid type-I ELMs completely~\cite{labit2019}. Using the simulation set-up for the type-I ELM cycles described above, but with a lower heating power, an operational regime with small ELM-like behaviour has been also obtained~\cite{Cathey2020C}, albeit not yet at ITER relevant conditions which remain a topic for future research. These simulations highlight the importance of the stabilising effects of plasma flow onto small ELMs~\cite{Harrer2020}.

\begin{figure}
\centering
  \includegraphics[width=0.7\textwidth]{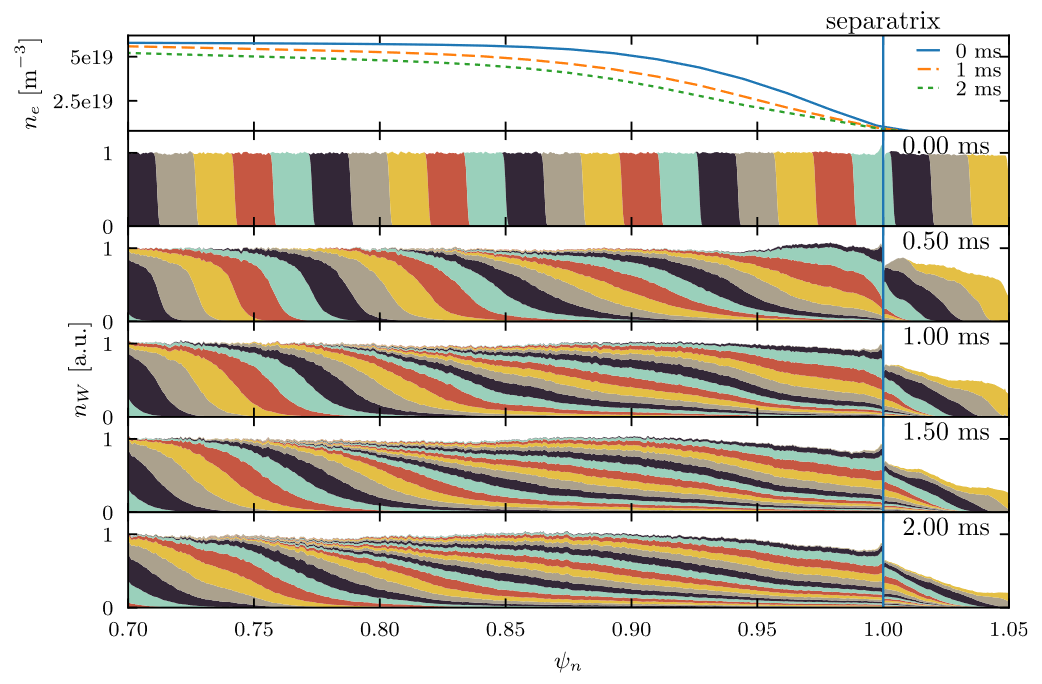}
\caption{During an ELM crash in ASDEX Upgrade (see Ref.~\cite{Hoelzl2018}), the transport of tungsten is investigated. The top figure shows the evolution of the density profile during the ELM crash, the figures below show the mixing of tungsten particles due to the MHD activity. Particles are colored according to their original location. A strong mixing is observed including a strong interchange transport across the separatrix. Re-print from Ref.~\cite{vanVugt2019}. \reprintaipOK}
\label{fig:ELMtungstentransport}
\end{figure}

Based on ELM simulations, fast ion losses were studied by particle tracing in the fields of the JOREK simulations confirming increased losses during the ELM crash~\cite{GarciaMunoz2013A,GarciaMunoz2013,vanVuuren2020}.
Tungsten transport during a large ELM crash in ASDEX Upgrade was investigated kinetically~\cite{vanVugt2019} (Figure~\ref{fig:ELMtungstentransport}) revealing that lower-dimensional diffusion-convection models cannot explain the far reaching interchange driven transport of the impurities and that high tungsten densities in the ITER scrape-off layer might cause a net inward transport of tungsten during an ELM crash.

\begin{figure}
\centering
  \includegraphics[width=0.7\textwidth]{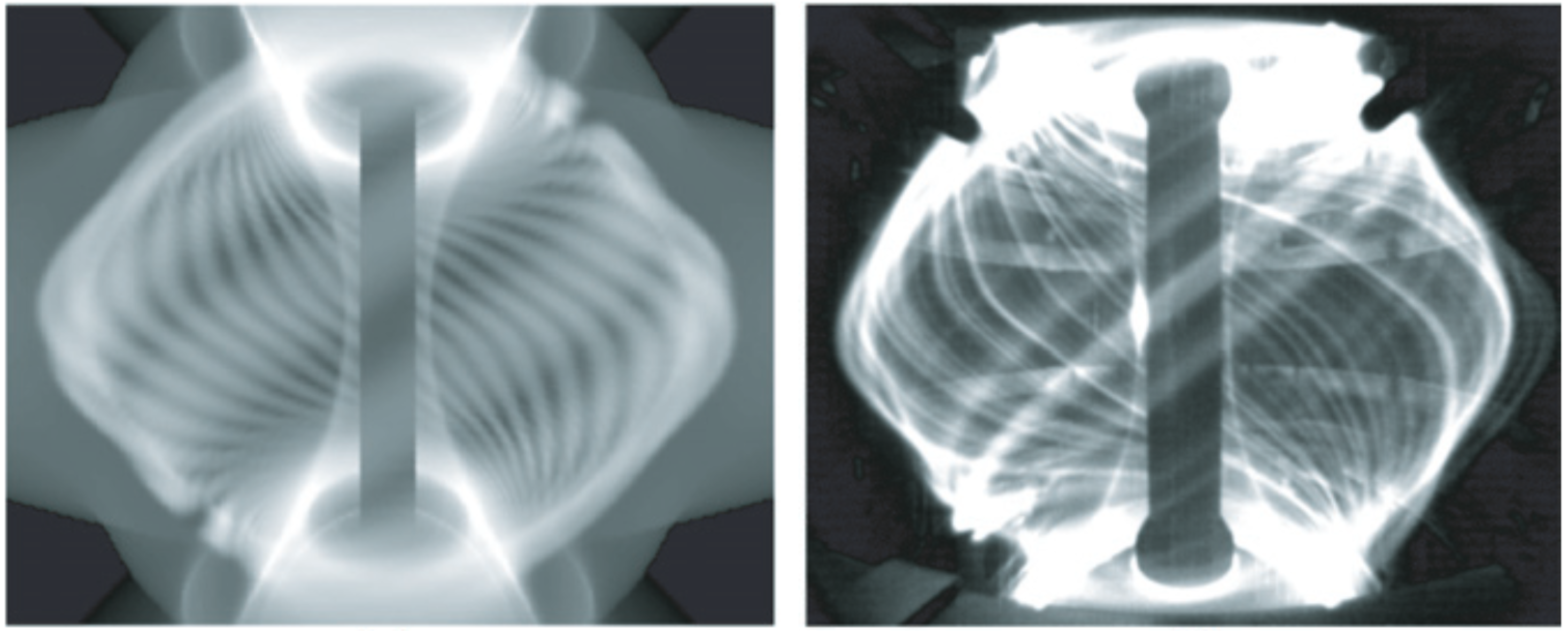}
\caption{The signature of ELM filaments in visible light is shown for ELMs in MAST. Left: virtual diagnostic data based on a JOREK simulation. Right: visible light picture in the experiment. Re-print from Ref.~\cite{Pamela2013}. \reprintiop}
\label{fig:mastfilaments}
\end{figure}

\revised{Simulations} for double X-point plasmas in the spherical tokamak MAST~\cite{Kirk2017} were reported in Ref.~\cite{Pamela2013}. In particular, the filament dynamics were compared to experimental observations revealing good qualitative agreement regarding structure and dynamics (Figure~\ref{fig:mastfilaments}). The general role of filaments in ELM dynamics and ELM energy losses was also addressed in Ref.~\cite{Ham2020A}. The ELM energy losses, and divertor heat flux profiles were found to agree reasonably well in spite of using a reduced MHD model in this low aspect-ratio configuration and neglecting diamagnetic drift effects. For the MAST Upgrade tokamak, \revised{predictive simulations} were reported in Refs.~\cite{Smith2018EPS,Smith2020,SmithEPS2019,SmithPhD}. Particular attention was paid to studying the effect of the Super-X divertor onto detachment in the inter-ELM phase, and burn-through during the ELM crash and resulting divertor heat fluxes. More details on this study are given in Section~\ref{:applic:edge:detachment} along with detachment and burn-through studies for ITER reported in Ref.~\cite{HuijsmansEPS2019}.

\subsection{Pellet ELM pacing}\label{:applic:edge:pellet}

\begin{figure}
\centering
  \includegraphics[width=0.5\textwidth]{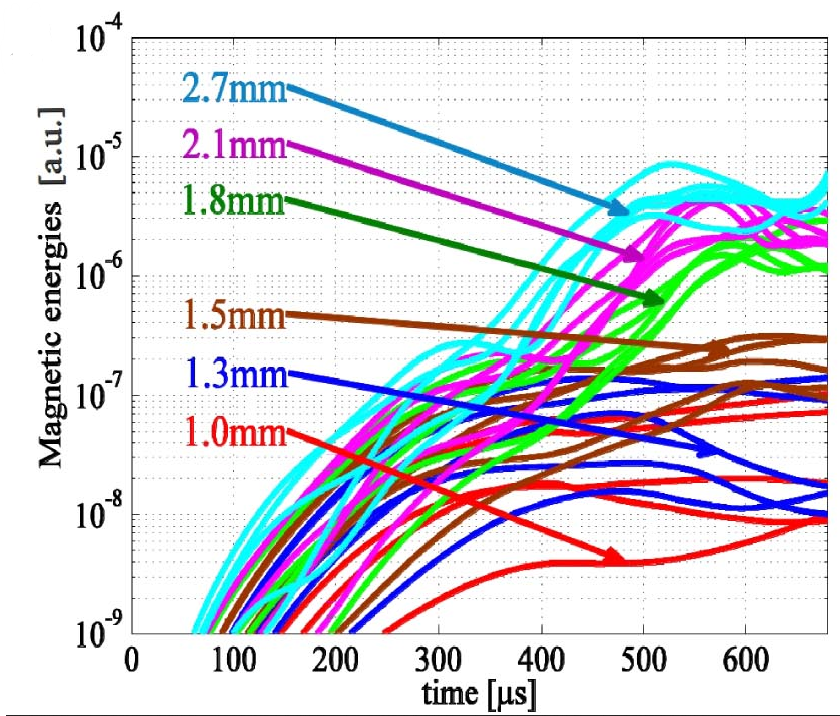}
\caption{The excitation of magnetic perturbations during pellet injection is shown. At a pellet size between 1.5 mm and 1.8 mm, a bifurcation is observed. For larger pellet sizes, a strong and sudden excitation of magnetic perturbation energies to higher amplitudes is observed, which is absent for the smaller pellet sizes. These strong perturbations are associated to a pellet triggered ELM crash and induce losses from the plasma. Re-print from Ref.~\cite{Futatani2014}. \reprintiaea}
\label{fig:pelletELM}
\end{figure}

Pellet injection into the pedestal of H-mode plasmas has experimentally shown to be capable of increasing the ELM frequency significantly above its natural value while decreasing ELM sizes, which may be beneficial for divertor lifetime and, in particular, for the control of impurity concentrations in the plasma. In non-linear simulations, such ELM triggering has first been demonstrated in Refs.~\cite{Huysmans2009,Huysmans2010IAEA}. Detailed comparisons to experimental observations for pellet triggered ELMs in the DIII-D tokamak were reported in Ref.~\cite{Futatani2014}. These studies revealed the detailed physical mechanisms for the destabilization of the ELM by a pellet: Ablation leads to a largely adiabatic increase of the plasma density and decrease of the plasma temperature in the vicinity of the pellet. Due to the high mobility of the electrons along field lines, the temperature within a flux surface is equilibrated again on an extremely fast time scale, while the density transport takes place on the far slower time scale of the ion sound speed. In combination, this leads to a helical structure with strongly increased pressure around the pellet location, which locally exceeds the ballooning threshold. In the further evolution, the initially localized perturbation spreads poloidally and toroidally such that the actual ELM crash leads to radial transport across the separatrix in a far less localized manner. A clear threshold was observed between a perturbation of the plasma and ELM triggering (Figure~\ref{fig:pelletELM}).

Simulations for ELM triggering in the JET tokamak revealed a toroidal asymmetry in the peak heat flux to the divertor~\cite{Futatani2019,Futatani2019EPS}, which constitutes a clear difference between the non-linear behaviour of a natural and pellet triggered ELM. The simulation shows good \revised{quantitative agreement (within 10-20\%)} against the divertor heat flux obtained with experimental thermography measurements at the single toroidal location where measurements are available. However, simulations predict a significantly larger divertor heat flux at a location which was not covered by measurements in the corresponding experiments.

\begin{figure}
\centering
  \includegraphics[width=0.7\textwidth]{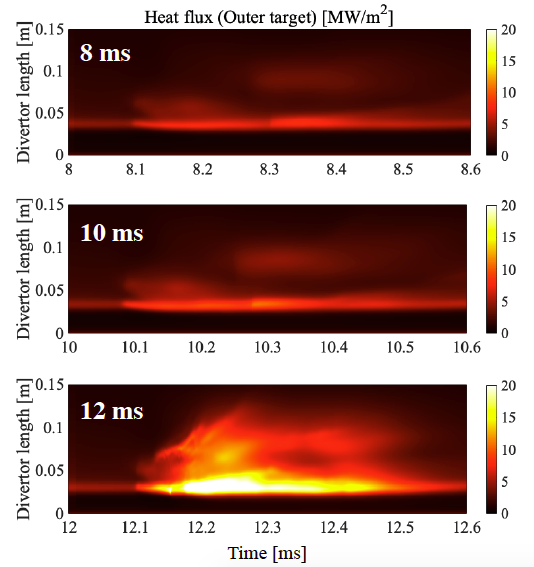} 
\caption{The time evolution of the heat flux onto the outer divertor targets which is caused by $0.8 \times 10^{20} D$ pellet injection with 560 m/s is shown. Three cases are compared, where the transition from no-ELM response (8 and 10 ms) to ELM triggering (12 ms) is observed. Re-print from Ref.~\cite{Futatani2020}. \reprintiaea}
\label{fig:heatflux_lagtime_08D}
\end{figure}

Recent simulations for pellet triggered ELMs in ASDEX Upgrade~\cite{Futatani2020} are studying the ELM triggering possibility at different phases of the pedestal build up, which provides additional insights into the evolution of the pedestal stability in the inter-ELM phase. These pellet simulations include self-consistent ExB and diamagnetic \revised{background plasma flows, use} realistic plasma parameters and are based on the type-I ELM cycle simulations of Ref.~\cite{Cathey2020}. The injection of pellets during the pedestal build-up is studied varying injection time, pellet size, and pellet velocity. Fig.~\ref{fig:heatflux_lagtime_08D} shows the divertor heat flux versus time and the divertor length (at the toroidal position of the pellet injection) for three cases: Pellets with $0.8 \times 10^{20} D$ atoms are injected into the plasma at 8, 10, and 12 ms during the pedestal build-up with an injection velocity of 560 m/s, while the natural ELM crash would occur here at around 16 ms. Note that the number of atoms reaching the plasma is given here; thus the simulation corresponds to a pellet with approximately $1.5\times10^{20}$ atoms in the experiment assuming 50\% losses in the guide tube. A sharp transition is observed between 10 and 12 ms from no-ELM response to ELM triggering. Thus, the experimentally observed~\cite{Lang_2014} lag-time has been reproduced qualitatively here, during which no ELM triggering is possible by means of pellet injection. In case of ELM triggering, large transient heat fluxes are observed in the range of $\sim 20~\mathrm{MW/m^{2}}$ for about 0.4 ms.  Furthermore, the ELM triggering cases show a far broader mode spectrum compared to the no-ELM response and the triggered ELMs features a toroidally asymmetric heat deposition with a strong n=1 component. A detailed comparison of the non-linear properties of triggered and spontaneous ELMs, e.g., regarding the wetted area, are shown in Ref.~\cite{Cathey2021}.

\subsection{ELM control by magnetic kicks}\label{:applic:edge:kicks}

\begin{figure}
\centering
  \includegraphics[width=0.7\textwidth]{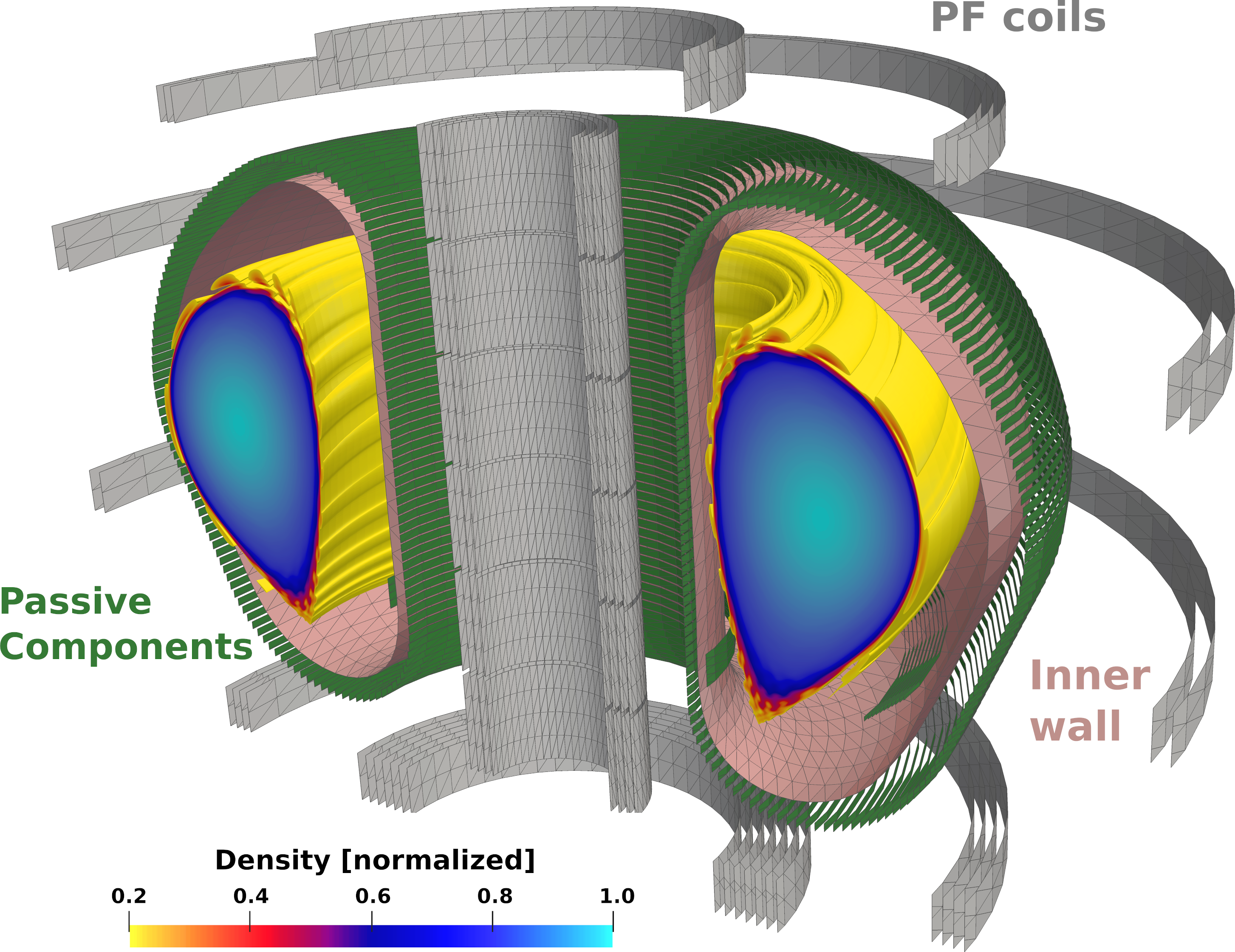}
\caption{The perturbed density during an ELM crash induced by a vertical magnetic kick is shown. All relevant conducting structures were included in the JOREK-STARWALL simulation. The figure was generated based on the simulation data published in Ref.~\cite{Artola2018}. \noreprint}
\label{fig:kick}
\end{figure}

Besides pellet injection discussed in the previous Section, ELM pacing has experimentally also been demonstrated via so-called ``vertical magnetic kicks'', during which the current in one or several poloidal field coils is evolved in time in such a way, that the plasma undergoes an excursion above and below the original location. 

A benchmark of the plasma excursion caused by a variation of the poloidal field coils (Figure~\ref{fig:DINA_kick_bench}), \revised{and fully self-consistent} simulations of ELM triggering by vertical magnetic kicks were shown in Refs.~\cite{Artola2018,Artola2018EPS,ArtolaPhD} for an ITER 7.5 MA plasma (Figure~\ref{fig:kick}). Consistently with the observations from several fusion experiments, it was demonstrated that peeling-ballooning modes (PBMs) are triggered during a downward (towards the X-point) excursion of the plasma leading to an ELM crash in an otherwise stable plasma configuration, while an upward excursion of the plasma does not give rise to such edge instabilities. The edge instabilities were shown to always appear at a particular vertical displacement of the plasma independently of the actual time scale of the applied oscillation, like also seen experimentally. Detailed analysis of the simulations and comparisons to analytical considerations allowed to confirm that an increase of the plasma edge current density during the downward motion of the plasma in the inhomogeneous magnetic field is responsible for the destabilization of the PBMs. A detailed analytical picture of the mechanisms of the edge current evolution was obtained consistent with the simulation results.

\subsection{ELM control by RMPs}\label{:applic:edge:rmp}

\begin{figure}
\centering
  \includegraphics[width=0.45\textwidth]{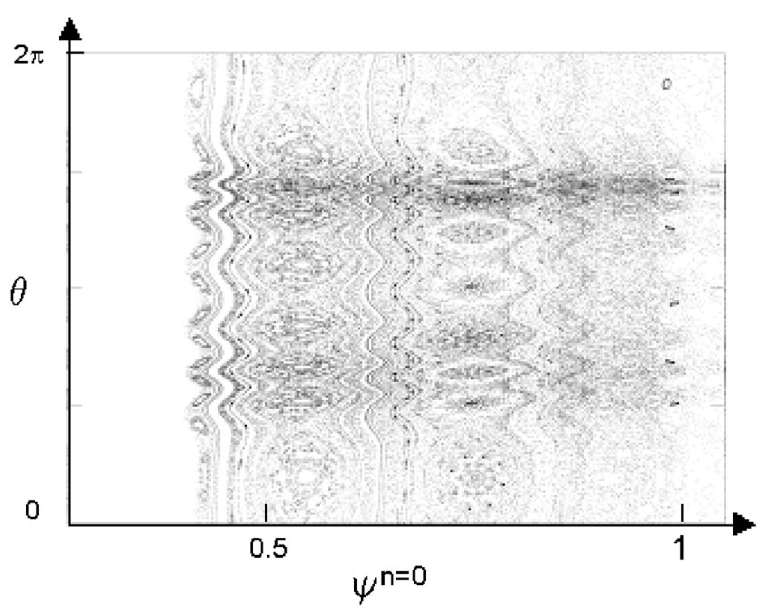}
  \includegraphics[width=0.45\textwidth]{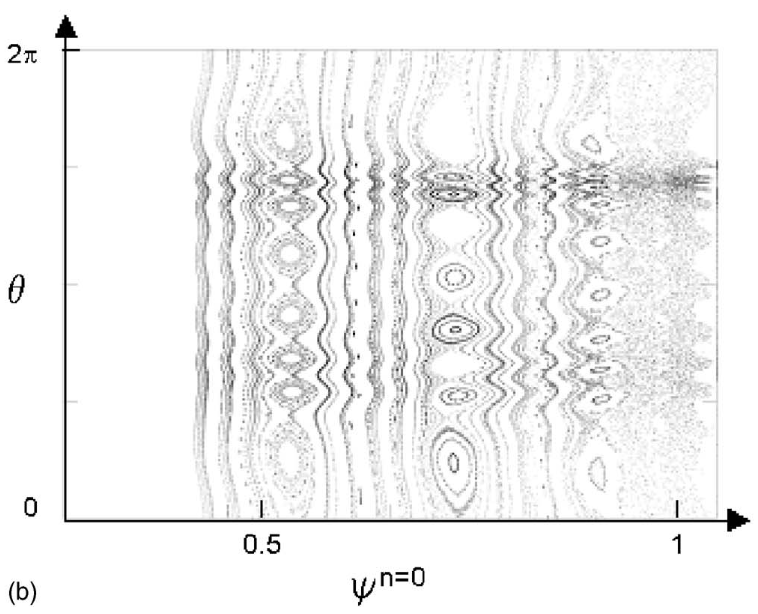}
\caption{The magnetic topology caused by resonant magnetic perturbations is shown. In the left figure, representing the ``vacuum field'' perturbation many island chains and stochastic regions are visible. In the right figure which includes the plasma response, most resonant components are shielded by the rotating plasma, and only a few penetrated island chains are visible. Re-print from Ref.~\cite{Nardon2007}. \reprintaipOK}
\label{fig:RMP}
\end{figure}

Simulations for the penetration of external magnetic perturbations (MPs) into plasmas of the DIII-D tokamak were shown in Refs.~\cite{Nardon2007,NardonPhD} for the first time, demonstrating that the magnetic topology obtained in simulations with plasma rotation considerably differs from the so-called ``vacuum approximation'', in which the vacuum magnetic field of the perturbation coils is simply added to the equilibrium magnetic field of the plasma (Figure~\ref{fig:RMP}). A strong suppression of magnetic islands and stochastic field regions by the plasma was shown in these simulations carried out in the ``zero-beta'' limit and with a rigid-body rotation. A radial $E\times B$ convective transport was observed in the presence of the MP fields and an important role of it for the experimentally observed density pump-out was proposed. However, note that the density pump-out due to this mechanism was not strong enough compared to the experiment. 

\begin{figure}
\centering
  \includegraphics[width=0.5\textwidth]{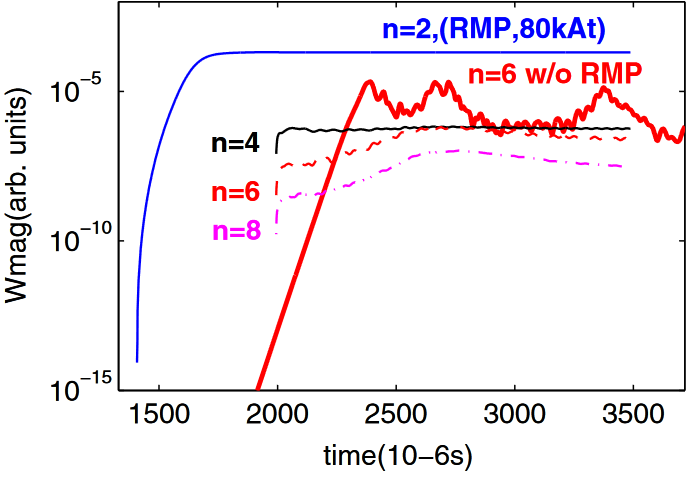}
\caption{The magnetic perturbation energies of a natural ELM crash (simulated with a single toroidal mode number $n=6$ are compared to a simulation with applied resonant magnetic perturbations. The $n=6$ magnetic perturbation energies are significantly reduced in the presence of the RMP fields. Instead of the bursts corresponding to an ELM crash, stationary modes are formed. Re-print from Ref.~\cite{Becoulet2014}. \reprintaps}
\label{fig:ELMRMP}
\end{figure}

With the implementation of neoclassical, diamagnetic and toroidal background plasma flows and taking into account also the self-consistent evolution of the plasma temperature, Refs.~\cite{Orain2013,Orain2014,Becoulet2014} were able to show in detail the dynamics of the penetration of the resonant magnetic perturbations (RMPs) into the plasma. In particular, the screening effect by the rotating plasma was studied in detail. Based on such simulations, Ref.~\cite{Cahyna2016} investigated the three dimensional lobe-structure of the homoclinic tangles formed in the presence of the 3D magnetic field perturbations in detail revealing a strike-line splitting like it is also observed in many experiments~\cite{Thornton_2014}. Refs.~\cite{Orain2014,Becoulet2014} also investigated the effect of the RMP fields onto the edge instabilities. In particular, it was shown that strong RMP fields can suppress peeling-ballooning instabilities in the simulations replacing them by saturated modes (Figure~\ref{fig:ELMRMP}), and that uneven-$n$ modes remain strongly sub-dominant in the presence of an $n=2$ magnetic perturbation, even if these are linearly unstable in the absence of the perturbation fields. 

\begin{figure}
\centering
  \includegraphics[width=0.45\textwidth]{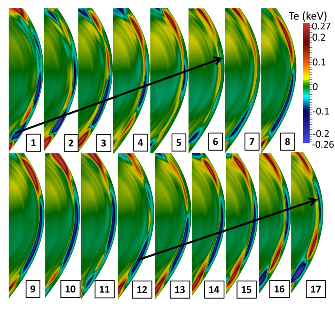}
  \includegraphics[width=0.45\textwidth]{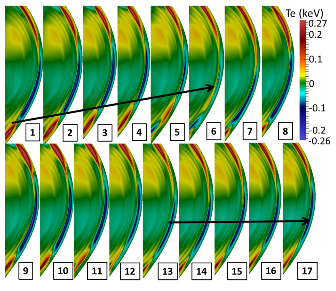}
\caption{Edge temperature perturbations (non-axisymmetric component) for ASDEX Upgrade simulations of a peeling-ballooning unstable plasma with different RMP amplitudes. The left panel shows different time points during a simulation with moderate RMP amplitude. Edge instabilities form rotating structures of reduced amplitudes compared to the natural ELM crash (mitigation regime). The right panel shows, that a higher RMP amplitude causes a locking of the instabilities. In this case, losses from the plasma are strongly reduced and stationary mode activity is seen (suppression regime). Re-print from Ref.~\cite{Orain2019}. \reprintaipOK}
\label{fig:ELMRMProt}
\end{figure}

Furthermore, simulations for RMP experiments in ASDEX Upgrade showed very good agreement for the penetration of the external fields into the plasma and the resulting corrugation of the separatrix for several investigated ``coil current phases'', i.e., different perturbation spectra~\cite{Orain2017}. The configuration leading to the strongest ELM mitigation effect in the experiments was identified in the simulation as the one with the largest kink response of the plasma near the X-point. In further studies shown in Ref.~\cite{Orain2018AAPPS,Orain2019}, the interaction of the RMP fields with plasma edge instabilities was investigated in a plasma configuration leading to an ELM crash in simulations without RMP fields. When increasing the RMP amplitude at given plasma rotation, and also when reducing the plasma rotation at fixed RMP amplitude, a transition was observed from an unmitigated ELM regime into a mitigated ELM regime with reduced perturbation amplitudes and losses, and further into an ELM suppressed state. While the 3D perturbations observed in the unmitigated and mitigated state show a ``bursty'' behaviour and rotate in the electron diamagnetic direction, the suppressed state is characterized by saturated modes, which are not rotating in the lab frame due to a locking with the external perturbation fields (Figure~\ref{fig:ELMRMProt}). Non-linear mode coupling is shown to be crucial for the ELM suppression.

Recent simulations for KSTAR~\cite{KimSK2020,KimSKPhD} investigate a plasma configuration unstable with respect to peeling-ballooning modes (PBMs). When $n=2$ RMPs are applied, a formation of islands and a stochastic field layer is observed at the plasma boundary and a density pump-out is observed, while the effect onto the temperature is weaker. The article shows, that the PBMs can be mitigated and eventually suppressed by the application of the RMP fields of sufficient amplitude. Consistently with Ref.~\cite{Orain2019}, a breaking of the mode rotation is seen in the suppressed state. Non-linear mode coupling plays an important role in the suppression, since the PBMs would be still unstable in the degraded plasma pedestal, when the 3D fields are removed from the simulation.

\subsection{ELM free regimes}\label{:applic:edge:elmfree}

\begin{figure}
\centering
  \includegraphics[width=0.5\textwidth]{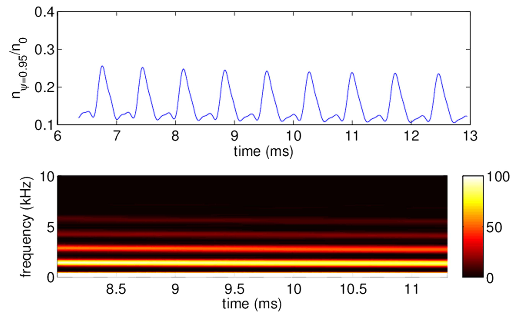}
\caption{The edge density perturbation during a QH-mode simulation is shown. The upper figure contains the time trace of the perturbation in the outer midplane induced by a stationary and rotating edge mode, the lower figure contains the spectrum analysis of this signal. The results are in very good agreement to the experimentally observed edge harmonic oscillations. Re-print from Ref.~\cite{LiuF2017}. \reprintiop}
\label{fig:EHO}
\end{figure}

Non-linearly saturated external kink modes in X-point plasma simulations including the scrape-off layer (SOL) were described for the first time in Ref.~\cite{Huysmans2005}. Dedicated simulations for quiescent H-mode (QH) plasmas from DIII-D were carried out in Ref.~\cite{LiuF2015,LiuF2017}. These showed the development of saturated kink-peeling modes (KPM) leading to the characteristic edge harmonic oscillation (EHO), which is observed experimentally in the density evolution and is caused by the rotating saturated modes (Figure~\ref{fig:EHO}). The influence of shear flows and resistive wall effects onto the development of the QH-mode was studied for DIII-D and predictively for ITER. Linear stability studies for the DIII-D case showed that the case is close to the peeling stability boundary. It was demonstrated that the saturated KPMs are replaced by an ELM crash, if the edge current density is decreased and the pedestal pressure increased.

Recently, simulations for experiments in ASDEX Upgrade have also been performed~\cite{LiuF2018IAEA}. A QH-mode experiment of the carbon wall ASDEX Upgrade was studied and saturated low-n kink-peeling modes were found to be non-linearly dominant at the edge of the plasma forming a helically perturbed structure. This is in spite of an initial small crash triggered by high-n ballooning modes, confirming that the simulations predict the formation of a QH-mode like state for this plasma configuration. In the same reference, also simulations for the ITER Q=10 baseline scenario were performed. Although further investigations may be necessary, these simulations indicate, that the bootstrap current may be high enough in this scenario to enter a QH-mode regime.
The ExB flows are shown to have a stabilizing influence on high-n modes in the ITER plasma. In the ITER relevant range of ExB rotation around 20km/s at the pedestal, the simulations show that low-n modes n=1-3 are destabilized and become dominant in the non-linear saturated state that establishes, while higher-n modes n=7-10 are significantly suppressed. The saturated low-n modes lead to density oscillations in the pedestal which are typical for EHO characteristics.
\revised{Also, simulations} with the free boundary extension (Section~\ref{:code:models:freebound}) were shown in order to investigate the effect of the true geometry of the vacuum vessel onto the instabilities. Related to the effect of resistive walls onto the KPM, simplified resistive wall mode studies had already been performed and benchmarked in Refs.~\cite{Hoelzl2014,McAdamsPhD}.
 
\subsection{Detachment physics}\label{:applic:edge:detachment}

\begin{figure}
\centering
  \includegraphics[width=0.8\textwidth]{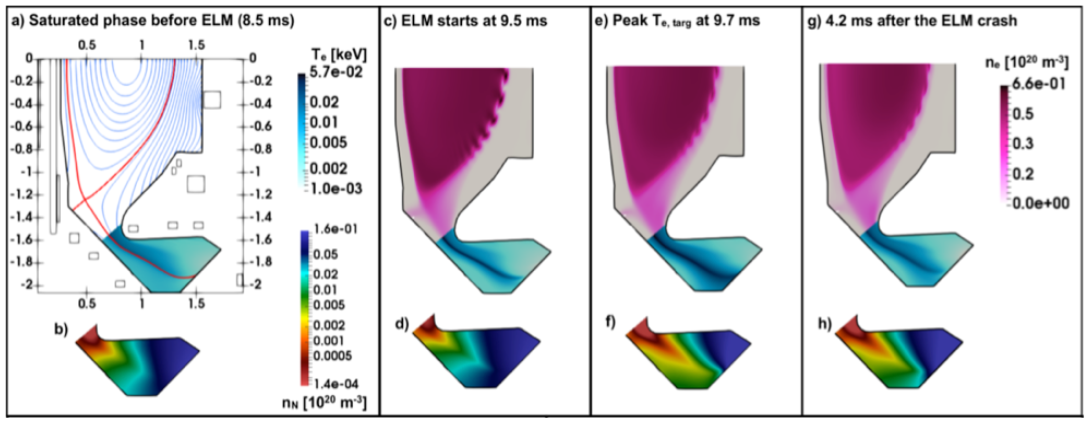}
\caption{The simulation of an ELM crash in MAST Upgrade is shown. The simulations are performed using the neutrals model available in JOREK. The originally detached state with low target temperatures (left plot) is re-attaching during the ELM crash in the Super-X divertor (right plots). Re-print from Ref.~\cite{SmithEPS2019}. \reprintccby}
\label{fig:MASTUburnthrough}
\end{figure}

\begin{figure}
\centering
  \includegraphics[width=0.98\textwidth]{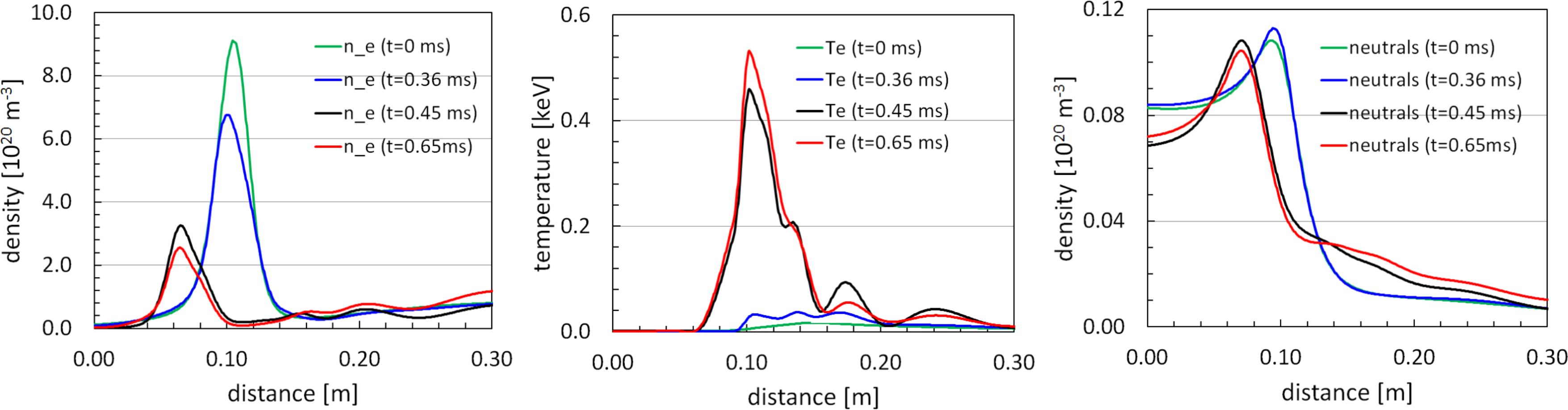}
\caption{Density, temperature and neutral density profiles along the ITER outer divertor. Re-print from Ref.~\cite{HuijsmansEPS2019}. \reprintccby}
\label{fig:ITER_ELM_neutrals}
\end{figure}

The simulations presented in Refs.~\cite{SmithEPS2019,HuijsmansEPS2019} \revised{study detachment/burn-through during an ELM crash in a non-linear MHD simulation}. Using the neutrals fluid model (Section~\ref{:code:models:neutrals}), simulations of plasma detachment for the MAST Upgrade tokamak were carried out~\cite{Smith2020,SmithEPS2019} and reasonable agreement with SOLPS simulations could be shown, although some discrepancies will require further investigations. In particular, the characteristic ``roll-over'' could be qualitative reproduced, where the target particle flux first increases and then decreases for an increasing midplane separatrix density, leading to a target temperature of only a few electron Volts (eV) in the detached state. Simulations of large type-I ELM crashes were carried out in such a detached plasma state, showing the ``burn-through'' and re-attachment in the Super-X divertor leg. The ELM energy fluence was found to be significantly lower than that predicted from the empirical scaling for the detached Super-X divertor~\cite{Smith2020}. Similar physics was studied for ITER~\cite{HuijsmansEPS2019}. It could be shown that small amplitude ELMs are sufficient to re-attach the plasma in the ITER high recycling divertor transiently, increasing the electron temperature at the divertor target from a few eV to several hundred eV within a fraction of a millisecond (Figure~\ref{fig:ITER_ELM_neutrals}). In this Reference, also an outlook to refined modelling of detachment and burn-through using a kinetic neutrals model (based on the particles framework described in Section~\ref{:code:models:particles}) is given, which will allow to capture the scrape-off layer dynamics even more accurately.

\subsection{Outlook}\label{:applic:edge:outlook}

Regarding the modelling of edge instabilities, further emphasis will be put onto a more accurate modelling of the scrape-off layer and divertor physics like shown for instance in Section~\ref{:applic:edge:detachment}. Simulations of the edge plasmas including such effects, further model enhancements, and based on further numerical improvements, aim to extend the modelling of ELM cycles to fully realistic plasma parameters and describe the physics mechanisms relevant for the transitions between the regimes of different ELM types and ELM free regimes.

%% file: 06_disruptions.tex
\section{Applications to disruptions and their control}\label{:applic:core}

In view of ITER, disruptions are presently the highest priority topic when it comes to large-scale plasma instabilities. Unmitigated disruptions are considered intolerable above modest values (by ITER standards) of the plasma current and thermal energy. The ITER Disruption Mitigation System (DMS), which is planned to rely on Shattered Pellet Injection (SPI), needs in particular to avoid substantial damage from heat loads, RE impacts and electromagnetic forces~\cite{Lehnen2015}. The current strategy to achieve this consists in: 1) mitigating heat loads by dissipating most of the plasma energy through uniform radiation, 2) avoiding, if possible, the formation of a RE beam by raising the electron density, 3) should a RE beam appear anyway, using SPI into the beam in order to make its impact as benign as possible, and 4) controlling (\textit{via} the plasma impurity content) the $I_p$ decay rate in order to mitigate electromagnetic loads. Achieving these goals simultaneously requires a deep understanding of SPI and disruption physics, and this motivates the many disruption-related investigations with JOREK which are described in this Section. In particular, \ref{:applic:core:predisruption_mech} discusses pre-disruption physics, i.e. the mechanisms leading to disruptions, \ref{:applic:core:disr} covers the dynamics of disruptions triggered by Massive Material Injection (MMI), \ref{:applic:core:vdes} deals with Vertical Displacement Events (VDEs) and halo current dynamics, and finally \ref{:applic:core:res} describes RE studies.

\subsection{Pre-disruption physics}\label{:applic:core:predisruption_mech}

\subsubsection{Tearing mode dynamics and mode locking}\label{:applic:core:islands}

\begin{figure}
\centering
  \includegraphics[width=0.6\textwidth]{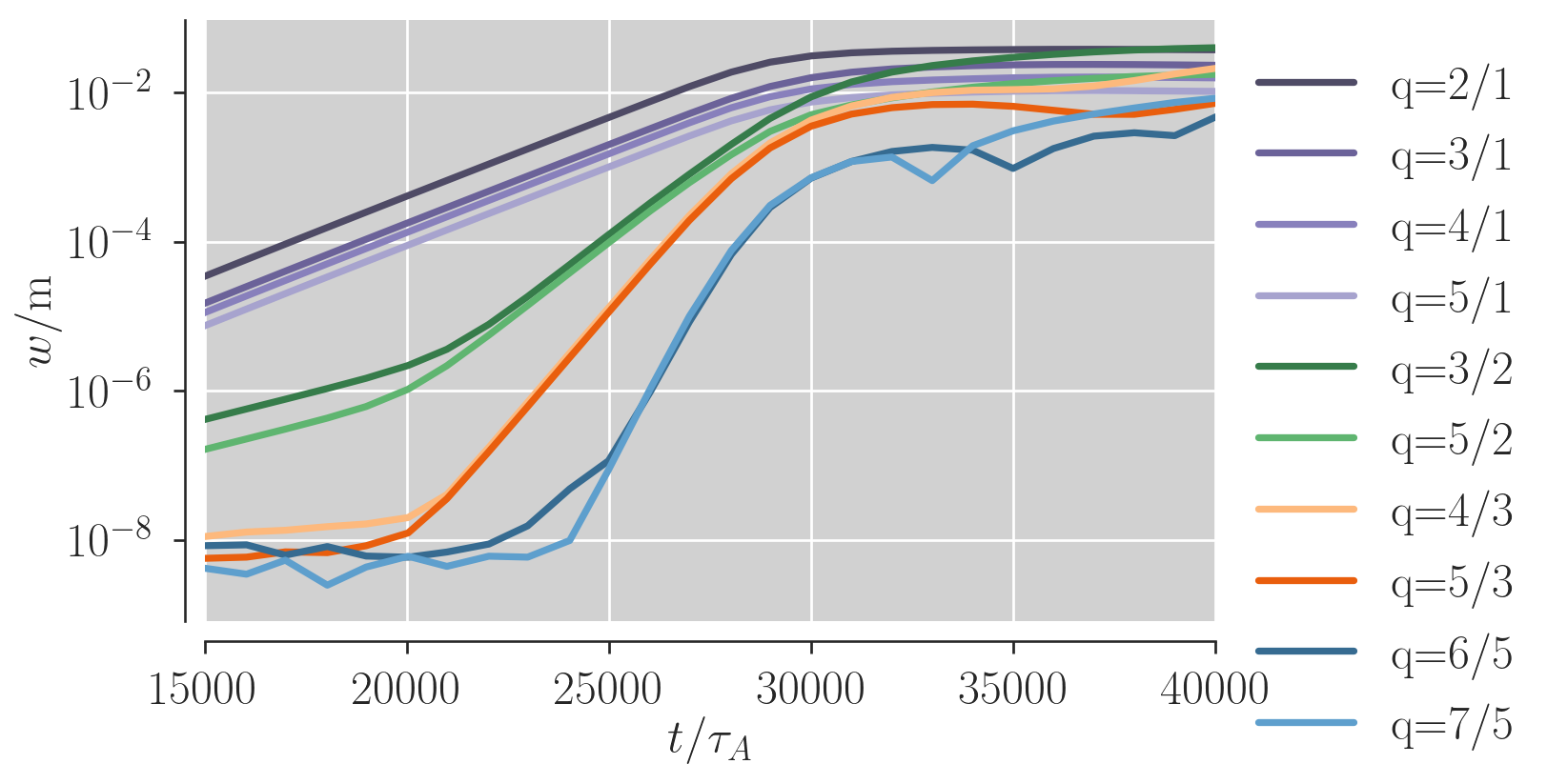}
\caption{Island width as a function of time for various islands, showing that higher $n$ islands are driven by mode coupling. \noreprint}
\label{fig:islands_width}
\end{figure}

Simulations in ASDEX Upgrade-like geometry have been run in which the current profile was tailored to make a dominantly $m/n=2/1$ tearing mode strongly unstable~\cite{WieschollekMaster}. Due to the inhomogeneous magnetic field as well as geometrical effects, the $n=1$ mode contains sidebands, i.e. components of the type $m/1$ with $m\neq2$. In the linear phase, Poincar{\'e} cross-sections thus show the presence of not only a $2/1$ island, but also $3/1$, $4/1$ and $5/1$ islands. In the non-linear phase, $n>1$ islands (in particular $3/2$ and $5/2$) grow fast. An analysis of the growth rate of the islands width, shown in Figure~\ref{fig:islands_width}, suggests that $n>1$ islands result from mode coupling. For example, the growth rate of $n=2$ islands is twice that of $n=1$ islands. Island overlapping leads to a stochastization of the magnetic field over roughly the outer half of the plasma. Due to the fast parallel heat transport, the temperature profile flattens across this region, reproducing a key feature of so-called `partial thermal quenches' in experiments~\cite{Sweeney_2018}. 

\begin{figure}
\centering
  \includegraphics[width=0.6\textwidth]{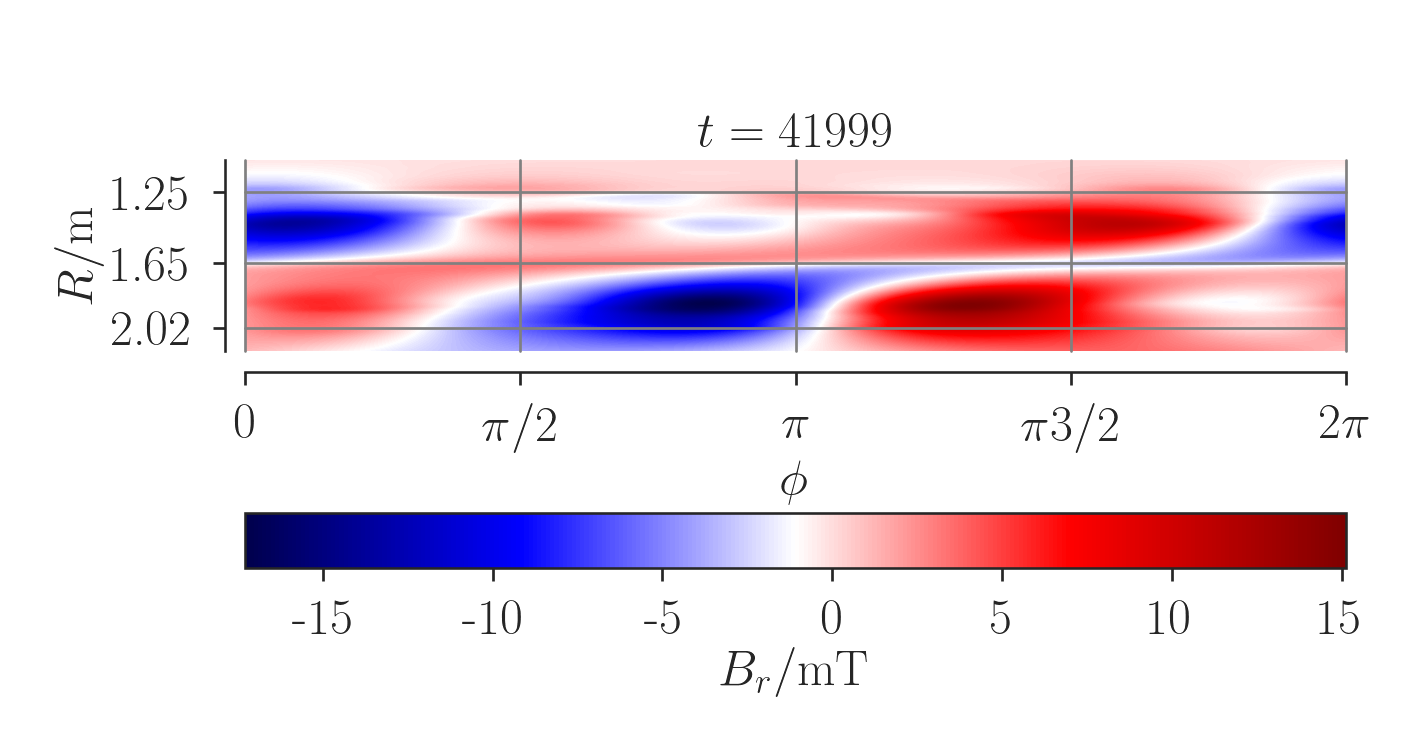}
\caption{Perturbed radial magnetic field $B_r$ at the midplane as a function of the toroidal angle $\phi$ and major radius $R$. The $q = 2$ surface (both on the HFS and LFS) and the magnetic axis are marked by grey lines. A HFS-LFS asymmetry clearly appears. \noreprint}
\label{fig:TM_midplane_Br}
\end{figure}

The amplitude of the magnetic field perturbation at the onset of the partial thermal quench, as measured by synthetic magnetic sensors localized in the midplane, was compared to the empirical scaling law identified by de Vries et al.~\cite{deVries2015}. Agreement within the error bars is found for Low Field Side (LFS) sensors. On the other hand, the High Field Side (HFS) synthetic sensors measure a magnetic perturbation much weaker than expected from the de Vries scaling. This LFS-HFS asymmetry, which is clearly visible in Figure~\ref{fig:TM_midplane_Br}, seems related to the fact that the various $m/1$ components interfere constructively on the LFS but destructively on the HFS. A related observation is that the O-points of the various $m/1$ islands align on the LFS (a feature also observed experimentally~\cite{Sweeney_2018}) while on the HFS, the alignment involves either the O-point or the X-point depending on the parity of $m$. These observations could help extract improved scaling laws in the future. 

Furthermore, the locking of a slowly rotating magnetic island to the vacuum vessel of ASDEX Upgrade was demonstrated~\cite{WieschollekMaster}. Scanning the vessel conductivity artificially shows that mode locking is fastest and most complete when $\tau_v \simeq m/\omega$, where $\tau_v$ is the vessel resistive time and $\omega$ is the mode frequency, as expected from theory~\cite{Nave_1990}.

\subsubsection{Effect of impurities on island growth and relation to the Greenwald limit}\label{:applic:core:impurities}

Using the fluid impurity model (Section~\ref{:code:models:impurities}), JOREK simulations show a strong growth of islands when the local radiation exceeds the Ohmic heating \cite{vanOverveldMaster}. The transition to strong island growth occurs in a range of densities near the Greenwald limit. For a single impurity species, the critical density depends on the impurity fraction and the temperature, in contrast to the scaling of the Greenwald density limit. However, a mixture of impurity species is found to remove some of these dependencies, consistently with~\cite{Teng_2016}.


\subsubsection{Tearing mode seeding via Resonant Magnetic Perturbations (RMPs)}\label{:applic:core:predisruption_mech:RMP_seeding}

\begin{figure}
\centering
  \includegraphics[width=0.5\textwidth]{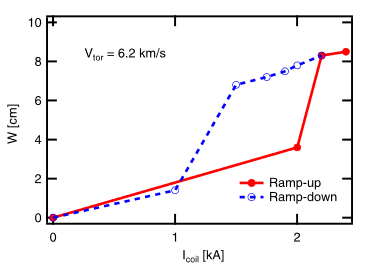}
\caption{$2/1$ island width versus RMP coils current, showing the hysteresis effect: mode penetration occurs slightly above 2kA in the RMP ramp-up phase, while mode expulsion occurs below 1.5kA in the ramp-down phase. Re-print from Ref.~\cite{Meshcheriakov2019}. \reprintaipOK}
\label{fig:TMhysteresis}
\end{figure}

In ASDEX Upgrade, experiments have been conducted using the RMP coils (the same as those used for ELM mitigation/suppression, see Section~\ref{:applic:edge:rmp}) with a coil configuration optimized to produce $m/n=2/1$ perturbations. Corresponding JOREK simulations show good qualitative agreement with the experiments and with analytical predictions regarding mode penetration~\cite{Meshcheriakov2019}. In particular, thresholds for the mode penetration in the RMP amplitude and plasma rotation frequency were observed, as well as a fast transition between the shielded and penetrated states and a hysteresis of the island size and plasma rotation between the ramp-up and ramp-down of the RMP. The latter is illustrated in Figure~\ref{fig:TMhysteresis}.

\subsubsection{Tearing mode control with Electron Cyclotron Current Drive (ECCD)}\label{:applic:core:eccd}

A useful tool to control tearing modes and help avoid disruptions is Electron Cyclotron Current Drive (ECCD). A fluid closure reproducing the dominant Fisch-Boozer current generation mechanism in ECCD has been developed and validated against full Fokker-Planck calculations of ECCD~\cite{Westerhof2014}. This closure model is implemented in JOREK and the stabilizing influence of ECCD onto a tearing mode has been demonstrated in simplified geometry~\cite{Pratt2016}.
 
\subsection{Dynamics of disruptions triggered by Massive Material Injection}\label{:applic:core:disr}

\revised{This section summarizes work performed with JOREK on the thermal quench triggering mechanisms (Section~\ref{:applic:core:disr:TQ_trigg}), the thermal quench dynamics and plasma current spike (Section~\ref{:applic:core:disr:TQ_dyn}), the assimilation and mixing of injected material (Section~\ref{:applic:core:disr:fuell_dens}) and the radiation fraction and asymmetry (Section~\ref{:applic:core:disr:rad_frac}). For work with other codes on these topics, see Refs.~\cite{Strauss2000,Izzo2008,Izzo2013,Izzo2015,Ferraro2018,Kim2019} and references therein.}

\subsubsection{Thermal Quench triggering mechanisms}\label{:applic:core:disr:TQ_trigg}


\begin{figure}
\centering
 \includegraphics[width=0.6\textwidth]{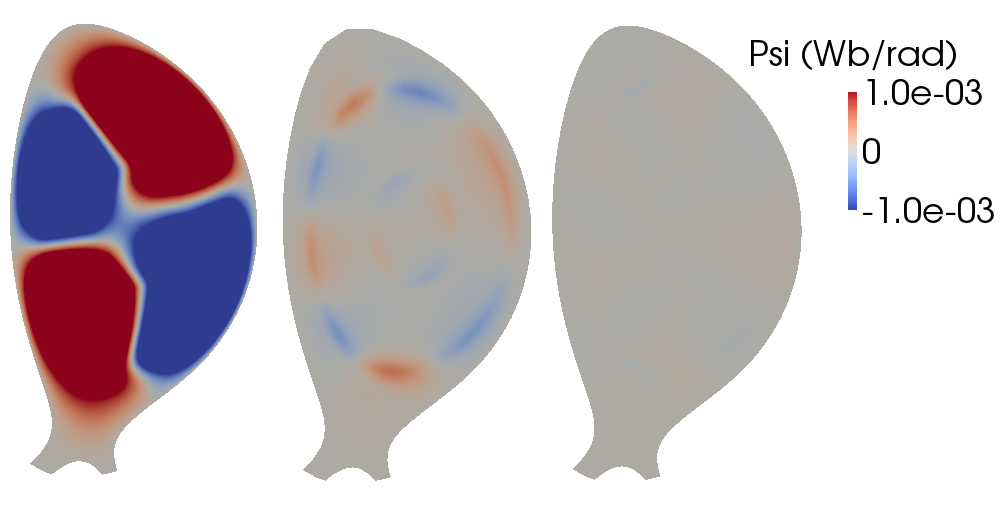}
\includegraphics[width=0.6\textwidth]{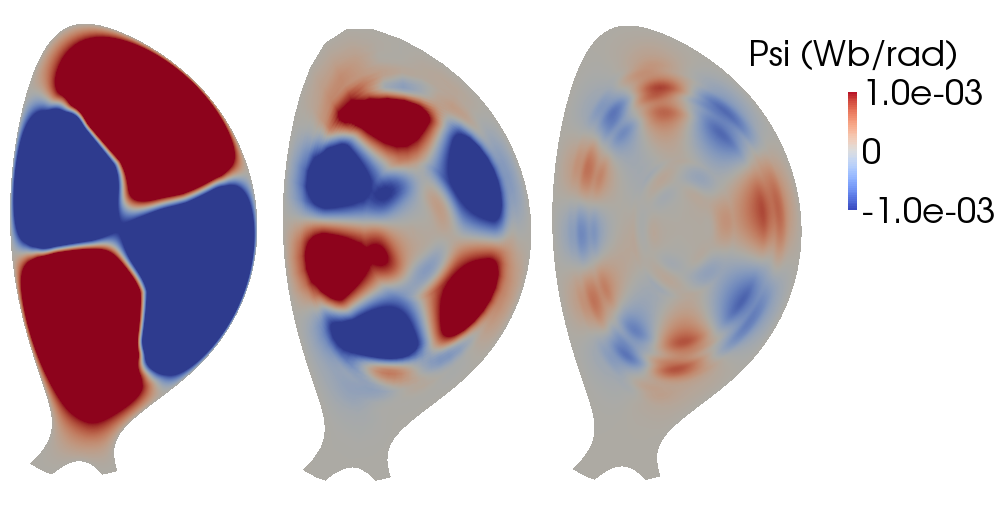}
\includegraphics[width=0.6\textwidth]{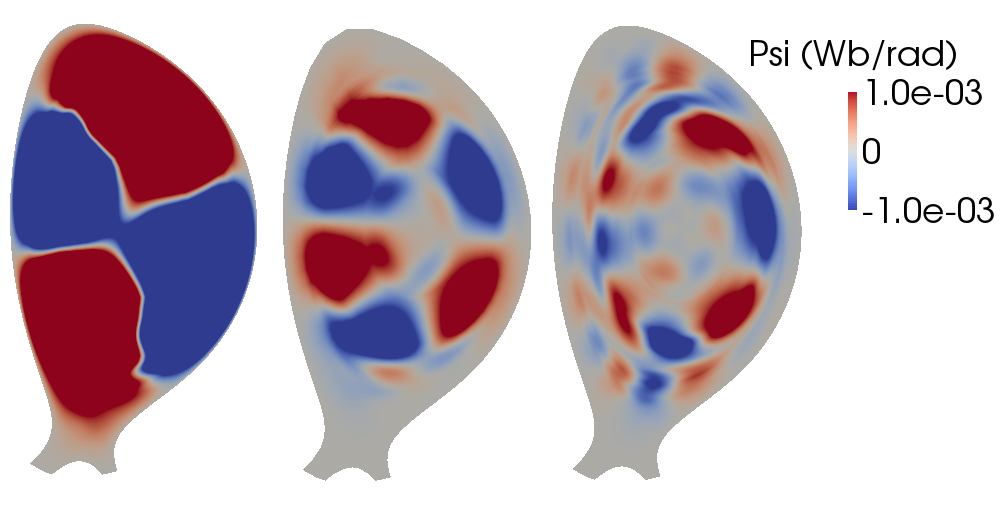}
\caption{Poloidal cross-sections at the toroidal position of the MGI of (from left to right) the $n = 1$, $n = 2$ and $n = 3$ cosine component of the poloidal flux $\psi$ at times (from top to bottom) t=4.1 ms, t=5.1 ms and t=5.7 ms, for the same simulation of deuterium MGI in JET as in Figure~\ref{fig:MGI_jprof}. The color scale is the same for all plots (note the saturation for the $n = 1$ mode which has a large amplitude compared to the other modes). Re-print from Ref.~\cite{Nardon2017}. \reprintiop}
\label{fig:MGI_psi_n123}
\end{figure}

\begin{figure}
\centering
  \includegraphics[width=0.6\textwidth]{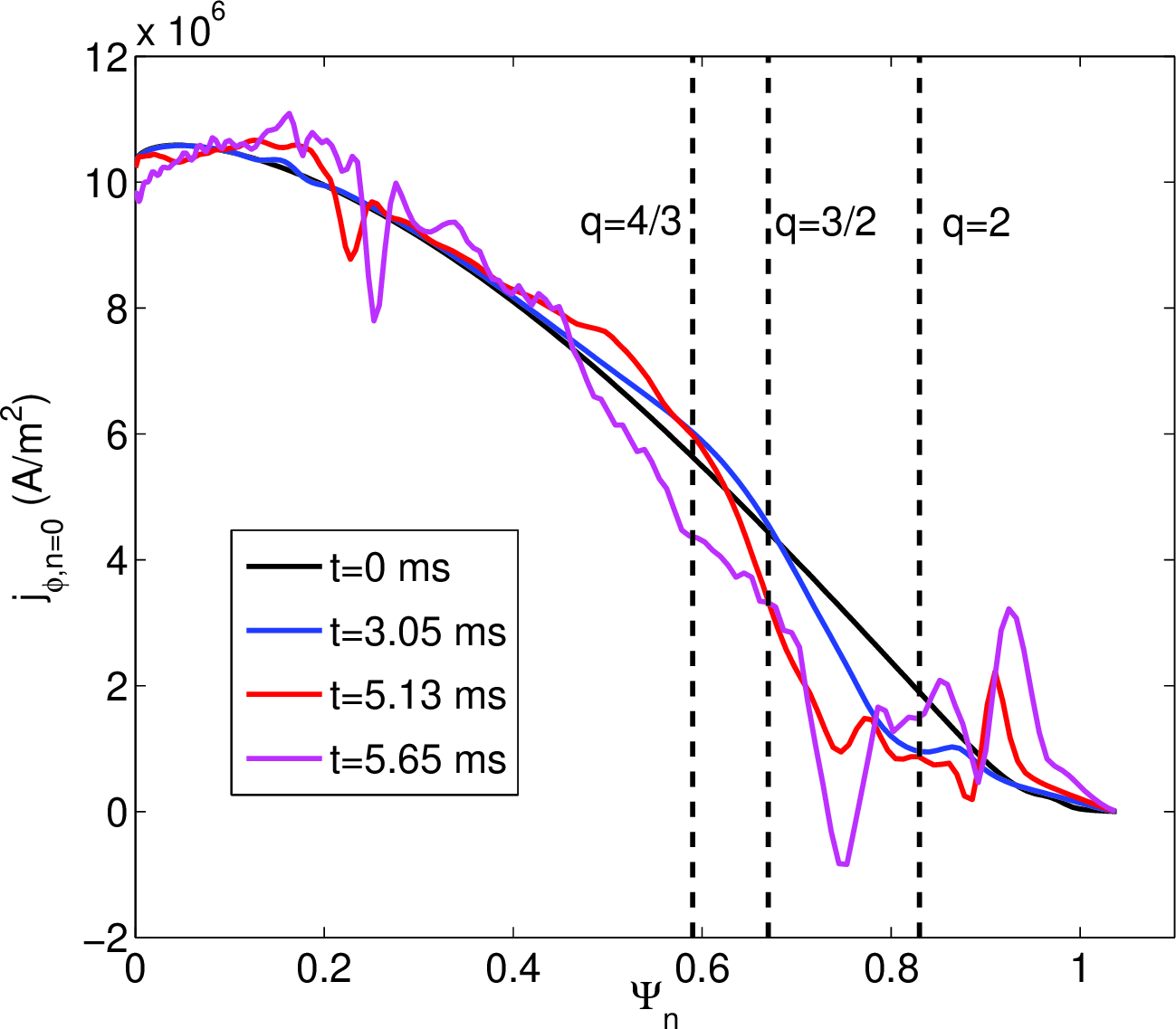}
\caption{Toroidal current density profiles at the midplane (low field side) at different times for the same simulation of deuterium MGI in JET as in Figure~\ref{fig:MGI_psi_n123}. The red and magenta profiles correspond to the last two rows of Figure~\ref{fig:MGI_psi_n123}. \revised{Note that, for clarity, the position of rational surfaces is indicated (by vertical dashed lines) referring to their location at the beginning of the simulation.} Re-print from Ref.~\cite{Nardon2017}. \reprintiop}
\label{fig:MGI_jprof}
\end{figure}


Disruptions triggered by a Massive Gas Injection (MGI) have been studied intensively with JOREK~\cite{Reux38thEPS,Fil2015,FilPhD,Nardon2017}. An analysis of simulations of Deuterium MGI in JET, illustrated by Figures~\ref{fig:MGI_psi_n123} and~\ref{fig:MGI_jprof}, suggests that the Thermal Quench (TQ) is triggered through a current profile avalanche effect. The avalanche is started by the penetration of an MGI-driven cold front up to the $q=2$ surface, at which point a large $m/n=2/1$ tearing mode is destabilized (first row of Figure~\ref{fig:MGI_psi_n123}). The $2/1$ mode, whose growth is boosted by an MGI-induced helical cooling effect inside the $2/1$ island, flattens the toroidal current density ($j_\phi$) profile around the $q=2$ surface, which results in a steepening of the $j_\phi$ gradient more inward, as evident in the blue and red profiles in Figure~\ref{fig:MGI_jprof}. When the steep $j_\phi$ gradient passes across the $q=3/2$ surface, a $3/2$ tearing mode is destabilized (second row of Figure~\ref{fig:MGI_psi_n123}). The $3/2$ mode in turns flattens $j_\phi$ locally and propagates the steep $j_\phi$ gradient more inward, which destabilizes a $4/3$ tearing mode (magenta curve in Figure~\ref{fig:MGI_jprof} and last row of Figure~\ref{fig:MGI_psi_n123}). These combined modes generate global magnetic stochasticity and thereby provoke the TQ. It is interesting to note that, for these MGI simulations, the $n>1$ modes seem to be destabilized via a current profile effect while in the study presented in Section~\ref{:applic:core:islands} they were destabilized by mode beating. The reason for this difference remains to be clarified.

SPI is also being investigated in detail with JOREK. Simulations of Deuterium SPI in JET~\cite{Hu2018,HuEPS2018} and ASDEX Upgrade~\cite{Hoelzl2020A} have shown that pre-TQ dynamics can vary drastically depending on parameters. If shards travel across the plasma relatively fast compared to the current decay time in the SPI-cooled region, then the main MHD destabilization mechanisms are 1) the helical cooling on low order rational $q$ surfaces (whereby shards may generate magnetic islands as they pass across these surfaces, possibly leading to a stochastization front progressing with the shards), and 2) for target plasmas with a central safety factor $q_0<1$, the excitation of the $1/1$ internal kink mode when shards reach the $q=1$ surface. This behaviour has been observed in JET Deuterium SPI simulations ~\cite{Hu2018,HuEPS2018}. In the opposite limit of slow shards with respect to the current decay time in the SPI-cooled region, the dynamics resemble that described above for MGI: the current profile contracts due to the SPI-induced cooling of the edge, which destabilizes tearing modes in cascade from the edge to the core, leading to a stochastization front progressing faster than the shards. In this regime, the TQ is typically triggered when shards reach the $q=2$ surface, but it can be triggered even before for large and slow pellets. This behaviour has been observed in ASDEX Upgrade Deuterium SPI simulations~\cite{Hoelzl2020A} and is also typically observed for any tokamak when simulating the injection of shattered pellets containing impurities~\cite{Hu2020}.

The key role of the ordering between the shards penetration time and the current decay time in the SPI-cooled region has a number of implications. First, it can be noticed that the resistive current decay time, $\sim \mu_0 l^2 / \eta$, is proportional to the square of the length scale of interest $l$. Hence, considering that $l$ is proportional to the machine size, the current decay time should grow like the machine size squared, while the shards penetration time, for a given velocity, increases linearly with machine size. This consideration is important for comparisons or extrapolations between machines of different sizes. It explains partly the different behaviour found in JET~\cite{Hu2018,HuEPS2018} and ASDEX Upgrade~\cite{Hoelzl2020A} Deuterium SPI simulations. Second, simulations in which the resistivity is artificially increased (which is often done for numerical reasons) should be considered with caution because the current decay time will be artificially shortened. To avoid this issue, present JOREK simulations use a realistic resistivity at low $T_e$ with a cut-off above a certain $T_e$ threshold. A third important remark is that the current decay time depends on the characteristics of the cooling, i.e. on 1) the timescale of the cooling and 2) the post-cooling temperature. 

This last observation explains the large differences observed between simulations of pure Deuterium versus impurity-containing MMI, and suggests that Deuterium SPI may be used as a means to promptly and strongly dilute ITER plasmas without immediately triggering a TQ, thanks to a relatively high post-cooling temperature. Such a strategy may be instrumental in avoiding RE generation, motivating its investigation with JOREK \cite{Nardon2020}. It was found that in the absence of pre-existing islands and provided that the density of background impurities is low enough, the desired effect can indeed be obtained. Simulations are on their way to assess whether pre-existing magnetic islands can have a detrimental influence for realizing this strategy (first considering ASDEX Upgrade plasmas). It is assessed in particular whether the presence of the $2/1$ island makes it harder to get SPI shards across the $q=2$ rational surface without triggering a TQ~\cite{Wieschollek2020}.

\subsubsection{Thermal Quench dynamics and plasma current spike}\label{:applic:core:disr:TQ_dyn}

\begin{figure}
\begin{center}
\centering
\includegraphics[trim=110 0 150 0, clip, width=0.24\textwidth]{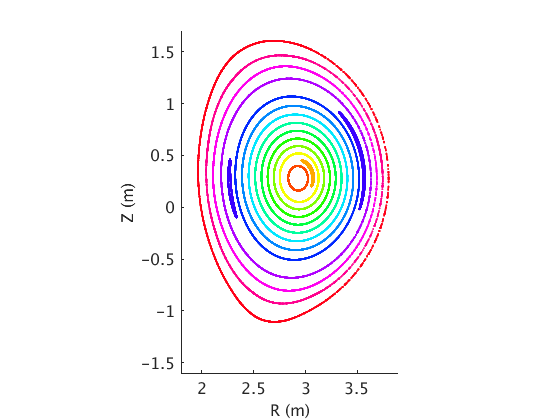}
\includegraphics[trim=110 0 150 0, clip, width=0.24\textwidth]{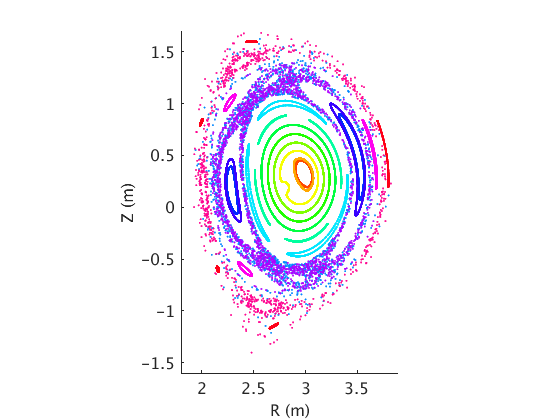}
\includegraphics[trim=110 0 150 0, clip, width=0.24\textwidth]{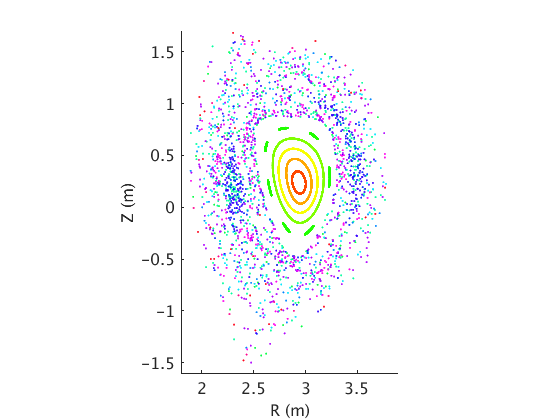}
\includegraphics[trim=110 0 150 0, clip, width=0.24\textwidth]{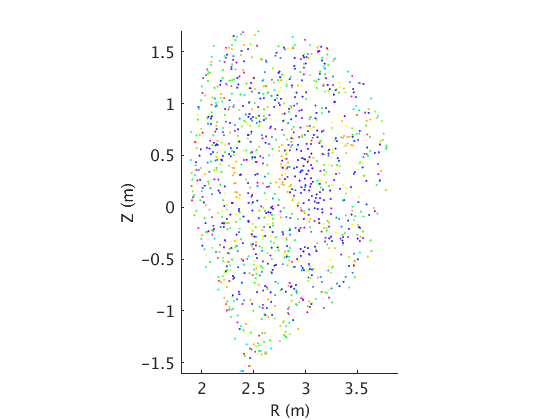}
\caption{Poincar{\'e} cross sections at times 1.19ms, 1.91ms, 2.29ms and 2.46ms in a simulation of massive Argon injection in JET (pulse 85943). The corresponding times are indicated by vertical lines in Figure~\ref{fig:Ip_spike}. \noreprint}
\label{fig:JET_MGI_Poinc}
\end{center}
\end{figure}

\begin{figure}
\centering
  \includegraphics[width=0.6\textwidth]{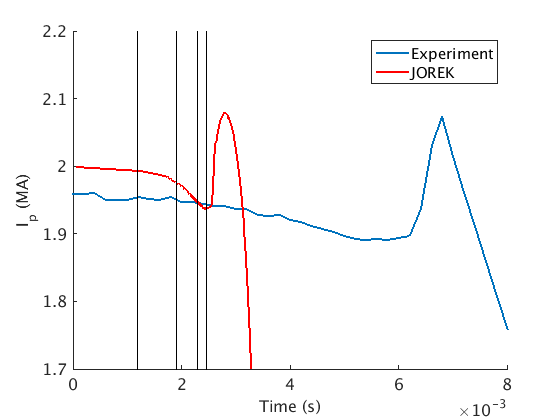}
\caption{Simulated and experimental time traces of the plasma current during a massive Argon injection in JET (pulse 85943). Vertical lines indicate the time of the Poincar{\'e} cross sections shown in Figure~\ref{fig:JET_MGI_Poinc}. \noreprint}
\label{fig:Ip_spike}
\end{figure}

The presence of an $I_p$ spike is a robust experimental observation associated to the TQ, but a difficult one to reproduce quantitatively with 3D non-linear MHD simulations. $I_p$ spikes in simulations have been reported in the past~\cite{Izzo_2010,Nardon2017} but with a significantly smaller amplitude than experimental spikes. However, in recent JOREK MGI simulations, pushing the parameters towards realistic experimental conditions and studying Argon MGI in JET, an $I_p$ spike of comparable magnitude to the experimental one has been obtained, as can be seen in Figure~\ref{fig:Ip_spike}. In these simulations, the \revised{mechanisms leading to the TQ are partly the same as described above for Deuterium MGI} but modes are more violently destabilized, leading to stronger stochasticity throughout the plasma (a series of Poincar{\'e} plots, at times indicated by vertical lines in Figure~\ref{fig:Ip_spike}, are shown in Figure~\ref{fig:JET_MGI_Poinc}). The fact that modes are more destabilized is a result of a stronger cooling with Argon than Deuterium MGI and of the more realistic (i.e. lower) resistivity used in the simulation, which generates sharper skin currents and thus a more unstable current profile. \revised{Also, a seemingly critical feature associated to a large $I_p$ spike \cite{Nardon2021} is a radiative cooling strong enough to persist near the O-point of the $2/1$ island, even as the island gets destroyed by magnetic stochasticity. This promotes a local collapse of the current density which drives the $2/1$ mode to a very large amplitude.} According to theory~\cite{Boozer_PPCF_2018}, the $I_p$ spike results from a relaxation of the current density profile at approximately fixed magnetic helicity. This relaxation is caused by the magnetic field stochastization and may be modelled, in a mean-field approach, by a hyper-resistivity term in the poloidal flux evolution equation~\cite{Boozer_PPCF_2018}. Boozer predicts a relation between the mean-field hyper-resistivity and the field line stochastic diffusivity (Eq. 68 in \cite{Boozer_PPCF_2018}). JOREK mean-field simulations, i.e. axisymmetric simulations with an \textit{ad hoc} hyper-resistivity term of the order of that predicted by Boozer, indeed match 3D simulations~\cite{Nardon2019}. The $I_p$ spike magnitude and the modes amplitude (which govern the field line stochastic diffusivity), thus appear to be strongly connected. In order to progress in the validation of JOREK simulations, ongoing work aims at comparing both the $I_p$ spike magnitude (and more generally the time evolution of $I_p$) and the modes amplitude with experimental data, in particular by using synthetic saddle coils in JET simulations. 

Another important question related to RE generation, is that of the decay of MHD modes and magnetic stochasticity after the TQ. This point is actually also related to the $I_p$ spike. Indeed, according to robust theoretical estimates done by Boozer, if the current profile relaxation during the TQ was complete, and assuming good magnetic helicity conservation, the $I_p$ spike would be much larger than experimentally observed~\cite{Boozer_PPCF_2018}. Thus, either helicity dissipation is non negligible, or the relaxation is incomplete. JOREK simulations seem to point to the latter explanation. Indeed, they usually display good helicity conservation during the TQ but a clear decay of stochasicity (starting from the core of the plasma) before the current profile has fully flattened, as visible for example in Figures 11 and 12 of \cite{Hoelzl2020A}.

\subsubsection{Assimilation and mixing of injected material}\label{:applic:core:disr:fuell_dens}

The question of the assimilation and mixing of the injected material is important, both for reducing localized heat fluxes and for avoiding REs. Indeed, the present ITER RE avoidance strategy relies on raising $n_e$ by more than one order of magnitude throughout the plasma ~\cite{Martin_Solis_2017}. This density rise should be uniform, otherwise REs could be generated in low density regions.

It is generally observed in JOREK MGI and SPI simulations that the deposited material initially expands in the parallel direction at the speed of sound, as also seen in simulations of pellet ELM pacing (see Section \ref{:applic:edge:pellet}). After some time, the part of the cloud expanding in one direction may run into the part expanding in the other direction, leading to substantial viscous dissipation.

\begin{figure}
\centering
  \includegraphics[width=0.6\textwidth]{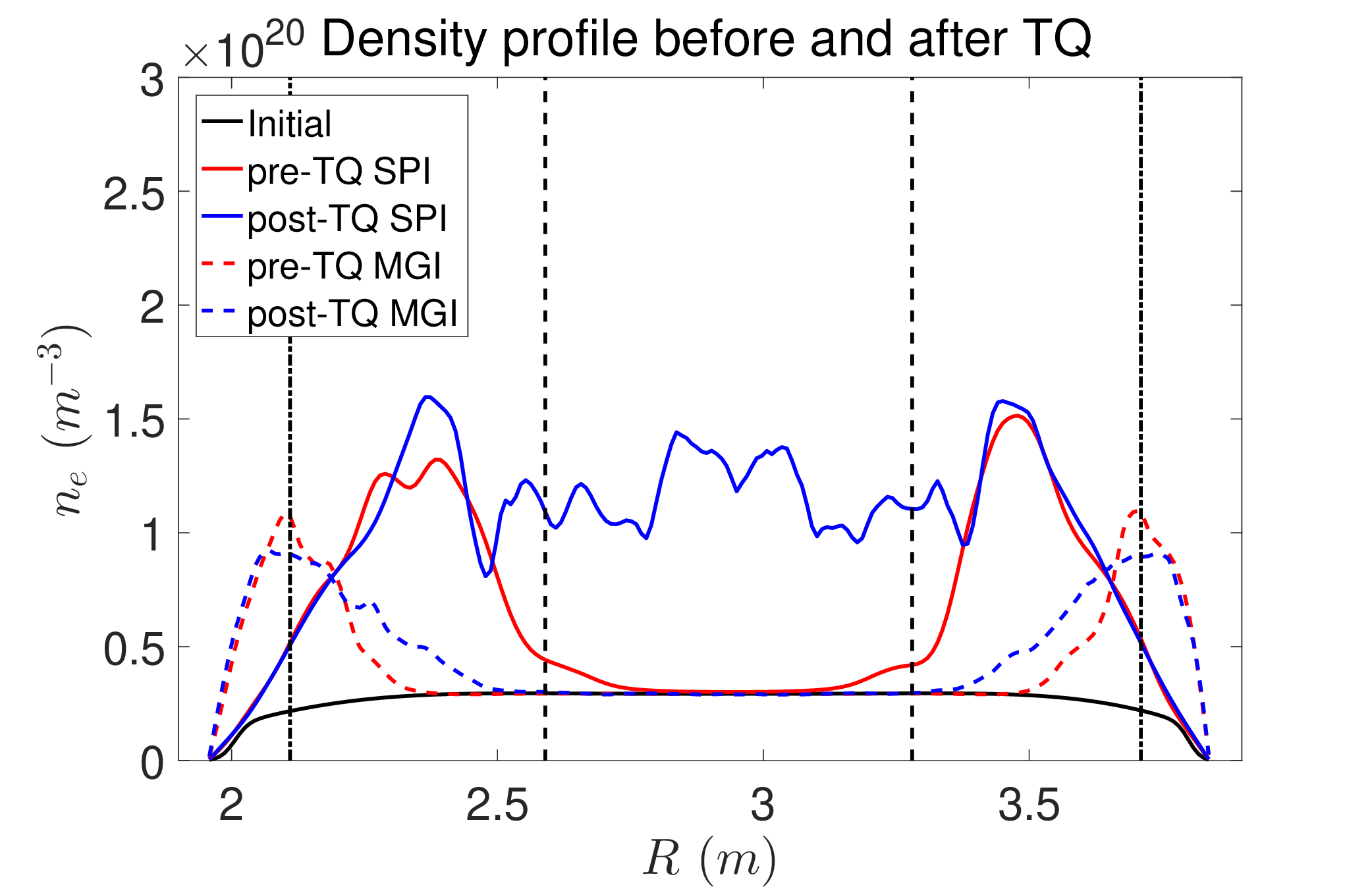}
\caption{Electron density profile just before (red) and just after (blue) the TQ for simulations of Deuterium SPI (plain) and Deuterium MGI (dashed) in JET. The black plain and dashed vertical lines indicate the position of the $q=2$ and $q=1$ surfaces, respectively. Re-print from Ref.~\cite{Hu2018}. \reprintiaea}
\label{fig:Di_fig8a_SPI_mixing}
\end{figure}

JOREK simulations of Deuterium MGI and SPI in JET~\cite{Hu2018,HuEPS2018} show that SPI is superior regarding material assimilation and mixing, as illustrated in Figure~\ref{fig:Di_fig8a_SPI_mixing}, thanks to the deeper penetration of solid shards compared to gas. The same simulations also suggest, for target plasmas with $q_0$ substantially below 1 and thus with a large $q=1$ radius, a key role of the $1/1$ internal kink mode for material mixing into the plasma core during the TQ. In this respect, the position of the shards at the time of the internal kink mode crash appears critical: if shards are `within reach' of the mode's flow structure, substantial material mixing into the core happens; otherwise, core mixing is poor. These observations are related to the 1/1 mode structure, which exhibits uniform and strong displacement within the $q=1$ surface, but a weak displacement outside of that surface. Simulations in which $q_0$ was elevated above 1 so as to remove the internal kink mode, also show poor core mixing early in the TQ.~\cite{White2001Book}. Recent impurity SPI simulations of ITER plasmas show similar results~\cite{Hu2020}.

\begin{figure}
\centering
  \includegraphics[width=0.9\textwidth]{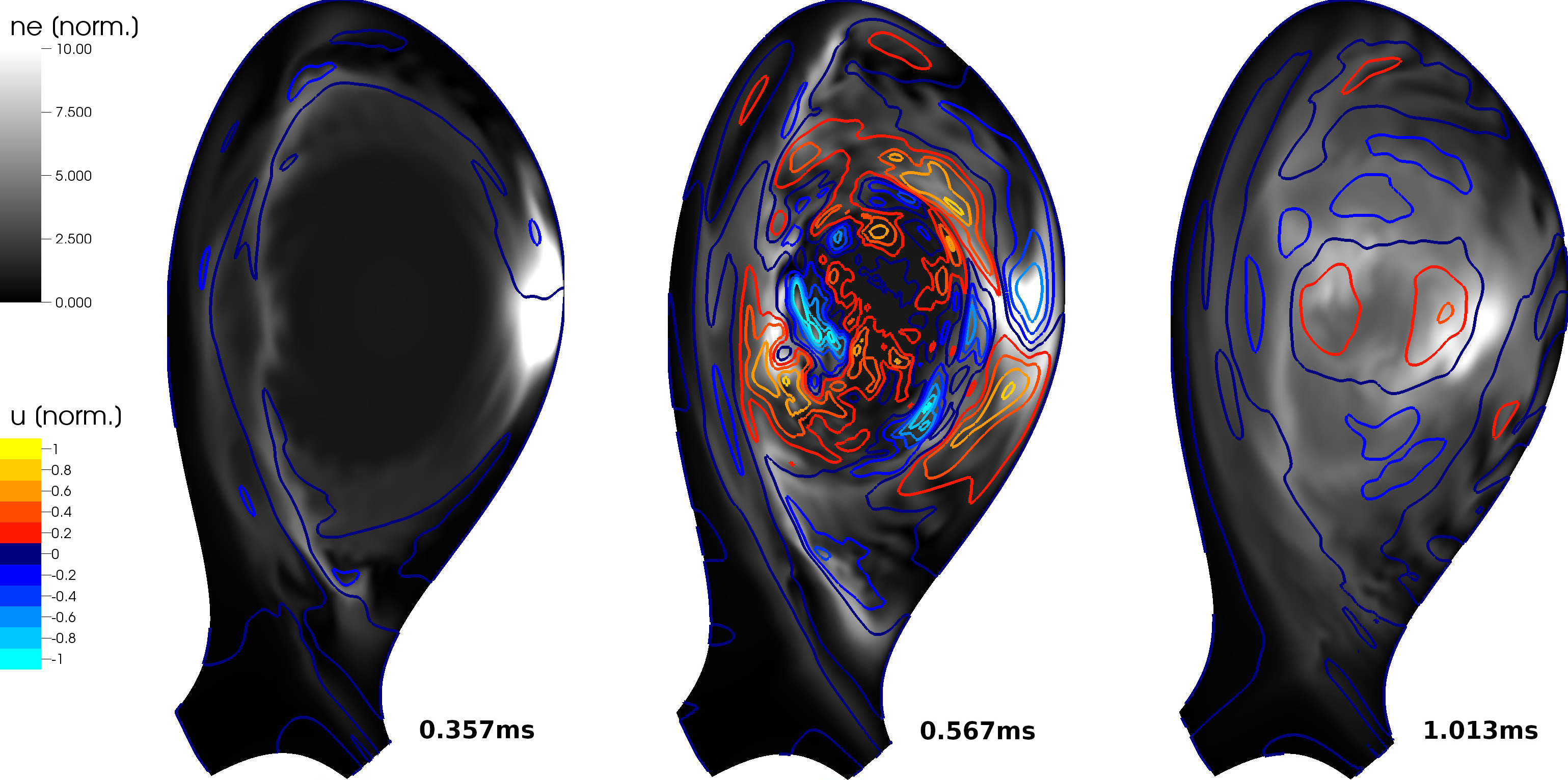}
\caption{Colormap of the electron density distribution (black and white) with overlaid iso-contours of the stream function for the perpendicular plasma velocity (colours) during a simulation of Deuterium SPI in ASDEX Upgrade. The 3 plots correspond, from left to right, to the pre-TQ, TQ, and early CQ phase. Re-print from Ref.~\cite{Hoelzl2020A}. \reprintaipOK}
\label{fig:AUG_SPIdisr}
\end{figure}

However, the absence of a $1/1$ mode does not necessarily imply bad mixing. Indeed, simulations of Deuterium SPI in ASDEX Upgrade~\cite{Hoelzl2020A}, for which the target plasma has $q_0>1$, found good core mixing even if the TQ happens while shards are still far from the core (e.g. near the $q=2$ surface). More precisely, as shown in Figure~\ref{fig:AUG_SPIdisr}, the density profile is usually hollow during and shortly after the TQ, but by the time flux surfaces reappear, the density profile has typically strongly flattened.

The more pronounced core mixing later into the TQ for the ASDEX Upgrade simulations is likely related to the much larger mode amplitudes and stronger magnetic stochasticity (as described in \ref{:applic:core:disr:TQ_trigg}), which lead to both stronger $\mathbf{E} \times \mathbf{B}$ flows and a larger radial transport by parallel flows. An important observation is that parallel flows are strongly driven during the TQ due to the heating of the relatively dense and cold region where the material has been deposited by the heat flux coming from the core along stochastic field lines.


\subsubsection{Radiated fraction and radiation asymmetry}\label{:applic:core:disr:rad_frac}

A uniform deposition onto the plasma facing components of the pre-TQ thermal energy $W_{th}$ is critical for efficient thermal load mitigation during ITER disruptions \cite{Lehnen201539}. To achieve this, both a large radiated thermal energy fraction ($\ge 90 \%$ for the baseline ITER scenario with $W_{th} = 350$ MJ) and a low radiation asymmetry are required. It is usual to characterize the radiation asymmetry by a Toroidal Peaking Factor (TPF) and a Poloidal Peaking Factor (PPF). In this section, we discuss the TPF, which is defined as the maximal (over the toroidal angle) over mean radiated power per unit radian. The target for ITER is a TPF lower than 2. The PPF shall be the subject of future studies.

\begin{figure}
\centering
  \includegraphics[width=0.55\textwidth]{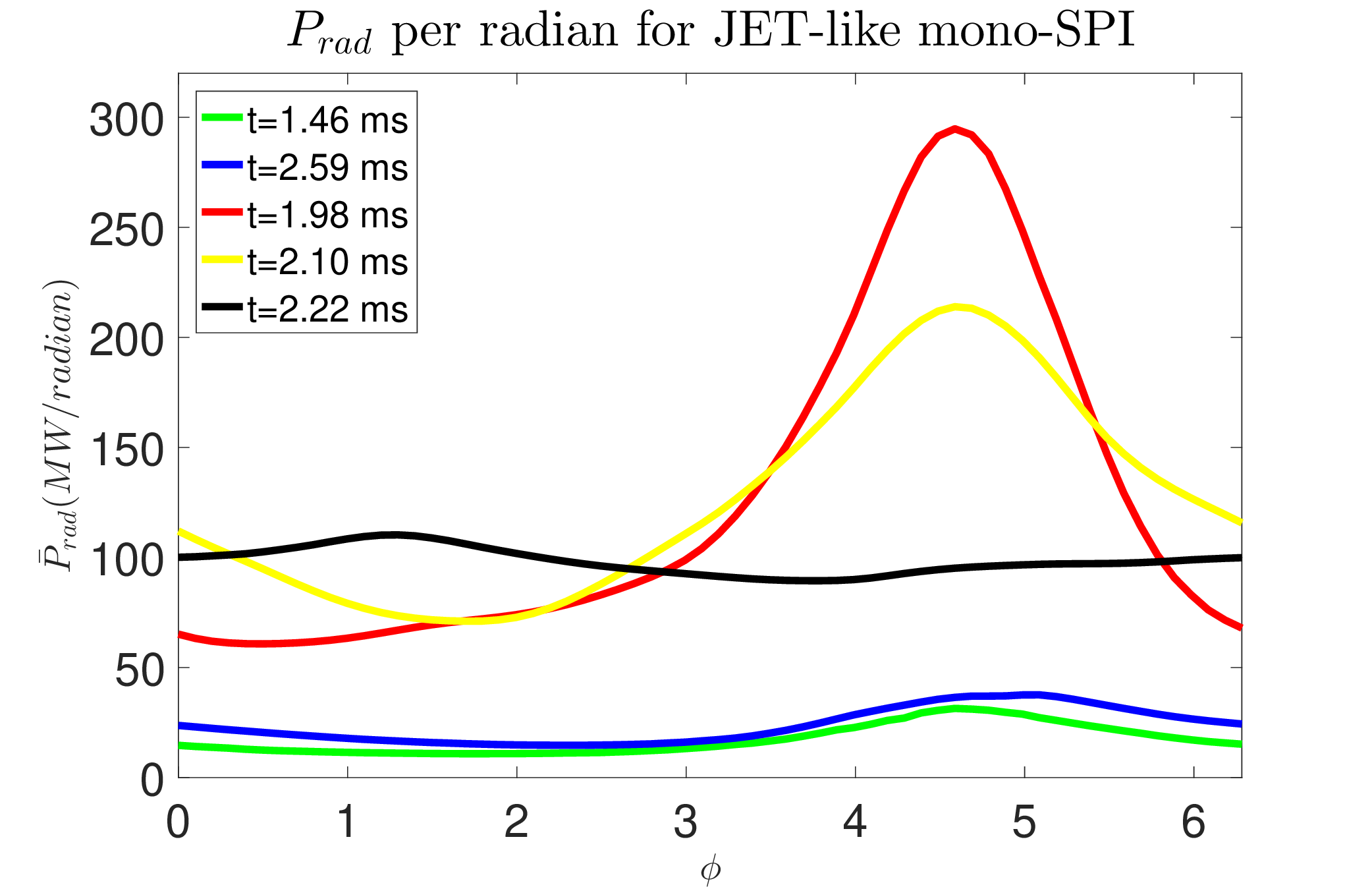}
\caption{The relative and absolute radiation power within each poloidal plane for Neon SPI into a JET L-mode plasma. The TQ occurs at $t=1.98ms$. The fragments are injected at a toroidal angle of $4.51$ radian.  \noreprint}.
\label{fig:mono_SPI_radiation_asym}
\end{figure}

With SPI from a single toroidal location, the TPF may be above the ITER target during the pre-TQ and early TQ stages. An example of such behaviour is shown in Fig. \ref{fig:mono_SPI_radiation_asym}. In this JET simulation, the injected fragments consist of pure Neon and are flying along the poloidal plane. It is evident that the peak radiation power is located close to the toroidal location of the fragments, although there is some drift along the course of the injection. The TPF reaches its peak value, which is larger than 2, at the time of the TQ onset. It then gradually relaxes over the course of the TQ, until the radiation power becomes relatively toroidally uniform. The cause of such asymmetry can be traced back to the unrelaxed impurity density along the field lines, which is related to the finite ion sound velocity. 

\begin{figure}
\centering
  \includegraphics[width=0.48\textwidth]{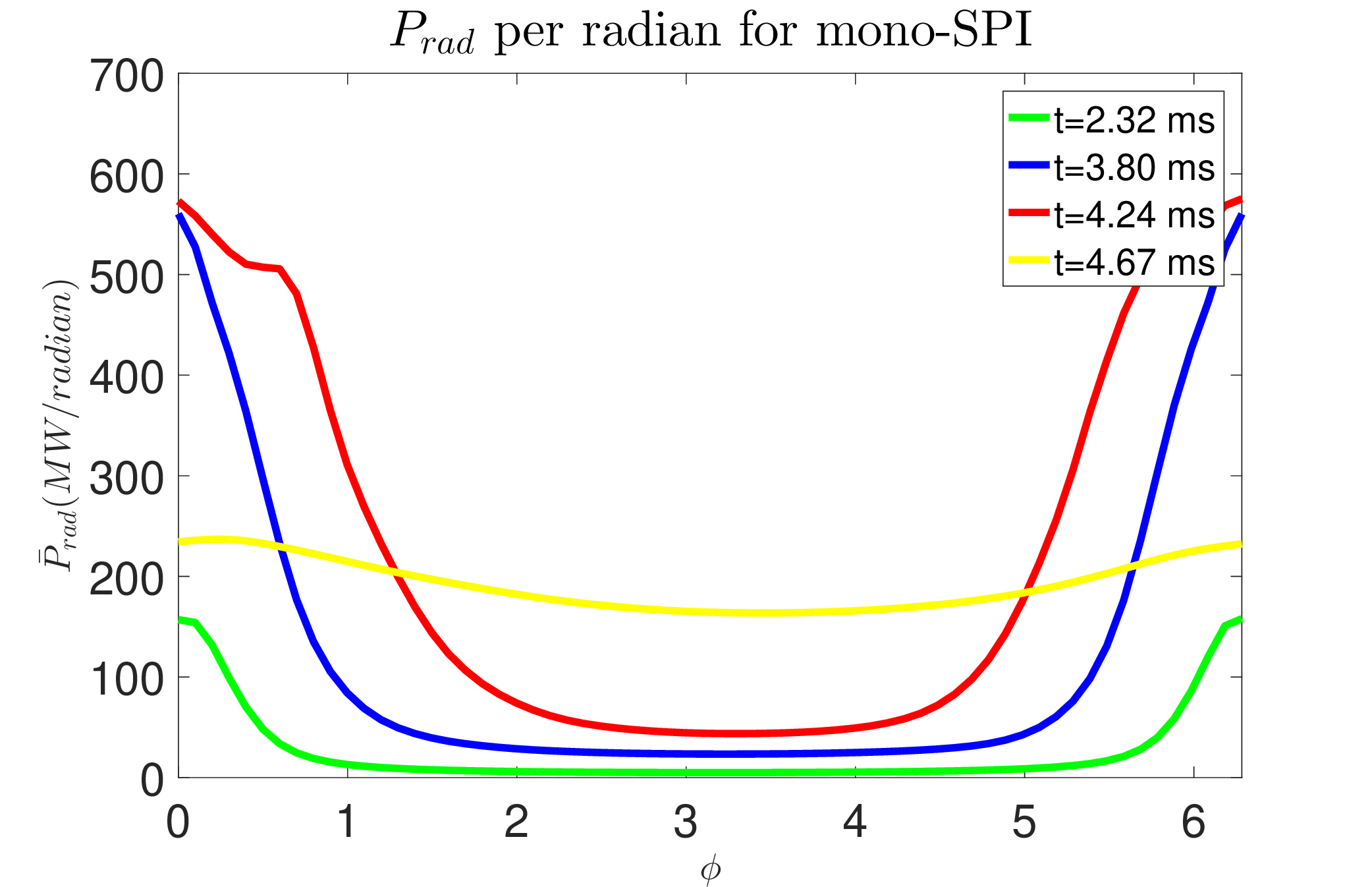}
  \includegraphics[width=0.48\textwidth]{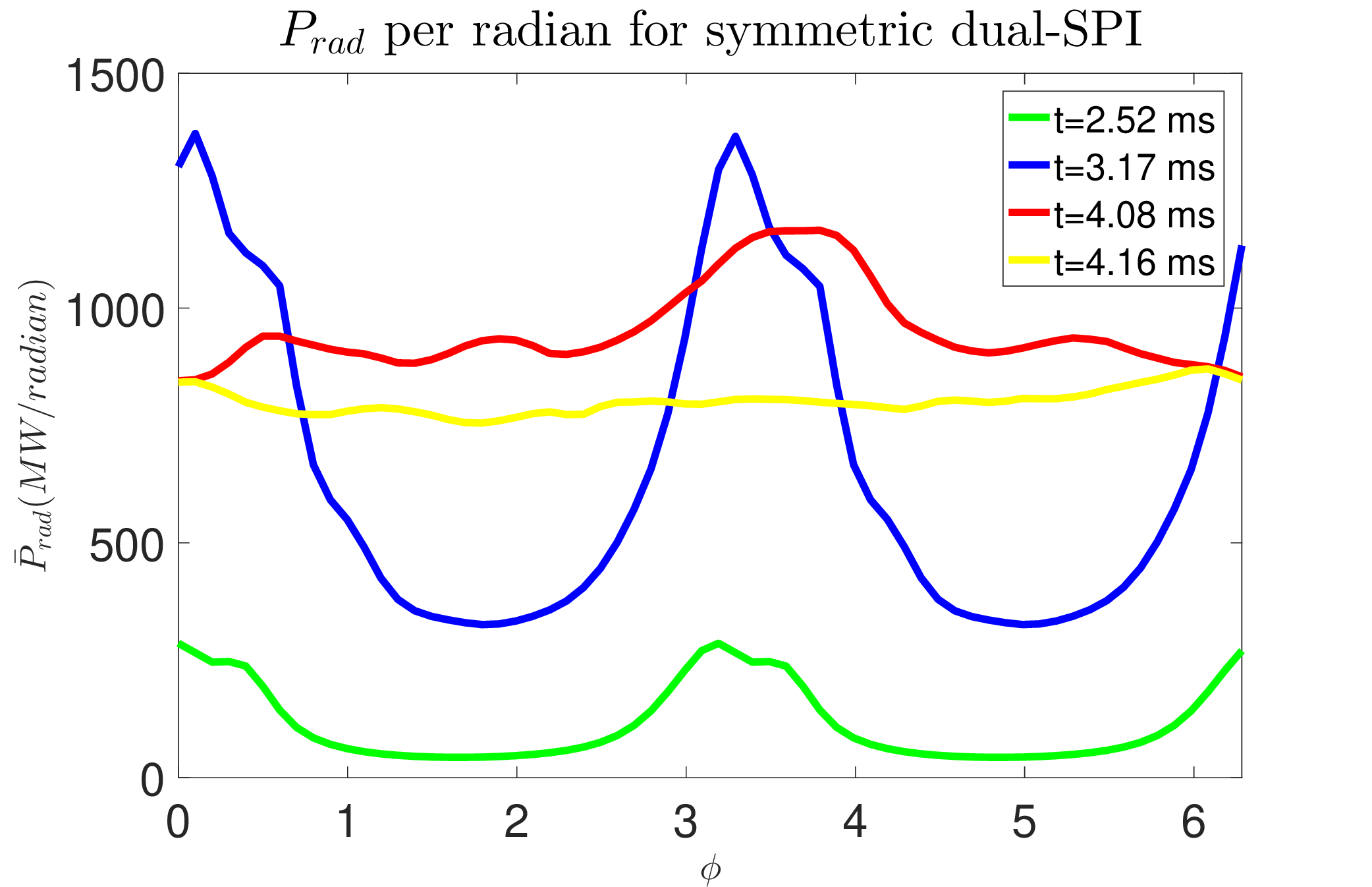}
\caption{Comparison of radiation asymmetries between Neon single-SPI (left) and dual-SPI (right) into an ITER L-mode plasma. Re-print from Ref.~\cite{Hu2020}. \reprintiaea}
\label{fig:compare_SPI_radiation_asym}
\end{figure}

JOREK simulations show that multiple injections may significantly mitigate the TPF, as illustrated in Fig. \ref{fig:compare_SPI_radiation_asym}. Here, we are comparing the toroidal radiation distribution of a single-SPI and a symmetric dual-SPI, both injecting $2.6\times 10^{22}$ Neon atoms along with $2.1\times 10^{24}$ Deuterium atoms into an ITER L-mode plasma. For the single-SPI, the injection is done at $0$ radian, while for the dual-SPI injections are done at $0$ and $\pi$ radian. A significant TPF is again exhibited in the single-SPI case, even during the early stage of TQ at $t=4.24$ ms. On the other hand, for the dual-SPI case, in spite of a strong asymmetry in the early pre-TQ stage, the asymmetry is strongly mitigated by the time of TQ at $t=4.08$ ms.

\subsection{Vertical Displacement Events and halo current dynamics}\label{:applic:core:vdes}

In the case where the vertical control of an elongated plasma is lost, the plasma column undergoes an axisymmetric instability referred to as a Vertical Displacement Event (VDE). The loss of vertical control can be due to a large MHD perturbation or to a technical failure of the vertical control system. As the plasma drifts vertically towards the wall, large heat and electromagnetic loads are deposited into the PFCs and surrounding structures. The extrapolation of the magnitude and distribution of these loads to larger machines, in particular the 3D distribution, is still not well established. In this respect the simulation of these events with 3D MHD codes is crucial. \revised{For other non-linear MHD codes investigating VDEs, refer, e.g., to Refs.~\cite{Strauss2020,Bunkers2020,Clauser2021,Marx2017} and references therein.}

The simulation of VDEs requires free-boundary conditions for the magnetic field and a self-consistent evolution of currents in the structures surrounding the plasma. This is possible (including with 3D conducting structures) with the free-boundary extension of JOREK (see Section~\ref{:code:models:freebound}). Based on this extension, a first benchmark of the growth rates for axisymmetric VDEs in a simplified ITER-like plasma had been carried out between JOREK and the CEDRES++ code in~\cite{Hoelzl2014} revealing \revised{good quantitative agreement (within 10-20\%)} over a large range of wall resistivities (including fully realistic values). First demonstrations of 3D VDEs \revised{using JOREK} were shown in~\cite{AleynikovaUnpublished,Hoelzl2014}.

\begin{figure}
\centering
  \includegraphics[width=0.75\textwidth]{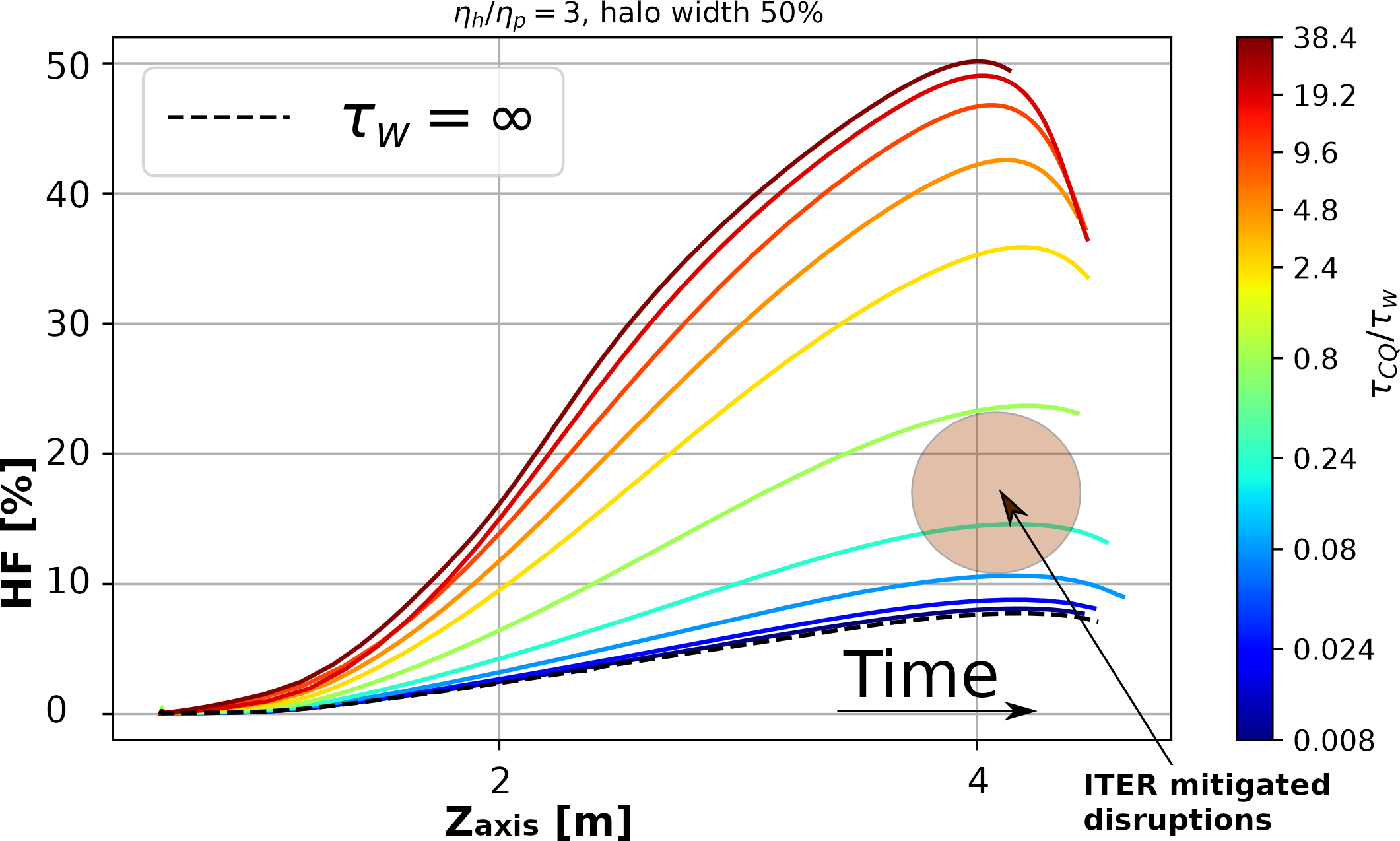}
\caption{Halo current fraction versus vertical position of the magnetic axis for different $\tau_\text{CQ}/\tau_w$ ratios~\cite{artola2020_param}. The halo current fraction is defined as $\text{HF}\equiv  (1/I_{p,0})\int |\mathbf{J}\cdot\mathbf{n}|/2 dS_{wall}$. \noreprint}
\label{fig:VDE_halo_fraction}
\end{figure}


An important area of research connected to VDEs and disruptions is the halo current, i.e. the current that flows partly in the plasma SOL, closing its path \textit{via} conducting structures in contact with the SOL~\cite{Gruber_1993}\cite{Humphreys_1999}\cite{Boozer_halo_2013}. Axisymmetric parametric scans for an ITER 15 MA upward VDE were performed in \cite{artola2020_param} by fixing the halo region temperature and width. Scans in the plasma resistivity showed that when the CQ time is much smaller than the decay time of the wall currents ($\tau_\text{CQ}\ll \tau_w$), the vertical position is a monotonic function of the plasma current $Z(I_p)$ as predicted by Kiramov's wire model \cite{Kiramov2017Ip}. In this limit the shape of the $Z(I_p)$ curve does not depend on specific time scales but only on the shape of the toroidal current profile. The implication is that ITER plasmas will transition from an X-point to a limiter configuration at a large total current ($\sim 10$ MA) regardless of the mitigation scheme used, which could have important consequences regarding the wall damage caused by a vertically unstable RE beam, if RE mitigation by SPI does not achieve its goal in ITER.

The latter scans show that the expected Halo current Fraction (HF) depends strongly on the $\tau_\text{CQ}/ \tau_w$ as shown in Fig. \ref{fig:VDE_halo_fraction}. A minimum HF of $\sim 10\%$ is found at $\tau_\text{CQ}/ \tau_w \ll 1$  and maximum fractions of $\sim 50\%$ have been found in the limit $\tau_\text{CQ}/ \tau_w \gg 1$, similar to the maximum values obtained with the DINA code \cite{Lehnen201539}. In the fast CQ limit the maximum HF is below $10\%$ regardless of halo width and temperature assumptions. Mitigated disruptions in ITER will target CQ times in the range of 50 to 150 ms (to be compared to $\tau_w=0.5$ s). The corresponding predicted HF range is $10-20\%$. This suggests that the ITER DMS system could be used to reduce the halo current electromagnetic loads by more than a factor 2 with respect to their maximum possible value.

\begin{figure}
\centering
  \includegraphics[width=0.65\textwidth]{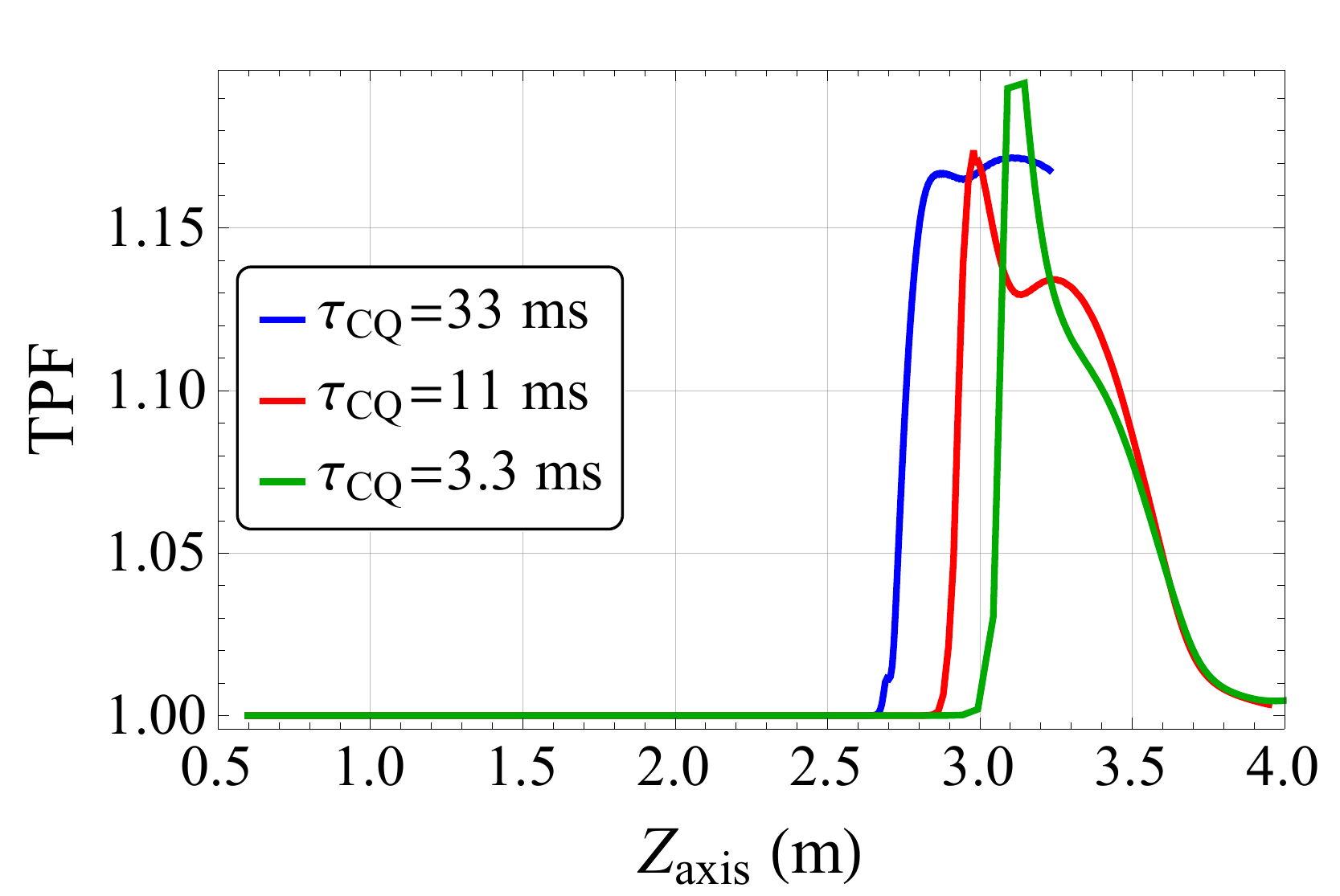}
\caption{Toroidal peaking factor of the halo current as a function of the vertical position of the magnetic axis for an ITER 15 MA upward VDE \cite{ArtolaPhD}. \noreprint}
\label{fig:TPF_halo}
\end{figure}

3D simulations for an ITER 15 MA plasma were conducted in the fast CQ limit \cite{ArtolaPhD}. The Toroidal Peaking Factor (TPF) of the poloidal halo currents is shown in Fig. \ref{fig:TPF_halo} for 3 different values of the plasma resistivity leading to different CQ times. It can be seen that the TPF (given by an $n=1$ kink mode) is very low in the fast CQ limit (while in experiments the TPF can reach values of 4). Moreover the maximum TPF does not seem to depend strongly on the CQ time if $\tau_\text{CQ}/ \tau_w \ll 1$. Note however that CQ times in these simulations are below 50 ms, which is not allowed in ITER due to eddy current constraints on the blanket modules \cite{Lehnen201539}. These results are therefore presented only for their physical interest and future work will be devoted to realistic ITER predictions. 

The change of the edge current density and edge safety factor during a VDE when the plasma moves into the PFCs is discussed in~\cite{Artola2020A}. Analytical theory and JOREK non-linear simulations show excellent agreement. When currents are lost in the scraped-off region of the plasma, a significant fraction of these currents is re-induced in the edge hot core region of the plasma. The same mechanisms also apply in case of a loss of the plasma edge current due to a cooling of the plasma edge caused by an MMI.

Predictive simulations of VDEs must include several effects that are relevant for the evolution of the density and of the temperature in the SOL region. The energy balance given by perpendicular and parallel transport, Ohmic heating, radiation, ionisation and other effects determines the plasma temperature, which in turn has a large impact on the current density and the VDE dynamics through the plasma resistivity. In~\cite{artola2020COMPASS}, a complete set of sheath boundary conditions was taken into account together with a neutral fluid model and realistic physical parameters, e.g., Spitzer resistivity and Spitzer-Härm parallel conduction. It was found that it is particularly important to couple the electrical current density to the particle density and temperature through a boundary condition that limits the maximum current density to the ion saturation current. Otherwise large currents can be induced in regions of very low particle density, leading to non-physical dynamics. Although the effect of impurity radiation was not included, these simulations are the first to provide a fully self-consistent evolution of halo currents. 

\subsection{Runaway electron physics}\label{:applic:core:res}

\subsubsection{Test electron dynamics during the thermal quench}\label{:applic:core:res:particles}

\begin{figure}
\centering
  \includegraphics[width=0.7\textwidth]{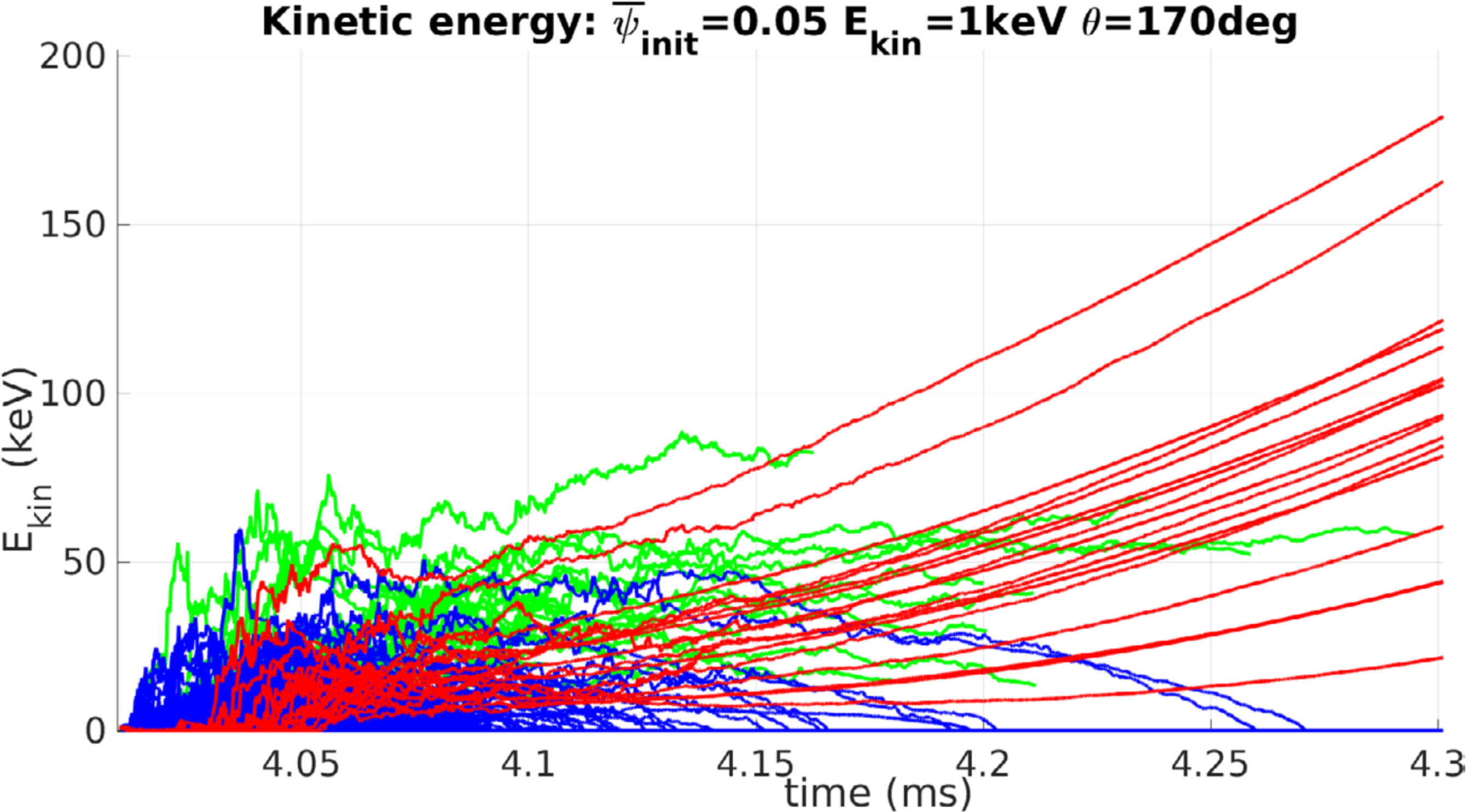}
\caption{Early evolution of the kinetic energy of a set of electrons initialized in the plasma core with a kinetic energy of 1 keV. Green lines correspond to lost electrons while red (resp.\ blue) lines correspond to electrons which remain confined and have a final energy above (resp.\ below) 1 MeV. Re-print from Ref.~\cite{Sommariva2018}. \reprintiaea}
\label{fig:REaccel}
\end{figure}

RE generation during disruptions is a major concern for ITER~\cite{Breizman2019}. A key question in this area is that of fast electron losses along stochastic field lines during a disruption. 
To study this question, a relativistic test particle tracer (with 2 options: full orbit or guiding center) has been implemented in JOREK (see Section~\ref{:code:models:particles}). It has first been applied to study electron losses at approximately fixed energy in the electromagnetic fields of MGI-triggered disruptions (see Section~\ref{:applic:core:disr})~\cite{Sommariva2017,SommarivaPhD}. Substantial electron losses were found during the strongly stochastic TQ phase, which increased with the electron energy until reaching a saturation as the electron velocity approaches the speed of light. However, a few tens of $\%$ of electrons at typical pre-TQ thermal energies ($\sim1$ keV) were found to remain in the plasma by the time flux surfaces start reforming after the TQ. This is by far enough to give birth to a large RE beam, but this number should be taken with caution since the magnetic field stochasticity may have been underestimated in these simulations, as suggested by the under-predicted plasma current spike. In a second study, the full dynamics of test electrons, including acceleration or braking by the parallel electric field and by collisions, were investigated in the same JOREK simulations~\cite{Sommariva2018,SommarivaPhD}. This revealed that the very large ($\sim 1$ kV/m) parallel electric field fluctuations during the TQ can have a key role in RE generation. These fluctuations indeed rapidly broaden the electron energy distribution, with a typical energy reaching a few tens of keV after a few tens of $\mu$s, as can be seen in Fig.~\ref{fig:REaccel}. This makes electrons less collisional and therefore promotes RE generation. This is a new effect, not captured by typical lower-dimensional studies where only the axisymmetric component of the electric field is considered. Finally, the magnetic field from a JOREK disruption simulation has been used in a recent test electron study with the ASCOT code~\cite{Sarkimaki2020}.

\subsubsection{Runaway electron fluid model applied to VDEs and RE beam termination}\label{:applic:core:res:fluid}

Complementary to the passive test particles approach, a RE fluid model has been implemented, which allows to describe the full non-linear interaction of the REs with the MHD activity~\cite{Bandaru2019A}. While various applications of the model are on their way, some studies have already been performed. \revised{Similar approaches are followed by other codes, see e.g., Refs.~\cite{Cai2015,Matsuyama2017,Zhao2020,Sainterme2020APS}.}

\begin{figure}
\centering
  \includegraphics[width=0.75\textwidth]{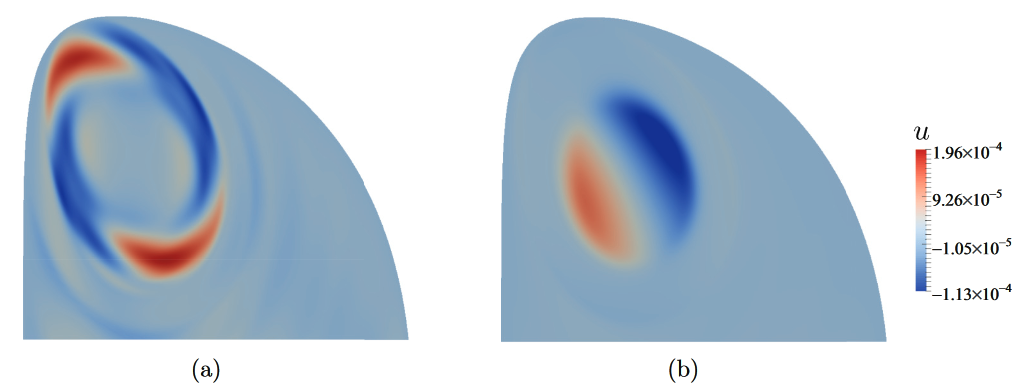}
\caption{Structure in the perturbed (normalized) electric potential $u$ during the non-linear phase of the mode growth. (a) Without REs, the $n=2$ component dominates while (b) with REs an $n=1$ mode is developing due to the different q-profile (1/1 internal kink). Re-print from Ref.~\cite{Bandaru2019A}. \reprintaps}
\label{fig:REVDE}
\end{figure}

Axisymmetric and non-axisymmetric VDEs have been studied with and without the presence of REs~\cite{Bandaru2018IAEA}, showing that the vertical motion of the plasma can be slowed down by the presence of REs if the decay of the plasma current is influencing the dynamics of the vertical instability significantly. It was also demonstrated that the RE current, peaking off-axis in this case, may lead to the destabilization of a $1/1$ internal kink instability in a plasma that would otherwise have a dominant $2/1$ helical perturbation, as shown in Fig.~\ref{fig:REVDE}. Such effects may have important consequences for the termination of RE beams. 

\begin{figure}
\centering
  \includegraphics[width=0.9\textwidth]{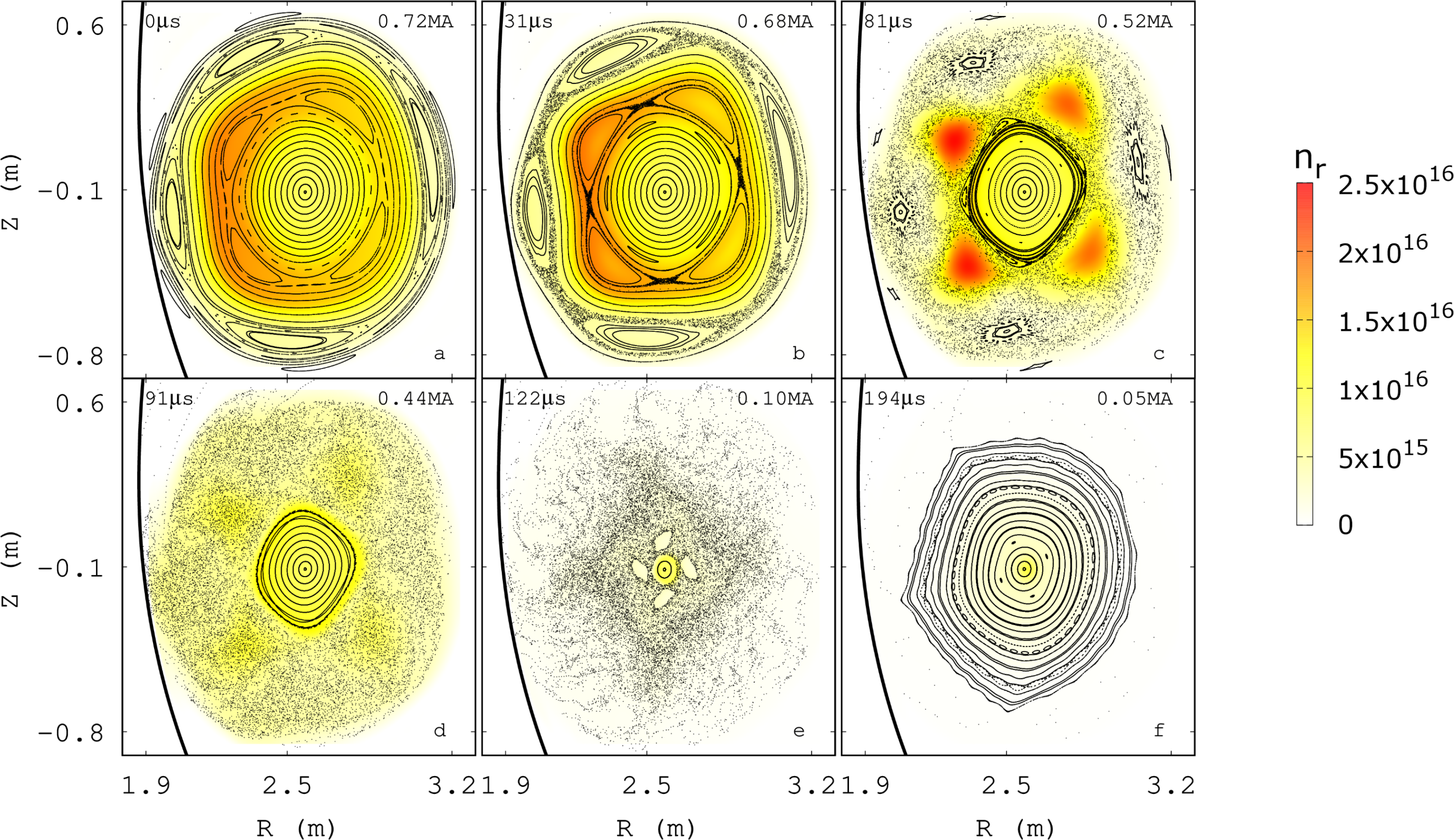}
\caption{Poincar{\'e} plots along with the RE number density (background color) during the MHD activity leading to the stochastization of the magnetic field and subsequent termination of the RE beam. Remaining RE current is also shown in each panel. Re-print from Ref.~\cite{Bandaru2020A}. \reprintiop}
\label{fig:re-termination}
\end{figure}

In a recent study, the benign termination of a RE in a JET experiment has been simulated~\cite{Reux2020,Bandaru2020A}. Based on experimental data, a plateau-phase RE beam with a hollow current density profile in a low density background plasma with negligible content of high-Z impurities is considered. It is found that an $n=1$ double tearing mode associated with the presence of two $q=4$ surfaces grow rapidly, with subsequent non-linear mode interaction causing global magnetic stochastization. This leads to the loss of REs on a timescale of $\leq100$ $\mu$s. Poincar{\'e} plots as well as the RE number density are shown in Figure~\ref{fig:re-termination}. The observed behaviour is in very good agreement with the experiment \revised{as it shows comparable mode structures and time scales}. Simulations also indicate a significant toroidal variation in the RE flux on the wall dominated by an $n=1$ structure and a poloidally broad RE deposition zone, partly explaining why no material damage is observed experimentally.

\subsection{Outlook}\label{:applic:core:outlook}

Concerning current and future work related to disruptions, like mentioned in Section~\ref{:code:models:outlook}, the development of a more realistic impurity model going beyond the coronal equilibrium assumption, is ongoing to improve the accuracy of MMI simulations.

A validation effort on MMI modelling is underway, involving in particular the simulation of MGI and SPI experiments in JET. Synthetic diagnostics such as interferometry, bolometry and saddle loops have been implemented for this purpose.

As described above, a growing effort is also devoted to testing disruption mitigation strategies for ITER. In particular, the study on the possibility of promptly diluting ITER plasmas with pure Deuterium SPI, described in Section ~\ref{:applic:core:disr:TQ_trigg}, is being refined, investigating e.g. the effect of pre-existing islands. Also, the work aimed at optimizing thermal load mitigation by radiation, described in Section ~\ref{:applic:core:disr:rad_frac}, is being pursued, considering more ITER scenarios and refining the treatment of impurities. The RE fluid model will be applied to further test RE avoidance or mitigation schemes for ITER. 

In the field of VDEs, current and future work involves experimental validation, implementation of sheath boundary conditions, a free-boundary extension for full MHD and a coupling with 3D volumetric wall codes.

%% file: 07_other.tex
\section{Further applications}\label{:applic:other}

\subsection{ITGs}\label{:applic:other:itgturbulence}

As described in Section~\ref{:code:models:itg}, an electrostatic model for ion temperature gradient (ITG) turbulence has been recently implemented. The fluid model and first benchmark results were shown in Ref.~\cite{Zielinski44thEPS}. The effect of magnetic shear onto the characteristics of global ITG modes are investigated in Ref.~\cite{Zielinski2020} in global simulations in simplified circular geometry similar to ~\cite{Dimits2000}. Typical up-down asymmetrical eigenmode structure with non zero ballooning angle is typically observed in the regimes of relatively large magnetic shear due to the increased coupling between rational surfaces ~\cite{Zielinski2020}). In the low shear regime,the unstable eigenmodes become narrowly localized on the corresponding rational surfaces  and exhibit no up-down asymmetry. The structure of the generated mean poloidal flow via Reynolds stress is investigated in more detail in Ref.~\cite{Zielinski2020}. 

\begin{figure}
\centering
  \includegraphics[width=0.48\textwidth]{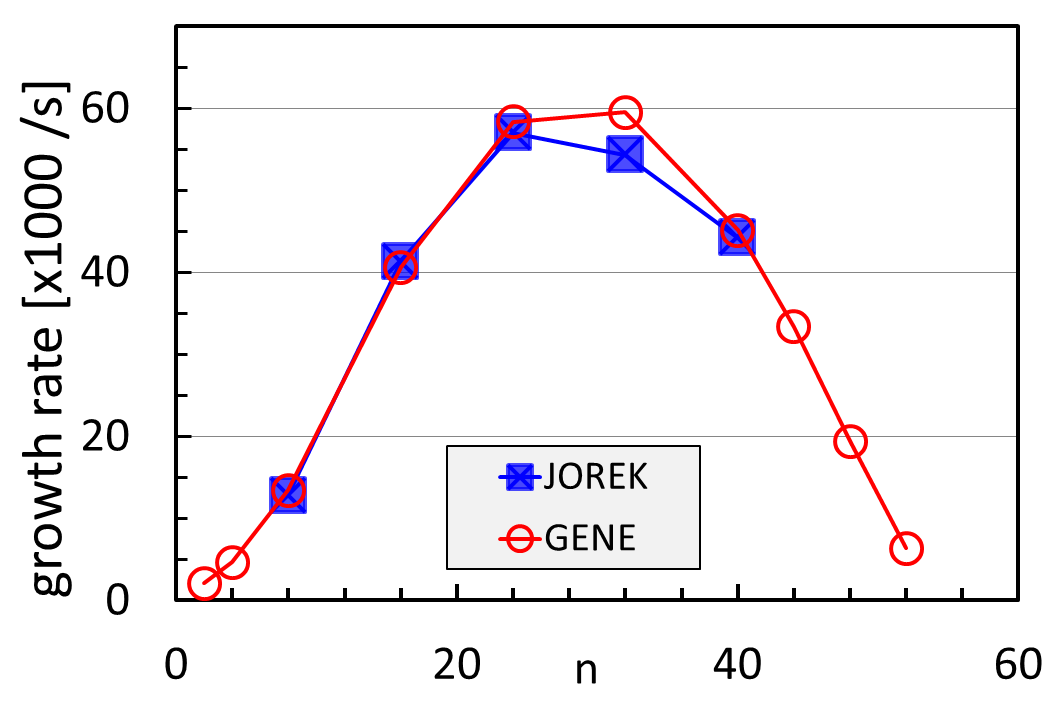}
  \includegraphics[width=0.315\textwidth]{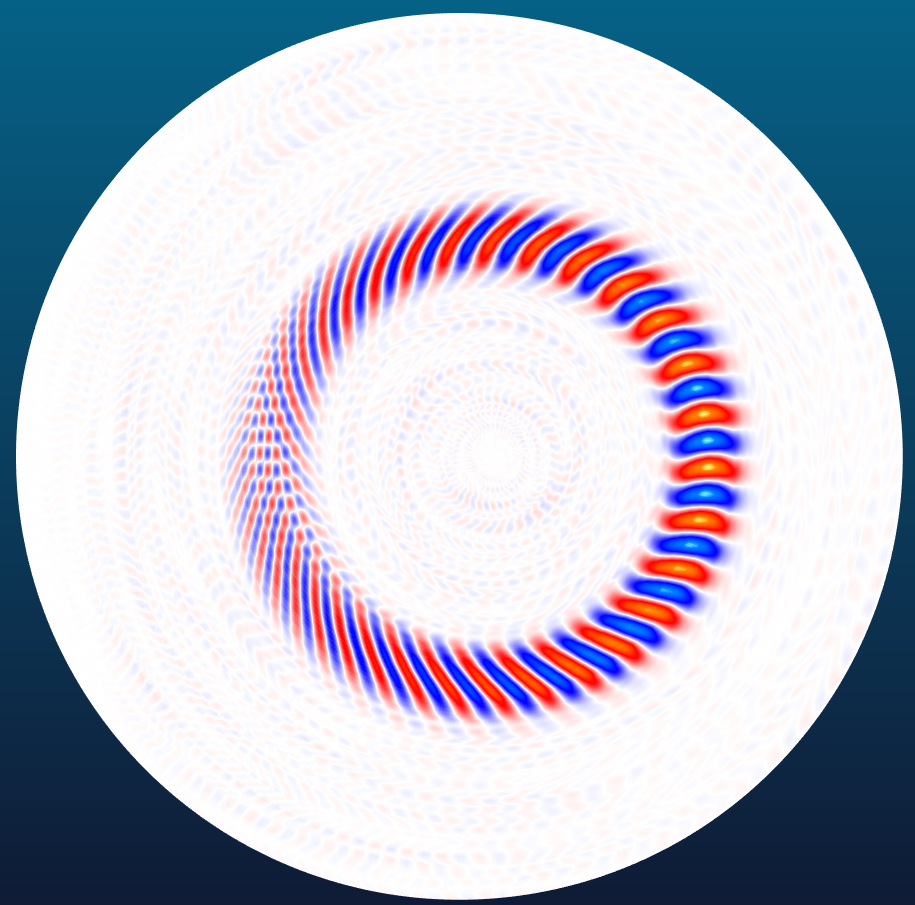}
  \includegraphics[width=0.18\textwidth]{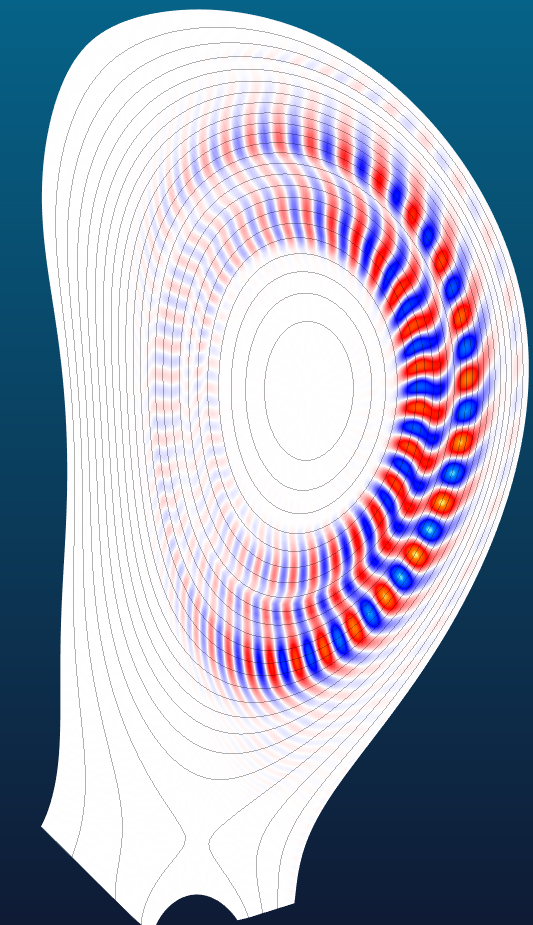}
\caption{\textbf{(Left)} Growth rate of ITG modes as a function of the toroidal mode number from the JOREK kinetic model compared to the GENE results from the benchmark \cite{Merlo2018}. \textbf{(Middle)} The ITG mode structure of the potential for the $n=24$ case. \textbf{(Right)} Example of the mode structure of an $n=20$ ITG mode in COMPASS X-point geometry L-mode plasma. \noreprint}
\label{fig:ITG_kinetic}
\end{figure}

The first benchmark results for linear ITG modes of the full-f electrostatic kinetic JOREK model with full orbit ions and adiabatic electrons are shown in Fig.~\ref{fig:ITG_kinetic}. For these cases, the full orbit of $10^9$ ions are traced with a time step of $5\cdot10^{-9} \mathrm{s}$ in a polar flux surface-aligned finite element grid of 101 radial and 256 poloidal elements. The full orbit ITG growth rates from JOREK are in good agreement with the gyrokinetic results from GENE, XGC and ORB5 presented in Ref.~\cite{Merlo2018}. 
Since the kinetic model uses the same grids and equilibria as the fluid models in JOREK, the kinetic model can also be applied in X-point geometry, including open field lines. As an illustration of this, Fig.~\ref{fig:ITG_kinetic}(right) shows an $n=20$ ITG mode in COMPASS X-point geometry, in the non-linear phase.

\subsection{Toroidal Alfv{\'e}n Eigenmodes and fast particles}\label{:applic:other:taes}
The excitation of TAE modes with an external antenna was studied in JET X-point plasmas was studied in Refs.~\cite{Dvornova2018EPS,Dvornova2020A} using the freeboundary extension described in Section~\ref{:code:models:freebound}. The paper addresses the question why, in JET experiments, TAE modes are more difficult to excite in X-point geometry as opposed to limiter geometry. The JOREK simulations show the same behavior as observed experimentally and point to the importance of the SOL with open field lines with a wider SOL leading to a less efficient antenna excitation.

\begin{figure}
\centering
  \includegraphics[width=0.5\textwidth]{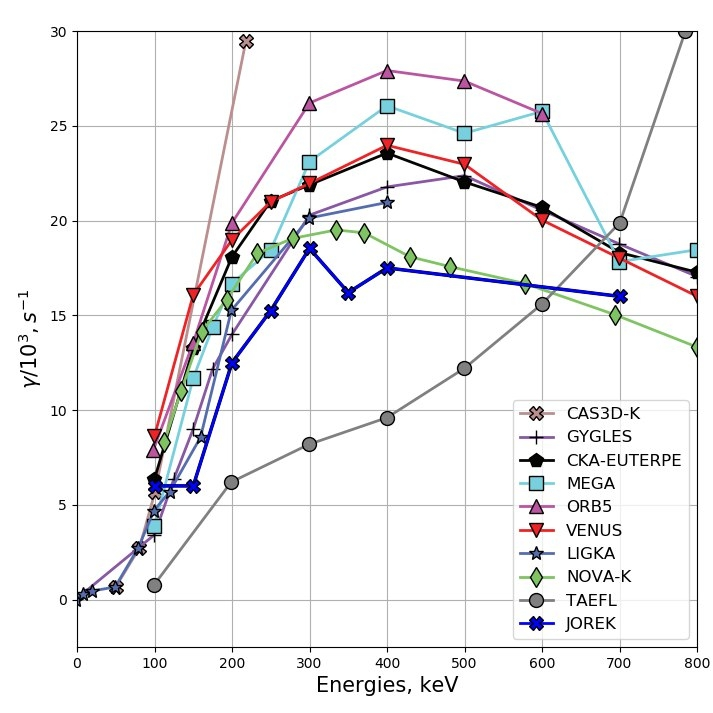}
\caption{The linear growth rate of the $n=6$ TAE mode as a function of the temperature of the fast particle population, comparing JOREK results to other codes (including FLR effects). The data is taken from Ref.~\cite{Konies2018} with the JOREK results from Ref.~\cite{DvornovaPhD} added. \noreprint}
\label{fig:TAE_benchmark_Dvornova}
\end{figure}

The particles framework (Section~\ref{:code:models:particles}) includes the  interaction of fast particles on the MHD fluid. The fast particle driven TAE instability has benchmarked against the well-established ITPA TAE benchmark as described in \cite{Konies2018}. In the JOREK simulations the pressure coupling scheme was used, the number of particles was $10^8$. Figure~\ref{fig:TAE_benchmark_Dvornova} shows a good agreement in the comparison of the JOREK results with several other codes on the $n=6$ TAE mode growth rate as a function of the fast particle temperature~\cite{DvornovaPhD}. Note that JOREK is the only code in this benchmark using full orbit fast ions.

\subsection{3D configurations}\label{:applic:other:3D-configs}

3D plasma configurations are becoming an important research area for JOREK, in particular quasi-axisymmetric (QA) stellarators are of interest. A first approach is to simulate an axisymmetric configuration with properties as close as possible to the true quasi-axisymmetric configuration. Linearly, such an approach has been followed in Ref.~\cite{Strumberger2019} and is presently improved by including ``virtual coil currents'' in the simulations, which provide an externally driven rotational transform like it is present in the QA-stellarator also for the axisymmetric simulation~\cite{Ramasamy2018Stell,Ramasamy2020,Ramasamy2021}. This allows to incorporate both the influence of the external rotational transform onto non-axisymmetric modes and the stabilizing effect onto vertical displacement events. In parallel to this axisymmetric approach to the QA stellarator, an extension of JOREK to 3D configurations is presently ongoing. As a first step, a hierarchy of reduced and full MHD models has been derived in a form suitable for stellarator devices~\cite{Nikulsin2019,Nikulsin2020}; implementation is presently on the way. The extension of the code to 3D grids taking place in parallel is briefly mentioned Section~\ref{:code:models:outlook} and~\ref{:code:numerics:outlook}. First simulations in simple stellarator geometry are expected to become possible very soon.

%% file: 08_summary.tex
\section{Conclusions and Summary}\label{:summary:conclusions}

A comprehensive summary of the JOREK simulation code was given covering the available physics models, the numerical methods employed and the verification, as well as the broad range of applications to magnetically confined fusion plasmas. An outlook was provided in each section onto further developments of the code and onto future applications.

It was shown that JOREK provides a framework which contains a number of different physics models ranging from reduced and full MHD models to a fully kinetic treatment of the plasma (so far electrostatic). Various extensions are available including separate electron and ion temperatures, diamagnetic drift, neoclassical effects, fluid and kinetic neutrals, a runaway electron fluid and test particle model, fluid and kinetic impurities, pellets, free boundary and conducting structures. \revised{While the single fluid reduced and full MHD models are energy conserving on the equation level, errors can arise from gyro-viscous cancellation, temporal discretization and too low toroidal resolution. Diagnostics running automatically for each simulation allow to confirm that energy is conserved reasonably well in practice. Momentum conservation is exact on the equation level for the full MHD model, but not for the reduced MHD model. The error has a low order as seen from analysis in Ref.~\cite{Nikulsin2020} and confirmed by the good linear and non-linear agreement between reduced and full MHD models shown in direct comparisons.}

The spatial discretization is based on a 2D $G^1$ continuous finite element formulation combined with a toroidal Fourier decomposition allowing to accurately align to flux surfaces improving numerical accuracy and to extend the computational domain across the separatrix up to divertor and plasma facing components. An extension of the axisymmetric grid to 3D stellarator configurations is under development. The robust fully implicit time advance allows to use large time steps where the physics processes allow for it. An iterative solver with a physics based preconditioner is applied to the large sparse matrix system, for which various options and library interfaces exist and further improvements are on their way. The code is very actively developed in an international community with automatic regression tests and code reviewing.

A large number of verification activities have been carried out over the years, while only selected ones could be shown in this article. This covers basic convergence properties of the discretization, tests for highly anisotropic heat transport, a verification of energy conservation, linear and non-linear benchmarks on core and edge instabilities, comparisons on the SOL models, benchmarks for the runaway electron fluid model, a variety of simpler and very advanced benchmarks for the free boundary extension, and detailed tests of the kinetic particle framework.

Regarding pedestal/edge simulations, a variety of results regarding natural ELMs was shown demonstrating that the JOREK code is able to reproduce key experimental observations qualitatively and even quantitatively. This is in particular the case for the divertor heat fluency scaling that is recovered reasonably well and for the explosive onset of type-I ELM crashes that has recently been reproduced for the first time in simulations of full ELM cycles. Regarding the control of ELMs, pellet ELM triggering has been studied extensively and, for instance, the lag-time experimentally observed in the ELM cycle has been reproduced recently in simulations with fully realistic parameters and flows. ELM triggering via vertical magnetic kicks has been demonstrated as well. The penetration of error fields into the plasma as well as the mitigation and suppression of ELMs via RMP fields was investigated in detail including realistic ExB and diamagnetic flows. ELM free QH-mode regimes were obtained and the EHO was explained by a saturated kink-peeling mode. Using a neutral fluid model, the ELM burn-through in detached conditions was \revised{demonstrated in non-linear simulations} and a more precise description of the SOL/divertor processes is on the way including a kinetic treatment of impurities and neutrals.

Regarding disruptions, results were presented on pre-disruption physics, the dynamics of massive material injection triggered disruptions, vertical displacement events, halo currents, and runaway electrons. Most prominently, JOREK is able to capture, at least in a qualitative sense, experimentally observed features of disruptions like the triggering of a thermal quench \textit{via} the destabilization of a $2/1$ tearing mode and the ensuing non-linear dynamics, and the $I_p$ spike. It can also describe accurately 3D vertical displacement events and halo current dynamics in spite of using a reduced MHD model, as benchmarks with full MHD codes have shown. While work is ongoing towards quantitative validation of disruption simulations, JOREK is already being used to test and optimize disruption mitigation strategies for ITER, and these efforts will intensify in the coming years. 

Further applications of JOREK include ITG turbulence simulations, the interaction of TAE modes with fast particles, and the ongoing extension towards stellarators.

%% file: 09_acknowledgements.tex
\section{Acknowledgements}\label{:ack}

Part of this work has been carried out within the framework of the EUROfusion Consortium and has received funding from the Euratom research and training programme 2014-2018 and 2019-2020 under grant agreement No 633053. Part of this work has been carried out in collaboration with the ITER Organization. Parts of this work was carried out using the Marconi-Fusion supercomputer. ITER is the Nuclear Facility INB no. 174. The views and opinions expressed herein do not necessarily reflect those of the ITER Organization. This publication is provided for scientific purposes only. Its contents should not be considered as commitments from the ITER Organization as a nuclear operator in the frame of the licensing process.

JOREK code developments and physics studies have over the years profited from national fusion programs, EFDA, the EUROfusion consortium, various collaborations with the ITER Organisation, and further funding sources (see acknowledgements in the original publications). Within EUROfusion, in particular the funding for three Enabling Research Projects (2014, 2015--2018, 2019--2020), \revised{a newly started Theory and Simulation Verification and Validation (TSVV) project (2021-2025), the European HPC infrastructure (presently Marconi-Fusion; also JFRS-1 in Japan via the Broader Approach) provided essential support. Fruitful collaborations with the Work Packages Medium Size Tokamaks (MST), JET and the new Tokamak Exploitation (TE) have lead to joint work on experiment interpretation and code validation. Several code optimization projects with the High Level Support Team (HLST) have contributed to the code development and helped to make the physics studies possible. For further information on HPC infrastructure used, please refer to the original publications.

The authors would like to thank the following persons for fruitful discussions (alphabetical): P Cahyna, B Dudson, M Dunne, X Garbet, S G{\"u}nter, R Hatzky, F Hindenlang, K Lackner, A Loarte, D Penko, S Pinches, M Rampp, E Viezzer, E Wolfrum. Please also refer to the author lists and acknowledgements of the original publications. We finally would like to thank the referees and editorial team of Nuclear Fusion for their substantial support in bringing this review paper to its final version.}

%% file: 10_appendix.tex
\appendix

\section{Coordinate systems}\label{:app:coord}

\paragraph{Cylindrical coordinates}
As already mentioned in Section~\ref{:code:models:equil}, the basic cylindrical coordinate system $(u^1,u^2,u^3)=(R,Z,\phi)$ of JOREK is given by
$x = R~\mathrm{cos} \phi$,
$y = -R~\mathrm{sin} \phi$, and
$z = Z$, where $(x,y,z)$ denotes Cartesian coordinates (Fig.~\ref{fig:cylcoord}).
The tokamak coordinate convention number COCOS~\cite{Sauter2013} used by JOREK is 8. This implies the COCOS coefficients $e_{\mathrm{Bp}}=0$, $\sigma_{\mathrm{Bp}}=-1$, $\sigma_{R\phi Z}=-1$, $\sigma_{\rho \theta \phi}=1$. The covariant basis vectors $\mathbf{a}_\alpha = \partial\mathbf{X}/\partial u^{\alpha}$ are given by 
\begin{equation*}
\mathbf{a}_1 = \begin{pmatrix} \mathrm{cos}\phi \\ -\mathrm{sin}\phi \\ 0\end{pmatrix},
~~~~~~~~
\mathbf{a}_2 = \begin{pmatrix} 0 \\ 0 \\ 1\end{pmatrix},
~~~~~~~~
\mathbf{a}_3 = \begin{pmatrix} -R~\mathrm{sin}\phi \\ -R~\mathrm{cos}\phi \\ 0\end{pmatrix}.
\end{equation*}
We use both $1,2,3$ and $R,Z,\phi$ synonymously as sub- or superscripts in our notation to identify the co- and contravariant components such that, e.g., $\mathbf{a}_\phi\equiv\mathbf{e}_3$. 
The cross products between these basis vectors are
$\mathbf{a}_1 \times \mathbf{a}_2 = \mathbf{a}_3/R$,
$\mathbf{a}_1 \times \mathbf{a}_3 = -R~\mathbf{a}_2$,
$\mathbf{a}_2 \times \mathbf{a}_3 = R~\mathbf{a}_1$,
and $\mathbf{a}_\alpha \times \mathbf{a}_\alpha = 0$ as well as $\mathbf{a}_\alpha \times \mathbf{a}_\beta = -\mathbf{a}_\beta \times \mathbf{a}_\alpha$. The contravariant basis vectors are given by $\mathbf{a}^1 = \mathbf{a}_1$, $\mathbf{a}^2 = \mathbf{a}_2$, $\mathbf{a}^3 = \mathbf{a}_3 / R^2$.
Of course, $\mathbf{a}_\alpha \cdot \mathbf{a}^\beta = \delta_\alpha^\beta$ and
$\mathbf{a}^1 = \nabla R$, $\mathbf{a}^2 = \nabla Z$, $\mathbf{a}^3 = \nabla \phi$.
Furthermore,
$\mathbf{a}_1 = J \nabla Z \times \nabla \phi$,
$\mathbf{a}_2 = J \nabla \phi \times \nabla R$,
$\mathbf{a}_3 = J \nabla R \times \nabla Z$.
And $\mathbf{a}^1 \times \mathbf{a}^2 = R\mathbf{a}^3$,
$\mathbf{a}^1 \times \mathbf{a}^3 = -\mathbf{a}^2/R$,
$\mathbf{a}^2 \times \mathbf{a}^3 = \mathbf{a}^1/R$.
Here the Jacobian is $J= \mathbf{a}_1\cdot(\mathbf{a_2}\times\mathbf{a}_3) = R$.
Normalized basis vectors are given by
$\mathbf{e}_1 = \mathbf{e}^1 = \mathbf{a}_1 = \mathbf{a}^1$,
$\mathbf{e}_2 = \mathbf{e}^2 = \mathbf{a}_2 = \mathbf{a}^2$,
$\mathbf{e}_3 = \mathbf{e}^3 = \mathbf{a}_3 / R = R~\mathbf{a}^3$.
The co- and contravariant metric tensors are given by
$g_{\alpha\beta} = \mathbf{a}_\alpha\cdot\mathbf{a}_\beta = \mathrm{diag}(1,~ 1, ~ R^2)$,
$g^{\alpha\beta} = \mathbf{a}^\alpha\cdot\mathbf{a}^\beta = \mathrm{diag}(1,~ 1, ~ 1/R^2)$.
The determinant of the covariant metric tensor is
$g \equiv J^2 =\det{\left(g_{\alpha\beta}\right)}=R^2$.
Differential Operators are given in the cylindrical coordinates of JOREK by
\begin{align*}
  \nabla U                &=\partial_1 U~\mathbf{a}^1
    + \partial_2 U~\mathbf{a}^2 + \partial_3 U~\mathbf{a}^3 \\
  \nabla_\text{pol} U                &=\partial_1 U~\mathbf{a}^1
    + \partial_2 U~\mathbf{a}^2 \\
  \nabla\cdot\mathbf{V}   &=\frac{1}{R}\partial_1 (R V^1) + \partial_2 V^2 + \partial_3 V^3 \\
\nabla \times \mathbf{V} &=  \frac{1}{R}\left(\partial_2 V_3 - \partial_3 V_2 \right)~\mathbf{a}_1
 + \frac{1}{R}\left(\partial_3 V_1 - \partial_1 V_3 \right)~\mathbf{a}_2
+ \frac{1}{R}\left(\partial_1 V_2 - \partial_2 V_1 \right)~\mathbf{a}_3 \\
  \Delta U &= \nabla\cdot\nabla U = \frac{1}{R}\partial_1(R \partial_1 U) + \partial_{2,2}U + \frac{1}{R^2}\partial_{3,3}U \\
  \Delta_\text{pol}U &= \nabla\cdot\nabla_\text{pol} U = \frac{1}{R}\partial_1(R \partial_1 U) + \partial_{2,2}U \\
  \Delta^*U &= R^2 \nabla\cdot\left(\frac{1}{R^2}\nabla_\text{pol} U\right)=R~\partial_1\left(\frac{1}{R}\partial_1 U\right)+\partial_{2,2} U \\
  [A,B] &= e_3 \cdot (\nabla A \times\nabla B) = \partial_1 A~\partial_2 B - \partial_2 A~\partial_1 B
\end{align*}
The Christoffel symbols defined by
\begin{align*}
   \mathbf{A} \cdot \nabla \mathbf{B} &= A^i \partial_i (B^j a_j) = A^i (\partial B^j)a_j + A^i B^j (\partial_i a_j
) = A^i(\partial_i B^j + \Gamma^j_{ik} B^k) a_j\\  (\mathbf{A} \cdot \nabla \mathbf{B})^j &= A^i (\partial_i B^j + \Gamma^j_{ik} B^k)
\end{align*}
are given by the following expressions in the JOREK cylindrical coordinate system:
\begin{equation*}
\partial_3 \mathbf{a}^1 = R \nabla \phi = R \mathbf{a}^3,
~~~~~~~~
\partial_3 \mathbf{a}^3 = -\frac{1}{R} \mathbf{a}^1,
~~~~~~~~
\partial_1 \mathbf{a}^3 = -\frac{1}{R} \mathbf{a}^3.
\end{equation*}

\paragraph{Element local coordinates}
Inside each of the quadrangular finite elements (see Section~\ref{:code:numerics:discretization}), a local coordinate system (s,t,$\phi$) is defined with s and t taking values in the range $[0,1]$. Both s and t are orthogonal with respect to $\phi$, but not with respect to each other. The local coordinates are defined by the cylindrical coordinates
$R(s,t)$ and $Z(s,t)$ being expressed in terms of the local coordinates in the same finite element basis as the physical variables. When the Poisson bracket $[a,b]$ is expressed in terms of the element-local coordinates, we get
$[a,b]=(a_{,s}b_{,t}-a_{,t}b_{,s})/J_2$
where $J_2=R_{,s}Z_{,t}-R_{,t}Z_{,s}$. Note, that the singularity at the grid center, where $J_2=0$ does not break the code, since the integration is carried out on the Gaussian integration points, where $J_2\ne0$.

While this mapping does not have a $\phi$ dependency at present, an extension of the code to 3D grids is presently ongoing (see Section~\ref{:code:numerics:outlook}) such that the mapping will become $R(s,t,\tilde\phi)$ and $Z(s,t,\tilde\phi)$, where $\tilde\phi$ does not have to be equivalent to $\phi$, but can be a function of $(R,Z,\phi)$ (e.g., in the case of Boozer coordinates). The orthogonality assumption of $\phi$ with respect to s and t might be dropped in the context of this extension.

\section{Normalization}\label{:app:norm}

{\small
\begin{tabular}{ l l | l }
\hline
\multicolumn{2}{ c| }{Connection between SI and normalized units} & Description of the quantity  \\
\hline
 $R_\text{SI}~[m]$ & $=R$ & Major radius \\
 $Z_\text{SI}~[m]$ & $=Z$ & Vertical coordinate \\
 $\mathbf{B}_\text{SI}~[T]$ & $=\mathbf{B}$ & Magnetic field vector \\
 $\mathbf{E}_\text{SI}~[V m^{-1}]$ & $=\mathbf{E}/\sqrt{\mu_0\rho_0}$ & Electric field vector \\
  $\Psi_\text{SI}~[T m^2]$ & $=\Psi$ & Poloidal magnetic flux \\
 $j_{\phi, \text{SI}}~[A m^{-2}]$ & $=-j/(R\;\mu_0)$ & Toroidal current density; $j_{\phi, \text{SI}}=\mathbf{j}_{\text{SI}}\cdot\hat{e}_\phi$ \\
 $n_\text{SI}~[m^{-3}]$ & $=\rho\;n_0$ & Particle density \\
 $\rho_\text{SI}~[kg~m^{-3}]$ & $=\rho\;\rho_0$ & Mass density = ion mass $\times$ particle density \\
 $T_\text{SI}~[K]$ & $=T/(k_B\;\mu_0\;n_0)$ & Temperature = electron + ion temperature \\
 $T_{eV}~[eV]$ & $=T/(e\;\mu_0\;n_0)$ & Temperature = electron + ion temperature \\
 $FF'_{SI}~[T\;\text{rad}]$ & $=FF'$ & Poloidal current stream function $F=RB_{\phi}$ and $'=d/d\psi$\\
 $p_\text{SI}~[N m^{-2}]$ & $=\rho\;T/\mu_0$ & Plasma pressure \\
 $\mathbf{v}_\text{SI}~[m s^{-1}]$ & $=\mathbf{v}/\sqrt{\mu_0\rho_0}$ & Velocity vector \\
 $v_{||, \text{SI}}~[m s^{-1}]$ & $=v_{||}\cdot B_\text{SI}/\sqrt{\mu_0\rho_0}$ & Parallel velocity component, where $B_\text{SI}=|\mathbf{B}_\text{SI}|$ \\
 $u_\text{SI}~[m s^{-1}]$ & $=u/\sqrt{\mu_0\rho_0}$ & $R u$ is the velocity stream function, $F_0 u$ is the potential \\
 $\omega_{\phi, \text{SI}}~[m^{-1} s^{-1}]$ & $=\omega/\sqrt{\mu_0\rho_0}$ & Toroidal vorticity \\
  $t_\text{SI}~[s]$ & $=t\cdot\sqrt{\mu_0\rho_0}$ & Time \\
 $\gamma_\text{SI}~[s^{-1}]$ & $=\gamma/\sqrt{\mu_0\rho_0}$ & Growth rate of an MHD mode \\
 $\eta_\text{SI}~[\Omega m]$ & $=\eta\cdot\sqrt{\mu_0/\rho_0}$ & Resistivity \\
 $\nu_\text{SI}~[kg~m^{-1} s^{-1}]$ & $=\nu\cdot\sqrt{\rho_0/\mu_0}$ & Dynamic viscosity \\
 $\tilde{\nu}_\text{SI}~[m^2 s^{-1}]$ & $=\tilde{\nu}_\text{SI}/\rho_{SI}$ & Kinematic viscosity ($\rho_{SI}$ is the local value) \\
 $D_\text{SI}~[m^2 s^{-1}]$ & $=D/\sqrt{\mu_0\rho_0}$ & Particle diffusivity ($||$ or $\bot$); Usually, $D_{||}=0$ \\
 $K_\text{SI}~[kg~m^{-1} s^{-1}]$ & $=K\cdot\sqrt{\rho_0/\mu_0}/\left(\gamma-1\right)$ & Heat diffusivity ($||$ or $\bot$) \\
 $S_{T, \text{SI}}~[W m^{-3}]$ & $=S_T/\sqrt{\mu_0^3\rho_0}$ & Heat source \\
 $S_{\rho, \text{SI}}~[kg~s^{-1} m^{-3}]$ & $=S_\rho\cdot\sqrt{\rho_0/\mu_0}$ & Particle source \\
 $\eta_{wall, thin, \text{SI}}~[\Omega]$ & $=\eta_{wall,thin}\cdot\sqrt{\mu_0/\rho_0}$ & Resistivity of conducting structures \\
 $R_\text{ion/rec,SI}~[m^{-3} s^{-1}]$ & $= R_\text{ion/rec} / (\sqrt{\mu_0 \rho_0} n_0)$ & Ionisation and recombination rate \\
 $E_\text{ion,SI} [J]$  & $= \xi_\text{ion}/(\frac{2}{3} \mu_0 n_0)$ & Ionisation energy \\
 ${L}_\text{rad,SI} [W m^3]$ & $= L_\text{rad} / (\frac{2}{3} \sqrt{\rho_0/\mu_0} n_0^2 \frac{m_i}{m_{imp}})$ & Radiation rate (impurity fluid model) \\
 ${P}_\text{rad,SI} [W m^{-3}]$ & $= P_\text{rad} / (\frac{2}{3} \sqrt{\rho_0/\mu_0})$ & Radiation power density (impurity fluid model) \\
 $q_\text{SI}~[A s]$ & $=q \sqrt{\mu_0\rho_0}$ & Particle charge \\
 $\mu_{neo, \text{SI}}~[s^{-1}]$ & $=\mu_{neo}/\sqrt{\rho_0 \mu_0}$ & Neoclassical friction rate \\
\end{tabular}
}
 
\begin{itemize}
\item The growth rate $\gamma_\text{n, SI}$ for harmonic n is calculated by $\ln[E_n(t_2)/E_n(t_1)]/[2(t_2-t_1)]$ with $E_n$ the mode energy and $t_2$ and $t_1$ the corresponding time points at which the energies are determined. The factor two appears in the denominator to obtain growth rates for amplitudes instead of energies and be comparable to linear codes.
\item Heat diffusion coefficients in JOREK are defined absorbing the factor $\gamma-1$, which would normally appear in the energy equation for diffusion terms. This means, JOREK diffusivities need to be multiplied by a factor $1.5$ (at $\gamma=5/3$) when comparing them to the usual definitions.
\item Note that $\chi_\text{SI}~[m^2 s^{-1}]=K_\text{SI}/\rho_\text{SI}$ where $\kappa_\text{SI}~[m^{-1}s^{-1}]=n_\text{SI}\,\chi_\text{SI}$
\item Thin wall resistivity: $\eta_{wall, thin, SI}~[\Omega] = \eta_{wall, SI}~[\Omega~m]/d_{wall} [m]$ with wall thickness $d_{wall}$; E.g. ITER: $\eta_{wall, thin, SI}=\eta_{wall, SI}/d_{wall}=8\cdot10^{-7} \Omega~m/(6 cm)=1.33\cdot10^{-5}\Omega$
\end{itemize}

\section{Time stepping scheme}\label{:app:tstep}

The time-integration of a set of equations of the form 
\begin{equation}\label{eq:timestep1}
  \pderiv{\mathbf{A}(\mathbf{u})}{t}=\mathbf{B}(\mathbf{u},t)
\end{equation} can be discretized by the general form (refer to \cite{Beam1980,Hirsch1991})
\begin{equation}\label{eq:timestep2}
  (1+\xi)\mathbf{A}^{n+1}-(1+2\xi)\mathbf{A}^n+\xi \mathbf{A}^{n-1} =
    \Delta t\left[\theta \mathbf{B}^{n+1}+(1-\theta-\phi)\mathbf{B}^n-\phi \mathbf{B}^{n-1}\right],
\end{equation}
which guarantees second-order accuracy, if $\phi+\theta-\xi=1/2$.
Superscripts like $B^n$ indicate at which timestep the corresponding expression is evaluated. The linearization 
$\mathbf{H}^{n+1}\approx \mathbf{H}^n+\partial\mathbf{H}/\partial\mathbf{u}|^n\cdot\delta\mathbf{u}^n$,
with $\mathbf{H}=\mathbf{A}$ or $\mathbf{H}=\mathbf{B}$, which is described in Ref~\cite{Hirsch1991}, allows to rewrite Equation \eqref{eq:timestep2} in the following way, where $\phi=0$ has been chosen:
\begin{equation}\begin{split}
  (1+\xi)\left[\mathbf{A}^n+\left(\pderiv{\mathbf{A}}{\mathbf{u}}\right)^n\delta\mathbf{u}^n\right]
    -(1+\xi)\mathbf{A}^n-\xi\mathbf{A}^n+\xi\mathbf{A}^{n-1} \\
    =\Delta t\left[\theta\left(\mathbf{B}^n+\left(\pderiv{\mathbf{B}}{\mathbf{u}}\right)^n\delta\mathbf{u}^n\right)
      +(1-\theta)\mathbf{B}^n\right]
\end{split}\end{equation}
Here, $\delta\mathbf{u}^n\equiv \mathbf{u}^{n+1}-\mathbf{u}^n$.
After some simplifications, and using the backward linearization
$\mathbf{H}^{n-1}\approx \mathbf{H}^n-\partial\mathbf{H}/\partial\mathbf{u}|^n\cdot\delta\mathbf{u}^{n-1}$,
one obtains 
\begin{equation}\label{eq:timestep4}
    \left[(1+\xi)\left(\pderiv{\mathbf{A}}{\mathbf{u}}\right)^n-\Delta t\theta\left(\pderiv{\mathbf{B}}{\mathbf{u}}\right)^n\right]\delta\mathbf{u}^n
      =\Delta t\mathbf{B}^n+\xi\left(\pderiv{\mathbf{A}}{\mathbf{u}}\right)^n\delta\mathbf{u}^{n-1},
\end{equation} which is the time-integration scheme implemented in JOREK. Certain parameter choices correspond to well-known time integration methods: Crank-Nicolson is selected by $\theta=1/2$ and $\xi=0$, BDF2 (Gears) is selected by $\theta=1$ and $\xi=1/2$, and first order implicit Euler method (not used in production) corresponds to $\theta=1$ and $\xi=0$.
The linearization shown above can also be replaced by Newton iterations with a beneficial effect onto non-linear stability in certain situation, as demonstrated in Ref.~\cite{Franck2015}. However, this is not implemented in the present code version.

\section{Setup for testing anisotropic heat transport}\label{:app:anisotropy-test}

\revised{For the tests shown in Section~\ref{:verification:anisotropy}, the following setup was used. The plasma cross section is circular with a major radius of $R_\text{axis}=100\,\mathrm{m}$ and a minor radius of $1\,\mathrm{m}$. The magnetic configuration is initialized by the poloidal flux distribution $\Psi=1-[ (R-R_\text{axis})^2+Z^2 ]^2\,\mathrm{T/m^2}$ and $F_0=200\,\mathrm{Tm}$. The applied perturbation is given by $\tilde\Psi=0.01\,\mathrm{T}\,[Z\,\sin(\phi) + (R-R_\text{axis}\,\cos(\phi)]$. The simulation domain is rectangular $2\,\mathrm{m}$ by $2\,\mathrm{m}$ for the non-aligned grid and circular with radius $1\,\mathrm{m}$ for the aligned grid. The temperature at the boundary of the computational domain is fixed at zero via Dirichlet boundary conditions. For the following, (normalized) pre-factors are omitted since they do not affect the results due to the self-similarity of the solution and since only relative errors are discussed. To establish a steady state temperature distribution, a source $S_T=\{0.5-0.5\,\tanh[(\Psi_n-0.1)/0.1]\}$ is applied in the plasma center. The perpendicular and parallel heat conduction coefficients are spatially constant and their ratio is varied for the tests. The density distribution is spatially constant. Convergence is tested by assessing the steady state value of the axis temperature that establishes.}